\documentclass[floatfix,aps,prd,twocolumn,letterpaper,lengthcheck,superscriptaddress,amssymb,amsmath,amsfonts,nofootinbib,altaffilletter]{revtex4-2}
\usepackage{color}
\usepackage[dvipsnames]{xcolor}
\usepackage{calc}
\usepackage{mathtools,graphicx}
\usepackage{pifont}
\usepackage{bm}
\usepackage{microtype}
\usepackage{booktabs}
\usepackage{times}
\usepackage[varg]{txfonts}
\usepackage[colorlinks, pdfborder={0 0 0}]{hyperref}
\usepackage[utf8]{inputenc}
\usepackage[caption=false]{subfig}
\usepackage{paralist}
\usepackage[normalem]{ulem}
\usepackage{multirow}
\usepackage{etoolbox}
\usepackage{xstring}
\usepackage{xparse}
\usepackage{dcolumn}

\definecolor{LinkColor}{rgb}{0.75, 0, 0}
\definecolor{CiteColor}{rgb}{0, 0.5, 0.5}
\definecolor{UrlColor}{rgb}{0, 0, 0.75}
\hypersetup{linkcolor=LinkColor}
\hypersetup{citecolor=CiteColor}
\hypersetup{urlcolor=UrlColor}
\allowdisplaybreaks
\DeclareFontFamily{OT1}{pzc}{}
\DeclareFontShape{OT1}{pzc}{m}{it}{<-> s * [1.10] pzcmi7t}{}
\DeclareMathAlphabet{\mathpzc}{OT1}{pzc}{m}{it}

\AtBeginDocument{%
    \newwrite\bibnotes
    \def\bibnotesext{Notes.bib}
    \immediate\openout\bibnotes=\jobname\bibnotesext
    \immediate\write\bibnotes{@CONTROL{REVTEX41Control}}
    \immediate\write\bibnotes{@CONTROL{%
    apsrev41Control,author="08",editor="1",pages="0",title="0",year="1"}}
     \if@filesw
     \immediate\write\@auxout{\string\citation{apsrev41Control}}%
    \fi
}%

\newtoggle{fullauthorlist}
\toggletrue{fullauthorlist}
\newtoggle{endauthorlist}
\toggletrue{endauthorlist}

\graphicspath{{./fig/}}

\usepackage{scrextend}
\usepackage{textgreek}

\usepackage{graphicx}
\usepackage{makerobust}
\makeatletter
\@ifundefined{caption@xref}{}{%
  \MakeRobustCommand\caption@xref
}
\makeatother

\def\mnras{\ref@jnl{MNRAS}}             

\renewcommand{\today}{\number\day\space\ifcase\month\or
  January\or February\or March\or April\or May\or June\or
  July\or August\or September\or October\or November\or December\fi
  \space\number\year}

\newcommand{\loglikelihoodminus}[1]{\IfEqCase{#1}{{GW190930A}{8.4}{GW190929A}{10.9}{GW190924A}{8.6}{GW190915A}{8.1}{GW190910A}{4.8}{GW190909A}{4.5}{GW190828B}{5.3}{GW190828A}{5.0}{GW190814A}{4.9}{GW190803A}{4.4}{GW190731A}{4.0}{GW190728A}{48007.6}{GW190727A}{5.1}{GW190720A}{9.4}{GW190719A}{4.2}{GW190708A}{4.8}{GW190707A}{7.1}{GW190706A}{5.2}{GW190701A}{3.9}{GW190630A}{5.3}{GW190620A}{5.0}{GW190602A}{4.3}{GW190527A}{6.7}{GW190521B}{5.9}{GW190521A}{11.1}{GW190519A}{17.8}{GW190517A}{5.9}{GW190514A}{4.6}{GW190513A}{4.7}{GW190512A}{5.5}{GW190503A}{4.4}{GW190426A}{5.6}{GW190425A}{5.7}{GW190424A}{3.9}{GW190421A}{4.0}{GW190413B}{5.7}{GW190413A}{4.8}{GW190412A}{10.1}{GW190408A}{5.0}}}
\newcommand{\loglikelihoodmed}[1]{\IfEqCase{#1}{{GW190930A}{-15934.9}{GW190929A}{-11962.2}{GW190924A}{-97031.8}{GW190915A}{-2809.6}{GW190910A}{93.5}{GW190909A}{25.8}{GW190828B}{43.6}{GW190828A}{121.4}{GW190814A}{298.6}{GW190803A}{28.7}{GW190731A}{29.8}{GW190728A}{64.1}{GW190727A}{58.4}{GW190720A}{-23904.6}{GW190719A}{27.0}{GW190708A}{76.7}{GW190707A}{-15883.0}{GW190706A}{72.6}{GW190701A}{55.9}{GW190630A}{114.8}{GW190620A}{65.2}{GW190602A}{73.6}{GW190527A}{22.9}{GW190521B}{322.2}{GW190521A}{-11913.6}{GW190519A}{111.0}{GW190517A}{48.8}{GW190514A}{25.9}{GW190513A}{74.3}{GW190512A}{67.5}{GW190503A}{68.5}{GW190426A}{-389547.0}{GW190425A}{-500483.9}{GW190424A}{45.7}{GW190421A}{48.8}{GW190413B}{42.7}{GW190413A}{28.4}{GW190412A}{-22827.3}{GW190408A}{108.8}}}
\newcommand{\loglikelihoodplus}[1]{\IfEqCase{#1}{{GW190930A}{15973.8}{GW190929A}{12007.8}{GW190924A}{97091.0}{GW190915A}{2897.5}{GW190910A}{4.0}{GW190909A}{3.6}{GW190828B}{4.0}{GW190828A}{4.2}{GW190814A}{3.0}{GW190803A}{2.7}{GW190731A}{2.5}{GW190728A}{13.4}{GW190727A}{5.2}{GW190720A}{23953.2}{GW190719A}{2.8}{GW190708A}{3.3}{GW190707A}{15964.2}{GW190706A}{4.0}{GW190701A}{2.6}{GW190630A}{3.8}{GW190620A}{4.1}{GW190602A}{3.3}{GW190527A}{3.0}{GW190521B}{4.7}{GW190521A}{12013.1}{GW190519A}{9.5}{GW190517A}{4.4}{GW190514A}{2.7}{GW190513A}{4.2}{GW190512A}{3.8}{GW190503A}{3.4}{GW190426A}{4.6}{GW190425A}{4.5}{GW190424A}{2.8}{GW190421A}{2.6}{GW190413B}{3.6}{GW190413A}{4.0}{GW190412A}{23002.9}{GW190408A}{3.7}}}
\newcommand{\chieffminus}[1]{\IfEqCase{#1}{{GW190930A}{0.15}{GW190929A}{0.33}{GW190924A}{0.09}{GW190915A}{0.25}{GW190910A}{0.18}{GW190909A}{0.36}{GW190828B}{0.16}{GW190828A}{0.16}{GW190814A}{0.06}{GW190803A}{0.27}{GW190731A}{0.24}{GW190728A}{0.07}{GW190727A}{0.25}{GW190720A}{0.12}{GW190719A}{0.31}{GW190708A}{0.08}{GW190707A}{0.08}{GW190706A}{0.29}{GW190701A}{0.29}{GW190630A}{0.13}{GW190620A}{0.25}{GW190602A}{0.24}{GW190527A}{0.28}{GW190521B}{0.13}{GW190521A}{0.39}{GW190519A}{0.22}{GW190517A}{0.19}{GW190514A}{0.32}{GW190513A}{0.17}{GW190512A}{0.13}{GW190503A}{0.26}{GW190426A}{0.30}{GW190425A}{0.05}{GW190424A}{0.22}{GW190421A}{0.27}{GW190413B}{0.29}{GW190413A}{0.34}{GW190412A}{0.11}{GW190408A}{0.19}}}
\newcommand{\chieffmed}[1]{\IfEqCase{#1}{{GW190930A}{0.14}{GW190929A}{0.01}{GW190924A}{0.03}{GW190915A}{0.02}{GW190910A}{0.02}{GW190909A}{-0.06}{GW190828B}{0.08}{GW190828A}{0.19}{GW190814A}{0.00}{GW190803A}{-0.03}{GW190731A}{0.06}{GW190728A}{0.12}{GW190727A}{0.11}{GW190720A}{0.18}{GW190719A}{0.32}{GW190708A}{0.02}{GW190707A}{-0.05}{GW190706A}{0.28}{GW190701A}{-0.07}{GW190630A}{0.10}{GW190620A}{0.33}{GW190602A}{0.07}{GW190527A}{0.11}{GW190521B}{0.09}{GW190521A}{0.03}{GW190519A}{0.31}{GW190517A}{0.52}{GW190514A}{-0.19}{GW190513A}{0.11}{GW190512A}{0.03}{GW190503A}{-0.03}{GW190426A}{-0.03}{GW190425A}{0.06}{GW190424A}{0.13}{GW190421A}{-0.06}{GW190413B}{-0.03}{GW190413A}{-0.01}{GW190412A}{0.25}{GW190408A}{-0.03}}}
\newcommand{\chieffplus}[1]{\IfEqCase{#1}{{GW190930A}{0.31}{GW190929A}{0.34}{GW190924A}{0.30}{GW190915A}{0.20}{GW190910A}{0.18}{GW190909A}{0.37}{GW190828B}{0.16}{GW190828A}{0.15}{GW190814A}{0.06}{GW190803A}{0.24}{GW190731A}{0.24}{GW190728A}{0.20}{GW190727A}{0.26}{GW190720A}{0.14}{GW190719A}{0.29}{GW190708A}{0.10}{GW190707A}{0.10}{GW190706A}{0.26}{GW190701A}{0.23}{GW190630A}{0.12}{GW190620A}{0.22}{GW190602A}{0.25}{GW190527A}{0.28}{GW190521B}{0.10}{GW190521A}{0.32}{GW190519A}{0.20}{GW190517A}{0.19}{GW190514A}{0.29}{GW190513A}{0.28}{GW190512A}{0.12}{GW190503A}{0.20}{GW190426A}{0.32}{GW190425A}{0.11}{GW190424A}{0.22}{GW190421A}{0.22}{GW190413B}{0.25}{GW190413A}{0.29}{GW190412A}{0.08}{GW190408A}{0.14}}}
\newcommand{\totalmasssourceminus}[1]{\IfEqCase{#1}{{GW190930A}{1.5}{GW190929A}{25.2}{GW190924A}{1.0}{GW190915A}{6.4}{GW190910A}{9.1}{GW190909A}{17.6}{GW190828B}{4.4}{GW190828A}{4.8}{GW190814A}{0.9}{GW190803A}{9.0}{GW190731A}{11.3}{GW190728A}{1.3}{GW190727A}{8.0}{GW190720A}{2.3}{GW190719A}{10.7}{GW190708A}{1.8}{GW190707A}{1.3}{GW190706A}{13.9}{GW190701A}{9.5}{GW190630A}{4.8}{GW190620A}{13.1}{GW190602A}{15.6}{GW190527A}{9.8}{GW190521B}{4.8}{GW190521A}{23.5}{GW190519A}{14.8}{GW190517A}{9.6}{GW190514A}{10.8}{GW190513A}{5.9}{GW190512A}{3.5}{GW190503A}{8.3}{GW190426A}{1.5}{GW190425A}{0.1}{GW190424A}{10.7}{GW190421A}{9.2}{GW190413B}{11.9}{GW190413A}{9.7}{GW190412A}{3.7}{GW190408A}{3.0}}}
\newcommand{\totalmasssourcemed}[1]{\IfEqCase{#1}{{GW190930A}{20.3}{GW190929A}{104.3}{GW190924A}{13.9}{GW190915A}{59.9}{GW190910A}{79.6}{GW190909A}{75.0}{GW190828B}{34.4}{GW190828A}{58.0}{GW190814A}{25.8}{GW190803A}{64.5}{GW190731A}{70.1}{GW190728A}{20.6}{GW190727A}{67.1}{GW190720A}{21.5}{GW190719A}{57.8}{GW190708A}{30.9}{GW190707A}{20.1}{GW190706A}{104.1}{GW190701A}{94.3}{GW190630A}{59.1}{GW190620A}{92.1}{GW190602A}{116.3}{GW190527A}{59.1}{GW190521B}{74.7}{GW190521A}{163.9}{GW190519A}{106.6}{GW190517A}{63.5}{GW190514A}{67.2}{GW190513A}{53.9}{GW190512A}{35.9}{GW190503A}{71.7}{GW190426A}{7.2}{GW190425A}{3.4}{GW190424A}{72.6}{GW190421A}{72.9}{GW190413B}{78.8}{GW190413A}{58.6}{GW190412A}{38.4}{GW190408A}{43.0}}}
\newcommand{\totalmasssourceplus}[1]{\IfEqCase{#1}{{GW190930A}{8.9}{GW190929A}{34.9}{GW190924A}{5.1}{GW190915A}{7.5}{GW190910A}{9.3}{GW190909A}{55.9}{GW190828B}{5.4}{GW190828A}{7.7}{GW190814A}{1.0}{GW190803A}{12.6}{GW190731A}{15.8}{GW190728A}{4.5}{GW190727A}{11.7}{GW190720A}{4.3}{GW190719A}{18.3}{GW190708A}{2.5}{GW190707A}{1.9}{GW190706A}{20.2}{GW190701A}{12.1}{GW190630A}{4.6}{GW190620A}{18.5}{GW190602A}{19.0}{GW190527A}{21.3}{GW190521B}{7.0}{GW190521A}{39.2}{GW190519A}{13.5}{GW190517A}{9.6}{GW190514A}{18.7}{GW190513A}{8.6}{GW190512A}{3.8}{GW190503A}{9.4}{GW190426A}{3.5}{GW190425A}{0.3}{GW190424A}{13.3}{GW190421A}{13.4}{GW190413B}{17.4}{GW190413A}{13.3}{GW190412A}{3.8}{GW190408A}{4.2}}}
\newcommand{\chipminus}[1]{\IfEqCase{#1}{{GW190930A}{0.24}{GW190929A}{0.45}{GW190924A}{0.18}{GW190915A}{0.39}{GW190910A}{0.32}{GW190909A}{0.38}{GW190828B}{0.23}{GW190828A}{0.31}{GW190814A}{0.03}{GW190803A}{0.33}{GW190731A}{0.30}{GW190728A}{0.20}{GW190727A}{0.36}{GW190720A}{0.22}{GW190719A}{0.30}{GW190708A}{0.24}{GW190707A}{0.23}{GW190706A}{0.28}{GW190701A}{0.31}{GW190630A}{0.23}{GW190620A}{0.28}{GW190602A}{0.31}{GW190527A}{0.34}{GW190521B}{0.29}{GW190521A}{0.44}{GW190519A}{0.29}{GW190517A}{0.29}{GW190514A}{0.34}{GW190513A}{0.22}{GW190512A}{0.17}{GW190503A}{0.29}{GW190426A}{0.00}{GW190425A}{0.27}{GW190424A}{0.38}{GW190421A}{0.36}{GW190413B}{0.41}{GW190413A}{0.31}{GW190412A}{0.16}{GW190408A}{0.31}}}
\newcommand{\chipmed}[1]{\IfEqCase{#1}{{GW190930A}{0.34}{GW190929A}{0.59}{GW190924A}{0.24}{GW190915A}{0.55}{GW190910A}{0.40}{GW190909A}{0.52}{GW190828B}{0.31}{GW190828A}{0.43}{GW190814A}{0.04}{GW190803A}{0.43}{GW190731A}{0.39}{GW190728A}{0.29}{GW190727A}{0.47}{GW190720A}{0.33}{GW190719A}{0.43}{GW190708A}{0.29}{GW190707A}{0.29}{GW190706A}{0.39}{GW190701A}{0.42}{GW190630A}{0.32}{GW190620A}{0.43}{GW190602A}{0.41}{GW190527A}{0.44}{GW190521B}{0.40}{GW190521A}{0.68}{GW190519A}{0.44}{GW190517A}{0.49}{GW190514A}{0.47}{GW190513A}{0.30}{GW190512A}{0.22}{GW190503A}{0.38}{GW190426A}{0.00}{GW190425A}{0.34}{GW190424A}{0.52}{GW190421A}{0.48}{GW190413B}{0.56}{GW190413A}{0.41}{GW190412A}{0.30}{GW190408A}{0.39}}}
\newcommand{\chipplus}[1]{\IfEqCase{#1}{{GW190930A}{0.40}{GW190929A}{0.32}{GW190924A}{0.40}{GW190915A}{0.36}{GW190910A}{0.39}{GW190909A}{0.39}{GW190828B}{0.38}{GW190828A}{0.36}{GW190814A}{0.04}{GW190803A}{0.42}{GW190731A}{0.46}{GW190728A}{0.37}{GW190727A}{0.40}{GW190720A}{0.43}{GW190719A}{0.37}{GW190708A}{0.43}{GW190707A}{0.39}{GW190706A}{0.39}{GW190701A}{0.41}{GW190630A}{0.32}{GW190620A}{0.37}{GW190602A}{0.42}{GW190527A}{0.43}{GW190521B}{0.32}{GW190521A}{0.26}{GW190519A}{0.34}{GW190517A}{0.30}{GW190514A}{0.39}{GW190513A}{0.39}{GW190512A}{0.36}{GW190503A}{0.41}{GW190426A}{0.00}{GW190425A}{0.43}{GW190424A}{0.38}{GW190421A}{0.39}{GW190413B}{0.37}{GW190413A}{0.41}{GW190412A}{0.19}{GW190408A}{0.38}}}
\newcommand{\spinoneyminus}[1]{\IfEqCase{#1}{{GW190930A}{0.47}{GW190929A}{0.71}{GW190924A}{0.35}{GW190915A}{0.68}{GW190910A}{0.48}{GW190909A}{0.66}{GW190828B}{0.41}{GW190828A}{0.51}{GW190814A}{0.04}{GW190803A}{0.58}{GW190731A}{0.52}{GW190728A}{0.37}{GW190727A}{0.61}{GW190720A}{0.49}{GW190719A}{0.55}{GW190708A}{0.43}{GW190707A}{0.39}{GW190706A}{0.50}{GW190701A}{0.52}{GW190630A}{0.36}{GW190620A}{0.53}{GW190602A}{0.52}{GW190527A}{0.61}{GW190521B}{0.44}{GW190521A}{0.76}{GW190519A}{0.55}{GW190517A}{0.58}{GW190514A}{0.56}{GW190513A}{0.41}{GW190512A}{0.29}{GW190503A}{0.48}{GW190426A}{0.00}{GW190425A}{0.48}{GW190424A}{0.64}{GW190421A}{0.58}{GW190413B}{0.70}{GW190413A}{0.54}{GW190412A}{0.39}{GW190408A}{0.48}}}
\newcommand{\spinoneymed}[1]{\IfEqCase{#1}{{GW190930A}{0.002}{GW190929A}{0.005}{GW190924A}{0.0009}{GW190915A}{0.00}{GW190910A}{0.0008}{GW190909A}{-0.01}{GW190828B}{0.0010}{GW190828A}{0.00}{GW190814A}{0.0008}{GW190803A}{0.0007}{GW190731A}{0.0007}{GW190728A}{0.004}{GW190727A}{0.0002}{GW190720A}{0.0009}{GW190719A}{0.004}{GW190708A}{0.00}{GW190707A}{0.00}{GW190706A}{0.00}{GW190701A}{0.003}{GW190630A}{0.0002}{GW190620A}{-0.01}{GW190602A}{0.00}{GW190527A}{0.00}{GW190521B}{0.00}{GW190521A}{0.0003}{GW190519A}{0.00}{GW190517A}{-0.01}{GW190514A}{0.00}{GW190513A}{0.0005}{GW190512A}{0.00}{GW190503A}{0.00}{GW190426A}{0.00}{GW190425A}{0.003}{GW190424A}{0.00}{GW190421A}{0.0008}{GW190413B}{0.00}{GW190413A}{0.003}{GW190412A}{0.06}{GW190408A}{0.001}}}
\newcommand{\spinoneyplus}[1]{\IfEqCase{#1}{{GW190930A}{0.48}{GW190929A}{0.70}{GW190924A}{0.36}{GW190915A}{0.68}{GW190910A}{0.50}{GW190909A}{0.64}{GW190828B}{0.43}{GW190828A}{0.53}{GW190814A}{0.04}{GW190803A}{0.55}{GW190731A}{0.54}{GW190728A}{0.39}{GW190727A}{0.59}{GW190720A}{0.44}{GW190719A}{0.55}{GW190708A}{0.41}{GW190707A}{0.38}{GW190706A}{0.51}{GW190701A}{0.53}{GW190630A}{0.37}{GW190620A}{0.55}{GW190602A}{0.54}{GW190527A}{0.59}{GW190521B}{0.45}{GW190521A}{0.75}{GW190519A}{0.54}{GW190517A}{0.57}{GW190514A}{0.59}{GW190513A}{0.41}{GW190512A}{0.29}{GW190503A}{0.49}{GW190426A}{0.00}{GW190425A}{0.48}{GW190424A}{0.63}{GW190421A}{0.60}{GW190413B}{0.69}{GW190413A}{0.55}{GW190412A}{0.33}{GW190408A}{0.49}}}
\newcommand{\finalmassdetminus}[1]{\IfEqCase{#1}{{GW190930A}{0.9}{GW190929A}{24.5}{GW190924A}{0.8}{GW190915A}{7.4}{GW190910A}{7.1}{GW190909A}{21.3}{GW190828B}{4.2}{GW190828A}{5.2}{GW190814A}{1.0}{GW190803A}{10.9}{GW190731A}{12.8}{GW190728A}{0.7}{GW190727A}{9.8}{GW190720A}{1.2}{GW190719A}{14.1}{GW190708A}{0.7}{GW190707A}{0.5}{GW190706A}{23.7}{GW190701A}{13.4}{GW190630A}{3.3}{GW190620A}{16.2}{GW190602A}{18.3}{GW190527A}{9.5}{GW190521B}{4.8}{GW190521A}{30.4}{GW190519A}{15.4}{GW190517A}{6.4}{GW190514A}{13.9}{GW190513A}{6.7}{GW190512A}{2.8}{GW190503A}{10.8}{GW190424A}{10.0}{GW190421A}{11.3}{GW190413B}{16.7}{GW190413A}{14.0}{GW190412A}{4.7}{GW190408A}{3.4}}}
\newcommand{\finalmassdetmed}[1]{\IfEqCase{#1}{{GW190930A}{22.1}{GW190929A}{144.3}{GW190924A}{14.8}{GW190915A}{74.8}{GW190910A}{97.0}{GW190909A}{114.5}{GW190828B}{42.7}{GW190828A}{75.7}{GW190814A}{26.9}{GW190803A}{95.8}{GW190731A}{104.6}{GW190728A}{22.7}{GW190727A}{99.2}{GW190720A}{23.7}{GW190719A}{90.0}{GW190708A}{34.4}{GW190707A}{22.1}{GW190706A}{171.1}{GW190701A}{124.0}{GW190630A}{66.3}{GW190620A}{130.3}{GW190602A}{163.8}{GW190527A}{80.3}{GW190521B}{88.0}{GW190521A}{256.6}{GW190519A}{146.8}{GW190517A}{79.8}{GW190514A}{108.3}{GW190513A}{70.6}{GW190512A}{43.5}{GW190503A}{87.6}{GW190424A}{96.0}{GW190421A}{103.9}{GW190413B}{129.8}{GW190413A}{89.6}{GW190412A}{42.9}{GW190408A}{53.0}}}
\newcommand{\finalmassdetplus}[1]{\IfEqCase{#1}{{GW190930A}{10.8}{GW190929A}{36.4}{GW190924A}{5.9}{GW190915A}{7.9}{GW190910A}{9.3}{GW190909A}{92.0}{GW190828B}{6.6}{GW190828A}{6.0}{GW190814A}{1.1}{GW190803A}{13.1}{GW190731A}{12.8}{GW190728A}{5.5}{GW190727A}{10.7}{GW190720A}{5.2}{GW190719A}{22.5}{GW190708A}{2.7}{GW190707A}{1.9}{GW190706A}{20.0}{GW190701A}{15.1}{GW190630A}{4.2}{GW190620A}{17.7}{GW190602A}{20.7}{GW190527A}{51.0}{GW190521B}{4.3}{GW190521A}{36.6}{GW190519A}{14.7}{GW190517A}{8.8}{GW190514A}{16.6}{GW190513A}{11.5}{GW190512A}{4.0}{GW190503A}{10.2}{GW190424A}{13.0}{GW190421A}{14.1}{GW190413B}{16.4}{GW190413A}{16.3}{GW190412A}{4.6}{GW190408A}{3.2}}}
\newcommand{\phioneminus}[1]{\IfEqCase{#1}{{GW190930A}{2.79}{GW190929A}{2.78}{GW190924A}{2.82}{GW190915A}{2.86}{GW190910A}{2.79}{GW190909A}{2.98}{GW190828B}{2.78}{GW190828A}{2.84}{GW190814A}{2.65}{GW190803A}{2.80}{GW190731A}{2.83}{GW190728A}{2.76}{GW190727A}{2.83}{GW190720A}{2.81}{GW190719A}{2.77}{GW190708A}{2.90}{GW190707A}{2.85}{GW190706A}{2.84}{GW190701A}{2.77}{GW190630A}{2.80}{GW190620A}{2.89}{GW190602A}{2.85}{GW190527A}{2.88}{GW190521B}{2.87}{GW190521A}{2.79}{GW190519A}{2.84}{GW190517A}{2.90}{GW190514A}{2.82}{GW190513A}{2.82}{GW190512A}{2.84}{GW190503A}{2.83}{GW190426A}{0.00}{GW190425A}{2.73}{GW190424A}{2.84}{GW190421A}{2.82}{GW190413B}{2.83}{GW190413A}{2.76}{GW190412A}{2.36}{GW190408A}{2.77}}}
\newcommand{\phionemed}[1]{\IfEqCase{#1}{{GW190930A}{3.10}{GW190929A}{3.09}{GW190924A}{3.09}{GW190915A}{3.17}{GW190910A}{3.12}{GW190909A}{3.26}{GW190828B}{3.09}{GW190828A}{3.16}{GW190814A}{2.97}{GW190803A}{3.12}{GW190731A}{3.12}{GW190728A}{3.05}{GW190727A}{3.13}{GW190720A}{3.12}{GW190719A}{3.09}{GW190708A}{3.21}{GW190707A}{3.16}{GW190706A}{3.15}{GW190701A}{3.07}{GW190630A}{3.13}{GW190620A}{3.23}{GW190602A}{3.18}{GW190527A}{3.16}{GW190521B}{3.17}{GW190521A}{3.14}{GW190519A}{3.15}{GW190517A}{3.23}{GW190514A}{3.18}{GW190513A}{3.11}{GW190512A}{3.15}{GW190503A}{3.15}{GW190426A}{0.00}{GW190425A}{3.05}{GW190424A}{3.16}{GW190421A}{3.13}{GW190413B}{3.16}{GW190413A}{3.06}{GW190412A}{2.69}{GW190408A}{3.08}}}
\newcommand{\phioneplus}[1]{\IfEqCase{#1}{{GW190930A}{2.86}{GW190929A}{2.90}{GW190924A}{2.86}{GW190915A}{2.79}{GW190910A}{2.85}{GW190909A}{2.71}{GW190828B}{2.87}{GW190828A}{2.82}{GW190814A}{2.95}{GW190803A}{2.85}{GW190731A}{2.88}{GW190728A}{2.92}{GW190727A}{2.83}{GW190720A}{2.84}{GW190719A}{2.86}{GW190708A}{2.79}{GW190707A}{2.83}{GW190706A}{2.83}{GW190701A}{2.88}{GW190630A}{2.83}{GW190620A}{2.75}{GW190602A}{2.77}{GW190527A}{2.83}{GW190521B}{2.80}{GW190521A}{2.80}{GW190519A}{2.83}{GW190517A}{2.76}{GW190514A}{2.77}{GW190513A}{2.86}{GW190512A}{2.84}{GW190503A}{2.81}{GW190426A}{0.00}{GW190425A}{2.90}{GW190424A}{2.78}{GW190421A}{2.82}{GW190413B}{2.79}{GW190413A}{2.90}{GW190412A}{3.20}{GW190408A}{2.87}}}
\newcommand{\phitwominus}[1]{\IfEqCase{#1}{{GW190930A}{2.89}{GW190929A}{2.80}{GW190924A}{2.83}{GW190915A}{2.82}{GW190910A}{2.83}{GW190909A}{2.79}{GW190828B}{2.84}{GW190828A}{2.83}{GW190814A}{2.74}{GW190803A}{2.82}{GW190731A}{2.88}{GW190728A}{2.85}{GW190727A}{2.83}{GW190720A}{2.82}{GW190719A}{2.74}{GW190708A}{2.76}{GW190707A}{2.79}{GW190706A}{2.81}{GW190701A}{2.97}{GW190630A}{2.80}{GW190620A}{2.84}{GW190602A}{2.84}{GW190527A}{2.79}{GW190521B}{2.82}{GW190521A}{2.86}{GW190519A}{2.79}{GW190517A}{2.82}{GW190514A}{2.89}{GW190513A}{2.80}{GW190512A}{2.75}{GW190503A}{2.80}{GW190426A}{0.00}{GW190425A}{2.84}{GW190424A}{2.84}{GW190421A}{2.81}{GW190413B}{2.97}{GW190413A}{2.83}{GW190412A}{2.75}{GW190408A}{2.80}}}
\newcommand{\phitwomed}[1]{\IfEqCase{#1}{{GW190930A}{3.21}{GW190929A}{3.12}{GW190924A}{3.15}{GW190915A}{3.16}{GW190910A}{3.14}{GW190909A}{3.09}{GW190828B}{3.16}{GW190828A}{3.13}{GW190814A}{3.07}{GW190803A}{3.14}{GW190731A}{3.21}{GW190728A}{3.15}{GW190727A}{3.12}{GW190720A}{3.15}{GW190719A}{3.08}{GW190708A}{3.05}{GW190707A}{3.09}{GW190706A}{3.12}{GW190701A}{3.27}{GW190630A}{3.12}{GW190620A}{3.14}{GW190602A}{3.15}{GW190527A}{3.06}{GW190521B}{3.15}{GW190521A}{3.17}{GW190519A}{3.12}{GW190517A}{3.14}{GW190514A}{3.19}{GW190513A}{3.14}{GW190512A}{3.07}{GW190503A}{3.14}{GW190426A}{0.00}{GW190425A}{3.15}{GW190424A}{3.15}{GW190421A}{3.14}{GW190413B}{3.28}{GW190413A}{3.16}{GW190412A}{3.09}{GW190408A}{3.11}}}
\newcommand{\phitwoplus}[1]{\IfEqCase{#1}{{GW190930A}{2.74}{GW190929A}{2.83}{GW190924A}{2.82}{GW190915A}{2.80}{GW190910A}{2.82}{GW190909A}{2.86}{GW190828B}{2.84}{GW190828A}{2.84}{GW190814A}{2.86}{GW190803A}{2.87}{GW190731A}{2.77}{GW190728A}{2.78}{GW190727A}{2.88}{GW190720A}{2.84}{GW190719A}{2.90}{GW190708A}{2.90}{GW190707A}{2.88}{GW190706A}{2.88}{GW190701A}{2.70}{GW190630A}{2.84}{GW190620A}{2.81}{GW190602A}{2.79}{GW190527A}{2.91}{GW190521B}{2.82}{GW190521A}{2.82}{GW190519A}{2.84}{GW190517A}{2.83}{GW190514A}{2.75}{GW190513A}{2.85}{GW190512A}{2.89}{GW190503A}{2.81}{GW190426A}{0.00}{GW190425A}{2.83}{GW190424A}{2.83}{GW190421A}{2.81}{GW190413B}{2.72}{GW190413A}{2.82}{GW190412A}{2.85}{GW190408A}{2.88}}}
\newcommand{\phionetwominus}[1]{\IfEqCase{#1}{{GW190930A}{2.84}{GW190929A}{2.79}{GW190924A}{2.83}{GW190915A}{2.89}{GW190910A}{2.78}{GW190909A}{2.85}{GW190828B}{2.83}{GW190828A}{2.63}{GW190814A}{2.71}{GW190803A}{2.82}{GW190731A}{2.73}{GW190728A}{2.79}{GW190727A}{2.83}{GW190720A}{2.74}{GW190719A}{2.75}{GW190708A}{2.66}{GW190707A}{2.79}{GW190706A}{2.80}{GW190701A}{2.82}{GW190630A}{2.74}{GW190620A}{2.78}{GW190602A}{2.78}{GW190527A}{2.77}{GW190521B}{2.78}{GW190521A}{3.09}{GW190519A}{2.92}{GW190517A}{2.91}{GW190514A}{2.83}{GW190513A}{2.84}{GW190512A}{2.80}{GW190503A}{2.70}{GW190426A}{0.00}{GW190425A}{2.87}{GW190424A}{2.85}{GW190421A}{2.83}{GW190413B}{2.67}{GW190413A}{2.84}{GW190412A}{3.10}{GW190408A}{2.77}}}
\newcommand{\phionetwomed}[1]{\IfEqCase{#1}{{GW190930A}{3.17}{GW190929A}{3.11}{GW190924A}{3.17}{GW190915A}{3.18}{GW190910A}{3.15}{GW190909A}{3.17}{GW190828B}{3.16}{GW190828A}{2.94}{GW190814A}{3.01}{GW190803A}{3.12}{GW190731A}{3.03}{GW190728A}{3.13}{GW190727A}{3.13}{GW190720A}{3.09}{GW190719A}{3.13}{GW190708A}{3.08}{GW190707A}{3.13}{GW190706A}{3.11}{GW190701A}{3.15}{GW190630A}{3.06}{GW190620A}{3.16}{GW190602A}{3.09}{GW190527A}{3.08}{GW190521B}{3.16}{GW190521A}{3.35}{GW190519A}{3.23}{GW190517A}{3.22}{GW190514A}{3.15}{GW190513A}{3.15}{GW190512A}{3.14}{GW190503A}{3.02}{GW190426A}{0.00}{GW190425A}{3.18}{GW190424A}{3.15}{GW190421A}{3.17}{GW190413B}{3.00}{GW190413A}{3.15}{GW190412A}{3.44}{GW190408A}{3.11}}}
\newcommand{\phionetwoplus}[1]{\IfEqCase{#1}{{GW190930A}{2.77}{GW190929A}{2.87}{GW190924A}{2.80}{GW190915A}{2.78}{GW190910A}{2.80}{GW190909A}{2.82}{GW190828B}{2.78}{GW190828A}{3.01}{GW190814A}{2.95}{GW190803A}{2.83}{GW190731A}{2.88}{GW190728A}{2.82}{GW190727A}{2.86}{GW190720A}{2.84}{GW190719A}{2.80}{GW190708A}{2.83}{GW190707A}{2.81}{GW190706A}{2.86}{GW190701A}{2.81}{GW190630A}{2.89}{GW190620A}{2.77}{GW190602A}{2.84}{GW190527A}{2.89}{GW190521B}{2.73}{GW190521A}{2.68}{GW190519A}{2.71}{GW190517A}{2.77}{GW190514A}{2.80}{GW190513A}{2.79}{GW190512A}{2.82}{GW190503A}{2.94}{GW190426A}{0.00}{GW190425A}{2.76}{GW190424A}{2.84}{GW190421A}{2.81}{GW190413B}{2.95}{GW190413A}{2.82}{GW190412A}{2.54}{GW190408A}{2.84}}}
\newcommand{\raminus}[1]{\IfEqCase{#1}{{GW190930A}{5.23224}{GW190929A}{3.05568}{GW190924A}{0.14434}{GW190915A}{0.08865}{GW190910A}{2.79654}{GW190909A}{1.21034}{GW190828B}{0.25804}{GW190828A}{0.19728}{GW190814A}{0.02832}{GW190803A}{0.32018}{GW190731A}{2.05429}{GW190728A}{3.93966}{GW190727A}{0.28275}{GW190720A}{5.03310}{GW190719A}{1.97394}{GW190708A}{2.51716}{GW190707A}{1.47850}{GW190706A}{0.11777}{GW190701A}{0.02938}{GW190630A}{3.00747}{GW190620A}{3.79912}{GW190602A}{0.16873}{GW190527A}{4.81181}{GW190521B}{0.49181}{GW190521A}{3.83925}{GW190519A}{3.53823}{GW190517A}{0.11026}{GW190514A}{2.80859}{GW190513A}{0.16674}{GW190512A}{0.34154}{GW190503A}{0.07308}{GW190426A}{5.11703}{GW190425A}{1.14713}{GW190424A}{2.87584}{GW190421A}{1.90797}{GW190413B}{0.19894}{GW190413A}{2.38305}{GW190412A}{0.06390}{GW190408A}{0.23336}}}
\newcommand{\ramed}[1]{\IfEqCase{#1}{{GW190930A}{5.56800}{GW190929A}{4.57328}{GW190924A}{2.28446}{GW190915A}{3.41338}{GW190910A}{3.71091}{GW190909A}{1.53780}{GW190828B}{2.46668}{GW190828A}{2.55563}{GW190814A}{0.22230}{GW190803A}{1.64028}{GW190731A}{3.03870}{GW190728A}{5.47833}{GW190727A}{1.82934}{GW190720A}{5.19166}{GW190719A}{2.73410}{GW190708A}{2.97596}{GW190707A}{3.54979}{GW190706A}{2.59111}{GW190701A}{0.66145}{GW190630A}{5.86181}{GW190620A}{4.24971}{GW190602A}{1.30570}{GW190527A}{5.13405}{GW190521B}{4.87951}{GW190521A}{3.88408}{GW190519A}{3.58629}{GW190517A}{4.07092}{GW190514A}{3.58258}{GW190513A}{0.89431}{GW190512A}{4.37140}{GW190503A}{1.65900}{GW190426A}{5.27391}{GW190425A}{1.62833}{GW190424A}{3.13318}{GW190421A}{3.50423}{GW190413B}{2.70054}{GW190413A}{2.53690}{GW190412A}{3.81244}{GW190408A}{6.08868}}}
\newcommand{\raplus}[1]{\IfEqCase{#1}{{GW190930A}{0.45017}{GW190929A}{0.94968}{GW190924A}{0.18201}{GW190915A}{0.07354}{GW190910A}{1.06677}{GW190909A}{4.24495}{GW190828B}{3.42033}{GW190828A}{3.27852}{GW190814A}{0.16914}{GW190803A}{1.69538}{GW190731A}{0.37410}{GW190728A}{0.52075}{GW190727A}{4.30388}{GW190720A}{0.96175}{GW190719A}{3.20570}{GW190708A}{2.84449}{GW190707A}{2.17972}{GW190706A}{3.23517}{GW190701A}{0.02994}{GW190630A}{0.12425}{GW190620A}{0.37483}{GW190602A}{0.27462}{GW190527A}{0.80861}{GW190521B}{0.59768}{GW190521A}{2.35925}{GW190519A}{2.66341}{GW190517A}{1.75963}{GW190514A}{1.66161}{GW190513A}{4.13656}{GW190512A}{0.17389}{GW190503A}{0.07931}{GW190426A}{0.85181}{GW190425A}{3.12911}{GW190424A}{2.81897}{GW190421A}{0.15768}{GW190413B}{1.98722}{GW190413A}{0.89215}{GW190412A}{0.03095}{GW190408A}{0.08337}}}
\newcommand{\phijlminus}[1]{\IfEqCase{#1}{{GW190930A}{2.81}{GW190929A}{3.08}{GW190924A}{2.69}{GW190915A}{2.76}{GW190910A}{2.83}{GW190909A}{2.84}{GW190828B}{2.87}{GW190828A}{3.02}{GW190814A}{1.87}{GW190803A}{2.70}{GW190731A}{2.69}{GW190728A}{2.87}{GW190727A}{2.85}{GW190720A}{2.86}{GW190719A}{2.89}{GW190708A}{2.84}{GW190707A}{2.89}{GW190706A}{2.70}{GW190701A}{2.57}{GW190630A}{2.93}{GW190620A}{3.11}{GW190602A}{2.78}{GW190527A}{2.80}{GW190521B}{2.82}{GW190521A}{2.81}{GW190519A}{2.86}{GW190517A}{2.16}{GW190514A}{2.83}{GW190513A}{2.69}{GW190512A}{2.82}{GW190503A}{3.64}{GW190426A}{1.43}{GW190425A}{2.88}{GW190424A}{2.83}{GW190421A}{2.83}{GW190413B}{3.08}{GW190413A}{2.65}{GW190412A}{3.64}{GW190408A}{2.54}}}
\newcommand{\phijlmed}[1]{\IfEqCase{#1}{{GW190930A}{3.14}{GW190929A}{3.37}{GW190924A}{3.02}{GW190915A}{3.39}{GW190910A}{3.15}{GW190909A}{3.16}{GW190828B}{3.19}{GW190828A}{3.35}{GW190814A}{2.28}{GW190803A}{3.05}{GW190731A}{3.03}{GW190728A}{3.19}{GW190727A}{3.21}{GW190720A}{3.15}{GW190719A}{3.21}{GW190708A}{3.17}{GW190707A}{3.21}{GW190706A}{3.03}{GW190701A}{3.05}{GW190630A}{3.23}{GW190620A}{3.51}{GW190602A}{3.12}{GW190527A}{3.13}{GW190521B}{3.17}{GW190521A}{3.14}{GW190519A}{3.17}{GW190517A}{2.38}{GW190514A}{3.15}{GW190513A}{3.00}{GW190512A}{3.16}{GW190503A}{3.90}{GW190426A}{1.70}{GW190425A}{3.23}{GW190424A}{3.16}{GW190421A}{3.14}{GW190413B}{3.36}{GW190413A}{2.98}{GW190412A}{3.82}{GW190408A}{2.90}}}
\newcommand{\phijlplus}[1]{\IfEqCase{#1}{{GW190930A}{2.86}{GW190929A}{2.65}{GW190924A}{2.94}{GW190915A}{2.29}{GW190910A}{2.84}{GW190909A}{2.81}{GW190828B}{2.77}{GW190828A}{2.59}{GW190814A}{3.50}{GW190803A}{2.90}{GW190731A}{2.89}{GW190728A}{2.76}{GW190727A}{2.79}{GW190720A}{2.84}{GW190719A}{2.78}{GW190708A}{2.80}{GW190707A}{2.77}{GW190706A}{2.92}{GW190701A}{2.72}{GW190630A}{2.76}{GW190620A}{2.40}{GW190602A}{2.83}{GW190527A}{2.85}{GW190521B}{2.86}{GW190521A}{2.80}{GW190519A}{2.79}{GW190517A}{3.65}{GW190514A}{2.83}{GW190513A}{2.95}{GW190512A}{2.80}{GW190503A}{2.15}{GW190426A}{1.19}{GW190425A}{2.76}{GW190424A}{2.82}{GW190421A}{2.84}{GW190413B}{2.68}{GW190413A}{2.97}{GW190412A}{2.23}{GW190408A}{3.00}}}
\newcommand{\tilttwominus}[1]{\IfEqCase{#1}{{GW190930A}{0.90}{GW190929A}{1.09}{GW190924A}{1.01}{GW190915A}{1.03}{GW190910A}{1.01}{GW190909A}{1.23}{GW190828B}{0.96}{GW190828A}{0.84}{GW190814A}{1.02}{GW190803A}{1.12}{GW190731A}{1.02}{GW190728A}{0.85}{GW190727A}{0.97}{GW190720A}{0.90}{GW190719A}{0.81}{GW190708A}{1.00}{GW190707A}{1.18}{GW190706A}{0.88}{GW190701A}{1.14}{GW190630A}{0.87}{GW190620A}{0.78}{GW190602A}{0.97}{GW190527A}{1.01}{GW190521B}{0.85}{GW190521A}{1.02}{GW190519A}{0.77}{GW190517A}{0.67}{GW190514A}{1.21}{GW190513A}{0.94}{GW190512A}{0.99}{GW190503A}{1.06}{GW190426A}{0.00}{GW190425A}{0.87}{GW190424A}{0.96}{GW190421A}{1.10}{GW190413B}{1.18}{GW190413A}{1.13}{GW190412A}{0.90}{GW190408A}{1.06}}}
\newcommand{\tilttwomed}[1]{\IfEqCase{#1}{{GW190930A}{1.26}{GW190929A}{1.53}{GW190924A}{1.42}{GW190915A}{1.56}{GW190910A}{1.53}{GW190909A}{1.74}{GW190828B}{1.32}{GW190828A}{1.19}{GW190814A}{1.60}{GW190803A}{1.63}{GW190731A}{1.42}{GW190728A}{1.17}{GW190727A}{1.37}{GW190720A}{1.25}{GW190719A}{1.11}{GW190708A}{1.43}{GW190707A}{1.76}{GW190706A}{1.19}{GW190701A}{1.74}{GW190630A}{1.23}{GW190620A}{1.05}{GW190602A}{1.38}{GW190527A}{1.41}{GW190521B}{1.27}{GW190521A}{1.59}{GW190519A}{1.05}{GW190517A}{0.91}{GW190514A}{1.94}{GW190513A}{1.32}{GW190512A}{1.41}{GW190503A}{1.59}{GW190426A}{0.00}{GW190425A}{1.41}{GW190424A}{1.37}{GW190421A}{1.73}{GW190413B}{1.70}{GW190413A}{1.65}{GW190412A}{1.32}{GW190408A}{1.59}}}
\newcommand{\tilttwoplus}[1]{\IfEqCase{#1}{{GW190930A}{1.21}{GW190929A}{1.13}{GW190924A}{1.16}{GW190915A}{1.09}{GW190910A}{1.04}{GW190909A}{1.01}{GW190828B}{1.20}{GW190828A}{1.25}{GW190814A}{1.00}{GW190803A}{1.05}{GW190731A}{1.16}{GW190728A}{1.29}{GW190727A}{1.20}{GW190720A}{1.25}{GW190719A}{1.31}{GW190708A}{1.13}{GW190707A}{0.92}{GW190706A}{1.29}{GW190701A}{0.97}{GW190630A}{1.18}{GW190620A}{1.32}{GW190602A}{1.15}{GW190527A}{1.20}{GW190521B}{1.13}{GW190521A}{1.05}{GW190519A}{1.26}{GW190517A}{1.30}{GW190514A}{0.88}{GW190513A}{1.19}{GW190512A}{1.11}{GW190503A}{1.07}{GW190426A}{3.14}{GW190425A}{0.94}{GW190424A}{1.16}{GW190421A}{0.99}{GW190413B}{1.00}{GW190413A}{1.04}{GW190412A}{1.12}{GW190408A}{1.03}}}
\newcommand{\costhetajnminus}[1]{\IfEqCase{#1}{{GW190930A}{1.54}{GW190929A}{0.76}{GW190924A}{1.67}{GW190915A}{0.50}{GW190910A}{0.88}{GW190909A}{1.06}{GW190828B}{0.71}{GW190828A}{0.36}{GW190814A}{1.35}{GW190803A}{1.39}{GW190731A}{1.33}{GW190728A}{1.30}{GW190727A}{0.98}{GW190720A}{0.20}{GW190719A}{0.92}{GW190708A}{1.17}{GW190707A}{0.43}{GW190706A}{1.14}{GW190701A}{0.41}{GW190630A}{1.29}{GW190620A}{0.60}{GW190602A}{0.82}{GW190527A}{1.25}{GW190521B}{1.05}{GW190521A}{1.35}{GW190519A}{0.81}{GW190517A}{0.34}{GW190514A}{1.02}{GW190513A}{1.65}{GW190512A}{0.91}{GW190503A}{0.23}{GW190426A}{0.84}{GW190425A}{1.43}{GW190424A}{1.02}{GW190421A}{0.80}{GW190413B}{0.63}{GW190413A}{1.35}{GW190412A}{0.35}{GW190408A}{1.65}}}
\newcommand{\costhetajnmed}[1]{\IfEqCase{#1}{{GW190930A}{0.59}{GW190929A}{-0.13}{GW190924A}{0.74}{GW190915A}{-0.45}{GW190910A}{-0.05}{GW190909A}{0.12}{GW190828B}{-0.25}{GW190828A}{-0.62}{GW190814A}{0.65}{GW190803A}{0.44}{GW190731A}{0.37}{GW190728A}{0.33}{GW190727A}{0.02}{GW190720A}{-0.79}{GW190719A}{-0.04}{GW190708A}{0.20}{GW190707A}{-0.55}{GW190706A}{0.20}{GW190701A}{0.84}{GW190630A}{0.34}{GW190620A}{-0.36}{GW190602A}{-0.14}{GW190527A}{0.30}{GW190521B}{0.09}{GW190521A}{0.41}{GW190519A}{-0.01}{GW190517A}{-0.64}{GW190514A}{0.07}{GW190513A}{0.70}{GW190512A}{-0.04}{GW190503A}{-0.75}{GW190426A}{-0.13}{GW190425A}{0.47}{GW190424A}{0.05}{GW190421A}{-0.17}{GW190413B}{-0.33}{GW190413A}{0.41}{GW190412A}{0.75}{GW190408A}{0.70}}}
\newcommand{\costhetajnplus}[1]{\IfEqCase{#1}{{GW190930A}{0.39}{GW190929A}{1.01}{GW190924A}{0.25}{GW190915A}{1.32}{GW190910A}{0.96}{GW190909A}{0.82}{GW190828B}{1.20}{GW190828A}{1.57}{GW190814A}{0.16}{GW190803A}{0.53}{GW190731A}{0.59}{GW190728A}{0.65}{GW190727A}{0.94}{GW190720A}{1.68}{GW190719A}{1.01}{GW190708A}{0.78}{GW190707A}{1.51}{GW190706A}{0.75}{GW190701A}{0.15}{GW190630A}{0.62}{GW190620A}{1.27}{GW190602A}{1.10}{GW190527A}{0.66}{GW190521B}{0.87}{GW190521A}{0.55}{GW190519A}{0.81}{GW190517A}{1.38}{GW190514A}{0.89}{GW190513A}{0.27}{GW190512A}{0.99}{GW190503A}{0.48}{GW190426A}{1.09}{GW190425A}{0.50}{GW190424A}{0.91}{GW190421A}{1.12}{GW190413B}{1.27}{GW190413A}{0.55}{GW190412A}{0.14}{GW190408A}{0.28}}}
\newcommand{\spintwominus}[1]{\IfEqCase{#1}{{GW190930A}{0.37}{GW190929A}{0.44}{GW190924A}{0.32}{GW190915A}{0.43}{GW190910A}{0.33}{GW190909A}{0.43}{GW190828B}{0.38}{GW190828A}{0.37}{GW190814A}{0.46}{GW190803A}{0.40}{GW190731A}{0.40}{GW190728A}{0.35}{GW190727A}{0.41}{GW190720A}{0.45}{GW190719A}{0.49}{GW190708A}{0.28}{GW190707A}{0.28}{GW190706A}{0.44}{GW190701A}{0.40}{GW190630A}{0.34}{GW190620A}{0.50}{GW190602A}{0.45}{GW190527A}{0.45}{GW190521B}{0.37}{GW190521A}{0.52}{GW190519A}{0.48}{GW190517A}{0.52}{GW190514A}{0.48}{GW190513A}{0.39}{GW190512A}{0.32}{GW190503A}{0.40}{GW190426A}{0.009}{GW190425A}{0.25}{GW190424A}{0.42}{GW190421A}{0.42}{GW190413B}{0.45}{GW190413A}{0.41}{GW190412A}{0.43}{GW190408A}{0.32}}}
\newcommand{\spintwomed}[1]{\IfEqCase{#1}{{GW190930A}{0.42}{GW190929A}{0.49}{GW190924A}{0.35}{GW190915A}{0.48}{GW190910A}{0.37}{GW190909A}{0.49}{GW190828B}{0.42}{GW190828A}{0.41}{GW190814A}{0.52}{GW190803A}{0.45}{GW190731A}{0.45}{GW190728A}{0.39}{GW190727A}{0.45}{GW190720A}{0.51}{GW190719A}{0.55}{GW190708A}{0.30}{GW190707A}{0.31}{GW190706A}{0.49}{GW190701A}{0.44}{GW190630A}{0.38}{GW190620A}{0.56}{GW190602A}{0.50}{GW190527A}{0.50}{GW190521B}{0.42}{GW190521A}{0.58}{GW190519A}{0.54}{GW190517A}{0.58}{GW190514A}{0.54}{GW190513A}{0.43}{GW190512A}{0.36}{GW190503A}{0.44}{GW190426A}{0.009}{GW190425A}{0.28}{GW190424A}{0.47}{GW190421A}{0.46}{GW190413B}{0.50}{GW190413A}{0.45}{GW190412A}{0.49}{GW190408A}{0.36}}}
\newcommand{\spintwoplus}[1]{\IfEqCase{#1}{{GW190930A}{0.49}{GW190929A}{0.45}{GW190924A}{0.51}{GW190915A}{0.46}{GW190910A}{0.51}{GW190909A}{0.45}{GW190828B}{0.49}{GW190828A}{0.46}{GW190814A}{0.41}{GW190803A}{0.49}{GW190731A}{0.48}{GW190728A}{0.50}{GW190727A}{0.46}{GW190720A}{0.43}{GW190719A}{0.40}{GW190708A}{0.50}{GW190707A}{0.52}{GW190706A}{0.45}{GW190701A}{0.48}{GW190630A}{0.46}{GW190620A}{0.40}{GW190602A}{0.44}{GW190527A}{0.45}{GW190521B}{0.39}{GW190521A}{0.38}{GW190519A}{0.41}{GW190517A}{0.38}{GW190514A}{0.42}{GW190513A}{0.48}{GW190512A}{0.51}{GW190503A}{0.48}{GW190426A}{0.03}{GW190425A}{0.51}{GW190424A}{0.47}{GW190421A}{0.47}{GW190413B}{0.44}{GW190413A}{0.48}{GW190412A}{0.44}{GW190408A}{0.53}}}
\newcommand{\massonedetminus}[1]{\IfEqCase{#1}{{GW190930A}{2.6}{GW190929A}{28.9}{GW190924A}{2.2}{GW190915A}{7.7}{GW190910A}{6.2}{GW190909A}{17.5}{GW190828B}{8.7}{GW190828A}{4.7}{GW190814A}{1.0}{GW190803A}{9.0}{GW190731A}{10.5}{GW190728A}{2.5}{GW190727A}{8.2}{GW190720A}{3.5}{GW190719A}{16.4}{GW190708A}{2.4}{GW190707A}{1.8}{GW190706A}{17.6}{GW190701A}{10.9}{GW190630A}{6.4}{GW190620A}{15.4}{GW190602A}{15.7}{GW190527A}{11.0}{GW190521B}{5.4}{GW190521A}{20.8}{GW190519A}{12.9}{GW190517A}{8.4}{GW190514A}{10.8}{GW190513A}{12.3}{GW190512A}{6.8}{GW190503A}{9.9}{GW190426A}{2.5}{GW190425A}{0.4}{GW190424A}{8.0}{GW190421A}{8.8}{GW190413B}{14.6}{GW190413A}{12.0}{GW190412A}{6.2}{GW190408A}{4.0}}}
\newcommand{\massonedetmed}[1]{\IfEqCase{#1}{{GW190930A}{14.2}{GW190929A}{111.3}{GW190924A}{9.9}{GW190915A}{46.0}{GW190910A}{56.3}{GW190909A}{73.0}{GW190828B}{31.1}{GW190828A}{43.9}{GW190814A}{24.4}{GW190803A}{57.6}{GW190731A}{64.4}{GW190728A}{14.4}{GW190727A}{58.8}{GW190720A}{15.7}{GW190719A}{60.3}{GW190708A}{20.6}{GW190707A}{13.4}{GW190706A}{112.9}{GW190701A}{74.1}{GW190630A}{41.4}{GW190620A}{84.6}{GW190602A}{101.7}{GW190527A}{52.0}{GW190521B}{51.9}{GW190521A}{153.4}{GW190519A}{95.4}{GW190517A}{50.5}{GW190514A}{64.9}{GW190513A}{49.0}{GW190512A}{29.4}{GW190503A}{55.2}{GW190426A}{6.2}{GW190425A}{2.1}{GW190424A}{56.1}{GW190421A}{61.1}{GW190413B}{80.7}{GW190413A}{55.3}{GW190412A}{34.6}{GW190408A}{31.5}}}
\newcommand{\massonedetplus}[1]{\IfEqCase{#1}{{GW190930A}{14.3}{GW190929A}{39.3}{GW190924A}{7.8}{GW190915A}{11.6}{GW190910A}{8.8}{GW190909A}{84.2}{GW190828B}{8.8}{GW190828A}{7.5}{GW190814A}{1.2}{GW190803A}{14.2}{GW190731A}{13.4}{GW190728A}{8.4}{GW190727A}{12.8}{GW190720A}{7.7}{GW190719A}{26.5}{GW190708A}{5.7}{GW190707A}{3.7}{GW190706A}{20.5}{GW190701A}{15.1}{GW190630A}{8.3}{GW190620A}{20.0}{GW190602A}{19.9}{GW190527A}{32.9}{GW190521B}{7.1}{GW190521A}{45.9}{GW190519A}{14.9}{GW190517A}{14.7}{GW190514A}{17.6}{GW190513A}{12.5}{GW190512A}{6.5}{GW190503A}{10.6}{GW190426A}{4.2}{GW190425A}{0.6}{GW190424A}{14.9}{GW190421A}{14.7}{GW190413B}{19.0}{GW190413A}{17.1}{GW190412A}{5.5}{GW190408A}{6.3}}}
\newcommand{\massratiominus}[1]{\IfEqCase{#1}{{GW190930A}{0.46}{GW190929A}{0.16}{GW190924A}{0.37}{GW190915A}{0.26}{GW190910A}{0.23}{GW190909A}{0.39}{GW190828B}{0.16}{GW190828A}{0.23}{GW190814A}{0.009}{GW190803A}{0.31}{GW190731A}{0.31}{GW190728A}{0.38}{GW190727A}{0.32}{GW190720A}{0.30}{GW190719A}{0.29}{GW190708A}{0.28}{GW190707A}{0.27}{GW190706A}{0.25}{GW190701A}{0.30}{GW190630A}{0.22}{GW190620A}{0.27}{GW190602A}{0.33}{GW190527A}{0.32}{GW190521B}{0.21}{GW190521A}{0.34}{GW190519A}{0.19}{GW190517A}{0.29}{GW190514A}{0.33}{GW190513A}{0.18}{GW190512A}{0.18}{GW190503A}{0.23}{GW190426A}{0.15}{GW190425A}{0.25}{GW190424A}{0.29}{GW190421A}{0.30}{GW190413B}{0.31}{GW190413A}{0.28}{GW190412A}{0.06}{GW190408A}{0.25}}}
\newcommand{\massratiomed}[1]{\IfEqCase{#1}{{GW190930A}{0.64}{GW190929A}{0.30}{GW190924A}{0.57}{GW190915A}{0.69}{GW190910A}{0.82}{GW190909A}{0.62}{GW190828B}{0.42}{GW190828A}{0.82}{GW190814A}{0.112}{GW190803A}{0.75}{GW190731A}{0.72}{GW190728A}{0.66}{GW190727A}{0.79}{GW190720A}{0.59}{GW190719A}{0.58}{GW190708A}{0.75}{GW190707A}{0.72}{GW190706A}{0.58}{GW190701A}{0.76}{GW190630A}{0.68}{GW190620A}{0.62}{GW190602A}{0.71}{GW190527A}{0.64}{GW190521B}{0.78}{GW190521A}{0.75}{GW190519A}{0.61}{GW190517A}{0.68}{GW190514A}{0.75}{GW190513A}{0.50}{GW190512A}{0.54}{GW190503A}{0.65}{GW190426A}{0.25}{GW190425A}{0.67}{GW190424A}{0.81}{GW190421A}{0.79}{GW190413B}{0.69}{GW190413A}{0.69}{GW190412A}{0.28}{GW190408A}{0.76}}}
\newcommand{\massratioplus}[1]{\IfEqCase{#1}{{GW190930A}{0.30}{GW190929A}{0.52}{GW190924A}{0.36}{GW190915A}{0.27}{GW190910A}{0.15}{GW190909A}{0.33}{GW190828B}{0.38}{GW190828A}{0.15}{GW190814A}{0.008}{GW190803A}{0.22}{GW190731A}{0.25}{GW190728A}{0.29}{GW190727A}{0.18}{GW190720A}{0.36}{GW190719A}{0.37}{GW190708A}{0.21}{GW190707A}{0.24}{GW190706A}{0.34}{GW190701A}{0.21}{GW190630A}{0.27}{GW190620A}{0.33}{GW190602A}{0.25}{GW190527A}{0.32}{GW190521B}{0.19}{GW190521A}{0.23}{GW190519A}{0.28}{GW190517A}{0.27}{GW190514A}{0.21}{GW190513A}{0.42}{GW190512A}{0.37}{GW190503A}{0.29}{GW190426A}{0.41}{GW190425A}{0.29}{GW190424A}{0.17}{GW190421A}{0.18}{GW190413B}{0.28}{GW190413A}{0.28}{GW190412A}{0.12}{GW190408A}{0.21}}}
\newcommand{\spinoneminus}[1]{\IfEqCase{#1}{{GW190930A}{0.35}{GW190929A}{0.54}{GW190924A}{0.21}{GW190915A}{0.49}{GW190910A}{0.30}{GW190909A}{0.52}{GW190828B}{0.26}{GW190828A}{0.40}{GW190814A}{0.03}{GW190803A}{0.37}{GW190731A}{0.34}{GW190728A}{0.28}{GW190727A}{0.42}{GW190720A}{0.35}{GW190719A}{0.53}{GW190708A}{0.20}{GW190707A}{0.21}{GW190706A}{0.48}{GW190701A}{0.36}{GW190630A}{0.23}{GW190620A}{0.50}{GW190602A}{0.34}{GW190527A}{0.43}{GW190521B}{0.28}{GW190521A}{0.63}{GW190519A}{0.50}{GW190517A}{0.35}{GW190514A}{0.46}{GW190513A}{0.28}{GW190512A}{0.16}{GW190503A}{0.31}{GW190426A}{0.14}{GW190425A}{0.25}{GW190424A}{0.47}{GW190421A}{0.41}{GW190413B}{0.52}{GW190413A}{0.36}{GW190412A}{0.22}{GW190408A}{0.31}}}
\newcommand{\spinonemed}[1]{\IfEqCase{#1}{{GW190930A}{0.39}{GW190929A}{0.64}{GW190924A}{0.24}{GW190915A}{0.55}{GW190910A}{0.34}{GW190909A}{0.58}{GW190828B}{0.28}{GW190828A}{0.44}{GW190814A}{0.03}{GW190803A}{0.41}{GW190731A}{0.37}{GW190728A}{0.32}{GW190727A}{0.46}{GW190720A}{0.40}{GW190719A}{0.62}{GW190708A}{0.22}{GW190707A}{0.24}{GW190706A}{0.55}{GW190701A}{0.40}{GW190630A}{0.26}{GW190620A}{0.61}{GW190602A}{0.38}{GW190527A}{0.47}{GW190521B}{0.31}{GW190521A}{0.73}{GW190519A}{0.60}{GW190517A}{0.86}{GW190514A}{0.52}{GW190513A}{0.30}{GW190512A}{0.17}{GW190503A}{0.34}{GW190426A}{0.14}{GW190425A}{0.27}{GW190424A}{0.53}{GW190421A}{0.46}{GW190413B}{0.58}{GW190413A}{0.40}{GW190412A}{0.44}{GW190408A}{0.34}}}
\newcommand{\spinoneplus}[1]{\IfEqCase{#1}{{GW190930A}{0.40}{GW190929A}{0.32}{GW190924A}{0.43}{GW190915A}{0.39}{GW190910A}{0.50}{GW190909A}{0.37}{GW190828B}{0.43}{GW190828A}{0.45}{GW190814A}{0.05}{GW190803A}{0.51}{GW190731A}{0.54}{GW190728A}{0.37}{GW190727A}{0.47}{GW190720A}{0.40}{GW190719A}{0.34}{GW190708A}{0.52}{GW190707A}{0.47}{GW190706A}{0.39}{GW190701A}{0.50}{GW190630A}{0.37}{GW190620A}{0.34}{GW190602A}{0.51}{GW190527A}{0.47}{GW190521B}{0.42}{GW190521A}{0.25}{GW190519A}{0.33}{GW190517A}{0.13}{GW190514A}{0.43}{GW190513A}{0.51}{GW190512A}{0.44}{GW190503A}{0.51}{GW190426A}{0.40}{GW190425A}{0.51}{GW190424A}{0.42}{GW190421A}{0.47}{GW190413B}{0.38}{GW190413A}{0.51}{GW190412A}{0.16}{GW190408A}{0.47}}}
\newcommand{\costiltoneminus}[1]{\IfEqCase{#1}{{GW190930A}{1.08}{GW190929A}{0.83}{GW190924A}{0.97}{GW190915A}{0.82}{GW190910A}{0.93}{GW190909A}{0.78}{GW190828B}{0.99}{GW190828A}{1.09}{GW190814A}{0.90}{GW190803A}{0.80}{GW190731A}{0.99}{GW190728A}{1.13}{GW190727A}{1.02}{GW190720A}{0.97}{GW190719A}{1.02}{GW190708A}{0.88}{GW190707A}{0.68}{GW190706A}{0.99}{GW190701A}{0.70}{GW190630A}{1.02}{GW190620A}{0.84}{GW190602A}{0.99}{GW190527A}{1.02}{GW190521B}{0.97}{GW190521A}{0.89}{GW190519A}{0.74}{GW190517A}{0.37}{GW190514A}{0.50}{GW190513A}{1.10}{GW190512A}{0.93}{GW190503A}{0.75}{GW190426A}{0.00}{GW190425A}{0.65}{GW190424A}{1.00}{GW190421A}{0.75}{GW190413B}{0.76}{GW190413A}{0.90}{GW190412A}{0.47}{GW190408A}{0.73}}}
\newcommand{\costiltonemed}[1]{\IfEqCase{#1}{{GW190930A}{0.47}{GW190929A}{0.02}{GW190924A}{0.19}{GW190915A}{0.06}{GW190910A}{0.08}{GW190909A}{-0.10}{GW190828B}{0.26}{GW190828A}{0.51}{GW190814A}{0.01}{GW190803A}{-0.07}{GW190731A}{0.16}{GW190728A}{0.49}{GW190727A}{0.30}{GW190720A}{0.54}{GW190719A}{0.66}{GW190708A}{0.07}{GW190707A}{-0.19}{GW190706A}{0.66}{GW190701A}{-0.22}{GW190630A}{0.29}{GW190620A}{0.68}{GW190602A}{0.18}{GW190527A}{0.28}{GW190521B}{0.17}{GW190521A}{0.05}{GW190519A}{0.65}{GW190517A}{0.83}{GW190514A}{-0.45}{GW190513A}{0.41}{GW190512A}{0.07}{GW190503A}{-0.16}{GW190426A}{-1.00}{GW190425A}{0.26}{GW190424A}{0.36}{GW190421A}{-0.15}{GW190413B}{-0.06}{GW190413A}{0.01}{GW190412A}{0.70}{GW190408A}{-0.17}}}
\newcommand{\costiltoneplus}[1]{\IfEqCase{#1}{{GW190930A}{0.49}{GW190929A}{0.65}{GW190924A}{0.76}{GW190915A}{0.73}{GW190910A}{0.79}{GW190909A}{0.96}{GW190828B}{0.63}{GW190828A}{0.44}{GW190814A}{0.87}{GW190803A}{0.91}{GW190731A}{0.74}{GW190728A}{0.47}{GW190727A}{0.61}{GW190720A}{0.42}{GW190719A}{0.31}{GW190708A}{0.80}{GW190707A}{0.97}{GW190706A}{0.31}{GW190701A}{1.01}{GW190630A}{0.63}{GW190620A}{0.29}{GW190602A}{0.72}{GW190527A}{0.65}{GW190521B}{0.71}{GW190521A}{0.78}{GW190519A}{0.32}{GW190517A}{0.16}{GW190514A}{1.08}{GW190513A}{0.53}{GW190512A}{0.82}{GW190503A}{0.96}{GW190426A}{2.00}{GW190425A}{0.61}{GW190424A}{0.56}{GW190421A}{0.94}{GW190413B}{0.82}{GW190413A}{0.87}{GW190412A}{0.20}{GW190408A}{0.94}}}
\newcommand{\finalmasssourceminus}[1]{\IfEqCase{#1}{{GW190930A}{1.5}{GW190929A}{25.3}{GW190924A}{1.0}{GW190915A}{6.0}{GW190910A}{8.6}{GW190909A}{16.8}{GW190828B}{4.5}{GW190828A}{4.3}{GW190814A}{0.9}{GW190803A}{8.5}{GW190731A}{10.8}{GW190728A}{1.3}{GW190727A}{7.5}{GW190720A}{2.2}{GW190719A}{10.2}{GW190708A}{1.8}{GW190707A}{1.3}{GW190706A}{13.5}{GW190701A}{8.9}{GW190630A}{4.6}{GW190620A}{12.1}{GW190602A}{14.9}{GW190527A}{9.3}{GW190521B}{4.4}{GW190521A}{22.4}{GW190519A}{13.8}{GW190517A}{8.9}{GW190514A}{10.4}{GW190513A}{5.8}{GW190512A}{3.5}{GW190503A}{7.7}{GW190424A}{10.1}{GW190421A}{8.7}{GW190413B}{11.4}{GW190413A}{9.2}{GW190412A}{3.8}{GW190408A}{2.8}}}
\newcommand{\finalmasssourcemed}[1]{\IfEqCase{#1}{{GW190930A}{19.4}{GW190929A}{101.5}{GW190924A}{13.3}{GW190915A}{57.2}{GW190910A}{75.8}{GW190909A}{72.0}{GW190828B}{33.1}{GW190828A}{54.9}{GW190814A}{25.6}{GW190803A}{61.7}{GW190731A}{67.0}{GW190728A}{19.6}{GW190727A}{63.8}{GW190720A}{20.4}{GW190719A}{54.9}{GW190708A}{29.5}{GW190707A}{19.2}{GW190706A}{99.0}{GW190701A}{90.2}{GW190630A}{56.4}{GW190620A}{87.2}{GW190602A}{110.9}{GW190527A}{56.4}{GW190521B}{71.0}{GW190521A}{156.3}{GW190519A}{101.0}{GW190517A}{59.3}{GW190514A}{64.5}{GW190513A}{51.6}{GW190512A}{34.5}{GW190503A}{68.6}{GW190424A}{68.9}{GW190421A}{69.7}{GW190413B}{75.5}{GW190413A}{56.0}{GW190412A}{37.3}{GW190408A}{41.1}}}
\newcommand{\finalmasssourceplus}[1]{\IfEqCase{#1}{{GW190930A}{9.2}{GW190929A}{33.6}{GW190924A}{5.2}{GW190915A}{7.1}{GW190910A}{8.5}{GW190909A}{54.9}{GW190828B}{5.5}{GW190828A}{7.2}{GW190814A}{1.1}{GW190803A}{11.8}{GW190731A}{14.6}{GW190728A}{4.7}{GW190727A}{10.9}{GW190720A}{4.5}{GW190719A}{17.3}{GW190708A}{2.5}{GW190707A}{1.9}{GW190706A}{18.3}{GW190701A}{11.3}{GW190630A}{4.4}{GW190620A}{16.8}{GW190602A}{17.7}{GW190527A}{20.2}{GW190521B}{6.5}{GW190521A}{36.8}{GW190519A}{12.4}{GW190517A}{9.1}{GW190514A}{17.9}{GW190513A}{8.2}{GW190512A}{3.8}{GW190503A}{8.8}{GW190424A}{12.4}{GW190421A}{12.5}{GW190413B}{16.4}{GW190413A}{12.5}{GW190412A}{3.9}{GW190408A}{3.9}}}
\newcommand{\phaseminus}[1]{\IfEqCase{#1}{{GW190930A}{2.91}{GW190929A}{2.79}{GW190924A}{2.77}{GW190915A}{2.96}{GW190910A}{2.81}{GW190909A}{2.80}{GW190828B}{2.90}{GW190828A}{2.89}{GW190814A}{2.76}{GW190803A}{2.85}{GW190731A}{2.83}{GW190728A}{2.80}{GW190727A}{2.88}{GW190720A}{2.81}{GW190719A}{2.84}{GW190708A}{2.83}{GW190707A}{2.91}{GW190706A}{2.45}{GW190701A}{2.60}{GW190630A}{3.46}{GW190620A}{2.68}{GW190602A}{2.80}{GW190527A}{2.85}{GW190521B}{1.76}{GW190521A}{2.87}{GW190519A}{2.80}{GW190517A}{2.79}{GW190514A}{2.81}{GW190513A}{2.78}{GW190512A}{2.87}{GW190503A}{2.88}{GW190426A}{2.79}{GW190425A}{2.82}{GW190424A}{2.86}{GW190421A}{2.89}{GW190413B}{2.82}{GW190413A}{2.81}{GW190412A}{1.89}{GW190408A}{2.79}}}
\newcommand{\phasemed}[1]{\IfEqCase{#1}{{GW190930A}{3.22}{GW190929A}{3.08}{GW190924A}{3.08}{GW190915A}{3.30}{GW190910A}{3.15}{GW190909A}{3.15}{GW190828B}{3.23}{GW190828A}{3.17}{GW190814A}{3.13}{GW190803A}{3.14}{GW190731A}{3.15}{GW190728A}{3.11}{GW190727A}{3.21}{GW190720A}{3.12}{GW190719A}{3.15}{GW190708A}{3.14}{GW190707A}{3.24}{GW190706A}{3.01}{GW190701A}{2.89}{GW190630A}{3.82}{GW190620A}{2.97}{GW190602A}{3.13}{GW190527A}{3.13}{GW190521B}{2.03}{GW190521A}{3.12}{GW190519A}{3.13}{GW190517A}{3.10}{GW190514A}{3.13}{GW190513A}{3.08}{GW190512A}{3.18}{GW190503A}{3.16}{GW190426A}{3.09}{GW190425A}{3.12}{GW190424A}{3.16}{GW190421A}{3.20}{GW190413B}{3.11}{GW190413A}{3.12}{GW190412A}{2.15}{GW190408A}{3.11}}}
\newcommand{\phaseplus}[1]{\IfEqCase{#1}{{GW190930A}{2.77}{GW190929A}{2.89}{GW190924A}{2.88}{GW190915A}{2.70}{GW190910A}{2.84}{GW190909A}{2.82}{GW190828B}{2.73}{GW190828A}{2.85}{GW190814A}{2.81}{GW190803A}{2.81}{GW190731A}{2.80}{GW190728A}{2.86}{GW190727A}{2.76}{GW190720A}{2.85}{GW190719A}{2.83}{GW190708A}{2.83}{GW190707A}{2.76}{GW190706A}{2.65}{GW190701A}{3.10}{GW190630A}{2.15}{GW190620A}{2.97}{GW190602A}{2.84}{GW190527A}{2.81}{GW190521B}{3.94}{GW190521A}{2.87}{GW190519A}{2.83}{GW190517A}{2.88}{GW190514A}{2.82}{GW190513A}{2.90}{GW190512A}{2.82}{GW190503A}{2.84}{GW190426A}{2.85}{GW190425A}{2.87}{GW190424A}{2.81}{GW190421A}{2.77}{GW190413B}{2.84}{GW190413A}{2.84}{GW190412A}{3.84}{GW190408A}{2.86}}}
\newcommand{\radiatedenergyminus}[1]{\IfEqCase{#1}{{GW190930A}{0.2}{GW190929A}{1.4}{GW190924A}{0.1}{GW190915A}{0.7}{GW190910A}{0.7}{GW190909A}{1.4}{GW190828B}{0.2}{GW190828A}{0.5}{GW190814A}{0.007}{GW190803A}{0.8}{GW190731A}{1.1}{GW190728A}{0.2}{GW190727A}{0.9}{GW190720A}{0.2}{GW190719A}{1.1}{GW190708A}{0.2}{GW190707A}{0.09}{GW190706A}{2.1}{GW190701A}{1.1}{GW190630A}{0.5}{GW190620A}{1.9}{GW190602A}{1.9}{GW190527A}{1.0}{GW190521B}{0.7}{GW190521A}{2.4}{GW190519A}{1.7}{GW190517A}{1.2}{GW190514A}{0.8}{GW190513A}{0.6}{GW190512A}{0.3}{GW190503A}{1.0}{GW190424A}{0.9}{GW190421A}{0.9}{GW190413B}{1.1}{GW190413A}{0.8}{GW190412A}{0.1}{GW190408A}{0.3}}}
\newcommand{\radiatedenergymed}[1]{\IfEqCase{#1}{{GW190930A}{0.9}{GW190929A}{2.7}{GW190924A}{0.6}{GW190915A}{2.7}{GW190910A}{3.8}{GW190909A}{3.0}{GW190828B}{1.2}{GW190828A}{3.1}{GW190814A}{0.2}{GW190803A}{2.9}{GW190731A}{3.2}{GW190728A}{1.0}{GW190727A}{3.3}{GW190720A}{1.0}{GW190719A}{2.9}{GW190708A}{1.4}{GW190707A}{0.9}{GW190706A}{5.3}{GW190701A}{4.1}{GW190630A}{2.8}{GW190620A}{4.9}{GW190602A}{5.4}{GW190527A}{2.7}{GW190521B}{3.7}{GW190521A}{7.6}{GW190519A}{5.6}{GW190517A}{4.1}{GW190514A}{2.7}{GW190513A}{2.2}{GW190512A}{1.5}{GW190503A}{3.1}{GW190424A}{3.6}{GW190421A}{3.3}{GW190413B}{3.4}{GW190413A}{2.6}{GW190412A}{1.1}{GW190408A}{1.9}}}
\newcommand{\radiatedenergyplus}[1]{\IfEqCase{#1}{{GW190930A}{0.1}{GW190929A}{2.8}{GW190924A}{0.06}{GW190915A}{0.7}{GW190910A}{0.9}{GW190909A}{2.2}{GW190828B}{0.3}{GW190828A}{0.7}{GW190814A}{0.006}{GW190803A}{0.9}{GW190731A}{1.2}{GW190728A}{0.09}{GW190727A}{1.1}{GW190720A}{0.1}{GW190719A}{1.7}{GW190708A}{0.1}{GW190707A}{0.08}{GW190706A}{2.3}{GW190701A}{1.1}{GW190630A}{0.5}{GW190620A}{2.0}{GW190602A}{1.8}{GW190527A}{1.5}{GW190521B}{0.6}{GW190521A}{2.9}{GW190519A}{1.7}{GW190517A}{1.3}{GW190514A}{1.1}{GW190513A}{1.1}{GW190512A}{0.3}{GW190503A}{0.9}{GW190424A}{1.2}{GW190421A}{1.0}{GW190413B}{1.1}{GW190413A}{1.0}{GW190412A}{0.2}{GW190408A}{0.3}}}
\newcommand{\masstwodetminus}[1]{\IfEqCase{#1}{{GW190930A}{3.9}{GW190929A}{16.4}{GW190924A}{2.1}{GW190915A}{8.5}{GW190910A}{9.1}{GW190909A}{23.3}{GW190828B}{2.7}{GW190828A}{6.5}{GW190814A}{0.09}{GW190803A}{13.5}{GW190731A}{16.4}{GW190728A}{3.0}{GW190727A}{13.8}{GW190720A}{2.6}{GW190719A}{12.7}{GW190708A}{3.1}{GW190707A}{1.9}{GW190706A}{27.0}{GW190701A}{17.5}{GW190630A}{5.5}{GW190620A}{19.7}{GW190602A}{27.8}{GW190527A}{12.3}{GW190521B}{7.6}{GW190521A}{39.7}{GW190519A}{16.9}{GW190517A}{9.5}{GW190514A}{15.9}{GW190513A}{6.0}{GW190512A}{2.9}{GW190503A}{10.6}{GW190426A}{0.5}{GW190425A}{0.3}{GW190424A}{10.3}{GW190421A}{13.2}{GW190413B}{20.0}{GW190413A}{11.5}{GW190412A}{1.0}{GW190408A}{4.6}}}
\newcommand{\masstwodetmed}[1]{\IfEqCase{#1}{{GW190930A}{9.1}{GW190929A}{33.9}{GW190924A}{5.6}{GW190915A}{31.9}{GW190910A}{45.9}{GW190909A}{47.1}{GW190828B}{13.3}{GW190828A}{36.1}{GW190814A}{2.72}{GW190803A}{42.7}{GW190731A}{45.6}{GW190728A}{9.5}{GW190727A}{46.0}{GW190720A}{9.2}{GW190719A}{34.5}{GW190708A}{15.5}{GW190707A}{9.7}{GW190706A}{66.6}{GW190701A}{56.2}{GW190630A}{28.0}{GW190620A}{53.1}{GW190602A}{71.5}{GW190527A}{32.8}{GW190521B}{40.5}{GW190521A}{114.8}{GW190519A}{59.3}{GW190517A}{34.4}{GW190514A}{48.0}{GW190513A}{24.7}{GW190512A}{15.8}{GW190503A}{36.2}{GW190426A}{1.6}{GW190425A}{1.4}{GW190424A}{44.8}{GW190421A}{47.8}{GW190413B}{55.2}{GW190413A}{38.0}{GW190412A}{9.6}{GW190408A}{23.7}}}
\newcommand{\masstwodetplus}[1]{\IfEqCase{#1}{{GW190930A}{1.8}{GW190929A}{37.2}{GW190924A}{1.5}{GW190915A}{7.0}{GW190910A}{7.0}{GW190909A}{20.9}{GW190828B}{4.9}{GW190828A}{4.6}{GW190814A}{0.08}{GW190803A}{9.8}{GW190731A}{11.8}{GW190728A}{1.8}{GW190727A}{8.4}{GW190720A}{2.4}{GW190719A}{13.3}{GW190708A}{2.0}{GW190707A}{1.4}{GW190706A}{23.9}{GW190701A}{11.7}{GW190630A}{5.5}{GW190620A}{17.1}{GW190602A}{18.6}{GW190527A}{19.4}{GW190521B}{5.7}{GW190521A}{26.0}{GW190519A}{16.1}{GW190517A}{8.2}{GW190514A}{11.1}{GW190513A}{10.6}{GW190512A}{4.7}{GW190503A}{10.4}{GW190426A}{0.9}{GW190425A}{0.3}{GW190424A}{8.3}{GW190421A}{9.4}{GW190413B}{14.9}{GW190413A}{10.7}{GW190412A}{1.7}{GW190408A}{3.6}}}
\newcommand{\masstwosourceminus}[1]{\IfEqCase{#1}{{GW190930A}{3.3}{GW190929A}{10.6}{GW190924A}{1.9}{GW190915A}{6.1}{GW190910A}{7.2}{GW190909A}{12.7}{GW190828B}{2.1}{GW190828A}{4.8}{GW190814A}{0.09}{GW190803A}{8.2}{GW190731A}{9.5}{GW190728A}{2.6}{GW190727A}{8.4}{GW190720A}{2.2}{GW190719A}{7.2}{GW190708A}{2.7}{GW190707A}{1.7}{GW190706A}{13.3}{GW190701A}{12.0}{GW190630A}{5.1}{GW190620A}{12.3}{GW190602A}{17.4}{GW190527A}{8.1}{GW190521B}{6.4}{GW190521A}{23.1}{GW190519A}{11.1}{GW190517A}{7.3}{GW190514A}{8.8}{GW190513A}{4.1}{GW190512A}{2.5}{GW190503A}{8.0}{GW190426A}{0.5}{GW190425A}{0.3}{GW190424A}{7.7}{GW190421A}{8.8}{GW190413B}{10.8}{GW190413A}{6.7}{GW190412A}{0.9}{GW190408A}{3.6}}}
\newcommand{\masstwosourcemed}[1]{\IfEqCase{#1}{{GW190930A}{7.8}{GW190929A}{24.1}{GW190924A}{5.0}{GW190915A}{24.4}{GW190910A}{35.6}{GW190909A}{28.3}{GW190828B}{10.2}{GW190828A}{26.2}{GW190814A}{2.59}{GW190803A}{27.3}{GW190731A}{28.8}{GW190728A}{8.1}{GW190727A}{29.4}{GW190720A}{7.8}{GW190719A}{20.8}{GW190708A}{13.2}{GW190707A}{8.4}{GW190706A}{38.2}{GW190701A}{40.8}{GW190630A}{23.7}{GW190620A}{35.5}{GW190602A}{47.8}{GW190527A}{22.6}{GW190521B}{32.8}{GW190521A}{69.0}{GW190519A}{40.5}{GW190517A}{25.3}{GW190514A}{28.4}{GW190513A}{18.0}{GW190512A}{12.6}{GW190503A}{28.4}{GW190426A}{1.5}{GW190425A}{1.4}{GW190424A}{31.8}{GW190421A}{31.9}{GW190413B}{31.8}{GW190413A}{23.7}{GW190412A}{8.3}{GW190408A}{18.4}}}
\newcommand{\masstwosourceplus}[1]{\IfEqCase{#1}{{GW190930A}{1.7}{GW190929A}{19.3}{GW190924A}{1.4}{GW190915A}{5.6}{GW190910A}{6.3}{GW190909A}{13.4}{GW190828B}{3.6}{GW190828A}{4.6}{GW190814A}{0.08}{GW190803A}{7.8}{GW190731A}{9.7}{GW190728A}{1.7}{GW190727A}{7.1}{GW190720A}{2.3}{GW190719A}{9.0}{GW190708A}{2.0}{GW190707A}{1.4}{GW190706A}{14.6}{GW190701A}{8.7}{GW190630A}{5.2}{GW190620A}{12.2}{GW190602A}{14.3}{GW190527A}{10.5}{GW190521B}{5.4}{GW190521A}{22.7}{GW190519A}{11.0}{GW190517A}{7.0}{GW190514A}{9.3}{GW190513A}{7.7}{GW190512A}{3.6}{GW190503A}{7.7}{GW190426A}{0.8}{GW190425A}{0.3}{GW190424A}{7.6}{GW190421A}{8.0}{GW190413B}{11.7}{GW190413A}{7.3}{GW190412A}{1.6}{GW190408A}{3.3}}}
\newcommand{\decminus}[1]{\IfEqCase{#1}{{GW190930A}{0.66804}{GW190929A}{1.09453}{GW190924A}{0.31095}{GW190915A}{0.43700}{GW190910A}{0.78759}{GW190909A}{1.37456}{GW190828B}{0.42413}{GW190828A}{0.45822}{GW190814A}{0.12765}{GW190803A}{0.76320}{GW190731A}{0.54299}{GW190728A}{1.45903}{GW190727A}{0.53050}{GW190720A}{1.79814}{GW190719A}{1.52271}{GW190708A}{1.12280}{GW190707A}{0.66130}{GW190706A}{1.13851}{GW190701A}{0.08561}{GW190630A}{0.88066}{GW190620A}{1.15741}{GW190602A}{0.22445}{GW190527A}{0.63740}{GW190521B}{0.61963}{GW190521A}{0.40205}{GW190519A}{1.22807}{GW190517A}{0.23138}{GW190514A}{1.30610}{GW190513A}{1.20492}{GW190512A}{0.07267}{GW190503A}{0.08741}{GW190426A}{1.53579}{GW190425A}{0.89977}{GW190424A}{1.08941}{GW190421A}{0.52647}{GW190413B}{0.10088}{GW190413A}{1.21171}{GW190412A}{0.03938}{GW190408A}{0.33259}}}
\newcommand{\decmed}[1]{\IfEqCase{#1}{{GW190930A}{0.64489}{GW190929A}{0.12643}{GW190924A}{0.16308}{GW190915A}{0.64820}{GW190910A}{-0.19299}{GW190909A}{0.45798}{GW190828B}{-0.70225}{GW190828A}{-0.38234}{GW190814A}{-0.43746}{GW190803A}{0.56728}{GW190731A}{-0.84452}{GW190728A}{0.14876}{GW190727A}{-0.69575}{GW190720A}{0.61861}{GW190719A}{0.60387}{GW190708A}{0.31278}{GW190707A}{-0.26771}{GW190706A}{0.49342}{GW190701A}{-0.11450}{GW190630A}{-0.17794}{GW190620A}{0.40157}{GW190602A}{-0.60585}{GW190527A}{-0.67031}{GW190521B}{0.31697}{GW190521A}{-0.79351}{GW190519A}{0.62138}{GW190517A}{-0.77834}{GW190514A}{0.75342}{GW190513A}{0.66794}{GW190512A}{-0.46498}{GW190503A}{-0.88291}{GW190426A}{0.90282}{GW190425A}{-0.13006}{GW190424A}{-0.00051}{GW190421A}{-0.82328}{GW190413B}{-0.53598}{GW190413A}{0.45042}{GW190412A}{0.63309}{GW190408A}{0.92018}}}
\newcommand{\decplus}[1]{\IfEqCase{#1}{{GW190930A}{0.44937}{GW190929A}{0.92641}{GW190924A}{0.26435}{GW190915A}{0.49993}{GW190910A}{0.99066}{GW190909A}{0.80671}{GW190828B}{1.28711}{GW190828A}{1.25309}{GW190814A}{0.03175}{GW190803A}{0.63137}{GW190731A}{1.11036}{GW190728A}{0.43956}{GW190727A}{1.58271}{GW190720A}{0.05969}{GW190719A}{0.55220}{GW190708A}{0.86640}{GW190707A}{1.42731}{GW190706A}{0.43610}{GW190701A}{0.08803}{GW190630A}{0.78734}{GW190620A}{0.77559}{GW190602A}{0.57550}{GW190527A}{0.97030}{GW190521B}{0.25837}{GW190521A}{1.54032}{GW190519A}{0.40124}{GW190517A}{0.96906}{GW190514A}{0.60827}{GW190513A}{0.37574}{GW190512A}{0.32502}{GW190503A}{0.10637}{GW190426A}{0.61722}{GW190425A}{0.96811}{GW190424A}{1.08246}{GW190421A}{0.53382}{GW190413B}{1.13504}{GW190413A}{0.90354}{GW190412A}{0.02643}{GW190408A}{0.08290}}}
\newcommand{\psiminus}[1]{\IfEqCase{#1}{{GW190930A}{1.82}{GW190929A}{1.43}{GW190924A}{1.79}{GW190915A}{1.85}{GW190910A}{2.23}{GW190909A}{1.42}{GW190828B}{1.30}{GW190828A}{2.05}{GW190814A}{0.32}{GW190803A}{2.82}{GW190731A}{2.78}{GW190728A}{1.92}{GW190727A}{2.89}{GW190720A}{1.90}{GW190719A}{1.84}{GW190708A}{2.83}{GW190707A}{1.83}{GW190706A}{2.03}{GW190701A}{1.87}{GW190630A}{1.25}{GW190620A}{1.54}{GW190602A}{2.89}{GW190527A}{2.77}{GW190521B}{1.24}{GW190521A}{1.42}{GW190519A}{2.65}{GW190517A}{2.20}{GW190514A}{2.82}{GW190513A}{2.03}{GW190512A}{2.81}{GW190503A}{2.51}{GW190426A}{1.40}{GW190425A}{1.46}{GW190424A}{2.78}{GW190421A}{2.77}{GW190413B}{2.84}{GW190413A}{2.79}{GW190412A}{2.36}{GW190408A}{2.73}}}
\newcommand{\psimed}[1]{\IfEqCase{#1}{{GW190930A}{2.02}{GW190929A}{1.62}{GW190924A}{2.00}{GW190915A}{2.06}{GW190910A}{3.19}{GW190909A}{1.60}{GW190828B}{1.45}{GW190828A}{2.21}{GW190814A}{0.39}{GW190803A}{3.18}{GW190731A}{3.13}{GW190728A}{2.16}{GW190727A}{3.16}{GW190720A}{2.09}{GW190719A}{2.05}{GW190708A}{3.14}{GW190707A}{2.05}{GW190706A}{2.19}{GW190701A}{2.03}{GW190630A}{1.81}{GW190620A}{1.82}{GW190602A}{3.13}{GW190527A}{3.10}{GW190521B}{1.73}{GW190521A}{1.59}{GW190519A}{3.30}{GW190517A}{2.37}{GW190514A}{3.18}{GW190513A}{2.26}{GW190512A}{3.14}{GW190503A}{3.12}{GW190426A}{1.58}{GW190425A}{1.62}{GW190424A}{3.15}{GW190421A}{3.11}{GW190413B}{3.10}{GW190413A}{3.15}{GW190412A}{2.56}{GW190408A}{3.14}}}
\newcommand{\psiplus}[1]{\IfEqCase{#1}{{GW190930A}{3.53}{GW190929A}{1.36}{GW190924A}{3.51}{GW190915A}{3.52}{GW190910A}{2.39}{GW190909A}{1.40}{GW190828B}{1.52}{GW190828A}{3.56}{GW190814A}{2.62}{GW190803A}{2.75}{GW190731A}{2.84}{GW190728A}{3.57}{GW190727A}{2.83}{GW190720A}{3.47}{GW190719A}{3.54}{GW190708A}{2.87}{GW190707A}{3.56}{GW190706A}{3.34}{GW190701A}{3.64}{GW190630A}{3.38}{GW190620A}{3.56}{GW190602A}{2.94}{GW190527A}{2.81}{GW190521B}{3.41}{GW190521A}{1.38}{GW190519A}{2.15}{GW190517A}{3.46}{GW190514A}{2.79}{GW190513A}{3.41}{GW190512A}{2.71}{GW190503A}{2.59}{GW190426A}{1.36}{GW190425A}{1.38}{GW190424A}{2.77}{GW190421A}{2.85}{GW190413B}{2.87}{GW190413A}{2.74}{GW190412A}{0.44}{GW190408A}{2.70}}}
\newcommand{\totalmassdetminus}[1]{\IfEqCase{#1}{{GW190930A}{1.0}{GW190929A}{26.3}{GW190924A}{0.7}{GW190915A}{8.1}{GW190910A}{7.8}{GW190909A}{22.9}{GW190828B}{4.0}{GW190828A}{5.9}{GW190814A}{1.0}{GW190803A}{11.9}{GW190731A}{14.3}{GW190728A}{0.7}{GW190727A}{10.9}{GW190720A}{1.2}{GW190719A}{15.5}{GW190708A}{0.8}{GW190707A}{0.5}{GW190706A}{27.7}{GW190701A}{14.8}{GW190630A}{3.5}{GW190620A}{18.4}{GW190602A}{20.6}{GW190527A}{10.3}{GW190521B}{5.4}{GW190521A}{34.6}{GW190519A}{17.9}{GW190517A}{7.3}{GW190514A}{15.1}{GW190513A}{6.7}{GW190512A}{2.8}{GW190503A}{11.8}{GW190426A}{1.6}{GW190425A}{0.08}{GW190424A}{10.9}{GW190421A}{12.4}{GW190413B}{18.0}{GW190413A}{15.3}{GW190412A}{4.6}{GW190408A}{3.8}}}
\newcommand{\totalmassdetmed}[1]{\IfEqCase{#1}{{GW190930A}{23.2}{GW190929A}{148.8}{GW190924A}{15.5}{GW190915A}{78.3}{GW190910A}{101.9}{GW190909A}{119.7}{GW190828B}{44.4}{GW190828A}{79.9}{GW190814A}{27.1}{GW190803A}{100.3}{GW190731A}{109.7}{GW190728A}{23.9}{GW190727A}{104.4}{GW190720A}{24.9}{GW190719A}{94.9}{GW190708A}{36.1}{GW190707A}{23.1}{GW190706A}{180.3}{GW190701A}{129.7}{GW190630A}{69.6}{GW190620A}{137.6}{GW190602A}{171.8}{GW190527A}{84.1}{GW190521B}{92.6}{GW190521A}{269.4}{GW190519A}{155.1}{GW190517A}{85.4}{GW190514A}{112.9}{GW190513A}{73.6}{GW190512A}{45.3}{GW190503A}{91.6}{GW190426A}{7.8}{GW190425A}{3.50}{GW190424A}{101.1}{GW190421A}{108.7}{GW190413B}{135.4}{GW190413A}{93.7}{GW190412A}{44.2}{GW190408A}{55.5}}}
\newcommand{\totalmassdetplus}[1]{\IfEqCase{#1}{{GW190930A}{10.5}{GW190929A}{38.6}{GW190924A}{5.7}{GW190915A}{8.4}{GW190910A}{10.4}{GW190909A}{95.3}{GW190828B}{6.4}{GW190828A}{6.9}{GW190814A}{1.1}{GW190803A}{14.1}{GW190731A}{14.3}{GW190728A}{5.3}{GW190727A}{11.9}{GW190720A}{5.0}{GW190719A}{24.4}{GW190708A}{2.5}{GW190707A}{1.8}{GW190706A}{23.3}{GW190701A}{16.4}{GW190630A}{4.2}{GW190620A}{20.1}{GW190602A}{23.2}{GW190527A}{53.7}{GW190521B}{4.8}{GW190521A}{39.8}{GW190519A}{16.7}{GW190517A}{9.6}{GW190514A}{17.8}{GW190513A}{12.7}{GW190512A}{3.9}{GW190503A}{11.2}{GW190426A}{3.7}{GW190425A}{0.3}{GW190424A}{14.4}{GW190421A}{15.3}{GW190413B}{17.9}{GW190413A}{17.8}{GW190412A}{4.5}{GW190408A}{3.5}}}
\newcommand{\thetajnminus}[1]{\IfEqCase{#1}{{GW190930A}{0.74}{GW190929A}{1.20}{GW190924A}{0.57}{GW190915A}{1.52}{GW190910A}{1.19}{GW190909A}{1.12}{GW190828B}{1.51}{GW190828A}{1.92}{GW190814A}{0.24}{GW190803A}{0.87}{GW190731A}{0.93}{GW190728A}{1.02}{GW190727A}{1.28}{GW190720A}{2.01}{GW190719A}{1.34}{GW190708A}{1.15}{GW190707A}{1.87}{GW190706A}{1.06}{GW190701A}{0.42}{GW190630A}{0.97}{GW190620A}{1.52}{GW190602A}{1.43}{GW190527A}{0.99}{GW190521B}{1.20}{GW190521A}{0.87}{GW190519A}{0.94}{GW190517A}{1.52}{GW190514A}{1.23}{GW190513A}{0.58}{GW190512A}{1.30}{GW190503A}{0.57}{GW190426A}{1.43}{GW190425A}{0.85}{GW190424A}{1.26}{GW190421A}{1.44}{GW190413B}{1.55}{GW190413A}{0.87}{GW190412A}{0.25}{GW190408A}{0.59}}}
\newcommand{\thetajnmed}[1]{\IfEqCase{#1}{{GW190930A}{0.94}{GW190929A}{1.70}{GW190924A}{0.74}{GW190915A}{2.03}{GW190910A}{1.62}{GW190909A}{1.45}{GW190828B}{1.83}{GW190828A}{2.24}{GW190814A}{0.86}{GW190803A}{1.12}{GW190731A}{1.19}{GW190728A}{1.23}{GW190727A}{1.55}{GW190720A}{2.48}{GW190719A}{1.61}{GW190708A}{1.37}{GW190707A}{2.15}{GW190706A}{1.37}{GW190701A}{0.58}{GW190630A}{1.22}{GW190620A}{1.94}{GW190602A}{1.71}{GW190527A}{1.26}{GW190521B}{1.48}{GW190521A}{1.15}{GW190519A}{1.58}{GW190517A}{2.26}{GW190514A}{1.50}{GW190513A}{0.79}{GW190512A}{1.61}{GW190503A}{2.41}{GW190426A}{1.70}{GW190425A}{1.08}{GW190424A}{1.52}{GW190421A}{1.74}{GW190413B}{1.90}{GW190413A}{1.15}{GW190412A}{0.72}{GW190408A}{0.79}}}
\newcommand{\thetajnplus}[1]{\IfEqCase{#1}{{GW190930A}{1.90}{GW190929A}{0.97}{GW190924A}{2.04}{GW190915A}{0.78}{GW190910A}{1.13}{GW190909A}{1.34}{GW190828B}{1.02}{GW190828A}{0.70}{GW190814A}{1.48}{GW190803A}{1.73}{GW190731A}{1.66}{GW190728A}{1.66}{GW190727A}{1.31}{GW190720A}{0.50}{GW190719A}{1.25}{GW190708A}{1.54}{GW190707A}{0.77}{GW190706A}{1.43}{GW190701A}{0.55}{GW190630A}{1.60}{GW190620A}{0.90}{GW190602A}{1.17}{GW190527A}{1.55}{GW190521B}{1.37}{GW190521A}{1.65}{GW190519A}{0.95}{GW190517A}{0.64}{GW190514A}{1.33}{GW190513A}{2.02}{GW190512A}{1.22}{GW190503A}{0.52}{GW190426A}{1.19}{GW190425A}{1.77}{GW190424A}{1.36}{GW190421A}{1.13}{GW190413B}{0.95}{GW190413A}{1.64}{GW190412A}{0.44}{GW190408A}{2.03}}}
\newcommand{\redshiftminus}[1]{\IfEqCase{#1}{{GW190930A}{0.06}{GW190929A}{0.17}{GW190924A}{0.04}{GW190915A}{0.10}{GW190910A}{0.10}{GW190909A}{0.33}{GW190828B}{0.10}{GW190828A}{0.15}{GW190814A}{0.010}{GW190803A}{0.24}{GW190731A}{0.26}{GW190728A}{0.07}{GW190727A}{0.22}{GW190720A}{0.06}{GW190719A}{0.29}{GW190708A}{0.07}{GW190707A}{0.07}{GW190706A}{0.27}{GW190701A}{0.12}{GW190630A}{0.07}{GW190620A}{0.20}{GW190602A}{0.17}{GW190527A}{0.20}{GW190521B}{0.10}{GW190521A}{0.28}{GW190519A}{0.14}{GW190517A}{0.14}{GW190514A}{0.31}{GW190513A}{0.13}{GW190512A}{0.10}{GW190503A}{0.11}{GW190426A}{0.03}{GW190425A}{0.02}{GW190424A}{0.19}{GW190421A}{0.21}{GW190413B}{0.30}{GW190413A}{0.24}{GW190412A}{0.03}{GW190408A}{0.10}}}
\newcommand{\redshiftmed}[1]{\IfEqCase{#1}{{GW190930A}{0.15}{GW190929A}{0.38}{GW190924A}{0.12}{GW190915A}{0.30}{GW190910A}{0.28}{GW190909A}{0.62}{GW190828B}{0.30}{GW190828A}{0.38}{GW190814A}{0.05}{GW190803A}{0.55}{GW190731A}{0.55}{GW190728A}{0.18}{GW190727A}{0.55}{GW190720A}{0.16}{GW190719A}{0.64}{GW190708A}{0.18}{GW190707A}{0.16}{GW190706A}{0.71}{GW190701A}{0.37}{GW190630A}{0.18}{GW190620A}{0.49}{GW190602A}{0.47}{GW190527A}{0.44}{GW190521B}{0.24}{GW190521A}{0.64}{GW190519A}{0.44}{GW190517A}{0.34}{GW190514A}{0.67}{GW190513A}{0.37}{GW190512A}{0.27}{GW190503A}{0.27}{GW190426A}{0.08}{GW190425A}{0.03}{GW190424A}{0.39}{GW190421A}{0.49}{GW190413B}{0.71}{GW190413A}{0.59}{GW190412A}{0.15}{GW190408A}{0.29}}}
\newcommand{\redshiftplus}[1]{\IfEqCase{#1}{{GW190930A}{0.06}{GW190929A}{0.49}{GW190924A}{0.04}{GW190915A}{0.11}{GW190910A}{0.16}{GW190909A}{0.41}{GW190828B}{0.10}{GW190828A}{0.10}{GW190814A}{0.009}{GW190803A}{0.26}{GW190731A}{0.31}{GW190728A}{0.05}{GW190727A}{0.21}{GW190720A}{0.12}{GW190719A}{0.33}{GW190708A}{0.06}{GW190707A}{0.07}{GW190706A}{0.32}{GW190701A}{0.11}{GW190630A}{0.10}{GW190620A}{0.23}{GW190602A}{0.25}{GW190527A}{0.34}{GW190521B}{0.07}{GW190521A}{0.28}{GW190519A}{0.25}{GW190517A}{0.24}{GW190514A}{0.33}{GW190513A}{0.13}{GW190512A}{0.09}{GW190503A}{0.11}{GW190426A}{0.04}{GW190425A}{0.01}{GW190424A}{0.23}{GW190421A}{0.19}{GW190413B}{0.31}{GW190413A}{0.29}{GW190412A}{0.03}{GW190408A}{0.06}}}
\newcommand{\iotaminus}[1]{\IfEqCase{#1}{{GW190930A}{0.73}{GW190929A}{1.24}{GW190924A}{0.56}{GW190915A}{1.73}{GW190910A}{1.19}{GW190909A}{1.11}{GW190828B}{1.55}{GW190828A}{1.97}{GW190814A}{0.27}{GW190803A}{0.85}{GW190731A}{0.90}{GW190728A}{1.01}{GW190727A}{1.26}{GW190720A}{2.01}{GW190719A}{1.34}{GW190708A}{1.14}{GW190707A}{1.88}{GW190706A}{1.04}{GW190701A}{0.43}{GW190630A}{0.98}{GW190620A}{1.69}{GW190602A}{1.53}{GW190527A}{0.97}{GW190521B}{1.19}{GW190521A}{0.87}{GW190519A}{0.93}{GW190517A}{1.40}{GW190514A}{1.22}{GW190513A}{0.58}{GW190512A}{1.29}{GW190503A}{0.57}{GW190426A}{1.43}{GW190425A}{0.85}{GW190424A}{1.26}{GW190421A}{1.48}{GW190413B}{1.61}{GW190413A}{0.86}{GW190412A}{0.35}{GW190408A}{0.59}}}
\newcommand{\iotamed}[1]{\IfEqCase{#1}{{GW190930A}{0.94}{GW190929A}{1.70}{GW190924A}{0.74}{GW190915A}{2.13}{GW190910A}{1.62}{GW190909A}{1.47}{GW190828B}{1.84}{GW190828A}{2.27}{GW190814A}{0.85}{GW190803A}{1.09}{GW190731A}{1.15}{GW190728A}{1.23}{GW190727A}{1.53}{GW190720A}{2.47}{GW190719A}{1.61}{GW190708A}{1.37}{GW190707A}{2.15}{GW190706A}{1.33}{GW190701A}{0.59}{GW190630A}{1.24}{GW190620A}{2.03}{GW190602A}{1.79}{GW190527A}{1.24}{GW190521B}{1.46}{GW190521A}{1.15}{GW190519A}{1.59}{GW190517A}{2.21}{GW190514A}{1.48}{GW190513A}{0.79}{GW190512A}{1.61}{GW190503A}{2.43}{GW190426A}{1.70}{GW190425A}{1.09}{GW190424A}{1.52}{GW190421A}{1.76}{GW190413B}{1.97}{GW190413A}{1.13}{GW190412A}{0.84}{GW190408A}{0.78}}}
\newcommand{\iotaplus}[1]{\IfEqCase{#1}{{GW190930A}{1.89}{GW190929A}{1.03}{GW190924A}{2.04}{GW190915A}{0.74}{GW190910A}{1.14}{GW190909A}{1.30}{GW190828B}{1.01}{GW190828A}{0.67}{GW190814A}{1.51}{GW190803A}{1.76}{GW190731A}{1.71}{GW190728A}{1.66}{GW190727A}{1.34}{GW190720A}{0.49}{GW190719A}{1.27}{GW190708A}{1.55}{GW190707A}{0.78}{GW190706A}{1.49}{GW190701A}{0.57}{GW190630A}{1.55}{GW190620A}{0.85}{GW190602A}{1.11}{GW190527A}{1.57}{GW190521B}{1.40}{GW190521A}{1.65}{GW190519A}{0.92}{GW190517A}{0.66}{GW190514A}{1.37}{GW190513A}{2.03}{GW190512A}{1.22}{GW190503A}{0.51}{GW190426A}{1.19}{GW190425A}{1.77}{GW190424A}{1.38}{GW190421A}{1.11}{GW190413B}{0.88}{GW190413A}{1.66}{GW190412A}{0.38}{GW190408A}{2.06}}}
\newcommand{\spinonexminus}[1]{\IfEqCase{#1}{{GW190930A}{0.44}{GW190929A}{0.71}{GW190924A}{0.35}{GW190915A}{0.67}{GW190910A}{0.51}{GW190909A}{0.63}{GW190828B}{0.42}{GW190828A}{0.52}{GW190814A}{0.04}{GW190803A}{0.57}{GW190731A}{0.56}{GW190728A}{0.37}{GW190727A}{0.58}{GW190720A}{0.43}{GW190719A}{0.55}{GW190708A}{0.42}{GW190707A}{0.39}{GW190706A}{0.53}{GW190701A}{0.52}{GW190630A}{0.36}{GW190620A}{0.55}{GW190602A}{0.52}{GW190527A}{0.59}{GW190521B}{0.44}{GW190521A}{0.74}{GW190519A}{0.54}{GW190517A}{0.57}{GW190514A}{0.57}{GW190513A}{0.42}{GW190512A}{0.30}{GW190503A}{0.49}{GW190426A}{0.00}{GW190425A}{0.50}{GW190424A}{0.64}{GW190421A}{0.59}{GW190413B}{0.68}{GW190413A}{0.53}{GW190412A}{0.33}{GW190408A}{0.47}}}
\newcommand{\spinonexmed}[1]{\IfEqCase{#1}{{GW190930A}{0.002}{GW190929A}{0.007}{GW190924A}{0.0001}{GW190915A}{0.00}{GW190910A}{0.00}{GW190909A}{0.002}{GW190828B}{0.00}{GW190828A}{0.00}{GW190814A}{0.00}{GW190803A}{0.00}{GW190731A}{0.0007}{GW190728A}{0.0008}{GW190727A}{0.002}{GW190720A}{0.003}{GW190719A}{0.004}{GW190708A}{0.004}{GW190707A}{0.003}{GW190706A}{0.00}{GW190701A}{0.00}{GW190630A}{0.00}{GW190620A}{0.00}{GW190602A}{0.00}{GW190527A}{0.002}{GW190521B}{0.001}{GW190521A}{-0.02}{GW190519A}{0.004}{GW190517A}{0.0009}{GW190514A}{0.00007}{GW190513A}{0.0006}{GW190512A}{0.0010}{GW190503A}{0.00}{GW190426A}{0.00}{GW190425A}{0.00}{GW190424A}{0.00}{GW190421A}{0.00005}{GW190413B}{0.00}{GW190413A}{0.0005}{GW190412A}{-0.02}{GW190408A}{0.003}}}
\newcommand{\spinonexplus}[1]{\IfEqCase{#1}{{GW190930A}{0.47}{GW190929A}{0.69}{GW190924A}{0.36}{GW190915A}{0.66}{GW190910A}{0.51}{GW190909A}{0.68}{GW190828B}{0.43}{GW190828A}{0.51}{GW190814A}{0.04}{GW190803A}{0.55}{GW190731A}{0.51}{GW190728A}{0.40}{GW190727A}{0.58}{GW190720A}{0.45}{GW190719A}{0.57}{GW190708A}{0.47}{GW190707A}{0.42}{GW190706A}{0.53}{GW190701A}{0.52}{GW190630A}{0.36}{GW190620A}{0.53}{GW190602A}{0.53}{GW190527A}{0.59}{GW190521B}{0.45}{GW190521A}{0.75}{GW190519A}{0.55}{GW190517A}{0.57}{GW190514A}{0.60}{GW190513A}{0.43}{GW190512A}{0.28}{GW190503A}{0.50}{GW190426A}{0.00}{GW190425A}{0.47}{GW190424A}{0.63}{GW190421A}{0.60}{GW190413B}{0.71}{GW190413A}{0.53}{GW190412A}{0.39}{GW190408A}{0.49}}}
\newcommand{\chirpmassdetminus}[1]{\IfEqCase{#1}{{GW190930A}{0.2}{GW190929A}{15.4}{GW190924A}{0.03}{GW190915A}{3.9}{GW190910A}{3.6}{GW190909A}{12.4}{GW190828B}{0.7}{GW190828A}{2.8}{GW190814A}{0.02}{GW190803A}{6.1}{GW190731A}{8.2}{GW190728A}{0.08}{GW190727A}{5.7}{GW190720A}{0.1}{GW190719A}{6.6}{GW190708A}{0.2}{GW190707A}{0.09}{GW190706A}{17.5}{GW190701A}{8.1}{GW190630A}{1.5}{GW190620A}{11.2}{GW190602A}{13.7}{GW190527A}{5.5}{GW190521B}{3.0}{GW190521A}{17.6}{GW190519A}{10.3}{GW190517A}{3.4}{GW190514A}{7.7}{GW190513A}{2.5}{GW190512A}{0.8}{GW190503A}{6.0}{GW190426A}{0.01}{GW190425A}{0.0006}{GW190424A}{4.8}{GW190421A}{6.0}{GW190413B}{9.8}{GW190413A}{6.6}{GW190412A}{0.2}{GW190408A}{1.7}}}
\newcommand{\chirpmassdetmed}[1]{\IfEqCase{#1}{{GW190930A}{9.8}{GW190929A}{52.2}{GW190924A}{6.44}{GW190915A}{33.1}{GW190910A}{43.9}{GW190909A}{49.8}{GW190828B}{17.4}{GW190828A}{34.5}{GW190814A}{6.41}{GW190803A}{42.7}{GW190731A}{46.6}{GW190728A}{10.1}{GW190727A}{44.7}{GW190720A}{10.4}{GW190719A}{38.7}{GW190708A}{15.5}{GW190707A}{9.89}{GW190706A}{75.1}{GW190701A}{55.5}{GW190630A}{29.4}{GW190620A}{57.5}{GW190602A}{72.9}{GW190527A}{34.9}{GW190521B}{39.8}{GW190521A}{114.8}{GW190519A}{65.1}{GW190517A}{35.9}{GW190514A}{48.1}{GW190513A}{29.5}{GW190512A}{18.6}{GW190503A}{38.6}{GW190426A}{2.60}{GW190425A}{1.49}{GW190424A}{43.4}{GW190421A}{46.6}{GW190413B}{57.0}{GW190413A}{39.4}{GW190412A}{15.2}{GW190408A}{23.7}}}
\newcommand{\chirpmassdetplus}[1]{\IfEqCase{#1}{{GW190930A}{0.2}{GW190929A}{19.9}{GW190924A}{0.04}{GW190915A}{3.3}{GW190910A}{4.6}{GW190909A}{32.2}{GW190828B}{0.6}{GW190828A}{2.9}{GW190814A}{0.02}{GW190803A}{6.3}{GW190731A}{6.8}{GW190728A}{0.09}{GW190727A}{5.3}{GW190720A}{0.2}{GW190719A}{9.2}{GW190708A}{0.3}{GW190707A}{0.1}{GW190706A}{11.0}{GW190701A}{7.3}{GW190630A}{1.6}{GW190620A}{9.0}{GW190602A}{10.8}{GW190527A}{21.7}{GW190521B}{2.2}{GW190521A}{15.2}{GW190519A}{7.7}{GW190517A}{4.0}{GW190514A}{7.5}{GW190513A}{5.6}{GW190512A}{0.9}{GW190503A}{5.3}{GW190426A}{0.01}{GW190425A}{0.0008}{GW190424A}{6.0}{GW190421A}{6.6}{GW190413B}{8.6}{GW190413A}{7.7}{GW190412A}{0.2}{GW190408A}{1.4}}}
\newcommand{\cosiotaminus}[1]{\IfEqCase{#1}{{GW190930A}{1.54}{GW190929A}{0.79}{GW190924A}{1.67}{GW190915A}{0.43}{GW190910A}{0.88}{GW190909A}{1.03}{GW190828B}{0.69}{GW190828A}{0.33}{GW190814A}{1.37}{GW190803A}{1.42}{GW190731A}{1.37}{GW190728A}{1.30}{GW190727A}{1.00}{GW190720A}{0.20}{GW190719A}{0.93}{GW190708A}{1.18}{GW190707A}{0.43}{GW190706A}{1.19}{GW190701A}{0.44}{GW190630A}{1.26}{GW190620A}{0.52}{GW190602A}{0.75}{GW190527A}{1.27}{GW190521B}{1.07}{GW190521A}{1.35}{GW190519A}{0.79}{GW190517A}{0.37}{GW190514A}{1.05}{GW190513A}{1.65}{GW190512A}{0.91}{GW190503A}{0.22}{GW190426A}{0.84}{GW190425A}{1.42}{GW190424A}{1.02}{GW190421A}{0.78}{GW190413B}{0.57}{GW190413A}{1.37}{GW190412A}{0.32}{GW190408A}{1.67}}}
\newcommand{\cosiotamed}[1]{\IfEqCase{#1}{{GW190930A}{0.59}{GW190929A}{-0.13}{GW190924A}{0.74}{GW190915A}{-0.53}{GW190910A}{-0.05}{GW190909A}{0.10}{GW190828B}{-0.27}{GW190828A}{-0.65}{GW190814A}{0.66}{GW190803A}{0.47}{GW190731A}{0.41}{GW190728A}{0.34}{GW190727A}{0.04}{GW190720A}{-0.78}{GW190719A}{-0.04}{GW190708A}{0.20}{GW190707A}{-0.55}{GW190706A}{0.24}{GW190701A}{0.83}{GW190630A}{0.32}{GW190620A}{-0.45}{GW190602A}{-0.22}{GW190527A}{0.32}{GW190521B}{0.11}{GW190521A}{0.41}{GW190519A}{-0.02}{GW190517A}{-0.59}{GW190514A}{0.09}{GW190513A}{0.71}{GW190512A}{-0.04}{GW190503A}{-0.76}{GW190426A}{-0.13}{GW190425A}{0.46}{GW190424A}{0.05}{GW190421A}{-0.19}{GW190413B}{-0.39}{GW190413A}{0.42}{GW190412A}{0.67}{GW190408A}{0.71}}}
\newcommand{\cosiotaplus}[1]{\IfEqCase{#1}{{GW190930A}{0.39}{GW190929A}{1.03}{GW190924A}{0.24}{GW190915A}{1.45}{GW190910A}{0.96}{GW190909A}{0.84}{GW190828B}{1.23}{GW190828A}{1.60}{GW190814A}{0.17}{GW190803A}{0.51}{GW190731A}{0.56}{GW190728A}{0.64}{GW190727A}{0.93}{GW190720A}{1.68}{GW190719A}{1.00}{GW190708A}{0.77}{GW190707A}{1.51}{GW190706A}{0.72}{GW190701A}{0.16}{GW190630A}{0.65}{GW190620A}{1.39}{GW190602A}{1.18}{GW190527A}{0.64}{GW190521B}{0.85}{GW190521A}{0.55}{GW190519A}{0.81}{GW190517A}{1.29}{GW190514A}{0.88}{GW190513A}{0.27}{GW190512A}{0.99}{GW190503A}{0.47}{GW190426A}{1.09}{GW190425A}{0.51}{GW190424A}{0.91}{GW190421A}{1.15}{GW190413B}{1.32}{GW190413A}{0.54}{GW190412A}{0.22}{GW190408A}{0.27}}}
\newcommand{\comovingdistminus}[1]{\IfEqCase{#1}{{GW190930A}{257}{GW190929A}{650}{GW190924A}{184}{GW190915A}{401}{GW190910A}{396}{GW190909A}{1128}{GW190828B}{395}{GW190828A}{567}{GW190814A}{41}{GW190803A}{824}{GW190731A}{906}{GW190728A}{285}{GW190727A}{774}{GW190720A}{254}{GW190719A}{965}{GW190708A}{307}{GW190707A}{295}{GW190706A}{859}{GW190701A}{438}{GW190630A}{289}{GW190620A}{723}{GW190602A}{621}{GW190527A}{725}{GW190521B}{409}{GW190521A}{943}{GW190519A}{518}{GW190517A}{538}{GW190514A}{1034}{GW190513A}{484}{GW190512A}{378}{GW190503A}{431}{GW190426A}{143}{GW190425A}{67}{GW190424A}{717}{GW190421A}{759}{GW190413B}{957}{GW190413A}{829}{GW190412A}{134}{GW190408A}{399}}}
\newcommand{\comovingdistmed}[1]{\IfEqCase{#1}{{GW190930A}{658}{GW190929A}{1541}{GW190924A}{507}{GW190915A}{1244}{GW190910A}{1139}{GW190909A}{2329}{GW190828B}{1228}{GW190828A}{1539}{GW190814A}{229}{GW190803A}{2108}{GW190731A}{2120}{GW190728A}{742}{GW190727A}{2119}{GW190720A}{678}{GW190719A}{2399}{GW190708A}{748}{GW190707A}{667}{GW190706A}{2594}{GW190701A}{1498}{GW190630A}{752}{GW190620A}{1893}{GW190602A}{1832}{GW190527A}{1729}{GW190521B}{1003}{GW190521A}{2390}{GW190519A}{1754}{GW190517A}{1389}{GW190514A}{2475}{GW190513A}{1501}{GW190512A}{1120}{GW190503A}{1133}{GW190426A}{344}{GW190425A}{151}{GW190424A}{1578}{GW190421A}{1923}{GW190413B}{2603}{GW190413A}{2232}{GW190412A}{640}{GW190408A}{1198}}}
\newcommand{\comovingdistplus}[1]{\IfEqCase{#1}{{GW190930A}{258}{GW190929A}{1537}{GW190924A}{174}{GW190915A}{403}{GW190910A}{588}{GW190909A}{1139}{GW190828B}{359}{GW190828A}{342}{GW190814A}{37}{GW190803A}{778}{GW190731A}{925}{GW190728A}{182}{GW190727A}{627}{GW190720A}{472}{GW190719A}{915}{GW190708A}{235}{GW190707A}{272}{GW190706A}{867}{GW190701A}{400}{GW190630A}{379}{GW190620A}{725}{GW190602A}{784}{GW190527A}{1068}{GW190521B}{253}{GW190521A}{795}{GW190519A}{816}{GW190517A}{816}{GW190514A}{915}{GW190513A}{453}{GW190512A}{330}{GW190503A}{406}{GW190426A}{154}{GW190425A}{64}{GW190424A}{755}{GW190421A}{598}{GW190413B}{831}{GW190413A}{856}{GW190412A}{105}{GW190408A}{240}}}
\newcommand{\spintwoyminus}[1]{\IfEqCase{#1}{{GW190930A}{0.54}{GW190929A}{0.57}{GW190924A}{0.48}{GW190915A}{0.61}{GW190910A}{0.52}{GW190909A}{0.61}{GW190828B}{0.54}{GW190828A}{0.50}{GW190814A}{0.61}{GW190803A}{0.57}{GW190731A}{0.58}{GW190728A}{0.51}{GW190727A}{0.59}{GW190720A}{0.55}{GW190719A}{0.55}{GW190708A}{0.44}{GW190707A}{0.47}{GW190706A}{0.51}{GW190701A}{0.58}{GW190630A}{0.48}{GW190620A}{0.56}{GW190602A}{0.60}{GW190527A}{0.61}{GW190521B}{0.53}{GW190521A}{0.68}{GW190519A}{0.55}{GW190517A}{0.54}{GW190514A}{0.60}{GW190513A}{0.54}{GW190512A}{0.51}{GW190503A}{0.57}{GW190426A}{0.00}{GW190425A}{0.48}{GW190424A}{0.60}{GW190421A}{0.59}{GW190413B}{0.59}{GW190413A}{0.57}{GW190412A}{0.57}{GW190408A}{0.52}}}
\newcommand{\spintwoymed}[1]{\IfEqCase{#1}{{GW190930A}{0.00}{GW190929A}{0.0008}{GW190924A}{0.00}{GW190915A}{0.00}{GW190910A}{0.00}{GW190909A}{0.003}{GW190828B}{0.00}{GW190828A}{0.0004}{GW190814A}{0.005}{GW190803A}{0.0002}{GW190731A}{0.00}{GW190728A}{0.00}{GW190727A}{0.0009}{GW190720A}{0.00}{GW190719A}{0.003}{GW190708A}{0.003}{GW190707A}{0.002}{GW190706A}{0.001}{GW190701A}{-0.01}{GW190630A}{0.0006}{GW190620A}{0.00010}{GW190602A}{0.00}{GW190527A}{0.004}{GW190521B}{0.00}{GW190521A}{0.00}{GW190519A}{0.001}{GW190517A}{0.00}{GW190514A}{0.00}{GW190513A}{0.00}{GW190512A}{0.002}{GW190503A}{0.00}{GW190426A}{0.00}{GW190425A}{0.00}{GW190424A}{0.00}{GW190421A}{0.00}{GW190413B}{-0.01}{GW190413A}{0.00}{GW190412A}{0.004}{GW190408A}{0.0008}}}
\newcommand{\spintwoyplus}[1]{\IfEqCase{#1}{{GW190930A}{0.51}{GW190929A}{0.59}{GW190924A}{0.49}{GW190915A}{0.60}{GW190910A}{0.52}{GW190909A}{0.57}{GW190828B}{0.54}{GW190828A}{0.50}{GW190814A}{0.61}{GW190803A}{0.58}{GW190731A}{0.55}{GW190728A}{0.50}{GW190727A}{0.56}{GW190720A}{0.56}{GW190719A}{0.55}{GW190708A}{0.46}{GW190707A}{0.48}{GW190706A}{0.53}{GW190701A}{0.55}{GW190630A}{0.49}{GW190620A}{0.54}{GW190602A}{0.58}{GW190527A}{0.59}{GW190521B}{0.52}{GW190521A}{0.69}{GW190519A}{0.55}{GW190517A}{0.55}{GW190514A}{0.60}{GW190513A}{0.55}{GW190512A}{0.55}{GW190503A}{0.58}{GW190426A}{0.00}{GW190425A}{0.48}{GW190424A}{0.60}{GW190421A}{0.59}{GW190413B}{0.62}{GW190413A}{0.56}{GW190412A}{0.58}{GW190408A}{0.53}}}
\newcommand{\tiltoneminus}[1]{\IfEqCase{#1}{{GW190930A}{0.79}{GW190929A}{0.72}{GW190924A}{1.05}{GW190915A}{0.85}{GW190910A}{0.97}{GW190909A}{1.14}{GW190828B}{0.83}{GW190828A}{0.72}{GW190814A}{1.08}{GW190803A}{1.06}{GW190731A}{0.96}{GW190728A}{0.77}{GW190727A}{0.85}{GW190720A}{0.72}{GW190719A}{0.59}{GW190708A}{0.98}{GW190707A}{1.08}{GW190706A}{0.61}{GW190701A}{1.12}{GW190630A}{0.87}{GW190620A}{0.59}{GW190602A}{0.94}{GW190527A}{0.89}{GW190521B}{0.90}{GW190521A}{0.93}{GW190519A}{0.60}{GW190517A}{0.42}{GW190514A}{1.15}{GW190513A}{0.81}{GW190512A}{1.03}{GW190503A}{1.10}{GW190426A}{0.00}{GW190425A}{0.80}{GW190424A}{0.80}{GW190421A}{1.06}{GW190413B}{0.93}{GW190413A}{1.06}{GW190412A}{0.35}{GW190408A}{1.06}}}
\newcommand{\tiltonemed}[1]{\IfEqCase{#1}{{GW190930A}{1.08}{GW190929A}{1.55}{GW190924A}{1.38}{GW190915A}{1.51}{GW190910A}{1.49}{GW190909A}{1.67}{GW190828B}{1.31}{GW190828A}{1.04}{GW190814A}{1.56}{GW190803A}{1.64}{GW190731A}{1.41}{GW190728A}{1.06}{GW190727A}{1.26}{GW190720A}{1.00}{GW190719A}{0.85}{GW190708A}{1.50}{GW190707A}{1.77}{GW190706A}{0.85}{GW190701A}{1.79}{GW190630A}{1.28}{GW190620A}{0.82}{GW190602A}{1.39}{GW190527A}{1.29}{GW190521B}{1.40}{GW190521A}{1.52}{GW190519A}{0.87}{GW190517A}{0.59}{GW190514A}{2.03}{GW190513A}{1.14}{GW190512A}{1.50}{GW190503A}{1.73}{GW190426A}{0.00}{GW190425A}{1.31}{GW190424A}{1.20}{GW190421A}{1.72}{GW190413B}{1.63}{GW190413A}{1.56}{GW190412A}{0.80}{GW190408A}{1.74}}}
\newcommand{\tiltoneplus}[1]{\IfEqCase{#1}{{GW190930A}{1.14}{GW190929A}{0.97}{GW190924A}{1.09}{GW190915A}{0.92}{GW190910A}{1.09}{GW190909A}{0.99}{GW190828B}{1.08}{GW190828A}{1.15}{GW190814A}{1.11}{GW190803A}{0.99}{GW190731A}{1.13}{GW190728A}{1.20}{GW190727A}{1.11}{GW190720A}{1.01}{GW190719A}{1.09}{GW190708A}{1.00}{GW190707A}{0.86}{GW190706A}{1.06}{GW190701A}{0.95}{GW190630A}{1.11}{GW190620A}{0.91}{GW190602A}{1.12}{GW190527A}{1.11}{GW190521B}{1.09}{GW190521A}{1.04}{GW190519A}{0.79}{GW190517A}{0.50}{GW190514A}{0.78}{GW190513A}{1.18}{GW190512A}{1.11}{GW190503A}{0.97}{GW190426A}{3.14}{GW190425A}{0.66}{GW190424A}{1.06}{GW190421A}{0.97}{GW190413B}{0.91}{GW190413A}{1.11}{GW190412A}{0.54}{GW190408A}{0.94}}}
\newcommand{\spintwozminus}[1]{\IfEqCase{#1}{{GW190930A}{0.41}{GW190929A}{0.55}{GW190924A}{0.36}{GW190915A}{0.51}{GW190910A}{0.36}{GW190909A}{0.62}{GW190828B}{0.42}{GW190828A}{0.35}{GW190814A}{0.53}{GW190803A}{0.55}{GW190731A}{0.47}{GW190728A}{0.38}{GW190727A}{0.45}{GW190720A}{0.53}{GW190719A}{0.46}{GW190708A}{0.33}{GW190707A}{0.36}{GW190706A}{0.45}{GW190701A}{0.54}{GW190630A}{0.31}{GW190620A}{0.47}{GW190602A}{0.46}{GW190527A}{0.49}{GW190521B}{0.32}{GW190521A}{0.54}{GW190519A}{0.43}{GW190517A}{0.46}{GW190514A}{0.59}{GW190513A}{0.41}{GW190512A}{0.33}{GW190503A}{0.51}{GW190426A}{0.03}{GW190425A}{0.18}{GW190424A}{0.44}{GW190421A}{0.53}{GW190413B}{0.55}{GW190413A}{0.54}{GW190412A}{0.44}{GW190408A}{0.37}}}
\newcommand{\spintwozmed}[1]{\IfEqCase{#1}{{GW190930A}{0.08}{GW190929A}{0.008}{GW190924A}{0.02}{GW190915A}{0.003}{GW190910A}{0.006}{GW190909A}{-0.04}{GW190828B}{0.06}{GW190828A}{0.11}{GW190814A}{-0.01}{GW190803A}{-0.01}{GW190731A}{0.03}{GW190728A}{0.11}{GW190727A}{0.04}{GW190720A}{0.11}{GW190719A}{0.16}{GW190708A}{0.03}{GW190707A}{-0.03}{GW190706A}{0.11}{GW190701A}{-0.04}{GW190630A}{0.09}{GW190620A}{0.21}{GW190602A}{0.05}{GW190527A}{0.04}{GW190521B}{0.09}{GW190521A}{-0.01}{GW190519A}{0.21}{GW190517A}{0.29}{GW190514A}{-0.13}{GW190513A}{0.06}{GW190512A}{0.03}{GW190503A}{0.00}{GW190426A}{0.00}{GW190425A}{0.03}{GW190424A}{0.06}{GW190421A}{-0.04}{GW190413B}{-0.03}{GW190413A}{-0.02}{GW190412A}{0.07}{GW190408A}{0.00}}}
\newcommand{\spintwozplus}[1]{\IfEqCase{#1}{{GW190930A}{0.50}{GW190929A}{0.57}{GW190924A}{0.48}{GW190915A}{0.47}{GW190910A}{0.39}{GW190909A}{0.51}{GW190828B}{0.54}{GW190828A}{0.48}{GW190814A}{0.52}{GW190803A}{0.47}{GW190731A}{0.54}{GW190728A}{0.48}{GW190727A}{0.53}{GW190720A}{0.54}{GW190719A}{0.61}{GW190708A}{0.39}{GW190707A}{0.34}{GW190706A}{0.62}{GW190701A}{0.42}{GW190630A}{0.44}{GW190620A}{0.57}{GW190602A}{0.56}{GW190527A}{0.60}{GW190521B}{0.36}{GW190521A}{0.53}{GW190519A}{0.56}{GW190517A}{0.53}{GW190514A}{0.44}{GW190513A}{0.54}{GW190512A}{0.45}{GW190503A}{0.46}{GW190426A}{0.03}{GW190425A}{0.30}{GW190424A}{0.52}{GW190421A}{0.41}{GW190413B}{0.49}{GW190413A}{0.48}{GW190412A}{0.57}{GW190408A}{0.38}}}
\newcommand{\massonesourceminus}[1]{\IfEqCase{#1}{{GW190930A}{2.3}{GW190929A}{33.2}{GW190924A}{2.0}{GW190915A}{6.4}{GW190910A}{6.1}{GW190909A}{13.3}{GW190828B}{7.2}{GW190828A}{4.0}{GW190814A}{1.0}{GW190803A}{7.0}{GW190731A}{9.0}{GW190728A}{2.2}{GW190727A}{6.2}{GW190720A}{3.0}{GW190719A}{10.3}{GW190708A}{2.3}{GW190707A}{1.7}{GW190706A}{16.2}{GW190701A}{8.0}{GW190630A}{5.6}{GW190620A}{12.7}{GW190602A}{13.0}{GW190527A}{9.0}{GW190521B}{4.8}{GW190521A}{18.9}{GW190519A}{12.0}{GW190517A}{7.6}{GW190514A}{8.2}{GW190513A}{9.2}{GW190512A}{5.8}{GW190503A}{8.1}{GW190426A}{2.3}{GW190425A}{0.3}{GW190424A}{7.3}{GW190421A}{6.9}{GW190413B}{10.7}{GW190413A}{8.1}{GW190412A}{5.1}{GW190408A}{3.4}}}
\newcommand{\massonesourcemed}[1]{\IfEqCase{#1}{{GW190930A}{12.3}{GW190929A}{80.8}{GW190924A}{8.9}{GW190915A}{35.3}{GW190910A}{43.9}{GW190909A}{45.8}{GW190828B}{24.1}{GW190828A}{32.1}{GW190814A}{23.2}{GW190803A}{37.3}{GW190731A}{41.5}{GW190728A}{12.3}{GW190727A}{38.0}{GW190720A}{13.4}{GW190719A}{36.5}{GW190708A}{17.6}{GW190707A}{11.6}{GW190706A}{67.0}{GW190701A}{53.9}{GW190630A}{35.1}{GW190620A}{57.1}{GW190602A}{69.1}{GW190527A}{36.5}{GW190521B}{42.2}{GW190521A}{95.3}{GW190519A}{66.0}{GW190517A}{37.4}{GW190514A}{39.0}{GW190513A}{35.7}{GW190512A}{23.3}{GW190503A}{43.3}{GW190426A}{5.7}{GW190425A}{2.0}{GW190424A}{40.5}{GW190421A}{41.3}{GW190413B}{47.5}{GW190413A}{34.7}{GW190412A}{30.1}{GW190408A}{24.6}}}
\newcommand{\massonesourceplus}[1]{\IfEqCase{#1}{{GW190930A}{12.4}{GW190929A}{33.0}{GW190924A}{7.0}{GW190915A}{9.5}{GW190910A}{7.6}{GW190909A}{52.7}{GW190828B}{7.0}{GW190828A}{5.8}{GW190814A}{1.1}{GW190803A}{10.6}{GW190731A}{12.2}{GW190728A}{7.2}{GW190727A}{9.5}{GW190720A}{6.7}{GW190719A}{18.0}{GW190708A}{4.7}{GW190707A}{3.3}{GW190706A}{14.6}{GW190701A}{11.8}{GW190630A}{6.9}{GW190620A}{16.0}{GW190602A}{15.7}{GW190527A}{16.4}{GW190521B}{5.9}{GW190521A}{28.7}{GW190519A}{10.7}{GW190517A}{11.7}{GW190514A}{14.7}{GW190513A}{9.5}{GW190512A}{5.3}{GW190503A}{9.2}{GW190426A}{3.9}{GW190425A}{0.6}{GW190424A}{11.1}{GW190421A}{10.4}{GW190413B}{13.5}{GW190413A}{12.6}{GW190412A}{4.7}{GW190408A}{5.1}}}
\newcommand{\geocenttimeminus}[1]{\IfEqCase{#1}{{GW190930A}{0.02}{GW190929A}{0.02}{GW190924A}{0.008}{GW190915A}{0.003}{GW190910A}{0.0}{GW190909A}{0.0}{GW190828B}{0.0}{GW190828A}{0.0}{GW190814A}{0.0009}{GW190803A}{0.0}{GW190731A}{0.0}{GW190728A}{0.03}{GW190727A}{0.0}{GW190720A}{0.01}{GW190719A}{0.0}{GW190708A}{0.0}{GW190707A}{0.03}{GW190706A}{0.0000002}{GW190701A}{0.0}{GW190630A}{0.0000002}{GW190620A}{0.0000002}{GW190602A}{0.0}{GW190527A}{0.0}{GW190521B}{0.0}{GW190521A}{0.04}{GW190519A}{0.0}{GW190517A}{0.0}{GW190514A}{0.0}{GW190513A}{0.010}{GW190512A}{0.0}{GW190503A}{0.0}{GW190426A}{0.03}{GW190425A}{0.009}{GW190424A}{0.0}{GW190421A}{0.0}{GW190413B}{0.0}{GW190413A}{0.0}{GW190412A}{0.001}{GW190408A}{0.0}}}
\newcommand{\geocenttimemed}[1]{\IfEqCase{#1}{{GW190930A}{1253885759.2}{GW190929A}{1253755327.5}{GW190924A}{1253326744.8}{GW190915A}{1252627040.7}{GW190910A}{1252150105.3}{GW190909A}{1252064527.7}{GW190828B}{1251010527.9}{GW190828A}{1251009263.8}{GW190814A}{1249852257.0}{GW190803A}{1248834439.9}{GW190731A}{1248617394.6}{GW190728A}{1248331528.6}{GW190727A}{1248242632.0}{GW190720A}{1247616534.7}{GW190719A}{1247608532.9}{GW190708A}{1246663515.4}{GW190707A}{1246527224.2}{GW190706A}{1246487219.3}{GW190701A}{1246048404.6}{GW190630A}{1245955943.2}{GW190620A}{1245035079.3}{GW190602A}{1243533585.1}{GW190527A}{1242984073.8}{GW190521B}{1242459857.5}{GW190521A}{1242442967.4}{GW190519A}{1242315362.4}{GW190517A}{1242107479.8}{GW190514A}{1241852074.8}{GW190513A}{1241816086.8}{GW190512A}{1241719652.4}{GW190503A}{1240944862.3}{GW190426A}{1240327333.4}{GW190425A}{1240215503.0}{GW190424A}{1240164426.1}{GW190421A}{1239917954.2}{GW190413B}{1239198206.7}{GW190413A}{1239168612.5}{GW190412A}{1239082262.2}{GW190408A}{1238782700.3}}}
\newcommand{\geocenttimeplus}[1]{\IfEqCase{#1}{{GW190930A}{0.002}{GW190929A}{0.03}{GW190924A}{0.008}{GW190915A}{0.002}{GW190910A}{0.0}{GW190909A}{0.0}{GW190828B}{0.0}{GW190828A}{0.0000005}{GW190814A}{0.004}{GW190803A}{0.0}{GW190731A}{0.0}{GW190728A}{0.0010}{GW190727A}{0.0}{GW190720A}{0.01}{GW190719A}{0.05}{GW190708A}{0.0}{GW190707A}{0.009}{GW190706A}{0.0}{GW190701A}{0.0000005}{GW190630A}{0.0}{GW190620A}{0.0}{GW190602A}{0.0}{GW190527A}{0.0}{GW190521B}{0.0}{GW190521A}{0.01}{GW190519A}{0.0}{GW190517A}{0.0}{GW190514A}{0.0}{GW190513A}{0.0}{GW190512A}{0.0}{GW190503A}{0.0}{GW190426A}{0.02}{GW190425A}{0.03}{GW190424A}{0.0}{GW190421A}{0.0}{GW190413B}{0.0}{GW190413A}{0.0}{GW190412A}{0.007}{GW190408A}{0.0}}}
\newcommand{\costilttwominus}[1]{\IfEqCase{#1}{{GW190930A}{1.09}{GW190929A}{0.93}{GW190924A}{1.00}{GW190915A}{0.90}{GW190910A}{0.88}{GW190909A}{0.75}{GW190828B}{1.06}{GW190828A}{1.14}{GW190814A}{0.83}{GW190803A}{0.83}{GW190731A}{0.99}{GW190728A}{1.16}{GW190727A}{1.04}{GW190720A}{1.11}{GW190719A}{1.19}{GW190708A}{0.97}{GW190707A}{0.71}{GW190706A}{1.16}{GW190701A}{0.74}{GW190630A}{1.08}{GW190620A}{1.22}{GW190602A}{1.01}{GW190527A}{1.02}{GW190521B}{1.03}{GW190521A}{0.86}{GW190519A}{1.17}{GW190517A}{1.21}{GW190514A}{0.59}{GW190513A}{1.05}{GW190512A}{0.98}{GW190503A}{0.87}{GW190426A}{0.00}{GW190425A}{0.87}{GW190424A}{1.02}{GW190421A}{0.76}{GW190413B}{0.77}{GW190413A}{0.82}{GW190412A}{1.01}{GW190408A}{0.85}}}
\newcommand{\costilttwomed}[1]{\IfEqCase{#1}{{GW190930A}{0.31}{GW190929A}{0.04}{GW190924A}{0.15}{GW190915A}{0.02}{GW190910A}{0.04}{GW190909A}{-0.17}{GW190828B}{0.24}{GW190828A}{0.37}{GW190814A}{-0.03}{GW190803A}{-0.06}{GW190731A}{0.15}{GW190728A}{0.39}{GW190727A}{0.20}{GW190720A}{0.31}{GW190719A}{0.44}{GW190708A}{0.14}{GW190707A}{-0.19}{GW190706A}{0.37}{GW190701A}{-0.17}{GW190630A}{0.33}{GW190620A}{0.50}{GW190602A}{0.19}{GW190527A}{0.16}{GW190521B}{0.29}{GW190521A}{-0.02}{GW190519A}{0.50}{GW190517A}{0.61}{GW190514A}{-0.36}{GW190513A}{0.25}{GW190512A}{0.16}{GW190503A}{-0.02}{GW190426A}{-1.00}{GW190425A}{0.16}{GW190424A}{0.20}{GW190421A}{-0.16}{GW190413B}{-0.13}{GW190413A}{-0.08}{GW190412A}{0.25}{GW190408A}{-0.02}}}
\newcommand{\costilttwoplus}[1]{\IfEqCase{#1}{{GW190930A}{0.63}{GW190929A}{0.86}{GW190924A}{0.77}{GW190915A}{0.85}{GW190910A}{0.83}{GW190909A}{1.04}{GW190828B}{0.69}{GW190828A}{0.57}{GW190814A}{0.87}{GW190803A}{0.93}{GW190731A}{0.77}{GW190728A}{0.56}{GW190727A}{0.72}{GW190720A}{0.62}{GW190719A}{0.51}{GW190708A}{0.77}{GW190707A}{1.02}{GW190706A}{0.58}{GW190701A}{1.00}{GW190630A}{0.60}{GW190620A}{0.47}{GW190602A}{0.73}{GW190527A}{0.76}{GW190521B}{0.62}{GW190521A}{0.86}{GW190519A}{0.46}{GW190517A}{0.36}{GW190514A}{1.11}{GW190513A}{0.68}{GW190512A}{0.75}{GW190503A}{0.88}{GW190426A}{2.00}{GW190425A}{0.70}{GW190424A}{0.72}{GW190421A}{0.96}{GW190413B}{1.00}{GW190413A}{0.95}{GW190412A}{0.67}{GW190408A}{0.88}}}
\newcommand{\finalspinminus}[1]{\IfEqCase{#1}{{GW190930A}{0.06}{GW190929A}{0.31}{GW190924A}{0.05}{GW190915A}{0.11}{GW190910A}{0.07}{GW190909A}{0.20}{GW190828B}{0.08}{GW190828A}{0.07}{GW190814A}{0.02}{GW190803A}{0.11}{GW190731A}{0.13}{GW190728A}{0.04}{GW190727A}{0.10}{GW190720A}{0.05}{GW190719A}{0.17}{GW190708A}{0.04}{GW190707A}{0.04}{GW190706A}{0.18}{GW190701A}{0.13}{GW190630A}{0.07}{GW190620A}{0.15}{GW190602A}{0.14}{GW190527A}{0.16}{GW190521B}{0.07}{GW190521A}{0.16}{GW190519A}{0.13}{GW190517A}{0.07}{GW190514A}{0.15}{GW190513A}{0.12}{GW190512A}{0.07}{GW190503A}{0.12}{GW190424A}{0.09}{GW190421A}{0.11}{GW190413B}{0.12}{GW190413A}{0.13}{GW190412A}{0.06}{GW190408A}{0.07}}}
\newcommand{\finalspinmed}[1]{\IfEqCase{#1}{{GW190930A}{0.72}{GW190929A}{0.66}{GW190924A}{0.67}{GW190915A}{0.70}{GW190910A}{0.70}{GW190909A}{0.66}{GW190828B}{0.65}{GW190828A}{0.75}{GW190814A}{0.28}{GW190803A}{0.68}{GW190731A}{0.70}{GW190728A}{0.71}{GW190727A}{0.73}{GW190720A}{0.72}{GW190719A}{0.78}{GW190708A}{0.69}{GW190707A}{0.66}{GW190706A}{0.78}{GW190701A}{0.66}{GW190630A}{0.70}{GW190620A}{0.79}{GW190602A}{0.70}{GW190527A}{0.71}{GW190521B}{0.72}{GW190521A}{0.71}{GW190519A}{0.79}{GW190517A}{0.87}{GW190514A}{0.63}{GW190513A}{0.68}{GW190512A}{0.65}{GW190503A}{0.66}{GW190424A}{0.74}{GW190421A}{0.67}{GW190413B}{0.68}{GW190413A}{0.68}{GW190412A}{0.67}{GW190408A}{0.67}}}
\newcommand{\finalspinplus}[1]{\IfEqCase{#1}{{GW190930A}{0.07}{GW190929A}{0.20}{GW190924A}{0.05}{GW190915A}{0.09}{GW190910A}{0.08}{GW190909A}{0.15}{GW190828B}{0.08}{GW190828A}{0.06}{GW190814A}{0.02}{GW190803A}{0.10}{GW190731A}{0.10}{GW190728A}{0.04}{GW190727A}{0.10}{GW190720A}{0.06}{GW190719A}{0.11}{GW190708A}{0.04}{GW190707A}{0.03}{GW190706A}{0.09}{GW190701A}{0.09}{GW190630A}{0.05}{GW190620A}{0.08}{GW190602A}{0.10}{GW190527A}{0.12}{GW190521B}{0.05}{GW190521A}{0.12}{GW190519A}{0.07}{GW190517A}{0.05}{GW190514A}{0.11}{GW190513A}{0.14}{GW190512A}{0.07}{GW190503A}{0.09}{GW190424A}{0.09}{GW190421A}{0.10}{GW190413B}{0.10}{GW190413A}{0.12}{GW190412A}{0.05}{GW190408A}{0.06}}}
\newcommand{\luminositydistanceminus}[1]{\IfEqCase{#1}{{GW190930A}{0.32}{GW190929A}{1.05}{GW190924A}{0.22}{GW190915A}{0.61}{GW190910A}{0.58}{GW190909A}{2.22}{GW190828B}{0.60}{GW190828A}{0.93}{GW190814A}{0.05}{GW190803A}{1.58}{GW190731A}{1.72}{GW190728A}{0.37}{GW190727A}{1.50}{GW190720A}{0.32}{GW190719A}{2.00}{GW190708A}{0.39}{GW190707A}{0.37}{GW190706A}{1.93}{GW190701A}{0.73}{GW190630A}{0.37}{GW190620A}{1.31}{GW190602A}{1.12}{GW190527A}{1.24}{GW190521B}{0.57}{GW190521A}{1.95}{GW190519A}{0.92}{GW190517A}{0.84}{GW190514A}{2.17}{GW190513A}{0.80}{GW190512A}{0.55}{GW190503A}{0.63}{GW190426A}{0.16}{GW190425A}{0.07}{GW190424A}{1.16}{GW190421A}{1.38}{GW190413B}{2.12}{GW190413A}{1.66}{GW190412A}{0.17}{GW190408A}{0.60}}}
\newcommand{\luminositydistancemed}[1]{\IfEqCase{#1}{{GW190930A}{0.76}{GW190929A}{2.13}{GW190924A}{0.57}{GW190915A}{1.62}{GW190910A}{1.46}{GW190909A}{3.77}{GW190828B}{1.60}{GW190828A}{2.13}{GW190814A}{0.24}{GW190803A}{3.27}{GW190731A}{3.30}{GW190728A}{0.87}{GW190727A}{3.30}{GW190720A}{0.79}{GW190719A}{3.94}{GW190708A}{0.88}{GW190707A}{0.77}{GW190706A}{4.42}{GW190701A}{2.06}{GW190630A}{0.89}{GW190620A}{2.81}{GW190602A}{2.69}{GW190527A}{2.49}{GW190521B}{1.24}{GW190521A}{3.92}{GW190519A}{2.53}{GW190517A}{1.86}{GW190514A}{4.13}{GW190513A}{2.06}{GW190512A}{1.43}{GW190503A}{1.45}{GW190426A}{0.37}{GW190425A}{0.16}{GW190424A}{2.20}{GW190421A}{2.88}{GW190413B}{4.45}{GW190413A}{3.55}{GW190412A}{0.74}{GW190408A}{1.55}}}
\newcommand{\luminositydistanceplus}[1]{\IfEqCase{#1}{{GW190930A}{0.36}{GW190929A}{3.65}{GW190924A}{0.22}{GW190915A}{0.71}{GW190910A}{1.03}{GW190909A}{3.27}{GW190828B}{0.62}{GW190828A}{0.66}{GW190814A}{0.04}{GW190803A}{1.95}{GW190731A}{2.39}{GW190728A}{0.26}{GW190727A}{1.54}{GW190720A}{0.69}{GW190719A}{2.59}{GW190708A}{0.33}{GW190707A}{0.38}{GW190706A}{2.59}{GW190701A}{0.76}{GW190630A}{0.56}{GW190620A}{1.68}{GW190602A}{1.79}{GW190527A}{2.48}{GW190521B}{0.40}{GW190521A}{2.19}{GW190519A}{1.83}{GW190517A}{1.62}{GW190514A}{2.65}{GW190513A}{0.88}{GW190512A}{0.55}{GW190503A}{0.69}{GW190426A}{0.18}{GW190425A}{0.07}{GW190424A}{1.58}{GW190421A}{1.37}{GW190413B}{2.48}{GW190413A}{2.27}{GW190412A}{0.14}{GW190408A}{0.40}}}
\newcommand{\spinonezminus}[1]{\IfEqCase{#1}{{GW190930A}{0.29}{GW190929A}{0.43}{GW190924A}{0.24}{GW190915A}{0.41}{GW190910A}{0.34}{GW190909A}{0.56}{GW190828B}{0.23}{GW190828A}{0.31}{GW190814A}{0.05}{GW190803A}{0.46}{GW190731A}{0.33}{GW190728A}{0.27}{GW190727A}{0.33}{GW190720A}{0.29}{GW190719A}{0.44}{GW190708A}{0.25}{GW190707A}{0.29}{GW190706A}{0.38}{GW190701A}{0.51}{GW190630A}{0.21}{GW190620A}{0.39}{GW190602A}{0.34}{GW190527A}{0.36}{GW190521B}{0.23}{GW190521A}{0.58}{GW190519A}{0.37}{GW190517A}{0.36}{GW190514A}{0.54}{GW190513A}{0.24}{GW190512A}{0.25}{GW190503A}{0.44}{GW190426A}{0.51}{GW190425A}{0.12}{GW190424A}{0.35}{GW190421A}{0.49}{GW190413B}{0.48}{GW190413A}{0.52}{GW190412A}{0.23}{GW190408A}{0.42}}}
\newcommand{\spinonezmed}[1]{\IfEqCase{#1}{{GW190930A}{0.15}{GW190929A}{0.008}{GW190924A}{0.02}{GW190915A}{0.02}{GW190910A}{0.01}{GW190909A}{-0.03}{GW190828B}{0.06}{GW190828A}{0.20}{GW190814A}{0.0001}{GW190803A}{-0.01}{GW190731A}{0.03}{GW190728A}{0.13}{GW190727A}{0.10}{GW190720A}{0.20}{GW190719A}{0.36}{GW190708A}{0.009}{GW190707A}{-0.03}{GW190706A}{0.33}{GW190701A}{-0.05}{GW190630A}{0.05}{GW190620A}{0.37}{GW190602A}{0.04}{GW190527A}{0.09}{GW190521B}{0.04}{GW190521A}{0.02}{GW190519A}{0.35}{GW190517A}{0.67}{GW190514A}{-0.18}{GW190513A}{0.09}{GW190512A}{0.005}{GW190503A}{-0.03}{GW190426A}{-0.03}{GW190425A}{0.06}{GW190424A}{0.15}{GW190421A}{-0.04}{GW190413B}{-0.02}{GW190413A}{0.001}{GW190412A}{0.30}{GW190408A}{-0.03}}}
\newcommand{\spinonezplus}[1]{\IfEqCase{#1}{{GW190930A}{0.41}{GW190929A}{0.45}{GW190924A}{0.39}{GW190915A}{0.40}{GW190910A}{0.39}{GW190909A}{0.53}{GW190828B}{0.24}{GW190828A}{0.41}{GW190814A}{0.04}{GW190803A}{0.39}{GW190731A}{0.45}{GW190728A}{0.30}{GW190727A}{0.48}{GW190720A}{0.29}{GW190719A}{0.43}{GW190708A}{0.24}{GW190707A}{0.20}{GW190706A}{0.43}{GW190701A}{0.34}{GW190630A}{0.28}{GW190620A}{0.42}{GW190602A}{0.46}{GW190527A}{0.50}{GW190521B}{0.32}{GW190521A}{0.53}{GW190519A}{0.37}{GW190517A}{0.25}{GW190514A}{0.39}{GW190513A}{0.46}{GW190512A}{0.21}{GW190503A}{0.31}{GW190426A}{0.36}{GW190425A}{0.18}{GW190424A}{0.46}{GW190421A}{0.40}{GW190413B}{0.40}{GW190413A}{0.44}{GW190412A}{0.12}{GW190408A}{0.26}}}
\newcommand{\chirpmasssourceminus}[1]{\IfEqCase{#1}{{GW190930A}{0.5}{GW190929A}{8.2}{GW190924A}{0.2}{GW190915A}{2.7}{GW190910A}{4.1}{GW190909A}{7.5}{GW190828B}{1.0}{GW190828A}{2.1}{GW190814A}{0.06}{GW190803A}{4.1}{GW190731A}{5.2}{GW190728A}{0.3}{GW190727A}{3.7}{GW190720A}{0.8}{GW190719A}{4.0}{GW190708A}{0.6}{GW190707A}{0.5}{GW190706A}{7.0}{GW190701A}{4.9}{GW190630A}{2.1}{GW190620A}{6.5}{GW190602A}{8.5}{GW190527A}{4.2}{GW190521B}{2.5}{GW190521A}{10.6}{GW190519A}{7.1}{GW190517A}{4.0}{GW190514A}{4.8}{GW190513A}{1.9}{GW190512A}{1.0}{GW190503A}{4.2}{GW190426A}{0.08}{GW190425A}{0.02}{GW190424A}{4.6}{GW190421A}{4.2}{GW190413B}{5.4}{GW190413A}{4.1}{GW190412A}{0.3}{GW190408A}{1.2}}}
\newcommand{\chirpmasssourcemed}[1]{\IfEqCase{#1}{{GW190930A}{8.5}{GW190929A}{35.8}{GW190924A}{5.8}{GW190915A}{25.3}{GW190910A}{34.3}{GW190909A}{30.9}{GW190828B}{13.3}{GW190828A}{25.0}{GW190814A}{6.09}{GW190803A}{27.3}{GW190731A}{29.5}{GW190728A}{8.6}{GW190727A}{28.6}{GW190720A}{8.9}{GW190719A}{23.5}{GW190708A}{13.2}{GW190707A}{8.5}{GW190706A}{42.7}{GW190701A}{40.3}{GW190630A}{24.9}{GW190620A}{38.3}{GW190602A}{49.1}{GW190527A}{24.3}{GW190521B}{32.1}{GW190521A}{69.2}{GW190519A}{44.5}{GW190517A}{26.6}{GW190514A}{28.5}{GW190513A}{21.6}{GW190512A}{14.6}{GW190503A}{30.2}{GW190426A}{2.41}{GW190425A}{1.44}{GW190424A}{31.0}{GW190421A}{31.2}{GW190413B}{33.0}{GW190413A}{24.6}{GW190412A}{13.3}{GW190408A}{18.3}}}
\newcommand{\chirpmasssourceplus}[1]{\IfEqCase{#1}{{GW190930A}{0.5}{GW190929A}{14.9}{GW190924A}{0.2}{GW190915A}{3.2}{GW190910A}{4.1}{GW190909A}{17.2}{GW190828B}{1.2}{GW190828A}{3.4}{GW190814A}{0.06}{GW190803A}{5.7}{GW190731A}{7.1}{GW190728A}{0.5}{GW190727A}{5.3}{GW190720A}{0.5}{GW190719A}{6.5}{GW190708A}{0.9}{GW190707A}{0.6}{GW190706A}{10.0}{GW190701A}{5.4}{GW190630A}{2.1}{GW190620A}{8.3}{GW190602A}{9.1}{GW190527A}{9.1}{GW190521B}{3.2}{GW190521A}{17.0}{GW190519A}{6.4}{GW190517A}{4.0}{GW190514A}{7.9}{GW190513A}{3.8}{GW190512A}{1.3}{GW190503A}{4.2}{GW190426A}{0.08}{GW190425A}{0.02}{GW190424A}{5.8}{GW190421A}{5.9}{GW190413B}{8.2}{GW190413A}{5.5}{GW190412A}{0.4}{GW190408A}{1.9}}}
\newcommand{\symmetricmassratiominus}[1]{\IfEqCase{#1}{{GW190930A}{0.11}{GW190929A}{0.07}{GW190924A}{0.09}{GW190915A}{0.03}{GW190910A}{0.01}{GW190909A}{0.09}{GW190828B}{0.04}{GW190828A}{0.01}{GW190814A}{0.006}{GW190803A}{0.03}{GW190731A}{0.04}{GW190728A}{0.07}{GW190727A}{0.03}{GW190720A}{0.06}{GW190719A}{0.06}{GW190708A}{0.03}{GW190707A}{0.03}{GW190706A}{0.05}{GW190701A}{0.03}{GW190630A}{0.03}{GW190620A}{0.04}{GW190602A}{0.04}{GW190527A}{0.06}{GW190521B}{0.01}{GW190521A}{0.04}{GW190519A}{0.03}{GW190517A}{0.04}{GW190514A}{0.04}{GW190513A}{0.04}{GW190512A}{0.03}{GW190503A}{0.03}{GW190426A}{0.08}{GW190425A}{0.03}{GW190424A}{0.02}{GW190421A}{0.03}{GW190413B}{0.04}{GW190413A}{0.04}{GW190412A}{0.02}{GW190408A}{0.02}}}
\newcommand{\symmetricmassratiomed}[1]{\IfEqCase{#1}{{GW190930A}{0.24}{GW190929A}{0.18}{GW190924A}{0.23}{GW190915A}{0.242}{GW190910A}{0.248}{GW190909A}{0.24}{GW190828B}{0.21}{GW190828A}{0.248}{GW190814A}{0.090}{GW190803A}{0.245}{GW190731A}{0.243}{GW190728A}{0.24}{GW190727A}{0.247}{GW190720A}{0.23}{GW190719A}{0.23}{GW190708A}{0.245}{GW190707A}{0.244}{GW190706A}{0.23}{GW190701A}{0.246}{GW190630A}{0.241}{GW190620A}{0.24}{GW190602A}{0.243}{GW190527A}{0.24}{GW190521B}{0.246}{GW190521A}{0.245}{GW190519A}{0.24}{GW190517A}{0.241}{GW190514A}{0.245}{GW190513A}{0.22}{GW190512A}{0.23}{GW190503A}{0.24}{GW190426A}{0.16}{GW190425A}{0.240}{GW190424A}{0.247}{GW190421A}{0.247}{GW190413B}{0.241}{GW190413A}{0.241}{GW190412A}{0.17}{GW190408A}{0.245}}}
\newcommand{\symmetricmassratioplus}[1]{\IfEqCase{#1}{{GW190930A}{0.01}{GW190929A}{0.07}{GW190924A}{0.02}{GW190915A}{0.008}{GW190910A}{0.002}{GW190909A}{0.01}{GW190828B}{0.04}{GW190828A}{0.002}{GW190814A}{0.005}{GW190803A}{0.005}{GW190731A}{0.007}{GW190728A}{0.01}{GW190727A}{0.003}{GW190720A}{0.02}{GW190719A}{0.02}{GW190708A}{0.005}{GW190707A}{0.006}{GW190706A}{0.02}{GW190701A}{0.004}{GW190630A}{0.009}{GW190620A}{0.01}{GW190602A}{0.007}{GW190527A}{0.01}{GW190521B}{0.004}{GW190521A}{0.005}{GW190519A}{0.01}{GW190517A}{0.009}{GW190514A}{0.005}{GW190513A}{0.03}{GW190512A}{0.02}{GW190503A}{0.01}{GW190426A}{0.08}{GW190425A}{0.010}{GW190424A}{0.003}{GW190421A}{0.003}{GW190413B}{0.008}{GW190413A}{0.008}{GW190412A}{0.03}{GW190408A}{0.005}}}
\newcommand{\spintwoxminus}[1]{\IfEqCase{#1}{{GW190930A}{0.52}{GW190929A}{0.60}{GW190924A}{0.49}{GW190915A}{0.59}{GW190910A}{0.52}{GW190909A}{0.58}{GW190828B}{0.55}{GW190828A}{0.50}{GW190814A}{0.62}{GW190803A}{0.58}{GW190731A}{0.57}{GW190728A}{0.46}{GW190727A}{0.55}{GW190720A}{0.58}{GW190719A}{0.56}{GW190708A}{0.46}{GW190707A}{0.46}{GW190706A}{0.53}{GW190701A}{0.55}{GW190630A}{0.49}{GW190620A}{0.54}{GW190602A}{0.60}{GW190527A}{0.61}{GW190521B}{0.51}{GW190521A}{0.66}{GW190519A}{0.57}{GW190517A}{0.53}{GW190514A}{0.60}{GW190513A}{0.53}{GW190512A}{0.50}{GW190503A}{0.55}{GW190426A}{0.00}{GW190425A}{0.47}{GW190424A}{0.59}{GW190421A}{0.59}{GW190413B}{0.59}{GW190413A}{0.58}{GW190412A}{0.57}{GW190408A}{0.53}}}
\newcommand{\spintwoxmed}[1]{\IfEqCase{#1}{{GW190930A}{0.00}{GW190929A}{0.0009}{GW190924A}{0.00}{GW190915A}{0.00}{GW190910A}{0.0003}{GW190909A}{0.0003}{GW190828B}{0.00}{GW190828A}{0.002}{GW190814A}{-0.01}{GW190803A}{0.006}{GW190731A}{0.003}{GW190728A}{0.001}{GW190727A}{0.002}{GW190720A}{0.002}{GW190719A}{0.00}{GW190708A}{0.002}{GW190707A}{0.0004}{GW190706A}{0.003}{GW190701A}{0.003}{GW190630A}{0.00}{GW190620A}{0.00006}{GW190602A}{0.00}{GW190527A}{0.005}{GW190521B}{0.00}{GW190521A}{0.002}{GW190519A}{0.0006}{GW190517A}{0.001}{GW190514A}{0.001}{GW190513A}{0.00}{GW190512A}{0.0009}{GW190503A}{0.001}{GW190426A}{0.00}{GW190425A}{0.0006}{GW190424A}{0.00}{GW190421A}{0.00}{GW190413B}{0.0007}{GW190413A}{0.00}{GW190412A}{-0.01}{GW190408A}{0.002}}}
\newcommand{\spintwoxplus}[1]{\IfEqCase{#1}{{GW190930A}{0.52}{GW190929A}{0.59}{GW190924A}{0.48}{GW190915A}{0.60}{GW190910A}{0.54}{GW190909A}{0.57}{GW190828B}{0.54}{GW190828A}{0.51}{GW190814A}{0.59}{GW190803A}{0.57}{GW190731A}{0.58}{GW190728A}{0.48}{GW190727A}{0.55}{GW190720A}{0.57}{GW190719A}{0.56}{GW190708A}{0.43}{GW190707A}{0.46}{GW190706A}{0.54}{GW190701A}{0.56}{GW190630A}{0.48}{GW190620A}{0.56}{GW190602A}{0.60}{GW190527A}{0.59}{GW190521B}{0.51}{GW190521A}{0.69}{GW190519A}{0.55}{GW190517A}{0.53}{GW190514A}{0.59}{GW190513A}{0.54}{GW190512A}{0.49}{GW190503A}{0.58}{GW190426A}{0.00}{GW190425A}{0.47}{GW190424A}{0.59}{GW190421A}{0.58}{GW190413B}{0.60}{GW190413A}{0.57}{GW190412A}{0.56}{GW190408A}{0.53}}}
\newcommand{\networkoptimalsnrminus}[1]{\IfEqCase{#1}{{GW190814A}{1.7}{GW190426A}{1.8}{GW190425A}{1.7}}}
\newcommand{\networkoptimalsnrmed}[1]{\IfEqCase{#1}{{GW190814A}{24.7}{GW190426A}{8.3}{GW190425A}{12.0}}}
\newcommand{\networkoptimalsnrplus}[1]{\IfEqCase{#1}{{GW190814A}{1.7}{GW190426A}{1.8}{GW190425A}{1.7}}}
\newcommand{\networkmatchedfiltersnrminus}[1]{\IfEqCase{#1}{{GW190814A}{0.2}{GW190426A}{0.6}{GW190425A}{0.4}{GW190412A}{0.4}}}
\newcommand{\networkmatchedfiltersnrmed}[1]{\IfEqCase{#1}{{GW190814A}{24.9}{GW190426A}{8.7}{GW190425A}{12.4}{GW190412A}{19.0}}}
\newcommand{\networkmatchedfiltersnrplus}[1]{\IfEqCase{#1}{{GW190814A}{0.1}{GW190426A}{0.5}{GW190425A}{0.3}{GW190412A}{0.2}}}
\newcommand{\logpriorminus}[1]{\IfEqCase{#1}{{GW190426A}{10.5}{GW190425A}{8.6}}}
\newcommand{\logpriormed}[1]{\IfEqCase{#1}{{GW190426A}{161.3}{GW190425A}{98.4}}}
\newcommand{\logpriorplus}[1]{\IfEqCase{#1}{{GW190426A}{8.6}{GW190425A}{6.7}}}
\newcommand{\PEpercentBNS}[1]{\IfEqCase{#1}{{GW190930A}{0}{GW190929A}{0}{GW190924A}{0}{GW190915A}{0}{GW190910A}{0}{GW190909A}{0}{GW190828B}{0}{GW190828A}{0}{GW190814A}{0}{GW190803A}{0}{GW190731A}{0}{GW190728A}{0}{GW190727A}{0}{GW190720A}{0}{GW190719A}{0}{GW190708A}{0}{GW190707A}{0}{GW190706A}{0}{GW190701A}{0}{GW190630A}{0}{GW190620A}{0}{GW190602A}{0}{GW190527A}{0}{GW190521B}{0}{GW190521A}{0}{GW190519A}{0}{GW190517A}{0}{GW190514A}{0}{GW190513A}{0}{GW190512A}{0}{GW190503A}{0}{GW190426A}{1}{GW190425A}{100}{GW190424A}{0}{GW190421A}{0}{GW190413B}{0}{GW190413A}{0}{GW190412A}{0}{GW190408A}{0}}}
\newcommand{\PEpercentNSBH}[1]{\IfEqCase{#1}{{GW190930A}{0}{GW190929A}{0}{GW190924A}{4}{GW190915A}{0}{GW190910A}{0}{GW190909A}{0}{GW190828B}{0}{GW190828A}{0}{GW190814A}{100}{GW190803A}{0}{GW190731A}{0}{GW190728A}{0}{GW190727A}{0}{GW190720A}{0}{GW190719A}{0}{GW190708A}{0}{GW190707A}{0}{GW190706A}{0}{GW190701A}{0}{GW190630A}{0}{GW190620A}{0}{GW190602A}{0}{GW190527A}{0}{GW190521B}{0}{GW190521A}{0}{GW190519A}{0}{GW190517A}{0}{GW190514A}{0}{GW190513A}{0}{GW190512A}{0}{GW190503A}{0}{GW190426A}{99}{GW190425A}{0}{GW190424A}{0}{GW190421A}{0}{GW190413B}{0}{GW190413A}{0}{GW190412A}{0}{GW190408A}{0}}}
\newcommand{\PEpercentBBH}[1]{\IfEqCase{#1}{{GW190930A}{100}{GW190929A}{100}{GW190924A}{96}{GW190915A}{100}{GW190910A}{100}{GW190909A}{100}{GW190828B}{100}{GW190828A}{100}{GW190814A}{0}{GW190803A}{100}{GW190731A}{100}{GW190728A}{100}{GW190727A}{100}{GW190720A}{100}{GW190719A}{100}{GW190708A}{100}{GW190707A}{100}{GW190706A}{100}{GW190701A}{100}{GW190630A}{100}{GW190620A}{100}{GW190602A}{100}{GW190527A}{100}{GW190521B}{100}{GW190521A}{100}{GW190519A}{100}{GW190517A}{100}{GW190514A}{100}{GW190513A}{100}{GW190512A}{100}{GW190503A}{100}{GW190426A}{0}{GW190425A}{0}{GW190424A}{100}{GW190421A}{100}{GW190413B}{100}{GW190413A}{100}{GW190412A}{100}{GW190408A}{100}}}
\newcommand{\PEpercentMassGap}[1]{\IfEqCase{#1}{{GW190930A}{0}{GW190929A}{0}{GW190924A}{0}{GW190915A}{0}{GW190910A}{0}{GW190909A}{0}{GW190828B}{0}{GW190828A}{0}{GW190814A}{0}{GW190803A}{0}{GW190731A}{0}{GW190728A}{0}{GW190727A}{0}{GW190720A}{0}{GW190719A}{0}{GW190708A}{0}{GW190707A}{0}{GW190706A}{0}{GW190701A}{0}{GW190630A}{0}{GW190620A}{0}{GW190602A}{0}{GW190527A}{0}{GW190521B}{0}{GW190521A}{0}{GW190519A}{0}{GW190517A}{0}{GW190514A}{0}{GW190513A}{0}{GW190512A}{0}{GW190503A}{0}{GW190426A}{0}{GW190425A}{0}{GW190424A}{0}{GW190421A}{0}{GW190413B}{0}{GW190413A}{0}{GW190412A}{0}{GW190408A}{0}}}

\newcommand{\costhetajnIMRminus}[1]{\IfEqCase{#1}{{GW190930A}{1.54}{GW190929A}{0.71}{GW190924A}{1.74}{GW190915A}{0.69}{GW190910A}{0.88}{GW190909A}{1.03}{GW190828B}{0.84}{GW190828A}{0.31}{GW190814A}{1.42}{GW190803A}{1.18}{GW190731A}{1.10}{GW190728A}{1.32}{GW190727A}{1.42}{GW190720A}{0.14}{GW190719A}{0.94}{GW190708A}{1.15}{GW190707A}{0.39}{GW190706A}{1.08}{GW190701A}{0.44}{GW190630A}{1.47}{GW190620A}{0.43}{GW190602A}{0.69}{GW190527A}{1.30}{GW190521B}{1.47}{GW190521A}{1.57}{GW190519A}{0.82}{GW190517A}{0.34}{GW190514A}{0.96}{GW190513A}{1.67}{GW190512A}{0.96}{GW190503A}{0.19}{GW190426A}{0.84}{GW190425A}{1.43}{GW190424A}{1.00}{GW190421A}{0.84}{GW190413B}{0.81}{GW190413A}{1.61}{GW190412A}{1.25}{GW190408A}{0.96}}}
\newcommand{\costhetajnIMRmed}[1]{\IfEqCase{#1}{{GW190930A}{0.58}{GW190929A}{-0.16}{GW190924A}{0.82}{GW190915A}{-0.22}{GW190910A}{-0.10}{GW190909A}{0.09}{GW190828B}{-0.13}{GW190828A}{-0.68}{GW190814A}{0.68}{GW190803A}{0.23}{GW190731A}{0.14}{GW190728A}{0.35}{GW190727A}{0.48}{GW190720A}{-0.85}{GW190719A}{-0.02}{GW190708A}{0.18}{GW190707A}{-0.59}{GW190706A}{0.13}{GW190701A}{0.77}{GW190630A}{0.51}{GW190620A}{-0.54}{GW190602A}{-0.27}{GW190527A}{0.37}{GW190521B}{0.53}{GW190521A}{0.62}{GW190519A}{-0.13}{GW190517A}{-0.62}{GW190514A}{0.01}{GW190513A}{0.75}{GW190512A}{-0.01}{GW190503A}{-0.80}{GW190426A}{-0.13}{GW190425A}{0.47}{GW190424A}{0.04}{GW190421A}{-0.11}{GW190413B}{-0.13}{GW190413A}{0.67}{GW190412A}{0.74}{GW190408A}{0.00}}}
\newcommand{\costhetajnIMRplus}[1]{\IfEqCase{#1}{{GW190930A}{0.40}{GW190929A}{0.97}{GW190924A}{0.17}{GW190915A}{1.09}{GW190910A}{1.06}{GW190909A}{0.86}{GW190828B}{1.10}{GW190828A}{1.63}{GW190814A}{0.15}{GW190803A}{0.72}{GW190731A}{0.82}{GW190728A}{0.63}{GW190727A}{0.49}{GW190720A}{1.63}{GW190719A}{0.97}{GW190708A}{0.80}{GW190707A}{1.56}{GW190706A}{0.82}{GW190701A}{0.21}{GW190630A}{0.47}{GW190620A}{1.47}{GW190602A}{1.22}{GW190527A}{0.58}{GW190521B}{0.43}{GW190521A}{0.35}{GW190519A}{1.07}{GW190517A}{1.36}{GW190514A}{0.94}{GW190513A}{0.23}{GW190512A}{0.98}{GW190503A}{0.52}{GW190426A}{1.09}{GW190425A}{0.50}{GW190424A}{0.91}{GW190421A}{1.06}{GW190413B}{1.05}{GW190413A}{0.30}{GW190412A}{0.16}{GW190408A}{0.97}}}
\newcommand{\loglikelihoodIMRminus}[1]{\IfEqCase{#1}{{GW190930A}{4.7}{GW190929A}{7.5}{GW190924A}{5.3}{GW190915A}{4.7}{GW190910A}{4.1}{GW190909A}{5.2}{GW190828B}{4.7}{GW190828A}{4.6}{GW190814A}{4.9}{GW190803A}{4.8}{GW190731A}{4.2}{GW190728A}{5.2}{GW190727A}{5.7}{GW190720A}{7.8}{GW190719A}{6.3}{GW190708A}{4.6}{GW190707A}{5.0}{GW190706A}{5.0}{GW190701A}{4.1}{GW190630A}{5.1}{GW190620A}{4.9}{GW190602A}{4.4}{GW190527A}{7.2}{GW190521B}{4.4}{GW190521A}{5.0}{GW190519A}{4.5}{GW190517A}{6.2}{GW190514A}{4.7}{GW190513A}{5.3}{GW190512A}{4.8}{GW190503A}{4.4}{GW190426A}{5.6}{GW190425A}{5.7}{GW190424A}{4.2}{GW190421A}{4.4}{GW190413B}{5.1}{GW190413A}{6.5}{GW190412A}{6.7}{GW190408A}{4.5}}}
\newcommand{\loglikelihoodIMRmed}[1]{\IfEqCase{#1}{{GW190930A}{-15939.8}{GW190929A}{-11966.5}{GW190924A}{-97036.5}{GW190915A}{-2814.1}{GW190910A}{-7980.5}{GW190909A}{-7987.2}{GW190828B}{-23976.5}{GW190828A}{-11929.3}{GW190814A}{298.0}{GW190803A}{-11706.7}{GW190731A}{-7957.8}{GW190728A}{-47939.6}{GW190727A}{-2700.5}{GW190720A}{-23907.6}{GW190719A}{-3904.4}{GW190708A}{-15917.4}{GW190707A}{-15886.3}{GW190706A}{-2802.0}{GW190701A}{-2790.2}{GW190630A}{-7866.2}{GW190620A}{-3893.6}{GW190602A}{-2793.2}{GW190527A}{-15670.8}{GW190521B}{-1871.5}{GW190521A}{-11921.7}{GW190519A}{-2814.6}{GW190517A}{-5836.2}{GW190514A}{-7948.9}{GW190513A}{-5849.7}{GW190512A}{-11692.6}{GW190503A}{-2783.0}{GW190426A}{-389547.0}{GW190425A}{-500483.9}{GW190424A}{-1961.6}{GW190421A}{-1851.9}{GW190413B}{-11957.9}{GW190413A}{-2782.0}{GW190412A}{-22832.7}{GW190408A}{-5836.3}}}
\newcommand{\loglikelihoodIMRplus}[1]{\IfEqCase{#1}{{GW190930A}{2.9}{GW190929A}{6.2}{GW190924A}{3.2}{GW190915A}{3.6}{GW190910A}{2.7}{GW190909A}{3.2}{GW190828B}{3.1}{GW190828A}{3.4}{GW190814A}{3.1}{GW190803A}{2.8}{GW190731A}{2.3}{GW190728A}{3.1}{GW190727A}{3.6}{GW190720A}{3.4}{GW190719A}{2.9}{GW190708A}{3.2}{GW190707A}{3.0}{GW190706A}{3.2}{GW190701A}{2.4}{GW190630A}{2.9}{GW190620A}{3.2}{GW190602A}{2.9}{GW190527A}{3.0}{GW190521B}{3.1}{GW190521A}{4.5}{GW190519A}{2.9}{GW190517A}{4.1}{GW190514A}{2.7}{GW190513A}{3.9}{GW190512A}{3.2}{GW190503A}{2.5}{GW190426A}{4.6}{GW190425A}{4.5}{GW190424A}{2.7}{GW190421A}{2.6}{GW190413B}{3.8}{GW190413A}{3.9}{GW190412A}{3.8}{GW190408A}{3.4}}}
\newcommand{\logpriorIMRminus}[1]{\IfEqCase{#1}{{GW190930A}{8.4}{GW190929A}{10.1}{GW190924A}{10.1}{GW190915A}{10.1}{GW190910A}{8.4}{GW190909A}{8.3}{GW190828B}{10.1}{GW190828A}{10.1}{GW190803A}{10.2}{GW190731A}{8.3}{GW190728A}{10.0}{GW190727A}{9.9}{GW190720A}{10.3}{GW190719A}{8.3}{GW190708A}{8.5}{GW190707A}{8.6}{GW190706A}{9.7}{GW190701A}{9.9}{GW190630A}{8.6}{GW190620A}{8.4}{GW190602A}{10.0}{GW190527A}{8.7}{GW190521B}{8.7}{GW190521A}{10.2}{GW190519A}{10.1}{GW190517A}{10.1}{GW190514A}{8.5}{GW190513A}{10.1}{GW190512A}{9.9}{GW190503A}{10.0}{GW190426A}{10.5}{GW190425A}{8.6}{GW190424A}{6.7}{GW190421A}{8.5}{GW190413B}{10.0}{GW190413A}{10.0}{GW190412A}{10.2}{GW190408A}{10.2}}}
\newcommand{\logpriorIMRmed}[1]{\IfEqCase{#1}{{GW190930A}{139.6}{GW190929A}{178.4}{GW190924A}{170.0}{GW190915A}{169.3}{GW190910A}{117.3}{GW190909A}{145.8}{GW190828B}{179.1}{GW190828A}{179.6}{GW190803A}{179.4}{GW190731A}{147.8}{GW190728A}{175.8}{GW190727A}{173.7}{GW190720A}{175.8}{GW190719A}{145.3}{GW190708A}{114.0}{GW190707A}{141.9}{GW190706A}{174.5}{GW190701A}{171.7}{GW190630A}{113.8}{GW190620A}{116.8}{GW190602A}{166.2}{GW190527A}{137.5}{GW190521B}{128.4}{GW190521A}{170.4}{GW190519A}{167.0}{GW190517A}{170.4}{GW190514A}{139.5}{GW190513A}{169.5}{GW190512A}{169.9}{GW190503A}{164.6}{GW190426A}{161.3}{GW190425A}{98.4}{GW190424A}{76.9}{GW190421A}{131.7}{GW190413B}{174.1}{GW190413A}{167.1}{GW190412A}{168.7}{GW190408A}{170.1}}}
\newcommand{\logpriorIMRplus}[1]{\IfEqCase{#1}{{GW190930A}{6.6}{GW190929A}{8.4}{GW190924A}{8.2}{GW190915A}{8.1}{GW190910A}{6.7}{GW190909A}{6.9}{GW190828B}{8.3}{GW190828A}{8.3}{GW190803A}{8.3}{GW190731A}{6.7}{GW190728A}{8.3}{GW190727A}{8.1}{GW190720A}{8.4}{GW190719A}{6.9}{GW190708A}{6.6}{GW190707A}{6.7}{GW190706A}{8.2}{GW190701A}{8.2}{GW190630A}{6.8}{GW190620A}{6.8}{GW190602A}{8.3}{GW190527A}{6.8}{GW190521B}{6.7}{GW190521A}{8.3}{GW190519A}{8.2}{GW190517A}{8.3}{GW190514A}{6.8}{GW190513A}{8.2}{GW190512A}{8.2}{GW190503A}{8.3}{GW190426A}{8.6}{GW190425A}{6.7}{GW190424A}{4.9}{GW190421A}{6.5}{GW190413B}{8.5}{GW190413A}{8.4}{GW190412A}{8.2}{GW190408A}{8.3}}}
\newcommand{\networkmatchedfiltersnrIMRminus}[1]{\IfEqCase{#1}{{GW190930A}{0.5}{GW190929A}{0.8}{GW190924A}{0.4}{GW190915A}{0.3}{GW190910A}{0.3}{GW190909A}{0.6}{GW190828B}{0.5}{GW190828A}{0.3}{GW190814A}{0.2}{GW190803A}{0.5}{GW190731A}{0.5}{GW190728A}{0.4}{GW190727A}{0.5}{GW190720A}{0.7}{GW190719A}{0.8}{GW190708A}{0.3}{GW190707A}{0.4}{GW190706A}{0.4}{GW190701A}{0.3}{GW190630A}{0.3}{GW190620A}{0.4}{GW190602A}{0.3}{GW190527A}{0.9}{GW190521B}{0.2}{GW190521A}{0.3}{GW190519A}{0.3}{GW190517A}{0.6}{GW190514A}{0.6}{GW190513A}{0.4}{GW190512A}{0.4}{GW190503A}{0.3}{GW190426A}{0.6}{GW190425A}{0.4}{GW190424A}{0.4}{GW190421A}{0.4}{GW190413B}{0.5}{GW190413A}{0.7}{GW190412A}{0.3}{GW190408A}{0.3}}}
\newcommand{\networkmatchedfiltersnrIMRmed}[1]{\IfEqCase{#1}{{GW190930A}{9.5}{GW190929A}{10.1}{GW190924A}{11.5}{GW190915A}{13.6}{GW190910A}{14.1}{GW190909A}{8.1}{GW190828B}{10.0}{GW190828A}{16.2}{GW190814A}{24.9}{GW190803A}{8.6}{GW190731A}{8.7}{GW190728A}{13.0}{GW190727A}{11.9}{GW190720A}{11.0}{GW190719A}{8.3}{GW190708A}{13.1}{GW190707A}{13.3}{GW190706A}{12.6}{GW190701A}{11.3}{GW190630A}{15.6}{GW190620A}{12.1}{GW190602A}{12.8}{GW190527A}{8.1}{GW190521B}{25.8}{GW190521A}{14.2}{GW190519A}{15.6}{GW190517A}{10.7}{GW190514A}{8.2}{GW190513A}{12.9}{GW190512A}{12.2}{GW190503A}{12.4}{GW190426A}{8.7}{GW190425A}{12.4}{GW190424A}{10.4}{GW190421A}{10.7}{GW190413B}{10.0}{GW190413A}{8.9}{GW190412A}{18.9}{GW190408A}{15.3}}}
\newcommand{\networkmatchedfiltersnrIMRplus}[1]{\IfEqCase{#1}{{GW190930A}{0.3}{GW190929A}{0.6}{GW190924A}{0.3}{GW190915A}{0.2}{GW190910A}{0.2}{GW190909A}{0.4}{GW190828B}{0.3}{GW190828A}{0.2}{GW190814A}{0.1}{GW190803A}{0.3}{GW190731A}{0.2}{GW190728A}{0.2}{GW190727A}{0.3}{GW190720A}{0.3}{GW190719A}{0.3}{GW190708A}{0.2}{GW190707A}{0.2}{GW190706A}{0.2}{GW190701A}{0.2}{GW190630A}{0.2}{GW190620A}{0.3}{GW190602A}{0.2}{GW190527A}{0.3}{GW190521B}{0.1}{GW190521A}{0.3}{GW190519A}{0.2}{GW190517A}{0.4}{GW190514A}{0.3}{GW190513A}{0.3}{GW190512A}{0.2}{GW190503A}{0.2}{GW190426A}{0.5}{GW190425A}{0.3}{GW190424A}{0.2}{GW190421A}{0.2}{GW190413B}{0.4}{GW190413A}{0.4}{GW190412A}{0.2}{GW190408A}{0.2}}}
\newcommand{\networkoptimalsnrIMRminus}[1]{\IfEqCase{#1}{{GW190930A}{1.7}{GW190929A}{1.8}{GW190924A}{1.7}{GW190915A}{1.7}{GW190910A}{1.7}{GW190909A}{1.8}{GW190828B}{1.7}{GW190828A}{1.7}{GW190814A}{1.7}{GW190803A}{1.8}{GW190731A}{1.8}{GW190728A}{1.7}{GW190727A}{1.7}{GW190720A}{1.8}{GW190719A}{1.8}{GW190708A}{1.7}{GW190707A}{1.7}{GW190706A}{1.7}{GW190701A}{1.7}{GW190630A}{1.7}{GW190620A}{1.7}{GW190602A}{1.7}{GW190527A}{1.8}{GW190521B}{1.6}{GW190521A}{1.7}{GW190519A}{1.6}{GW190517A}{1.7}{GW190514A}{1.8}{GW190513A}{1.7}{GW190512A}{1.7}{GW190503A}{1.7}{GW190426A}{1.8}{GW190425A}{1.7}{GW190424A}{1.7}{GW190421A}{1.7}{GW190413B}{1.7}{GW190413A}{1.8}{GW190412A}{1.7}{GW190408A}{1.7}}}
\newcommand{\networkoptimalsnrIMRmed}[1]{\IfEqCase{#1}{{GW190930A}{9.1}{GW190929A}{9.7}{GW190924A}{11.1}{GW190915A}{13.3}{GW190910A}{13.9}{GW190909A}{7.8}{GW190828B}{9.6}{GW190828A}{16.0}{GW190814A}{24.7}{GW190803A}{8.3}{GW190731A}{8.3}{GW190728A}{12.7}{GW190727A}{11.6}{GW190720A}{10.6}{GW190719A}{7.9}{GW190708A}{12.8}{GW190707A}{13.0}{GW190706A}{12.4}{GW190701A}{11.0}{GW190630A}{15.4}{GW190620A}{11.9}{GW190602A}{12.6}{GW190527A}{7.7}{GW190521B}{25.6}{GW190521A}{14.0}{GW190519A}{15.4}{GW190517A}{10.4}{GW190514A}{7.8}{GW190513A}{12.7}{GW190512A}{12.0}{GW190503A}{12.1}{GW190426A}{8.3}{GW190425A}{12.0}{GW190424A}{10.1}{GW190421A}{10.4}{GW190413B}{9.7}{GW190413A}{8.5}{GW190412A}{18.7}{GW190408A}{15.1}}}
\newcommand{\networkoptimalsnrIMRplus}[1]{\IfEqCase{#1}{{GW190930A}{1.7}{GW190929A}{1.8}{GW190924A}{1.7}{GW190915A}{1.7}{GW190910A}{1.7}{GW190909A}{1.7}{GW190828B}{1.7}{GW190828A}{1.7}{GW190814A}{1.6}{GW190803A}{1.7}{GW190731A}{1.8}{GW190728A}{1.7}{GW190727A}{1.7}{GW190720A}{1.7}{GW190719A}{1.7}{GW190708A}{1.7}{GW190707A}{1.7}{GW190706A}{1.7}{GW190701A}{1.7}{GW190630A}{1.7}{GW190620A}{1.7}{GW190602A}{1.7}{GW190527A}{1.8}{GW190521B}{1.7}{GW190521A}{1.7}{GW190519A}{1.7}{GW190517A}{1.7}{GW190514A}{1.7}{GW190513A}{1.7}{GW190512A}{1.7}{GW190503A}{1.7}{GW190426A}{1.8}{GW190425A}{1.7}{GW190424A}{1.7}{GW190421A}{1.7}{GW190413B}{1.7}{GW190413A}{1.8}{GW190412A}{1.7}{GW190408A}{1.7}}}
\newcommand{\phaseIMRminus}[1]{\IfEqCase{#1}{{GW190930A}{2.97}{GW190929A}{2.87}{GW190924A}{2.80}{GW190915A}{3.06}{GW190910A}{2.76}{GW190909A}{2.73}{GW190828B}{2.82}{GW190828A}{2.80}{GW190814A}{3.42}{GW190803A}{2.89}{GW190731A}{2.91}{GW190728A}{2.83}{GW190727A}{2.90}{GW190720A}{2.76}{GW190719A}{2.85}{GW190708A}{2.82}{GW190707A}{2.97}{GW190706A}{2.82}{GW190701A}{2.86}{GW190630A}{2.78}{GW190620A}{2.91}{GW190602A}{2.78}{GW190527A}{2.66}{GW190521B}{2.87}{GW190521A}{3.03}{GW190519A}{2.70}{GW190517A}{2.73}{GW190514A}{2.70}{GW190513A}{2.68}{GW190512A}{2.73}{GW190503A}{2.85}{GW190426A}{2.79}{GW190425A}{2.82}{GW190424A}{2.84}{GW190421A}{2.73}{GW190413B}{2.97}{GW190413A}{2.81}{GW190412A}{1.37}{GW190408A}{2.92}}}
\newcommand{\phaseIMRmed}[1]{\IfEqCase{#1}{{GW190930A}{3.29}{GW190929A}{3.16}{GW190924A}{3.10}{GW190915A}{3.44}{GW190910A}{3.02}{GW190909A}{3.05}{GW190828B}{3.15}{GW190828A}{3.13}{GW190814A}{3.61}{GW190803A}{3.19}{GW190731A}{3.28}{GW190728A}{3.14}{GW190727A}{3.23}{GW190720A}{3.07}{GW190719A}{3.20}{GW190708A}{3.09}{GW190707A}{3.33}{GW190706A}{3.09}{GW190701A}{3.18}{GW190630A}{3.06}{GW190620A}{3.21}{GW190602A}{3.09}{GW190527A}{2.97}{GW190521B}{3.23}{GW190521A}{3.37}{GW190519A}{3.02}{GW190517A}{3.03}{GW190514A}{3.01}{GW190513A}{2.96}{GW190512A}{3.03}{GW190503A}{3.16}{GW190426A}{3.09}{GW190425A}{3.12}{GW190424A}{3.12}{GW190421A}{3.03}{GW190413B}{3.29}{GW190413A}{3.14}{GW190412A}{1.62}{GW190408A}{3.26}}}
\newcommand{\phaseIMRplus}[1]{\IfEqCase{#1}{{GW190930A}{2.71}{GW190929A}{2.82}{GW190924A}{2.85}{GW190915A}{2.58}{GW190910A}{2.94}{GW190909A}{2.91}{GW190828B}{2.82}{GW190828A}{2.83}{GW190814A}{2.48}{GW190803A}{2.78}{GW190731A}{2.71}{GW190728A}{2.82}{GW190727A}{2.74}{GW190720A}{2.91}{GW190719A}{2.74}{GW190708A}{2.89}{GW190707A}{2.65}{GW190706A}{2.84}{GW190701A}{2.80}{GW190630A}{2.91}{GW190620A}{2.77}{GW190602A}{2.83}{GW190527A}{2.98}{GW190521B}{2.78}{GW190521A}{2.59}{GW190519A}{2.91}{GW190517A}{2.90}{GW190514A}{2.93}{GW190513A}{3.00}{GW190512A}{2.95}{GW190503A}{2.81}{GW190426A}{2.85}{GW190425A}{2.87}{GW190424A}{2.85}{GW190421A}{2.86}{GW190413B}{2.69}{GW190413A}{2.81}{GW190412A}{4.32}{GW190408A}{2.75}}}
\newcommand{\phionetwoIMRminus}[1]{\IfEqCase{#1}{{GW190930A}{2.78}{GW190929A}{2.83}{GW190924A}{2.84}{GW190915A}{2.82}{GW190910A}{2.84}{GW190909A}{2.80}{GW190828B}{2.71}{GW190828A}{2.74}{GW190814A}{2.79}{GW190803A}{2.73}{GW190731A}{2.85}{GW190728A}{2.84}{GW190727A}{2.76}{GW190720A}{2.60}{GW190719A}{2.85}{GW190708A}{2.83}{GW190707A}{2.85}{GW190706A}{2.86}{GW190701A}{2.82}{GW190630A}{2.67}{GW190620A}{2.91}{GW190602A}{2.84}{GW190527A}{2.96}{GW190521B}{2.74}{GW190521A}{3.06}{GW190519A}{2.78}{GW190517A}{2.81}{GW190514A}{2.74}{GW190513A}{3.07}{GW190512A}{2.83}{GW190503A}{2.95}{GW190426A}{0.00}{GW190425A}{2.87}{GW190424A}{2.70}{GW190421A}{2.92}{GW190413B}{2.85}{GW190413A}{2.71}{GW190412A}{3.11}{GW190408A}{2.80}}}
\newcommand{\phionetwoIMRmed}[1]{\IfEqCase{#1}{{GW190930A}{3.12}{GW190929A}{3.14}{GW190924A}{3.15}{GW190915A}{3.11}{GW190910A}{3.14}{GW190909A}{3.15}{GW190828B}{3.00}{GW190828A}{3.05}{GW190814A}{3.10}{GW190803A}{3.05}{GW190731A}{3.15}{GW190728A}{3.16}{GW190727A}{3.10}{GW190720A}{2.95}{GW190719A}{3.18}{GW190708A}{3.20}{GW190707A}{3.15}{GW190706A}{3.18}{GW190701A}{3.11}{GW190630A}{2.98}{GW190620A}{3.23}{GW190602A}{3.15}{GW190527A}{3.29}{GW190521B}{3.04}{GW190521A}{3.34}{GW190519A}{3.11}{GW190517A}{3.11}{GW190514A}{3.05}{GW190513A}{3.43}{GW190512A}{3.13}{GW190503A}{3.31}{GW190426A}{0.00}{GW190425A}{3.18}{GW190424A}{3.05}{GW190421A}{3.24}{GW190413B}{3.19}{GW190413A}{3.02}{GW190412A}{3.43}{GW190408A}{3.08}}}
\newcommand{\phionetwoIMRplus}[1]{\IfEqCase{#1}{{GW190930A}{2.85}{GW190929A}{2.83}{GW190924A}{2.82}{GW190915A}{2.88}{GW190910A}{2.82}{GW190909A}{2.82}{GW190828B}{2.93}{GW190828A}{2.86}{GW190814A}{2.87}{GW190803A}{2.91}{GW190731A}{2.81}{GW190728A}{2.82}{GW190727A}{2.85}{GW190720A}{2.99}{GW190719A}{2.81}{GW190708A}{2.75}{GW190707A}{2.83}{GW190706A}{2.84}{GW190701A}{2.85}{GW190630A}{3.00}{GW190620A}{2.74}{GW190602A}{2.82}{GW190527A}{2.69}{GW190521B}{2.94}{GW190521A}{2.65}{GW190519A}{2.83}{GW190517A}{2.85}{GW190514A}{2.90}{GW190513A}{2.56}{GW190512A}{2.86}{GW190503A}{2.65}{GW190426A}{0.00}{GW190425A}{2.76}{GW190424A}{2.93}{GW190421A}{2.74}{GW190413B}{2.78}{GW190413A}{2.93}{GW190412A}{2.56}{GW190408A}{2.88}}}
\newcommand{\phijlIMRminus}[1]{\IfEqCase{#1}{{GW190930A}{2.93}{GW190929A}{3.18}{GW190924A}{2.75}{GW190915A}{2.52}{GW190910A}{3.01}{GW190909A}{2.86}{GW190828B}{2.70}{GW190828A}{2.89}{GW190814A}{1.88}{GW190803A}{2.86}{GW190731A}{2.79}{GW190728A}{2.94}{GW190727A}{2.82}{GW190720A}{2.65}{GW190719A}{2.93}{GW190708A}{2.89}{GW190707A}{2.92}{GW190706A}{2.89}{GW190701A}{4.08}{GW190630A}{2.47}{GW190620A}{3.04}{GW190602A}{2.52}{GW190527A}{2.75}{GW190521B}{3.17}{GW190521A}{3.22}{GW190519A}{2.77}{GW190517A}{1.84}{GW190514A}{2.85}{GW190513A}{2.33}{GW190512A}{2.74}{GW190503A}{2.95}{GW190426A}{1.43}{GW190425A}{2.88}{GW190424A}{2.78}{GW190421A}{2.56}{GW190413B}{2.94}{GW190413A}{2.81}{GW190412A}{1.07}{GW190408A}{3.05}}}
\newcommand{\phijlIMRmed}[1]{\IfEqCase{#1}{{GW190930A}{3.26}{GW190929A}{3.42}{GW190924A}{3.06}{GW190915A}{3.39}{GW190910A}{3.33}{GW190909A}{3.26}{GW190828B}{3.03}{GW190828A}{3.16}{GW190814A}{2.29}{GW190803A}{3.19}{GW190731A}{3.17}{GW190728A}{3.28}{GW190727A}{3.15}{GW190720A}{2.95}{GW190719A}{3.22}{GW190708A}{3.19}{GW190707A}{3.24}{GW190706A}{3.31}{GW190701A}{4.47}{GW190630A}{2.70}{GW190620A}{3.30}{GW190602A}{2.97}{GW190527A}{3.02}{GW190521B}{3.54}{GW190521A}{3.58}{GW190519A}{3.02}{GW190517A}{2.26}{GW190514A}{3.17}{GW190513A}{2.88}{GW190512A}{3.06}{GW190503A}{3.13}{GW190426A}{1.70}{GW190425A}{3.23}{GW190424A}{3.11}{GW190421A}{3.02}{GW190413B}{3.22}{GW190413A}{3.13}{GW190412A}{1.15}{GW190408A}{3.33}}}
\newcommand{\phijlIMRplus}[1]{\IfEqCase{#1}{{GW190930A}{2.75}{GW190929A}{2.58}{GW190924A}{2.90}{GW190915A}{2.27}{GW190910A}{2.64}{GW190909A}{2.64}{GW190828B}{2.93}{GW190828A}{2.80}{GW190814A}{3.42}{GW190803A}{2.80}{GW190731A}{2.74}{GW190728A}{2.69}{GW190727A}{2.82}{GW190720A}{3.01}{GW190719A}{2.77}{GW190708A}{2.79}{GW190707A}{2.75}{GW190706A}{2.63}{GW190701A}{1.54}{GW190630A}{3.30}{GW190620A}{2.81}{GW190602A}{2.91}{GW190527A}{3.00}{GW190521B}{2.26}{GW190521A}{2.34}{GW190519A}{2.97}{GW190517A}{3.41}{GW190514A}{2.75}{GW190513A}{3.00}{GW190512A}{2.82}{GW190503A}{2.99}{GW190426A}{1.19}{GW190425A}{2.76}{GW190424A}{2.82}{GW190421A}{2.82}{GW190413B}{2.72}{GW190413A}{2.83}{GW190412A}{5.01}{GW190408A}{2.68}}}
\newcommand{\massratioIMRminus}[1]{\IfEqCase{#1}{{GW190930A}{0.46}{GW190929A}{0.14}{GW190924A}{0.36}{GW190915A}{0.24}{GW190910A}{0.25}{GW190909A}{0.37}{GW190828B}{0.20}{GW190828A}{0.26}{GW190814A}{0.010}{GW190803A}{0.32}{GW190731A}{0.33}{GW190728A}{0.41}{GW190727A}{0.35}{GW190720A}{0.30}{GW190719A}{0.26}{GW190708A}{0.30}{GW190707A}{0.29}{GW190706A}{0.33}{GW190701A}{0.30}{GW190630A}{0.19}{GW190620A}{0.25}{GW190602A}{0.29}{GW190527A}{0.32}{GW190521B}{0.22}{GW190521A}{0.29}{GW190519A}{0.29}{GW190517A}{0.36}{GW190514A}{0.34}{GW190513A}{0.18}{GW190512A}{0.24}{GW190503A}{0.29}{GW190426A}{0.15}{GW190425A}{0.25}{GW190424A}{0.30}{GW190421A}{0.31}{GW190413B}{0.35}{GW190413A}{0.30}{GW190412A}{0.08}{GW190408A}{0.26}}}
\newcommand{\massratioIMRmed}[1]{\IfEqCase{#1}{{GW190930A}{0.61}{GW190929A}{0.27}{GW190924A}{0.51}{GW190915A}{0.65}{GW190910A}{0.80}{GW190909A}{0.66}{GW190828B}{0.47}{GW190828A}{0.80}{GW190814A}{0.111}{GW190803A}{0.73}{GW190731A}{0.71}{GW190728A}{0.66}{GW190727A}{0.74}{GW190720A}{0.57}{GW190719A}{0.48}{GW190708A}{0.70}{GW190707A}{0.72}{GW190706A}{0.62}{GW190701A}{0.71}{GW190630A}{0.60}{GW190620A}{0.56}{GW190602A}{0.54}{GW190527A}{0.53}{GW190521B}{0.74}{GW190521A}{0.72}{GW190519A}{0.64}{GW190517A}{0.67}{GW190514A}{0.75}{GW190513A}{0.48}{GW190512A}{0.57}{GW190503A}{0.67}{GW190426A}{0.25}{GW190425A}{0.67}{GW190424A}{0.77}{GW190421A}{0.72}{GW190413B}{0.64}{GW190413A}{0.74}{GW190412A}{0.31}{GW190408A}{0.73}}}
\newcommand{\massratioIMRplus}[1]{\IfEqCase{#1}{{GW190930A}{0.34}{GW190929A}{0.48}{GW190924A}{0.42}{GW190915A}{0.30}{GW190910A}{0.18}{GW190909A}{0.30}{GW190828B}{0.41}{GW190828A}{0.18}{GW190814A}{0.009}{GW190803A}{0.24}{GW190731A}{0.26}{GW190728A}{0.31}{GW190727A}{0.23}{GW190720A}{0.39}{GW190719A}{0.44}{GW190708A}{0.26}{GW190707A}{0.25}{GW190706A}{0.34}{GW190701A}{0.25}{GW190630A}{0.34}{GW190620A}{0.37}{GW190602A}{0.40}{GW190527A}{0.40}{GW190521B}{0.23}{GW190521A}{0.25}{GW190519A}{0.32}{GW190517A}{0.29}{GW190514A}{0.22}{GW190513A}{0.43}{GW190512A}{0.37}{GW190503A}{0.29}{GW190426A}{0.41}{GW190425A}{0.29}{GW190424A}{0.20}{GW190421A}{0.25}{GW190413B}{0.32}{GW190413A}{0.23}{GW190412A}{0.12}{GW190408A}{0.24}}}
\newcommand{\geocenttimeIMRminus}[1]{\IfEqCase{#1}{{GW190930A}{0.009}{GW190929A}{0.04}{GW190924A}{0.004}{GW190915A}{0.004}{GW190910A}{0.04}{GW190909A}{0.03}{GW190828B}{0.002}{GW190828A}{0.002}{GW190814A}{0.003}{GW190803A}{0.009}{GW190731A}{0.009}{GW190728A}{0.04}{GW190727A}{0.004}{GW190720A}{0.02}{GW190719A}{0.03}{GW190708A}{0.04}{GW190707A}{0.01}{GW190706A}{0.04}{GW190701A}{0.003}{GW190630A}{0.005}{GW190620A}{0.04}{GW190602A}{0.006}{GW190527A}{0.02}{GW190521B}{0.007}{GW190521A}{0.01}{GW190519A}{0.04}{GW190517A}{0.02}{GW190514A}{0.04}{GW190513A}{0.002}{GW190512A}{0.001}{GW190503A}{0.002}{GW190426A}{0.03}{GW190425A}{0.009}{GW190424A}{0.02}{GW190421A}{0.003}{GW190413B}{0.004}{GW190413A}{0.006}{GW190412A}{0.005}{GW190408A}{0.004}}}
\newcommand{\geocenttimeIMRmed}[1]{\IfEqCase{#1}{{GW190930A}{1253885759.2}{GW190929A}{1253755327.5}{GW190924A}{1253326744.8}{GW190915A}{1252627040.7}{GW190910A}{1252150105.3}{GW190909A}{1252064527.7}{GW190828B}{1251010527.9}{GW190828A}{1251009263.7}{GW190814A}{1249852257.0}{GW190803A}{1248834439.9}{GW190731A}{1248617394.6}{GW190728A}{1248331528.5}{GW190727A}{1248242632.0}{GW190720A}{1247616534.7}{GW190719A}{1247608532.9}{GW190708A}{1246663515.4}{GW190707A}{1246527224.2}{GW190706A}{1246487219.4}{GW190701A}{1246048404.6}{GW190630A}{1245955943.2}{GW190620A}{1245035079.3}{GW190602A}{1243533585.1}{GW190527A}{1242984073.8}{GW190521B}{1242459857.5}{GW190521A}{1242442967.4}{GW190519A}{1242315362.4}{GW190517A}{1242107479.8}{GW190514A}{1241852074.9}{GW190513A}{1241816086.7}{GW190512A}{1241719652.4}{GW190503A}{1240944862.3}{GW190426A}{1240327333.4}{GW190425A}{1240215503.0}{GW190424A}{1240164426.1}{GW190421A}{1239917954.2}{GW190413B}{1239198206.7}{GW190413A}{1239168612.5}{GW190412A}{1239082262.2}{GW190408A}{1238782700.3}}}
\newcommand{\geocenttimeIMRplus}[1]{\IfEqCase{#1}{{GW190930A}{0.01}{GW190929A}{0.02}{GW190924A}{0.004}{GW190915A}{0.002}{GW190910A}{0.007}{GW190909A}{0.010}{GW190828B}{0.04}{GW190828A}{0.04}{GW190814A}{0.003}{GW190803A}{0.02}{GW190731A}{0.03}{GW190728A}{0.001}{GW190727A}{0.04}{GW190720A}{0.008}{GW190719A}{0.01}{GW190708A}{0.004}{GW190707A}{0.03}{GW190706A}{0.006}{GW190701A}{0.003}{GW190630A}{0.03}{GW190620A}{0.006}{GW190602A}{0.009}{GW190527A}{0.02}{GW190521B}{0.002}{GW190521A}{0.04}{GW190519A}{0.007}{GW190517A}{0.01}{GW190514A}{0.005}{GW190513A}{0.04}{GW190512A}{0.005}{GW190503A}{0.003}{GW190426A}{0.02}{GW190425A}{0.03}{GW190424A}{0.02}{GW190421A}{0.009}{GW190413B}{0.04}{GW190413A}{0.03}{GW190412A}{0.005}{GW190408A}{0.001}}}
\newcommand{\raIMRminus}[1]{\IfEqCase{#1}{{GW190930A}{5.30187}{GW190929A}{2.90716}{GW190924A}{0.11450}{GW190915A}{0.09186}{GW190910A}{1.32977}{GW190909A}{1.44176}{GW190828B}{0.24432}{GW190828A}{0.11562}{GW190814A}{0.03020}{GW190803A}{0.30342}{GW190731A}{1.62088}{GW190728A}{3.94631}{GW190727A}{1.00546}{GW190720A}{4.69082}{GW190719A}{2.52616}{GW190708A}{2.56101}{GW190707A}{2.80344}{GW190706A}{0.12785}{GW190701A}{0.02858}{GW190630A}{3.00962}{GW190620A}{3.89597}{GW190602A}{0.16106}{GW190527A}{4.87431}{GW190521B}{0.65316}{GW190521A}{3.53306}{GW190519A}{3.22456}{GW190517A}{0.09560}{GW190514A}{2.74884}{GW190513A}{4.24807}{GW190512A}{0.45841}{GW190503A}{0.06967}{GW190426A}{5.11703}{GW190425A}{1.14713}{GW190424A}{2.92902}{GW190421A}{1.98508}{GW190413B}{0.51111}{GW190413A}{0.86077}{GW190412A}{0.36956}{GW190408A}{5.78710}}}
\newcommand{\raIMRmed}[1]{\IfEqCase{#1}{{GW190930A}{5.56626}{GW190929A}{4.52976}{GW190924A}{2.23783}{GW190915A}{3.41447}{GW190910A}{1.88063}{GW190909A}{1.80555}{GW190828B}{2.45490}{GW190828A}{2.48782}{GW190814A}{0.22046}{GW190803A}{1.62682}{GW190731A}{2.62361}{GW190728A}{5.48446}{GW190727A}{2.50075}{GW190720A}{5.19392}{GW190719A}{2.85941}{GW190708A}{3.04830}{GW190707A}{3.69054}{GW190706A}{2.59840}{GW190701A}{0.66316}{GW190630A}{5.87324}{GW190620A}{4.31970}{GW190602A}{1.29270}{GW190527A}{5.16266}{GW190521B}{5.00715}{GW190521A}{3.56762}{GW190519A}{3.26337}{GW190517A}{4.05235}{GW190514A}{3.51453}{GW190513A}{4.99127}{GW190512A}{4.37724}{GW190503A}{1.65256}{GW190426A}{5.27391}{GW190425A}{1.62833}{GW190424A}{3.15377}{GW190421A}{3.46633}{GW190413B}{2.71115}{GW190413A}{1.17004}{GW190412A}{3.81292}{GW190408A}{5.99953}}}
\newcommand{\raIMRplus}[1]{\IfEqCase{#1}{{GW190930A}{0.49890}{GW190929A}{0.95496}{GW190924A}{0.21465}{GW190915A}{0.07280}{GW190910A}{2.90915}{GW190909A}{4.03198}{GW190828B}{3.50244}{GW190828A}{3.27136}{GW190814A}{0.17455}{GW190803A}{1.69156}{GW190731A}{0.76888}{GW190728A}{0.71196}{GW190727A}{3.68550}{GW190720A}{0.71951}{GW190719A}{3.21725}{GW190708A}{2.86189}{GW190707A}{2.10330}{GW190706A}{3.42235}{GW190701A}{0.03131}{GW190630A}{0.11225}{GW190620A}{0.28935}{GW190602A}{0.29864}{GW190527A}{0.82398}{GW190521B}{0.58797}{GW190521A}{2.68052}{GW190519A}{2.97976}{GW190517A}{1.58008}{GW190514A}{1.87892}{GW190513A}{0.11611}{GW190512A}{0.57197}{GW190503A}{0.07425}{GW190426A}{0.85181}{GW190425A}{3.12911}{GW190424A}{2.78982}{GW190421A}{0.17619}{GW190413B}{2.10904}{GW190413A}{2.27912}{GW190412A}{0.03668}{GW190408A}{0.15567}}}
\newcommand{\decIMRminus}[1]{\IfEqCase{#1}{{GW190930A}{0.69744}{GW190929A}{1.03996}{GW190924A}{0.37638}{GW190915A}{0.46141}{GW190910A}{0.98249}{GW190909A}{1.21610}{GW190828B}{0.48258}{GW190828A}{0.47136}{GW190814A}{0.13140}{GW190803A}{0.61258}{GW190731A}{0.76462}{GW190728A}{1.45373}{GW190727A}{0.21353}{GW190720A}{1.54766}{GW190719A}{1.34010}{GW190708A}{1.27055}{GW190707A}{0.78951}{GW190706A}{1.26217}{GW190701A}{0.08511}{GW190630A}{0.81825}{GW190620A}{1.14171}{GW190602A}{0.24139}{GW190527A}{0.73358}{GW190521B}{0.80120}{GW190521A}{0.31872}{GW190519A}{0.89619}{GW190517A}{0.21140}{GW190514A}{1.34361}{GW190513A}{0.06842}{GW190512A}{0.15610}{GW190503A}{0.08094}{GW190426A}{1.53579}{GW190425A}{0.89977}{GW190424A}{1.12293}{GW190421A}{0.57456}{GW190413B}{0.19980}{GW190413A}{0.08405}{GW190412A}{0.25558}{GW190408A}{0.61019}}}
\newcommand{\decIMRmed}[1]{\IfEqCase{#1}{{GW190930A}{0.65524}{GW190929A}{0.08602}{GW190924A}{0.24165}{GW190915A}{0.63294}{GW190910A}{0.05486}{GW190909A}{0.28574}{GW190828B}{-0.64372}{GW190828A}{-0.39759}{GW190814A}{-0.43687}{GW190803A}{0.40407}{GW190731A}{-0.63750}{GW190728A}{0.16190}{GW190727A}{-1.03131}{GW190720A}{0.62896}{GW190719A}{0.34485}{GW190708A}{0.32622}{GW190707A}{-0.14200}{GW190706A}{0.46456}{GW190701A}{-0.11203}{GW190630A}{-0.12269}{GW190620A}{0.33303}{GW190602A}{-0.58336}{GW190527A}{-0.57177}{GW190521B}{0.40716}{GW190521A}{-0.88872}{GW190519A}{0.24979}{GW190517A}{-0.76968}{GW190514A}{0.62000}{GW190513A}{-0.49061}{GW190512A}{-0.46654}{GW190503A}{-0.87887}{GW190426A}{0.90282}{GW190425A}{-0.13006}{GW190424A}{0.01445}{GW190421A}{-0.82553}{GW190413B}{-0.52708}{GW190413A}{-0.74413}{GW190412A}{0.63171}{GW190408A}{0.84614}}}
\newcommand{\decIMRplus}[1]{\IfEqCase{#1}{{GW190930A}{0.45251}{GW190929A}{0.94880}{GW190924A}{0.20954}{GW190915A}{0.51929}{GW190910A}{0.90077}{GW190909A}{0.95134}{GW190828B}{1.27320}{GW190828A}{1.21142}{GW190814A}{0.03517}{GW190803A}{0.78424}{GW190731A}{1.04803}{GW190728A}{0.44605}{GW190727A}{1.95957}{GW190720A}{0.04883}{GW190719A}{0.83400}{GW190708A}{0.86955}{GW190707A}{1.35253}{GW190706A}{0.48044}{GW190701A}{0.09559}{GW190630A}{0.67598}{GW190620A}{0.82034}{GW190602A}{0.59155}{GW190527A}{0.93448}{GW190521B}{0.19354}{GW190521A}{1.59538}{GW190519A}{0.76745}{GW190517A}{1.17654}{GW190514A}{0.76564}{GW190513A}{1.48376}{GW190512A}{0.48829}{GW190503A}{0.11333}{GW190426A}{0.61722}{GW190425A}{0.96811}{GW190424A}{1.09684}{GW190421A}{0.64278}{GW190413B}{1.84696}{GW190413A}{2.10974}{GW190412A}{0.02860}{GW190408A}{0.35802}}}
\newcommand{\luminositydistanceIMRminus}[1]{\IfEqCase{#1}{{GW190930A}{0.32}{GW190929A}{0.82}{GW190924A}{0.21}{GW190915A}{0.59}{GW190910A}{0.83}{GW190909A}{2.07}{GW190828B}{0.70}{GW190828A}{0.86}{GW190814A}{0.04}{GW190803A}{1.42}{GW190731A}{1.59}{GW190728A}{0.38}{GW190727A}{1.15}{GW190720A}{0.29}{GW190719A}{1.91}{GW190708A}{0.38}{GW190707A}{0.34}{GW190706A}{2.12}{GW190701A}{0.72}{GW190630A}{0.40}{GW190620A}{1.29}{GW190602A}{1.16}{GW190527A}{1.12}{GW190521B}{0.49}{GW190521A}{1.80}{GW190519A}{1.43}{GW190517A}{0.81}{GW190514A}{1.98}{GW190513A}{0.81}{GW190512A}{0.60}{GW190503A}{0.61}{GW190426A}{0.16}{GW190425A}{0.07}{GW190424A}{1.09}{GW190421A}{1.12}{GW190413B}{1.82}{GW190413A}{1.29}{GW190412A}{0.21}{GW190408A}{0.56}}}
\newcommand{\luminositydistanceIMRmed}[1]{\IfEqCase{#1}{{GW190930A}{0.76}{GW190929A}{1.85}{GW190924A}{0.56}{GW190915A}{1.60}{GW190910A}{1.80}{GW190909A}{3.60}{GW190828B}{1.67}{GW190828A}{2.03}{GW190814A}{0.25}{GW190803A}{2.96}{GW190731A}{3.11}{GW190728A}{0.88}{GW190727A}{2.75}{GW190720A}{0.74}{GW190719A}{3.41}{GW190708A}{0.84}{GW190707A}{0.73}{GW190706A}{4.56}{GW190701A}{1.87}{GW190630A}{0.93}{GW190620A}{2.63}{GW190602A}{2.33}{GW190527A}{2.24}{GW190521B}{1.13}{GW190521A}{4.25}{GW190519A}{3.45}{GW190517A}{1.83}{GW190514A}{3.68}{GW190513A}{1.92}{GW190512A}{1.42}{GW190503A}{1.46}{GW190426A}{0.37}{GW190425A}{0.16}{GW190424A}{1.96}{GW190421A}{2.43}{GW190413B}{3.69}{GW190413A}{3.15}{GW190412A}{0.74}{GW190408A}{1.40}}}
\newcommand{\luminositydistanceIMRplus}[1]{\IfEqCase{#1}{{GW190930A}{0.34}{GW190929A}{3.06}{GW190924A}{0.21}{GW190915A}{0.72}{GW190910A}{0.88}{GW190909A}{3.12}{GW190828B}{0.65}{GW190828A}{0.62}{GW190814A}{0.04}{GW190803A}{1.75}{GW190731A}{2.30}{GW190728A}{0.24}{GW190727A}{1.27}{GW190720A}{0.55}{GW190719A}{2.88}{GW190708A}{0.34}{GW190707A}{0.37}{GW190706A}{2.36}{GW190701A}{0.78}{GW190630A}{0.49}{GW190620A}{1.64}{GW190602A}{1.75}{GW190527A}{2.30}{GW190521B}{0.39}{GW190521A}{1.62}{GW190519A}{1.70}{GW190517A}{1.52}{GW190514A}{2.65}{GW190513A}{0.73}{GW190512A}{0.54}{GW190503A}{0.61}{GW190426A}{0.18}{GW190425A}{0.07}{GW190424A}{1.50}{GW190421A}{1.41}{GW190413B}{2.40}{GW190413A}{1.93}{GW190412A}{0.15}{GW190408A}{0.44}}}
\newcommand{\psiIMRminus}[1]{\IfEqCase{#1}{{GW190930A}{1.36}{GW190929A}{1.44}{GW190924A}{1.39}{GW190915A}{1.40}{GW190910A}{1.41}{GW190909A}{1.46}{GW190828B}{1.47}{GW190828A}{1.36}{GW190814A}{0.23}{GW190803A}{1.47}{GW190731A}{1.43}{GW190728A}{1.41}{GW190727A}{1.36}{GW190720A}{1.38}{GW190719A}{1.37}{GW190708A}{1.43}{GW190707A}{1.33}{GW190706A}{1.48}{GW190701A}{1.43}{GW190630A}{1.47}{GW190620A}{1.42}{GW190602A}{1.46}{GW190527A}{1.44}{GW190521B}{1.44}{GW190521A}{1.00}{GW190519A}{1.45}{GW190517A}{1.46}{GW190514A}{1.44}{GW190513A}{1.27}{GW190512A}{1.45}{GW190503A}{1.44}{GW190426A}{1.40}{GW190425A}{1.46}{GW190424A}{1.38}{GW190421A}{1.42}{GW190413B}{1.44}{GW190413A}{1.46}{GW190412A}{2.20}{GW190408A}{1.38}}}
\newcommand{\psiIMRmed}[1]{\IfEqCase{#1}{{GW190930A}{1.51}{GW190929A}{1.62}{GW190924A}{1.55}{GW190915A}{1.55}{GW190910A}{1.55}{GW190909A}{1.61}{GW190828B}{1.63}{GW190828A}{1.51}{GW190814A}{0.27}{GW190803A}{1.63}{GW190731A}{1.60}{GW190728A}{1.57}{GW190727A}{1.49}{GW190720A}{1.52}{GW190719A}{1.56}{GW190708A}{1.55}{GW190707A}{1.50}{GW190706A}{1.67}{GW190701A}{1.58}{GW190630A}{1.59}{GW190620A}{1.55}{GW190602A}{1.62}{GW190527A}{1.61}{GW190521B}{1.61}{GW190521A}{1.11}{GW190519A}{1.60}{GW190517A}{1.67}{GW190514A}{1.61}{GW190513A}{1.41}{GW190512A}{1.54}{GW190503A}{1.61}{GW190426A}{1.58}{GW190425A}{1.62}{GW190424A}{1.51}{GW190421A}{1.56}{GW190413B}{1.60}{GW190413A}{1.63}{GW190412A}{2.30}{GW190408A}{1.59}}}
\newcommand{\psiIMRplus}[1]{\IfEqCase{#1}{{GW190930A}{1.44}{GW190929A}{1.35}{GW190924A}{1.41}{GW190915A}{1.40}{GW190910A}{1.43}{GW190909A}{1.38}{GW190828B}{1.32}{GW190828A}{1.47}{GW190814A}{2.81}{GW190803A}{1.34}{GW190731A}{1.40}{GW190728A}{1.42}{GW190727A}{1.47}{GW190720A}{1.46}{GW190719A}{1.40}{GW190708A}{1.43}{GW190707A}{1.45}{GW190706A}{1.29}{GW190701A}{1.42}{GW190630A}{1.46}{GW190620A}{1.45}{GW190602A}{1.39}{GW190527A}{1.38}{GW190521B}{1.38}{GW190521A}{1.91}{GW190519A}{1.40}{GW190517A}{1.28}{GW190514A}{1.37}{GW190513A}{1.52}{GW190512A}{1.47}{GW190503A}{1.36}{GW190426A}{1.36}{GW190425A}{1.38}{GW190424A}{1.46}{GW190421A}{1.41}{GW190413B}{1.38}{GW190413A}{1.36}{GW190412A}{0.72}{GW190408A}{1.37}}}
\newcommand{\chirpmassdetIMRminus}[1]{\IfEqCase{#1}{{GW190930A}{0.3}{GW190929A}{13.9}{GW190924A}{0.03}{GW190915A}{3.6}{GW190910A}{3.9}{GW190909A}{11.1}{GW190828B}{0.8}{GW190828A}{2.6}{GW190814A}{0.02}{GW190803A}{6.5}{GW190731A}{8.6}{GW190728A}{0.08}{GW190727A}{7.0}{GW190720A}{0.1}{GW190719A}{7.0}{GW190708A}{0.3}{GW190707A}{0.08}{GW190706A}{17.0}{GW190701A}{8.7}{GW190630A}{1.6}{GW190620A}{12.0}{GW190602A}{19.3}{GW190527A}{6.5}{GW190521B}{2.7}{GW190521A}{19.2}{GW190519A}{11.7}{GW190517A}{4.2}{GW190514A}{8.1}{GW190513A}{2.2}{GW190512A}{0.9}{GW190503A}{6.7}{GW190426A}{0.01}{GW190425A}{0.0006}{GW190424A}{5.5}{GW190421A}{7.0}{GW190413B}{13.8}{GW190413A}{6.0}{GW190412A}{0.2}{GW190408A}{1.7}}}
\newcommand{\chirpmassdetIMRmed}[1]{\IfEqCase{#1}{{GW190930A}{9.8}{GW190929A}{49.5}{GW190924A}{6.44}{GW190915A}{33.2}{GW190910A}{43.7}{GW190909A}{47.6}{GW190828B}{17.3}{GW190828A}{33.9}{GW190814A}{6.41}{GW190803A}{42.8}{GW190731A}{47.0}{GW190728A}{10.1}{GW190727A}{44.7}{GW190720A}{10.4}{GW190719A}{38.0}{GW190708A}{15.4}{GW190707A}{9.87}{GW190706A}{77.3}{GW190701A}{55.1}{GW190630A}{28.9}{GW190620A}{56.6}{GW190602A}{69.4}{GW190527A}{34.1}{GW190521B}{39.0}{GW190521A}{112.5}{GW190519A}{67.4}{GW190517A}{35.9}{GW190514A}{48.7}{GW190513A}{28.7}{GW190512A}{18.5}{GW190503A}{38.4}{GW190426A}{2.60}{GW190425A}{1.49}{GW190424A}{43.0}{GW190421A}{45.7}{GW190413B}{57.0}{GW190413A}{39.0}{GW190412A}{15.2}{GW190408A}{23.7}}}
\newcommand{\chirpmassdetIMRplus}[1]{\IfEqCase{#1}{{GW190930A}{0.2}{GW190929A}{17.7}{GW190924A}{0.06}{GW190915A}{3.3}{GW190910A}{4.0}{GW190909A}{14.4}{GW190828B}{0.7}{GW190828A}{2.8}{GW190814A}{0.02}{GW190803A}{6.2}{GW190731A}{7.1}{GW190728A}{0.09}{GW190727A}{5.3}{GW190720A}{0.1}{GW190719A}{30.6}{GW190708A}{0.3}{GW190707A}{0.1}{GW190706A}{9.5}{GW190701A}{7.6}{GW190630A}{2.1}{GW190620A}{8.9}{GW190602A}{12.9}{GW190527A}{26.9}{GW190521B}{2.6}{GW190521A}{15.3}{GW190519A}{6.8}{GW190517A}{3.2}{GW190514A}{7.0}{GW190513A}{4.0}{GW190512A}{0.9}{GW190503A}{5.7}{GW190426A}{0.01}{GW190425A}{0.0008}{GW190424A}{5.6}{GW190421A}{6.0}{GW190413B}{9.5}{GW190413A}{6.7}{GW190412A}{0.3}{GW190408A}{1.3}}}
\newcommand{\spinoneIMRminus}[1]{\IfEqCase{#1}{{GW190930A}{0.36}{GW190929A}{0.53}{GW190924A}{0.27}{GW190915A}{0.49}{GW190910A}{0.34}{GW190909A}{0.44}{GW190828B}{0.21}{GW190828A}{0.37}{GW190814A}{0.03}{GW190803A}{0.39}{GW190731A}{0.37}{GW190728A}{0.31}{GW190727A}{0.46}{GW190720A}{0.38}{GW190719A}{0.51}{GW190708A}{0.21}{GW190707A}{0.22}{GW190706A}{0.54}{GW190701A}{0.37}{GW190630A}{0.27}{GW190620A}{0.50}{GW190602A}{0.36}{GW190527A}{0.42}{GW190521B}{0.31}{GW190521A}{0.57}{GW190519A}{0.51}{GW190517A}{0.34}{GW190514A}{0.46}{GW190513A}{0.28}{GW190512A}{0.20}{GW190503A}{0.31}{GW190426A}{0.14}{GW190425A}{0.25}{GW190424A}{0.52}{GW190421A}{0.42}{GW190413B}{0.49}{GW190413A}{0.37}{GW190412A}{0.25}{GW190408A}{0.29}}}
\newcommand{\spinoneIMRmed}[1]{\IfEqCase{#1}{{GW190930A}{0.41}{GW190929A}{0.66}{GW190924A}{0.30}{GW190915A}{0.55}{GW190910A}{0.38}{GW190909A}{0.50}{GW190828B}{0.23}{GW190828A}{0.41}{GW190814A}{0.03}{GW190803A}{0.44}{GW190731A}{0.41}{GW190728A}{0.35}{GW190727A}{0.50}{GW190720A}{0.45}{GW190719A}{0.58}{GW190708A}{0.23}{GW190707A}{0.25}{GW190706A}{0.69}{GW190701A}{0.42}{GW190630A}{0.30}{GW190620A}{0.61}{GW190602A}{0.40}{GW190527A}{0.47}{GW190521B}{0.34}{GW190521A}{0.64}{GW190519A}{0.75}{GW190517A}{0.86}{GW190514A}{0.51}{GW190513A}{0.31}{GW190512A}{0.22}{GW190503A}{0.35}{GW190426A}{0.14}{GW190425A}{0.27}{GW190424A}{0.59}{GW190421A}{0.47}{GW190413B}{0.54}{GW190413A}{0.41}{GW190412A}{0.40}{GW190408A}{0.32}}}
\newcommand{\spinoneIMRplus}[1]{\IfEqCase{#1}{{GW190930A}{0.44}{GW190929A}{0.29}{GW190924A}{0.45}{GW190915A}{0.40}{GW190910A}{0.53}{GW190909A}{0.44}{GW190828B}{0.46}{GW190828A}{0.49}{GW190814A}{0.06}{GW190803A}{0.48}{GW190731A}{0.51}{GW190728A}{0.40}{GW190727A}{0.44}{GW190720A}{0.42}{GW190719A}{0.38}{GW190708A}{0.47}{GW190707A}{0.48}{GW190706A}{0.27}{GW190701A}{0.49}{GW190630A}{0.46}{GW190620A}{0.34}{GW190602A}{0.50}{GW190527A}{0.46}{GW190521B}{0.52}{GW190521A}{0.32}{GW190519A}{0.22}{GW190517A}{0.12}{GW190514A}{0.43}{GW190513A}{0.52}{GW190512A}{0.48}{GW190503A}{0.51}{GW190426A}{0.40}{GW190425A}{0.51}{GW190424A}{0.37}{GW190421A}{0.47}{GW190413B}{0.40}{GW190413A}{0.51}{GW190412A}{0.22}{GW190408A}{0.52}}}
\newcommand{\spintwoIMRminus}[1]{\IfEqCase{#1}{{GW190930A}{0.41}{GW190929A}{0.45}{GW190924A}{0.36}{GW190915A}{0.46}{GW190910A}{0.37}{GW190909A}{0.46}{GW190828B}{0.36}{GW190828A}{0.37}{GW190814A}{0.45}{GW190803A}{0.41}{GW190731A}{0.41}{GW190728A}{0.37}{GW190727A}{0.42}{GW190720A}{0.47}{GW190719A}{0.47}{GW190708A}{0.30}{GW190707A}{0.31}{GW190706A}{0.52}{GW190701A}{0.41}{GW190630A}{0.36}{GW190620A}{0.48}{GW190602A}{0.42}{GW190527A}{0.45}{GW190521B}{0.36}{GW190521A}{0.48}{GW190519A}{0.49}{GW190517A}{0.61}{GW190514A}{0.46}{GW190513A}{0.39}{GW190512A}{0.37}{GW190503A}{0.39}{GW190426A}{0.009}{GW190425A}{0.25}{GW190424A}{0.44}{GW190421A}{0.42}{GW190413B}{0.47}{GW190413A}{0.41}{GW190412A}{0.45}{GW190408A}{0.38}}}
\newcommand{\spintwoIMRmed}[1]{\IfEqCase{#1}{{GW190930A}{0.46}{GW190929A}{0.50}{GW190924A}{0.40}{GW190915A}{0.52}{GW190910A}{0.41}{GW190909A}{0.51}{GW190828B}{0.40}{GW190828A}{0.41}{GW190814A}{0.50}{GW190803A}{0.46}{GW190731A}{0.45}{GW190728A}{0.41}{GW190727A}{0.47}{GW190720A}{0.55}{GW190719A}{0.52}{GW190708A}{0.33}{GW190707A}{0.34}{GW190706A}{0.58}{GW190701A}{0.46}{GW190630A}{0.40}{GW190620A}{0.53}{GW190602A}{0.47}{GW190527A}{0.50}{GW190521B}{0.40}{GW190521A}{0.53}{GW190519A}{0.55}{GW190517A}{0.70}{GW190514A}{0.52}{GW190513A}{0.43}{GW190512A}{0.41}{GW190503A}{0.43}{GW190426A}{0.009}{GW190425A}{0.28}{GW190424A}{0.49}{GW190421A}{0.46}{GW190413B}{0.52}{GW190413A}{0.45}{GW190412A}{0.50}{GW190408A}{0.42}}}
\newcommand{\spintwoIMRplus}[1]{\IfEqCase{#1}{{GW190930A}{0.46}{GW190929A}{0.44}{GW190924A}{0.51}{GW190915A}{0.43}{GW190910A}{0.52}{GW190909A}{0.43}{GW190828B}{0.52}{GW190828A}{0.50}{GW190814A}{0.43}{GW190803A}{0.47}{GW190731A}{0.47}{GW190728A}{0.50}{GW190727A}{0.47}{GW190720A}{0.40}{GW190719A}{0.43}{GW190708A}{0.58}{GW190707A}{0.53}{GW190706A}{0.37}{GW190701A}{0.48}{GW190630A}{0.50}{GW190620A}{0.41}{GW190602A}{0.46}{GW190527A}{0.44}{GW190521B}{0.51}{GW190521A}{0.42}{GW190519A}{0.40}{GW190517A}{0.26}{GW190514A}{0.42}{GW190513A}{0.49}{GW190512A}{0.49}{GW190503A}{0.49}{GW190426A}{0.03}{GW190425A}{0.51}{GW190424A}{0.45}{GW190421A}{0.47}{GW190413B}{0.42}{GW190413A}{0.47}{GW190412A}{0.45}{GW190408A}{0.49}}}
\newcommand{\tiltoneIMRminus}[1]{\IfEqCase{#1}{{GW190930A}{0.73}{GW190929A}{0.72}{GW190924A}{0.98}{GW190915A}{0.84}{GW190910A}{0.97}{GW190909A}{1.12}{GW190828B}{0.98}{GW190828A}{0.76}{GW190814A}{1.14}{GW190803A}{1.04}{GW190731A}{0.91}{GW190728A}{0.75}{GW190727A}{0.83}{GW190720A}{0.72}{GW190719A}{0.62}{GW190708A}{0.95}{GW190707A}{1.10}{GW190706A}{0.53}{GW190701A}{1.11}{GW190630A}{0.83}{GW190620A}{0.61}{GW190602A}{0.89}{GW190527A}{0.90}{GW190521B}{0.85}{GW190521A}{0.88}{GW190519A}{0.56}{GW190517A}{0.47}{GW190514A}{1.11}{GW190513A}{0.93}{GW190512A}{1.05}{GW190503A}{1.10}{GW190426A}{0.00}{GW190425A}{0.80}{GW190424A}{0.81}{GW190421A}{1.00}{GW190413B}{1.03}{GW190413A}{1.12}{GW190412A}{0.48}{GW190408A}{1.11}}}
\newcommand{\tiltoneIMRmed}[1]{\IfEqCase{#1}{{GW190930A}{0.99}{GW190929A}{1.65}{GW190924A}{1.24}{GW190915A}{1.51}{GW190910A}{1.53}{GW190909A}{1.89}{GW190828B}{1.46}{GW190828A}{1.09}{GW190814A}{1.56}{GW190803A}{1.62}{GW190731A}{1.37}{GW190728A}{1.02}{GW190727A}{1.27}{GW190720A}{0.99}{GW190719A}{0.87}{GW190708A}{1.48}{GW190707A}{1.78}{GW190706A}{0.75}{GW190701A}{1.80}{GW190630A}{1.32}{GW190620A}{0.87}{GW190602A}{1.34}{GW190527A}{1.39}{GW190521B}{1.40}{GW190521A}{1.46}{GW190519A}{0.80}{GW190517A}{0.66}{GW190514A}{1.92}{GW190513A}{1.35}{GW190512A}{1.57}{GW190503A}{1.69}{GW190426A}{0.00}{GW190425A}{1.31}{GW190424A}{1.24}{GW190421A}{1.77}{GW190413B}{1.67}{GW190413A}{1.70}{GW190412A}{0.91}{GW190408A}{1.75}}}
\newcommand{\tiltoneIMRplus}[1]{\IfEqCase{#1}{{GW190930A}{1.22}{GW190929A}{0.89}{GW190924A}{1.17}{GW190915A}{0.86}{GW190910A}{1.02}{GW190909A}{0.89}{GW190828B}{1.05}{GW190828A}{1.22}{GW190814A}{1.15}{GW190803A}{1.02}{GW190731A}{1.13}{GW190728A}{1.19}{GW190727A}{1.08}{GW190720A}{0.87}{GW190719A}{1.12}{GW190708A}{1.01}{GW190707A}{0.80}{GW190706A}{0.93}{GW190701A}{0.93}{GW190630A}{0.98}{GW190620A}{0.92}{GW190602A}{1.04}{GW190527A}{1.06}{GW190521B}{1.00}{GW190521A}{1.13}{GW190519A}{0.69}{GW190517A}{0.52}{GW190514A}{0.92}{GW190513A}{1.07}{GW190512A}{1.01}{GW190503A}{1.02}{GW190426A}{3.14}{GW190425A}{0.66}{GW190424A}{1.00}{GW190421A}{0.93}{GW190413B}{0.97}{GW190413A}{1.02}{GW190412A}{0.58}{GW190408A}{0.97}}}
\newcommand{\tilttwoIMRminus}[1]{\IfEqCase{#1}{{GW190930A}{0.97}{GW190929A}{1.11}{GW190924A}{1.06}{GW190915A}{1.00}{GW190910A}{1.11}{GW190909A}{1.22}{GW190828B}{1.09}{GW190828A}{0.90}{GW190814A}{1.07}{GW190803A}{1.11}{GW190731A}{1.02}{GW190728A}{0.90}{GW190727A}{0.99}{GW190720A}{1.03}{GW190719A}{0.86}{GW190708A}{1.02}{GW190707A}{1.19}{GW190706A}{0.82}{GW190701A}{1.18}{GW190630A}{0.99}{GW190620A}{0.85}{GW190602A}{0.98}{GW190527A}{1.03}{GW190521B}{1.00}{GW190521A}{1.02}{GW190519A}{0.84}{GW190517A}{0.63}{GW190514A}{1.17}{GW190513A}{1.01}{GW190512A}{1.03}{GW190503A}{1.10}{GW190426A}{0.00}{GW190425A}{0.87}{GW190424A}{0.92}{GW190421A}{1.15}{GW190413B}{1.06}{GW190413A}{1.15}{GW190412A}{0.87}{GW190408A}{1.07}}}
\newcommand{\tilttwoIMRmed}[1]{\IfEqCase{#1}{{GW190930A}{1.38}{GW190929A}{1.56}{GW190924A}{1.53}{GW190915A}{1.54}{GW190910A}{1.69}{GW190909A}{1.84}{GW190828B}{1.60}{GW190828A}{1.26}{GW190814A}{1.68}{GW190803A}{1.64}{GW190731A}{1.49}{GW190728A}{1.28}{GW190727A}{1.43}{GW190720A}{1.46}{GW190719A}{1.21}{GW190708A}{1.49}{GW190707A}{1.84}{GW190706A}{1.10}{GW190701A}{1.79}{GW190630A}{1.48}{GW190620A}{1.16}{GW190602A}{1.35}{GW190527A}{1.46}{GW190521B}{1.52}{GW190521A}{1.57}{GW190519A}{1.13}{GW190517A}{0.89}{GW190514A}{1.91}{GW190513A}{1.39}{GW190512A}{1.53}{GW190503A}{1.71}{GW190426A}{0.00}{GW190425A}{1.41}{GW190424A}{1.37}{GW190421A}{1.82}{GW190413B}{1.60}{GW190413A}{1.71}{GW190412A}{1.24}{GW190408A}{1.75}}}
\newcommand{\tilttwoIMRplus}[1]{\IfEqCase{#1}{{GW190930A}{1.16}{GW190929A}{1.12}{GW190924A}{1.08}{GW190915A}{1.08}{GW190910A}{0.98}{GW190909A}{0.94}{GW190828B}{1.06}{GW190828A}{1.24}{GW190814A}{1.00}{GW190803A}{1.04}{GW190731A}{1.14}{GW190728A}{1.25}{GW190727A}{1.16}{GW190720A}{1.13}{GW190719A}{1.27}{GW190708A}{1.07}{GW190707A}{0.88}{GW190706A}{1.28}{GW190701A}{0.97}{GW190630A}{1.11}{GW190620A}{1.23}{GW190602A}{1.20}{GW190527A}{1.20}{GW190521B}{1.03}{GW190521A}{1.09}{GW190519A}{1.26}{GW190517A}{1.15}{GW190514A}{0.89}{GW190513A}{1.21}{GW190512A}{1.07}{GW190503A}{1.01}{GW190426A}{3.14}{GW190425A}{0.94}{GW190424A}{1.15}{GW190421A}{0.92}{GW190413B}{1.03}{GW190413A}{1.04}{GW190412A}{1.31}{GW190408A}{0.93}}}
\newcommand{\thetajnIMRminus}[1]{\IfEqCase{#1}{{GW190930A}{0.76}{GW190929A}{1.11}{GW190924A}{0.47}{GW190915A}{1.27}{GW190910A}{1.41}{GW190909A}{1.15}{GW190828B}{1.46}{GW190828A}{2.00}{GW190814A}{0.23}{GW190803A}{1.04}{GW190731A}{1.15}{GW190728A}{1.02}{GW190727A}{0.82}{GW190720A}{1.91}{GW190719A}{1.28}{GW190708A}{1.17}{GW190707A}{1.94}{GW190706A}{1.13}{GW190701A}{0.50}{GW190630A}{0.82}{GW190620A}{1.77}{GW190602A}{1.53}{GW190527A}{0.89}{GW190521B}{0.74}{GW190521A}{0.66}{GW190519A}{1.36}{GW190517A}{1.50}{GW190514A}{1.24}{GW190513A}{0.54}{GW190512A}{1.33}{GW190503A}{0.64}{GW190426A}{1.43}{GW190425A}{0.85}{GW190424A}{1.23}{GW190421A}{1.35}{GW190413B}{1.29}{GW190413A}{0.62}{GW190412A}{0.29}{GW190408A}{1.31}}}
\newcommand{\thetajnIMRmed}[1]{\IfEqCase{#1}{{GW190930A}{0.95}{GW190929A}{1.73}{GW190924A}{0.61}{GW190915A}{1.79}{GW190910A}{1.67}{GW190909A}{1.48}{GW190828B}{1.70}{GW190828A}{2.32}{GW190814A}{0.83}{GW190803A}{1.34}{GW190731A}{1.43}{GW190728A}{1.21}{GW190727A}{1.08}{GW190720A}{2.59}{GW190719A}{1.59}{GW190708A}{1.39}{GW190707A}{2.21}{GW190706A}{1.44}{GW190701A}{0.69}{GW190630A}{1.04}{GW190620A}{2.14}{GW190602A}{1.84}{GW190527A}{1.19}{GW190521B}{1.01}{GW190521A}{0.91}{GW190519A}{1.70}{GW190517A}{2.24}{GW190514A}{1.56}{GW190513A}{0.72}{GW190512A}{1.58}{GW190503A}{2.50}{GW190426A}{1.70}{GW190425A}{1.08}{GW190424A}{1.53}{GW190421A}{1.68}{GW190413B}{1.70}{GW190413A}{0.83}{GW190412A}{0.74}{GW190408A}{1.57}}}
\newcommand{\thetajnIMRplus}[1]{\IfEqCase{#1}{{GW190930A}{1.91}{GW190929A}{0.90}{GW190924A}{2.14}{GW190915A}{0.92}{GW190910A}{1.24}{GW190909A}{1.31}{GW190828B}{1.21}{GW190828A}{0.64}{GW190814A}{1.59}{GW190803A}{1.48}{GW190731A}{1.42}{GW190728A}{1.70}{GW190727A}{1.73}{GW190720A}{0.41}{GW190719A}{1.25}{GW190708A}{1.52}{GW190707A}{0.73}{GW190706A}{1.37}{GW190701A}{0.54}{GW190630A}{1.83}{GW190620A}{0.75}{GW190602A}{1.00}{GW190527A}{1.58}{GW190521B}{1.79}{GW190521A}{1.92}{GW190519A}{1.12}{GW190517A}{0.61}{GW190514A}{1.26}{GW190513A}{2.00}{GW190512A}{1.32}{GW190503A}{0.47}{GW190426A}{1.19}{GW190425A}{1.77}{GW190424A}{1.33}{GW190421A}{1.14}{GW190413B}{1.08}{GW190413A}{1.94}{GW190412A}{1.37}{GW190408A}{1.31}}}
\newcommand{\massonedetIMRminus}[1]{\IfEqCase{#1}{{GW190930A}{2.9}{GW190929A}{30.5}{GW190924A}{2.8}{GW190915A}{8.1}{GW190910A}{6.7}{GW190909A}{14.3}{GW190828B}{7.9}{GW190828A}{5.1}{GW190814A}{1.2}{GW190803A}{9.6}{GW190731A}{10.7}{GW190728A}{2.6}{GW190727A}{9.0}{GW190720A}{3.8}{GW190719A}{19.3}{GW190708A}{3.2}{GW190707A}{1.9}{GW190706A}{20.7}{GW190701A}{11.6}{GW190630A}{8.2}{GW190620A}{16.6}{GW190602A}{20.8}{GW190527A}{14.4}{GW190521B}{6.5}{GW190521A}{20.9}{GW190519A}{16.8}{GW190517A}{8.4}{GW190514A}{10.4}{GW190513A}{12.7}{GW190512A}{6.3}{GW190503A}{9.4}{GW190426A}{2.5}{GW190425A}{0.4}{GW190424A}{9.0}{GW190421A}{9.9}{GW190413B}{15.2}{GW190413A}{10.1}{GW190412A}{5.3}{GW190408A}{4.6}}}
\newcommand{\massonedetIMRmed}[1]{\IfEqCase{#1}{{GW190930A}{14.4}{GW190929A}{112.7}{GW190924A}{10.5}{GW190915A}{47.5}{GW190910A}{56.9}{GW190909A}{69.2}{GW190828B}{29.4}{GW190828A}{43.9}{GW190814A}{24.4}{GW190803A}{58.6}{GW190731A}{65.3}{GW190728A}{14.4}{GW190727A}{60.7}{GW190720A}{15.9}{GW190719A}{63.9}{GW190708A}{21.2}{GW190707A}{13.4}{GW190706A}{113.9}{GW190701A}{76.1}{GW190630A}{43.3}{GW190620A}{86.9}{GW190602A}{109.6}{GW190527A}{56.3}{GW190521B}{52.6}{GW190521A}{153.2}{GW190519A}{97.8}{GW190517A}{50.9}{GW190514A}{65.6}{GW190513A}{48.8}{GW190512A}{28.6}{GW190503A}{54.7}{GW190426A}{6.2}{GW190425A}{2.1}{GW190424A}{57.1}{GW190421A}{62.8}{GW190413B}{83.2}{GW190413A}{53.0}{GW190412A}{32.4}{GW190408A}{32.1}}}
\newcommand{\massonedetIMRplus}[1]{\IfEqCase{#1}{{GW190930A}{17.1}{GW190929A}{36.8}{GW190924A}{10.7}{GW190915A}{10.9}{GW190910A}{9.6}{GW190909A}{30.8}{GW190828B}{9.3}{GW190828A}{8.6}{GW190814A}{1.3}{GW190803A}{14.5}{GW190731A}{16.0}{GW190728A}{10.1}{GW190727A}{14.7}{GW190720A}{8.3}{GW190719A}{69.1}{GW190708A}{7.4}{GW190707A}{4.1}{GW190706A}{29.7}{GW190701A}{16.0}{GW190630A}{8.7}{GW190620A}{19.4}{GW190602A}{23.6}{GW190527A}{41.5}{GW190521B}{7.9}{GW190521A}{38.0}{GW190519A}{22.3}{GW190517A}{17.1}{GW190514A}{19.1}{GW190513A}{12.1}{GW190512A}{8.9}{GW190503A}{12.2}{GW190426A}{4.2}{GW190425A}{0.6}{GW190424A}{14.4}{GW190421A}{15.5}{GW190413B}{21.4}{GW190413A}{14.7}{GW190412A}{5.5}{GW190408A}{7.2}}}
\newcommand{\masstwodetIMRminus}[1]{\IfEqCase{#1}{{GW190930A}{4.0}{GW190929A}{14.4}{GW190924A}{2.3}{GW190915A}{8.4}{GW190910A}{9.9}{GW190909A}{20.8}{GW190828B}{3.3}{GW190828A}{7.3}{GW190814A}{0.10}{GW190803A}{14.0}{GW190731A}{17.8}{GW190728A}{3.3}{GW190727A}{16.5}{GW190720A}{2.7}{GW190719A}{12.1}{GW190708A}{3.5}{GW190707A}{2.0}{GW190706A}{31.4}{GW190701A}{18.4}{GW190630A}{5.1}{GW190620A}{19.0}{GW190602A}{29.0}{GW190527A}{13.7}{GW190521B}{8.0}{GW190521A}{36.8}{GW190519A}{23.4}{GW190517A}{12.2}{GW190514A}{16.9}{GW190513A}{5.6}{GW190512A}{4.1}{GW190503A}{12.9}{GW190426A}{0.5}{GW190425A}{0.3}{GW190424A}{12.1}{GW190421A}{15.5}{GW190413B}{25.7}{GW190413A}{11.3}{GW190412A}{1.2}{GW190408A}{5.2}}}
\newcommand{\masstwodetIMRmed}[1]{\IfEqCase{#1}{{GW190930A}{8.9}{GW190929A}{30.8}{GW190924A}{5.3}{GW190915A}{31.1}{GW190910A}{45.0}{GW190909A}{45.5}{GW190828B}{13.8}{GW190828A}{34.9}{GW190814A}{2.72}{GW190803A}{42.3}{GW190731A}{46.0}{GW190728A}{9.5}{GW190727A}{44.6}{GW190720A}{9.0}{GW190719A}{32.0}{GW190708A}{14.9}{GW190707A}{9.6}{GW190706A}{71.3}{GW190701A}{54.2}{GW190630A}{26.0}{GW190620A}{49.6}{GW190602A}{59.7}{GW190527A}{29.7}{GW190521B}{38.8}{GW190521A}{110.7}{GW190519A}{62.5}{GW190517A}{33.9}{GW190514A}{48.7}{GW190513A}{23.1}{GW190512A}{16.2}{GW190503A}{36.6}{GW190426A}{1.6}{GW190425A}{1.4}{GW190424A}{43.8}{GW190421A}{45.0}{GW190413B}{53.3}{GW190413A}{38.6}{GW190412A}{10.1}{GW190408A}{23.3}}}
\newcommand{\masstwodetIMRplus}[1]{\IfEqCase{#1}{{GW190930A}{2.1}{GW190929A}{33.0}{GW190924A}{1.8}{GW190915A}{7.5}{GW190910A}{7.2}{GW190909A}{17.9}{GW190828B}{5.1}{GW190828A}{5.3}{GW190814A}{0.09}{GW190803A}{10.1}{GW190731A}{11.8}{GW190728A}{2.0}{GW190727A}{9.2}{GW190720A}{2.6}{GW190719A}{17.8}{GW190708A}{2.5}{GW190707A}{1.5}{GW190706A}{21.8}{GW190701A}{13.3}{GW190630A}{7.2}{GW190620A}{19.2}{GW190602A}{27.7}{GW190527A}{26.4}{GW190521B}{6.7}{GW190521A}{26.1}{GW190519A}{17.0}{GW190517A}{8.3}{GW190514A}{11.1}{GW190513A}{10.1}{GW190512A}{4.6}{GW190503A}{10.3}{GW190426A}{0.9}{GW190425A}{0.3}{GW190424A}{8.3}{GW190421A}{10.5}{GW190413B}{17.4}{GW190413A}{10.0}{GW190412A}{1.6}{GW190408A}{4.0}}}
\newcommand{\totalmassdetIMRminus}[1]{\IfEqCase{#1}{{GW190930A}{1.1}{GW190929A}{24.9}{GW190924A}{1.0}{GW190915A}{6.6}{GW190910A}{8.0}{GW190909A}{19.0}{GW190828B}{3.2}{GW190828A}{5.3}{GW190814A}{1.1}{GW190803A}{12.2}{GW190731A}{14.2}{GW190728A}{0.7}{GW190727A}{11.1}{GW190720A}{1.3}{GW190719A}{14.9}{GW190708A}{1.0}{GW190707A}{0.5}{GW190706A}{22.5}{GW190701A}{14.9}{GW190630A}{3.9}{GW190620A}{18.9}{GW190602A}{24.3}{GW190527A}{10.6}{GW190521B}{4.5}{GW190521A}{34.0}{GW190519A}{15.8}{GW190517A}{6.8}{GW190514A}{14.5}{GW190513A}{6.3}{GW190512A}{2.8}{GW190503A}{11.3}{GW190426A}{1.6}{GW190425A}{0.08}{GW190424A}{11.3}{GW190421A}{12.5}{GW190413B}{18.4}{GW190413A}{13.3}{GW190412A}{3.7}{GW190408A}{3.5}}}
\newcommand{\totalmassdetIMRmed}[1]{\IfEqCase{#1}{{GW190930A}{23.4}{GW190929A}{145.9}{GW190924A}{15.8}{GW190915A}{78.8}{GW190910A}{101.6}{GW190909A}{113.9}{GW190828B}{43.1}{GW190828A}{78.8}{GW190814A}{27.2}{GW190803A}{100.8}{GW190731A}{110.7}{GW190728A}{23.9}{GW190727A}{104.9}{GW190720A}{25.0}{GW190719A}{94.7}{GW190708A}{36.3}{GW190707A}{23.1}{GW190706A}{185.6}{GW190701A}{129.9}{GW190630A}{69.7}{GW190620A}{137.2}{GW190602A}{169.1}{GW190527A}{84.9}{GW190521B}{91.2}{GW190521A}{263.8}{GW190519A}{160.4}{GW190517A}{85.5}{GW190514A}{114.3}{GW190513A}{72.1}{GW190512A}{44.9}{GW190503A}{91.2}{GW190426A}{7.8}{GW190425A}{3.50}{GW190424A}{100.8}{GW190421A}{107.7}{GW190413B}{136.0}{GW190413A}{92.0}{GW190412A}{42.5}{GW190408A}{55.7}}}
\newcommand{\totalmassdetIMRplus}[1]{\IfEqCase{#1}{{GW190930A}{13.0}{GW190929A}{33.3}{GW190924A}{8.5}{GW190915A}{8.1}{GW190910A}{9.2}{GW190909A}{35.4}{GW190828B}{6.2}{GW190828A}{6.6}{GW190814A}{1.3}{GW190803A}{14.0}{GW190731A}{15.8}{GW190728A}{6.7}{GW190727A}{11.9}{GW190720A}{5.7}{GW190719A}{91.0}{GW190708A}{3.8}{GW190707A}{2.0}{GW190706A}{19.3}{GW190701A}{16.3}{GW190630A}{5.1}{GW190620A}{18.5}{GW190602A}{24.8}{GW190527A}{64.4}{GW190521B}{5.5}{GW190521A}{36.6}{GW190519A}{15.4}{GW190517A}{8.5}{GW190514A}{16.8}{GW190513A}{9.6}{GW190512A}{5.0}{GW190503A}{12.1}{GW190426A}{3.7}{GW190425A}{0.3}{GW190424A}{12.9}{GW190421A}{13.7}{GW190413B}{20.7}{GW190413A}{15.3}{GW190412A}{4.4}{GW190408A}{3.4}}}
\newcommand{\symmetricmassratioIMRminus}[1]{\IfEqCase{#1}{{GW190930A}{0.12}{GW190929A}{0.07}{GW190924A}{0.11}{GW190915A}{0.03}{GW190910A}{0.02}{GW190909A}{0.06}{GW190828B}{0.05}{GW190828A}{0.02}{GW190814A}{0.006}{GW190803A}{0.04}{GW190731A}{0.04}{GW190728A}{0.08}{GW190727A}{0.04}{GW190720A}{0.07}{GW190719A}{0.07}{GW190708A}{0.04}{GW190707A}{0.03}{GW190706A}{0.06}{GW190701A}{0.04}{GW190630A}{0.03}{GW190620A}{0.05}{GW190602A}{0.07}{GW190527A}{0.08}{GW190521B}{0.02}{GW190521A}{0.03}{GW190519A}{0.05}{GW190517A}{0.06}{GW190514A}{0.04}{GW190513A}{0.04}{GW190512A}{0.05}{GW190503A}{0.04}{GW190426A}{0.08}{GW190425A}{0.03}{GW190424A}{0.03}{GW190421A}{0.04}{GW190413B}{0.07}{GW190413A}{0.03}{GW190412A}{0.03}{GW190408A}{0.03}}}
\newcommand{\symmetricmassratioIMRmed}[1]{\IfEqCase{#1}{{GW190930A}{0.24}{GW190929A}{0.17}{GW190924A}{0.22}{GW190915A}{0.24}{GW190910A}{0.247}{GW190909A}{0.24}{GW190828B}{0.22}{GW190828A}{0.247}{GW190814A}{0.090}{GW190803A}{0.244}{GW190731A}{0.243}{GW190728A}{0.24}{GW190727A}{0.244}{GW190720A}{0.23}{GW190719A}{0.22}{GW190708A}{0.242}{GW190707A}{0.243}{GW190706A}{0.24}{GW190701A}{0.243}{GW190630A}{0.23}{GW190620A}{0.23}{GW190602A}{0.23}{GW190527A}{0.23}{GW190521B}{0.244}{GW190521A}{0.243}{GW190519A}{0.24}{GW190517A}{0.240}{GW190514A}{0.245}{GW190513A}{0.22}{GW190512A}{0.23}{GW190503A}{0.240}{GW190426A}{0.16}{GW190425A}{0.240}{GW190424A}{0.246}{GW190421A}{0.243}{GW190413B}{0.24}{GW190413A}{0.245}{GW190412A}{0.18}{GW190408A}{0.244}}}
\newcommand{\symmetricmassratioIMRplus}[1]{\IfEqCase{#1}{{GW190930A}{0.01}{GW190929A}{0.08}{GW190924A}{0.03}{GW190915A}{0.01}{GW190910A}{0.003}{GW190909A}{0.01}{GW190828B}{0.03}{GW190828A}{0.003}{GW190814A}{0.006}{GW190803A}{0.006}{GW190731A}{0.007}{GW190728A}{0.01}{GW190727A}{0.006}{GW190720A}{0.02}{GW190719A}{0.03}{GW190708A}{0.007}{GW190707A}{0.007}{GW190706A}{0.01}{GW190701A}{0.007}{GW190630A}{0.02}{GW190620A}{0.02}{GW190602A}{0.02}{GW190527A}{0.02}{GW190521B}{0.006}{GW190521A}{0.007}{GW190519A}{0.01}{GW190517A}{0.010}{GW190514A}{0.005}{GW190513A}{0.03}{GW190512A}{0.02}{GW190503A}{0.010}{GW190426A}{0.08}{GW190425A}{0.010}{GW190424A}{0.004}{GW190421A}{0.007}{GW190413B}{0.01}{GW190413A}{0.005}{GW190412A}{0.03}{GW190408A}{0.006}}}
\newcommand{\iotaIMRminus}[1]{\IfEqCase{#1}{{GW190930A}{0.76}{GW190929A}{1.35}{GW190924A}{0.46}{GW190915A}{1.52}{GW190910A}{1.42}{GW190909A}{1.14}{GW190828B}{1.46}{GW190828A}{2.02}{GW190814A}{0.26}{GW190803A}{1.03}{GW190731A}{1.14}{GW190728A}{1.01}{GW190727A}{0.83}{GW190720A}{1.88}{GW190719A}{1.25}{GW190708A}{1.16}{GW190707A}{1.96}{GW190706A}{1.14}{GW190701A}{0.56}{GW190630A}{0.82}{GW190620A}{1.70}{GW190602A}{1.30}{GW190527A}{0.90}{GW190521B}{0.73}{GW190521A}{0.80}{GW190519A}{1.31}{GW190517A}{1.36}{GW190514A}{1.20}{GW190513A}{0.51}{GW190512A}{1.33}{GW190503A}{0.67}{GW190426A}{1.43}{GW190425A}{0.85}{GW190424A}{1.18}{GW190421A}{1.26}{GW190413B}{1.21}{GW190413A}{0.62}{GW190412A}{0.32}{GW190408A}{1.36}}}
\newcommand{\iotaIMRmed}[1]{\IfEqCase{#1}{{GW190930A}{0.96}{GW190929A}{1.80}{GW190924A}{0.62}{GW190915A}{1.90}{GW190910A}{1.70}{GW190909A}{1.52}{GW190828B}{1.70}{GW190828A}{2.33}{GW190814A}{0.81}{GW190803A}{1.33}{GW190731A}{1.43}{GW190728A}{1.22}{GW190727A}{1.11}{GW190720A}{2.56}{GW190719A}{1.58}{GW190708A}{1.37}{GW190707A}{2.21}{GW190706A}{1.49}{GW190701A}{0.78}{GW190630A}{1.01}{GW190620A}{2.11}{GW190602A}{1.69}{GW190527A}{1.22}{GW190521B}{1.09}{GW190521A}{1.06}{GW190519A}{1.69}{GW190517A}{2.15}{GW190514A}{1.56}{GW190513A}{0.71}{GW190512A}{1.58}{GW190503A}{2.47}{GW190426A}{1.70}{GW190425A}{1.09}{GW190424A}{1.53}{GW190421A}{1.67}{GW190413B}{1.68}{GW190413A}{0.84}{GW190412A}{0.73}{GW190408A}{1.60}}}
\newcommand{\iotaIMRplus}[1]{\IfEqCase{#1}{{GW190930A}{1.89}{GW190929A}{0.95}{GW190924A}{2.12}{GW190915A}{0.95}{GW190910A}{1.19}{GW190909A}{1.22}{GW190828B}{1.20}{GW190828A}{0.63}{GW190814A}{1.63}{GW190803A}{1.48}{GW190731A}{1.40}{GW190728A}{1.68}{GW190727A}{1.68}{GW190720A}{0.42}{GW190719A}{1.25}{GW190708A}{1.56}{GW190707A}{0.73}{GW190706A}{1.27}{GW190701A}{0.66}{GW190630A}{1.87}{GW190620A}{0.77}{GW190602A}{1.09}{GW190527A}{1.51}{GW190521B}{1.63}{GW190521A}{1.74}{GW190519A}{1.08}{GW190517A}{0.64}{GW190514A}{1.24}{GW190513A}{1.96}{GW190512A}{1.30}{GW190503A}{0.49}{GW190426A}{1.19}{GW190425A}{1.77}{GW190424A}{1.29}{GW190421A}{1.10}{GW190413B}{1.06}{GW190413A}{1.95}{GW190412A}{1.40}{GW190408A}{1.31}}}
\newcommand{\spinonexIMRminus}[1]{\IfEqCase{#1}{{GW190930A}{0.44}{GW190929A}{0.71}{GW190924A}{0.39}{GW190915A}{0.68}{GW190910A}{0.55}{GW190909A}{0.59}{GW190828B}{0.38}{GW190828A}{0.50}{GW190814A}{0.04}{GW190803A}{0.59}{GW190731A}{0.57}{GW190728A}{0.41}{GW190727A}{0.62}{GW190720A}{0.46}{GW190719A}{0.56}{GW190708A}{0.41}{GW190707A}{0.36}{GW190706A}{0.55}{GW190701A}{0.54}{GW190630A}{0.44}{GW190620A}{0.55}{GW190602A}{0.56}{GW190527A}{0.59}{GW190521B}{0.51}{GW190521A}{0.71}{GW190519A}{0.61}{GW190517A}{0.59}{GW190514A}{0.60}{GW190513A}{0.47}{GW190512A}{0.37}{GW190503A}{0.48}{GW190426A}{0.00}{GW190425A}{0.50}{GW190424A}{0.65}{GW190421A}{0.64}{GW190413B}{0.64}{GW190413A}{0.53}{GW190412A}{0.31}{GW190408A}{0.48}}}
\newcommand{\spinonexIMRmed}[1]{\IfEqCase{#1}{{GW190930A}{0.002}{GW190929A}{0.00}{GW190924A}{0.0001}{GW190915A}{0.00}{GW190910A}{0.00}{GW190909A}{0.00}{GW190828B}{0.00005}{GW190828A}{0.0006}{GW190814A}{0.00}{GW190803A}{0.00}{GW190731A}{0.002}{GW190728A}{0.00}{GW190727A}{0.00}{GW190720A}{0.002}{GW190719A}{-0.01}{GW190708A}{0.0004}{GW190707A}{0.003}{GW190706A}{0.002}{GW190701A}{0.001}{GW190630A}{-0.01}{GW190620A}{0.009}{GW190602A}{0.0004}{GW190527A}{0.00}{GW190521B}{0.00}{GW190521A}{-0.04}{GW190519A}{0.006}{GW190517A}{-0.03}{GW190514A}{0.00}{GW190513A}{0.0007}{GW190512A}{0.0006}{GW190503A}{0.00}{GW190426A}{0.00}{GW190425A}{0.00}{GW190424A}{0.00}{GW190421A}{0.00}{GW190413B}{0.0002}{GW190413A}{0.0007}{GW190412A}{-0.04}{GW190408A}{0.0008}}}
\newcommand{\spinonexIMRplus}[1]{\IfEqCase{#1}{{GW190930A}{0.50}{GW190929A}{0.71}{GW190924A}{0.38}{GW190915A}{0.68}{GW190910A}{0.56}{GW190909A}{0.58}{GW190828B}{0.36}{GW190828A}{0.56}{GW190814A}{0.04}{GW190803A}{0.57}{GW190731A}{0.56}{GW190728A}{0.41}{GW190727A}{0.63}{GW190720A}{0.48}{GW190719A}{0.52}{GW190708A}{0.37}{GW190707A}{0.43}{GW190706A}{0.57}{GW190701A}{0.56}{GW190630A}{0.43}{GW190620A}{0.56}{GW190602A}{0.55}{GW190527A}{0.61}{GW190521B}{0.49}{GW190521A}{0.68}{GW190519A}{0.61}{GW190517A}{0.63}{GW190514A}{0.60}{GW190513A}{0.43}{GW190512A}{0.37}{GW190503A}{0.46}{GW190426A}{0.00}{GW190425A}{0.47}{GW190424A}{0.68}{GW190421A}{0.63}{GW190413B}{0.67}{GW190413A}{0.54}{GW190412A}{0.44}{GW190408A}{0.50}}}
\newcommand{\spinoneyIMRminus}[1]{\IfEqCase{#1}{{GW190930A}{0.46}{GW190929A}{0.73}{GW190924A}{0.39}{GW190915A}{0.69}{GW190910A}{0.55}{GW190909A}{0.59}{GW190828B}{0.37}{GW190828A}{0.52}{GW190814A}{0.04}{GW190803A}{0.57}{GW190731A}{0.55}{GW190728A}{0.41}{GW190727A}{0.64}{GW190720A}{0.55}{GW190719A}{0.57}{GW190708A}{0.38}{GW190707A}{0.41}{GW190706A}{0.55}{GW190701A}{0.54}{GW190630A}{0.46}{GW190620A}{0.54}{GW190602A}{0.56}{GW190527A}{0.58}{GW190521B}{0.52}{GW190521A}{0.71}{GW190519A}{0.61}{GW190517A}{0.61}{GW190514A}{0.59}{GW190513A}{0.46}{GW190512A}{0.37}{GW190503A}{0.49}{GW190426A}{0.00}{GW190425A}{0.48}{GW190424A}{0.67}{GW190421A}{0.60}{GW190413B}{0.66}{GW190413A}{0.53}{GW190412A}{0.40}{GW190408A}{0.46}}}
\newcommand{\spinoneyIMRmed}[1]{\IfEqCase{#1}{{GW190930A}{0.0009}{GW190929A}{0.01}{GW190924A}{0.0008}{GW190915A}{0.003}{GW190910A}{0.002}{GW190909A}{0.0010}{GW190828B}{0.0009}{GW190828A}{0.00}{GW190814A}{0.00}{GW190803A}{0.002}{GW190731A}{0.003}{GW190728A}{0.0008}{GW190727A}{0.00}{GW190720A}{0.0007}{GW190719A}{0.00}{GW190708A}{0.0010}{GW190707A}{0.0002}{GW190706A}{0.005}{GW190701A}{0.00}{GW190630A}{0.00}{GW190620A}{0.00}{GW190602A}{0.002}{GW190527A}{0.00}{GW190521B}{0.003}{GW190521A}{-0.01}{GW190519A}{0.00}{GW190517A}{0.00}{GW190514A}{0.003}{GW190513A}{0.00}{GW190512A}{0.00}{GW190503A}{0.00}{GW190426A}{0.00}{GW190425A}{0.003}{GW190424A}{-0.01}{GW190421A}{0.003}{GW190413B}{0.004}{GW190413A}{0.00}{GW190412A}{0.09}{GW190408A}{0.00}}}
\newcommand{\spinoneyIMRplus}[1]{\IfEqCase{#1}{{GW190930A}{0.48}{GW190929A}{0.71}{GW190924A}{0.39}{GW190915A}{0.67}{GW190910A}{0.58}{GW190909A}{0.59}{GW190828B}{0.37}{GW190828A}{0.51}{GW190814A}{0.04}{GW190803A}{0.59}{GW190731A}{0.57}{GW190728A}{0.41}{GW190727A}{0.62}{GW190720A}{0.49}{GW190719A}{0.54}{GW190708A}{0.40}{GW190707A}{0.39}{GW190706A}{0.55}{GW190701A}{0.55}{GW190630A}{0.44}{GW190620A}{0.55}{GW190602A}{0.57}{GW190527A}{0.62}{GW190521B}{0.51}{GW190521A}{0.70}{GW190519A}{0.61}{GW190517A}{0.60}{GW190514A}{0.58}{GW190513A}{0.49}{GW190512A}{0.37}{GW190503A}{0.48}{GW190426A}{0.00}{GW190425A}{0.48}{GW190424A}{0.69}{GW190421A}{0.60}{GW190413B}{0.66}{GW190413A}{0.54}{GW190412A}{0.34}{GW190408A}{0.47}}}
\newcommand{\spinonezIMRminus}[1]{\IfEqCase{#1}{{GW190930A}{0.31}{GW190929A}{0.43}{GW190924A}{0.23}{GW190915A}{0.38}{GW190910A}{0.39}{GW190909A}{0.57}{GW190828B}{0.24}{GW190828A}{0.31}{GW190814A}{0.06}{GW190803A}{0.46}{GW190731A}{0.34}{GW190728A}{0.28}{GW190727A}{0.35}{GW190720A}{0.30}{GW190719A}{0.39}{GW190708A}{0.22}{GW190707A}{0.28}{GW190706A}{0.47}{GW190701A}{0.50}{GW190630A}{0.22}{GW190620A}{0.40}{GW190602A}{0.32}{GW190527A}{0.36}{GW190521B}{0.26}{GW190521A}{0.61}{GW190519A}{0.46}{GW190517A}{0.38}{GW190514A}{0.55}{GW190513A}{0.28}{GW190512A}{0.30}{GW190503A}{0.43}{GW190426A}{0.51}{GW190425A}{0.12}{GW190424A}{0.38}{GW190421A}{0.46}{GW190413B}{0.51}{GW190413A}{0.55}{GW190412A}{0.22}{GW190408A}{0.36}}}
\newcommand{\spinonezIMRmed}[1]{\IfEqCase{#1}{{GW190930A}{0.18}{GW190929A}{-0.04}{GW190924A}{0.06}{GW190915A}{0.02}{GW190910A}{0.007}{GW190909A}{-0.11}{GW190828B}{0.01}{GW190828A}{0.16}{GW190814A}{0.00010}{GW190803A}{-0.01}{GW190731A}{0.05}{GW190728A}{0.16}{GW190727A}{0.11}{GW190720A}{0.23}{GW190719A}{0.32}{GW190708A}{0.01}{GW190707A}{-0.03}{GW190706A}{0.45}{GW190701A}{-0.06}{GW190630A}{0.05}{GW190620A}{0.35}{GW190602A}{0.06}{GW190527A}{0.05}{GW190521B}{0.03}{GW190521A}{0.04}{GW190519A}{0.48}{GW190517A}{0.64}{GW190514A}{-0.12}{GW190513A}{0.04}{GW190512A}{0.00005}{GW190503A}{-0.02}{GW190426A}{-0.03}{GW190425A}{0.06}{GW190424A}{0.14}{GW190421A}{-0.06}{GW190413B}{-0.03}{GW190413A}{-0.02}{GW190412A}{0.24}{GW190408A}{-0.03}}}
\newcommand{\spinonezIMRplus}[1]{\IfEqCase{#1}{{GW190930A}{0.42}{GW190929A}{0.46}{GW190924A}{0.43}{GW190915A}{0.40}{GW190910A}{0.40}{GW190909A}{0.38}{GW190828B}{0.26}{GW190828A}{0.42}{GW190814A}{0.05}{GW190803A}{0.44}{GW190731A}{0.48}{GW190728A}{0.32}{GW190727A}{0.50}{GW190720A}{0.31}{GW190719A}{0.46}{GW190708A}{0.26}{GW190707A}{0.24}{GW190706A}{0.38}{GW190701A}{0.35}{GW190630A}{0.34}{GW190620A}{0.42}{GW190602A}{0.44}{GW190527A}{0.51}{GW190521B}{0.34}{GW190521A}{0.50}{GW190519A}{0.34}{GW190517A}{0.26}{GW190514A}{0.41}{GW190513A}{0.39}{GW190512A}{0.24}{GW190503A}{0.33}{GW190426A}{0.36}{GW190425A}{0.18}{GW190424A}{0.48}{GW190421A}{0.37}{GW190413B}{0.43}{GW190413A}{0.41}{GW190412A}{0.17}{GW190408A}{0.31}}}
\newcommand{\spintwoxIMRminus}[1]{\IfEqCase{#1}{{GW190930A}{0.57}{GW190929A}{0.59}{GW190924A}{0.53}{GW190915A}{0.63}{GW190910A}{0.57}{GW190909A}{0.56}{GW190828B}{0.56}{GW190828A}{0.52}{GW190814A}{0.61}{GW190803A}{0.59}{GW190731A}{0.58}{GW190728A}{0.53}{GW190727A}{0.59}{GW190720A}{0.63}{GW190719A}{0.59}{GW190708A}{0.52}{GW190707A}{0.49}{GW190706A}{0.57}{GW190701A}{0.57}{GW190630A}{0.51}{GW190620A}{0.58}{GW190602A}{0.56}{GW190527A}{0.57}{GW190521B}{0.56}{GW190521A}{0.67}{GW190519A}{0.59}{GW190517A}{0.61}{GW190514A}{0.59}{GW190513A}{0.54}{GW190512A}{0.55}{GW190503A}{0.56}{GW190426A}{0.00}{GW190425A}{0.47}{GW190424A}{0.59}{GW190421A}{0.57}{GW190413B}{0.62}{GW190413A}{0.56}{GW190412A}{0.56}{GW190408A}{0.57}}}
\newcommand{\spintwoxIMRmed}[1]{\IfEqCase{#1}{{GW190930A}{0.00}{GW190929A}{0.00}{GW190924A}{0.0004}{GW190915A}{0.00}{GW190910A}{0.00}{GW190909A}{0.0007}{GW190828B}{0.00}{GW190828A}{-0.01}{GW190814A}{0.0009}{GW190803A}{0.00}{GW190731A}{0.0006}{GW190728A}{0.0003}{GW190727A}{0.00004}{GW190720A}{0.00}{GW190719A}{0.00}{GW190708A}{0.001}{GW190707A}{0.0002}{GW190706A}{0.007}{GW190701A}{0.00}{GW190630A}{0.004}{GW190620A}{0.002}{GW190602A}{0.00}{GW190527A}{0.00}{GW190521B}{0.00}{GW190521A}{-0.01}{GW190519A}{0.0001}{GW190517A}{0.00003}{GW190514A}{0.0009}{GW190513A}{0.00}{GW190512A}{0.0002}{GW190503A}{0.00}{GW190426A}{0.00}{GW190425A}{0.0006}{GW190424A}{0.002}{GW190421A}{-0.01}{GW190413B}{0.00}{GW190413A}{0.002}{GW190412A}{0.002}{GW190408A}{0.00}}}
\newcommand{\spintwoxIMRplus}[1]{\IfEqCase{#1}{{GW190930A}{0.57}{GW190929A}{0.60}{GW190924A}{0.54}{GW190915A}{0.63}{GW190910A}{0.58}{GW190909A}{0.60}{GW190828B}{0.53}{GW190828A}{0.49}{GW190814A}{0.61}{GW190803A}{0.58}{GW190731A}{0.55}{GW190728A}{0.53}{GW190727A}{0.59}{GW190720A}{0.59}{GW190719A}{0.58}{GW190708A}{0.52}{GW190707A}{0.51}{GW190706A}{0.58}{GW190701A}{0.58}{GW190630A}{0.53}{GW190620A}{0.58}{GW190602A}{0.59}{GW190527A}{0.59}{GW190521B}{0.55}{GW190521A}{0.61}{GW190519A}{0.56}{GW190517A}{0.60}{GW190514A}{0.57}{GW190513A}{0.55}{GW190512A}{0.55}{GW190503A}{0.57}{GW190426A}{0.00}{GW190425A}{0.47}{GW190424A}{0.58}{GW190421A}{0.56}{GW190413B}{0.63}{GW190413A}{0.57}{GW190412A}{0.58}{GW190408A}{0.54}}}
\newcommand{\spintwoyIMRminus}[1]{\IfEqCase{#1}{{GW190930A}{0.58}{GW190929A}{0.59}{GW190924A}{0.53}{GW190915A}{0.63}{GW190910A}{0.57}{GW190909A}{0.58}{GW190828B}{0.55}{GW190828A}{0.54}{GW190814A}{0.60}{GW190803A}{0.59}{GW190731A}{0.55}{GW190728A}{0.53}{GW190727A}{0.59}{GW190720A}{0.61}{GW190719A}{0.56}{GW190708A}{0.51}{GW190707A}{0.48}{GW190706A}{0.59}{GW190701A}{0.57}{GW190630A}{0.54}{GW190620A}{0.60}{GW190602A}{0.58}{GW190527A}{0.58}{GW190521B}{0.60}{GW190521A}{0.63}{GW190519A}{0.59}{GW190517A}{0.61}{GW190514A}{0.56}{GW190513A}{0.54}{GW190512A}{0.53}{GW190503A}{0.56}{GW190426A}{0.00}{GW190425A}{0.48}{GW190424A}{0.60}{GW190421A}{0.57}{GW190413B}{0.62}{GW190413A}{0.56}{GW190412A}{0.53}{GW190408A}{0.56}}}
\newcommand{\spintwoyIMRmed}[1]{\IfEqCase{#1}{{GW190930A}{0.00}{GW190929A}{0.0003}{GW190924A}{0.0007}{GW190915A}{0.00}{GW190910A}{0.002}{GW190909A}{0.002}{GW190828B}{0.004}{GW190828A}{0.00}{GW190814A}{0.003}{GW190803A}{0.00}{GW190731A}{0.0006}{GW190728A}{0.00}{GW190727A}{0.0008}{GW190720A}{0.004}{GW190719A}{0.0004}{GW190708A}{0.00007}{GW190707A}{0.00}{GW190706A}{0.0006}{GW190701A}{0.0008}{GW190630A}{0.002}{GW190620A}{0.002}{GW190602A}{0.005}{GW190527A}{0.002}{GW190521B}{0.003}{GW190521A}{0.01}{GW190519A}{0.00}{GW190517A}{0.00}{GW190514A}{0.003}{GW190513A}{0.00009}{GW190512A}{0.003}{GW190503A}{0.0007}{GW190426A}{0.00}{GW190425A}{0.00}{GW190424A}{0.00}{GW190421A}{0.0002}{GW190413B}{0.0009}{GW190413A}{0.0001}{GW190412A}{0.01}{GW190408A}{0.0008}}}
\newcommand{\spintwoyIMRplus}[1]{\IfEqCase{#1}{{GW190930A}{0.56}{GW190929A}{0.59}{GW190924A}{0.55}{GW190915A}{0.64}{GW190910A}{0.58}{GW190909A}{0.59}{GW190828B}{0.57}{GW190828A}{0.50}{GW190814A}{0.61}{GW190803A}{0.59}{GW190731A}{0.59}{GW190728A}{0.52}{GW190727A}{0.60}{GW190720A}{0.58}{GW190719A}{0.60}{GW190708A}{0.49}{GW190707A}{0.50}{GW190706A}{0.59}{GW190701A}{0.57}{GW190630A}{0.53}{GW190620A}{0.56}{GW190602A}{0.59}{GW190527A}{0.59}{GW190521B}{0.57}{GW190521A}{0.66}{GW190519A}{0.58}{GW190517A}{0.60}{GW190514A}{0.58}{GW190513A}{0.54}{GW190512A}{0.56}{GW190503A}{0.55}{GW190426A}{0.00}{GW190425A}{0.48}{GW190424A}{0.60}{GW190421A}{0.57}{GW190413B}{0.60}{GW190413A}{0.56}{GW190412A}{0.59}{GW190408A}{0.57}}}
\newcommand{\spintwozIMRminus}[1]{\IfEqCase{#1}{{GW190930A}{0.47}{GW190929A}{0.56}{GW190924A}{0.44}{GW190915A}{0.53}{GW190910A}{0.47}{GW190909A}{0.62}{GW190828B}{0.46}{GW190828A}{0.39}{GW190814A}{0.56}{GW190803A}{0.53}{GW190731A}{0.48}{GW190728A}{0.46}{GW190727A}{0.45}{GW190720A}{0.60}{GW190719A}{0.48}{GW190708A}{0.39}{GW190707A}{0.41}{GW190706A}{0.47}{GW190701A}{0.57}{GW190630A}{0.46}{GW190620A}{0.47}{GW190602A}{0.48}{GW190527A}{0.52}{GW190521B}{0.42}{GW190521A}{0.55}{GW190519A}{0.44}{GW190517A}{0.50}{GW190514A}{0.58}{GW190513A}{0.51}{GW190512A}{0.44}{GW190503A}{0.56}{GW190426A}{0.03}{GW190425A}{0.18}{GW190424A}{0.47}{GW190421A}{0.57}{GW190413B}{0.57}{GW190413A}{0.57}{GW190412A}{0.51}{GW190408A}{0.51}}}
\newcommand{\spintwozIMRmed}[1]{\IfEqCase{#1}{{GW190930A}{0.05}{GW190929A}{0.002}{GW190924A}{0.005}{GW190915A}{0.006}{GW190910A}{-0.02}{GW190909A}{-0.08}{GW190828B}{0.00}{GW190828A}{0.08}{GW190814A}{-0.03}{GW190803A}{-0.01}{GW190731A}{0.02}{GW190728A}{0.08}{GW190727A}{0.03}{GW190720A}{0.03}{GW190719A}{0.12}{GW190708A}{0.01}{GW190707A}{-0.06}{GW190706A}{0.19}{GW190701A}{-0.06}{GW190630A}{0.02}{GW190620A}{0.15}{GW190602A}{0.06}{GW190527A}{0.02}{GW190521B}{0.01}{GW190521A}{0.000002}{GW190519A}{0.18}{GW190517A}{0.39}{GW190514A}{-0.12}{GW190513A}{0.04}{GW190512A}{0.007}{GW190503A}{-0.03}{GW190426A}{0.00}{GW190425A}{0.03}{GW190424A}{0.05}{GW190421A}{-0.07}{GW190413B}{-0.01}{GW190413A}{-0.03}{GW190412A}{0.10}{GW190408A}{-0.04}}}
\newcommand{\spintwozIMRplus}[1]{\IfEqCase{#1}{{GW190930A}{0.53}{GW190929A}{0.57}{GW190924A}{0.51}{GW190915A}{0.51}{GW190910A}{0.40}{GW190909A}{0.46}{GW190828B}{0.48}{GW190828A}{0.54}{GW190814A}{0.47}{GW190803A}{0.46}{GW190731A}{0.52}{GW190728A}{0.51}{GW190727A}{0.53}{GW190720A}{0.59}{GW190719A}{0.61}{GW190708A}{0.46}{GW190707A}{0.35}{GW190706A}{0.59}{GW190701A}{0.44}{GW190630A}{0.44}{GW190620A}{0.60}{GW190602A}{0.57}{GW190527A}{0.58}{GW190521B}{0.39}{GW190521A}{0.50}{GW190519A}{0.59}{GW190517A}{0.43}{GW190514A}{0.44}{GW190513A}{0.51}{GW190512A}{0.46}{GW190503A}{0.42}{GW190426A}{0.03}{GW190425A}{0.30}{GW190424A}{0.51}{GW190421A}{0.42}{GW190413B}{0.49}{GW190413A}{0.47}{GW190412A}{0.61}{GW190408A}{0.41}}}
\newcommand{\phioneIMRminus}[1]{\IfEqCase{#1}{{GW190930A}{2.83}{GW190929A}{2.72}{GW190924A}{2.81}{GW190915A}{2.79}{GW190910A}{2.75}{GW190909A}{2.81}{GW190828B}{2.79}{GW190828A}{2.89}{GW190814A}{3.11}{GW190803A}{2.78}{GW190731A}{2.77}{GW190728A}{2.81}{GW190727A}{2.85}{GW190720A}{2.82}{GW190719A}{2.84}{GW190708A}{2.76}{GW190707A}{2.82}{GW190706A}{2.82}{GW190701A}{2.86}{GW190630A}{2.89}{GW190620A}{2.88}{GW190602A}{2.79}{GW190527A}{2.87}{GW190521B}{2.73}{GW190521A}{2.82}{GW190519A}{2.84}{GW190517A}{2.83}{GW190514A}{2.77}{GW190513A}{2.97}{GW190512A}{2.84}{GW190503A}{2.86}{GW190426A}{0.00}{GW190425A}{2.73}{GW190424A}{2.94}{GW190421A}{2.76}{GW190413B}{2.75}{GW190413A}{2.85}{GW190412A}{2.13}{GW190408A}{2.84}}}
\newcommand{\phioneIMRmed}[1]{\IfEqCase{#1}{{GW190930A}{3.12}{GW190929A}{3.05}{GW190924A}{3.11}{GW190915A}{3.09}{GW190910A}{3.08}{GW190909A}{3.12}{GW190828B}{3.10}{GW190828A}{3.19}{GW190814A}{3.43}{GW190803A}{3.09}{GW190731A}{3.05}{GW190728A}{3.12}{GW190727A}{3.16}{GW190720A}{3.13}{GW190719A}{3.22}{GW190708A}{3.08}{GW190707A}{3.13}{GW190706A}{3.10}{GW190701A}{3.19}{GW190630A}{3.24}{GW190620A}{3.17}{GW190602A}{3.09}{GW190527A}{3.18}{GW190521B}{3.06}{GW190521A}{3.22}{GW190519A}{3.16}{GW190517A}{3.16}{GW190514A}{3.08}{GW190513A}{3.29}{GW190512A}{3.15}{GW190503A}{3.19}{GW190426A}{0.00}{GW190425A}{3.05}{GW190424A}{3.28}{GW190421A}{3.07}{GW190413B}{3.07}{GW190413A}{3.17}{GW190412A}{2.50}{GW190408A}{3.15}}}
\newcommand{\phioneIMRplus}[1]{\IfEqCase{#1}{{GW190930A}{2.84}{GW190929A}{2.91}{GW190924A}{2.85}{GW190915A}{2.88}{GW190910A}{2.88}{GW190909A}{2.81}{GW190828B}{2.86}{GW190828A}{2.80}{GW190814A}{2.54}{GW190803A}{2.86}{GW190731A}{2.92}{GW190728A}{2.85}{GW190727A}{2.81}{GW190720A}{2.83}{GW190719A}{2.75}{GW190708A}{2.87}{GW190707A}{2.85}{GW190706A}{2.88}{GW190701A}{2.80}{GW190630A}{2.69}{GW190620A}{2.81}{GW190602A}{2.86}{GW190527A}{2.78}{GW190521B}{2.90}{GW190521A}{2.69}{GW190519A}{2.82}{GW190517A}{2.77}{GW190514A}{2.87}{GW190513A}{2.70}{GW190512A}{2.83}{GW190503A}{2.76}{GW190426A}{0.00}{GW190425A}{2.90}{GW190424A}{2.68}{GW190421A}{2.89}{GW190413B}{2.88}{GW190413A}{2.80}{GW190412A}{3.29}{GW190408A}{2.84}}}
\newcommand{\phitwoIMRminus}[1]{\IfEqCase{#1}{{GW190930A}{2.88}{GW190929A}{2.82}{GW190924A}{2.81}{GW190915A}{2.86}{GW190910A}{2.78}{GW190909A}{2.79}{GW190828B}{2.71}{GW190828A}{2.82}{GW190814A}{2.78}{GW190803A}{2.84}{GW190731A}{2.82}{GW190728A}{2.84}{GW190727A}{2.79}{GW190720A}{2.76}{GW190719A}{2.83}{GW190708A}{2.81}{GW190707A}{2.82}{GW190706A}{2.81}{GW190701A}{2.81}{GW190630A}{2.79}{GW190620A}{2.80}{GW190602A}{2.74}{GW190527A}{2.78}{GW190521B}{2.73}{GW190521A}{2.60}{GW190519A}{2.84}{GW190517A}{2.90}{GW190514A}{2.76}{GW190513A}{2.82}{GW190512A}{2.76}{GW190503A}{2.81}{GW190426A}{0.00}{GW190425A}{2.84}{GW190424A}{2.86}{GW190421A}{2.78}{GW190413B}{2.81}{GW190413A}{2.83}{GW190412A}{2.62}{GW190408A}{2.80}}}
\newcommand{\phitwoIMRmed}[1]{\IfEqCase{#1}{{GW190930A}{3.22}{GW190929A}{3.13}{GW190924A}{3.11}{GW190915A}{3.18}{GW190910A}{3.10}{GW190909A}{3.10}{GW190828B}{3.04}{GW190828A}{3.17}{GW190814A}{3.08}{GW190803A}{3.16}{GW190731A}{3.11}{GW190728A}{3.15}{GW190727A}{3.11}{GW190720A}{3.08}{GW190719A}{3.13}{GW190708A}{3.14}{GW190707A}{3.15}{GW190706A}{3.12}{GW190701A}{3.11}{GW190630A}{3.09}{GW190620A}{3.11}{GW190602A}{3.05}{GW190527A}{3.11}{GW190521B}{3.06}{GW190521A}{2.93}{GW190519A}{3.17}{GW190517A}{3.20}{GW190514A}{3.10}{GW190513A}{3.14}{GW190512A}{3.07}{GW190503A}{3.12}{GW190426A}{0.00}{GW190425A}{3.15}{GW190424A}{3.17}{GW190421A}{3.13}{GW190413B}{3.12}{GW190413A}{3.14}{GW190412A}{2.91}{GW190408A}{3.12}}}
\newcommand{\phitwoIMRplus}[1]{\IfEqCase{#1}{{GW190930A}{2.73}{GW190929A}{2.84}{GW190924A}{2.84}{GW190915A}{2.78}{GW190910A}{2.87}{GW190909A}{2.85}{GW190828B}{2.92}{GW190828A}{2.77}{GW190814A}{2.88}{GW190803A}{2.82}{GW190731A}{2.85}{GW190728A}{2.82}{GW190727A}{2.84}{GW190720A}{2.88}{GW190719A}{2.87}{GW190708A}{2.83}{GW190707A}{2.83}{GW190706A}{2.86}{GW190701A}{2.86}{GW190630A}{2.88}{GW190620A}{2.86}{GW190602A}{2.91}{GW190527A}{2.87}{GW190521B}{2.88}{GW190521A}{3.01}{GW190519A}{2.79}{GW190517A}{2.78}{GW190514A}{2.89}{GW190513A}{2.84}{GW190512A}{2.90}{GW190503A}{2.81}{GW190426A}{0.00}{GW190425A}{2.83}{GW190424A}{2.77}{GW190421A}{2.83}{GW190413B}{2.83}{GW190413A}{2.83}{GW190412A}{3.05}{GW190408A}{2.83}}}
\newcommand{\chieffIMRminus}[1]{\IfEqCase{#1}{{GW190930A}{0.16}{GW190929A}{0.33}{GW190924A}{0.11}{GW190915A}{0.23}{GW190910A}{0.20}{GW190909A}{0.35}{GW190828B}{0.15}{GW190828A}{0.16}{GW190814A}{0.06}{GW190803A}{0.28}{GW190731A}{0.24}{GW190728A}{0.07}{GW190727A}{0.25}{GW190720A}{0.11}{GW190719A}{0.31}{GW190708A}{0.08}{GW190707A}{0.07}{GW190706A}{0.29}{GW190701A}{0.29}{GW190630A}{0.14}{GW190620A}{0.27}{GW190602A}{0.24}{GW190527A}{0.28}{GW190521B}{0.12}{GW190521A}{0.43}{GW190519A}{0.21}{GW190517A}{0.22}{GW190514A}{0.32}{GW190513A}{0.17}{GW190512A}{0.16}{GW190503A}{0.24}{GW190426A}{0.30}{GW190425A}{0.05}{GW190424A}{0.25}{GW190421A}{0.25}{GW190413B}{0.35}{GW190413A}{0.33}{GW190412A}{0.10}{GW190408A}{0.18}}}
\newcommand{\chieffIMRmed}[1]{\IfEqCase{#1}{{GW190930A}{0.15}{GW190929A}{-0.03}{GW190924A}{0.05}{GW190915A}{0.02}{GW190910A}{-0.01}{GW190909A}{-0.13}{GW190828B}{0.01}{GW190828A}{0.16}{GW190814A}{-0.01}{GW190803A}{-0.02}{GW190731A}{0.06}{GW190728A}{0.12}{GW190727A}{0.11}{GW190720A}{0.16}{GW190719A}{0.28}{GW190708A}{0.01}{GW190707A}{-0.07}{GW190706A}{0.38}{GW190701A}{-0.08}{GW190630A}{0.06}{GW190620A}{0.30}{GW190602A}{0.09}{GW190527A}{0.07}{GW190521B}{0.04}{GW190521A}{0.03}{GW190519A}{0.37}{GW190517A}{0.54}{GW190514A}{-0.15}{GW190513A}{0.05}{GW190512A}{0.002}{GW190503A}{-0.04}{GW190426A}{-0.03}{GW190425A}{0.06}{GW190424A}{0.13}{GW190421A}{-0.09}{GW190413B}{-0.03}{GW190413A}{-0.05}{GW190412A}{0.22}{GW190408A}{-0.05}}}
\newcommand{\chieffIMRplus}[1]{\IfEqCase{#1}{{GW190930A}{0.35}{GW190929A}{0.37}{GW190924A}{0.37}{GW190915A}{0.20}{GW190910A}{0.18}{GW190909A}{0.29}{GW190828B}{0.16}{GW190828A}{0.15}{GW190814A}{0.06}{GW190803A}{0.24}{GW190731A}{0.25}{GW190728A}{0.24}{GW190727A}{0.26}{GW190720A}{0.16}{GW190719A}{0.31}{GW190708A}{0.13}{GW190707A}{0.12}{GW190706A}{0.21}{GW190701A}{0.23}{GW190630A}{0.15}{GW190620A}{0.22}{GW190602A}{0.24}{GW190527A}{0.29}{GW190521B}{0.13}{GW190521A}{0.31}{GW190519A}{0.17}{GW190517A}{0.15}{GW190514A}{0.30}{GW190513A}{0.23}{GW190512A}{0.14}{GW190503A}{0.21}{GW190426A}{0.32}{GW190425A}{0.11}{GW190424A}{0.23}{GW190421A}{0.23}{GW190413B}{0.27}{GW190413A}{0.27}{GW190412A}{0.08}{GW190408A}{0.14}}}
\newcommand{\chipIMRminus}[1]{\IfEqCase{#1}{{GW190930A}{0.25}{GW190929A}{0.46}{GW190924A}{0.19}{GW190915A}{0.38}{GW190910A}{0.34}{GW190909A}{0.32}{GW190828B}{0.19}{GW190828A}{0.30}{GW190814A}{0.03}{GW190803A}{0.34}{GW190731A}{0.31}{GW190728A}{0.21}{GW190727A}{0.38}{GW190720A}{0.24}{GW190719A}{0.30}{GW190708A}{0.22}{GW190707A}{0.23}{GW190706A}{0.29}{GW190701A}{0.31}{GW190630A}{0.24}{GW190620A}{0.29}{GW190602A}{0.30}{GW190527A}{0.34}{GW190521B}{0.30}{GW190521A}{0.43}{GW190519A}{0.33}{GW190517A}{0.29}{GW190514A}{0.33}{GW190513A}{0.23}{GW190512A}{0.21}{GW190503A}{0.26}{GW190426A}{0.00}{GW190425A}{0.27}{GW190424A}{0.41}{GW190421A}{0.34}{GW190413B}{0.37}{GW190413A}{0.31}{GW190412A}{0.17}{GW190408A}{0.28}}}
\newcommand{\chipIMRmed}[1]{\IfEqCase{#1}{{GW190930A}{0.34}{GW190929A}{0.60}{GW190924A}{0.25}{GW190915A}{0.55}{GW190910A}{0.43}{GW190909A}{0.45}{GW190828B}{0.24}{GW190828A}{0.41}{GW190814A}{0.04}{GW190803A}{0.45}{GW190731A}{0.42}{GW190728A}{0.30}{GW190727A}{0.50}{GW190720A}{0.35}{GW190719A}{0.40}{GW190708A}{0.28}{GW190707A}{0.29}{GW190706A}{0.45}{GW190701A}{0.42}{GW190630A}{0.31}{GW190620A}{0.44}{GW190602A}{0.39}{GW190527A}{0.43}{GW190521B}{0.39}{GW190521A}{0.59}{GW190519A}{0.52}{GW190517A}{0.54}{GW190514A}{0.47}{GW190513A}{0.30}{GW190512A}{0.26}{GW190503A}{0.35}{GW190426A}{0.00}{GW190425A}{0.34}{GW190424A}{0.55}{GW190421A}{0.46}{GW190413B}{0.51}{GW190413A}{0.41}{GW190412A}{0.30}{GW190408A}{0.37}}}
\newcommand{\chipIMRplus}[1]{\IfEqCase{#1}{{GW190930A}{0.42}{GW190929A}{0.30}{GW190924A}{0.46}{GW190915A}{0.36}{GW190910A}{0.43}{GW190909A}{0.42}{GW190828B}{0.43}{GW190828A}{0.41}{GW190814A}{0.04}{GW190803A}{0.41}{GW190731A}{0.44}{GW190728A}{0.42}{GW190727A}{0.39}{GW190720A}{0.46}{GW190719A}{0.40}{GW190708A}{0.41}{GW190707A}{0.42}{GW190706A}{0.36}{GW190701A}{0.43}{GW190630A}{0.40}{GW190620A}{0.36}{GW190602A}{0.45}{GW190527A}{0.43}{GW190521B}{0.43}{GW190521A}{0.33}{GW190519A}{0.32}{GW190517A}{0.26}{GW190514A}{0.41}{GW190513A}{0.46}{GW190512A}{0.42}{GW190503A}{0.44}{GW190426A}{0.00}{GW190425A}{0.43}{GW190424A}{0.35}{GW190421A}{0.43}{GW190413B}{0.39}{GW190413A}{0.42}{GW190412A}{0.25}{GW190408A}{0.45}}}
\newcommand{\costiltoneIMRminus}[1]{\IfEqCase{#1}{{GW190930A}{1.15}{GW190929A}{0.75}{GW190924A}{1.07}{GW190915A}{0.78}{GW190910A}{0.87}{GW190909A}{0.62}{GW190828B}{0.92}{GW190828A}{1.13}{GW190814A}{0.92}{GW190803A}{0.83}{GW190731A}{1.00}{GW190728A}{1.12}{GW190727A}{1.00}{GW190720A}{0.84}{GW190719A}{1.05}{GW190708A}{0.89}{GW190707A}{0.64}{GW190706A}{0.84}{GW190701A}{0.69}{GW190630A}{0.91}{GW190620A}{0.86}{GW190602A}{0.96}{GW190527A}{0.95}{GW190521B}{0.91}{GW190521A}{0.96}{GW190519A}{0.61}{GW190517A}{0.41}{GW190514A}{0.61}{GW190513A}{0.97}{GW190512A}{0.85}{GW190503A}{0.79}{GW190426A}{0.00}{GW190425A}{0.65}{GW190424A}{0.95}{GW190421A}{0.71}{GW190413B}{0.78}{GW190413A}{0.78}{GW190412A}{0.53}{GW190408A}{0.74}}}
\newcommand{\costiltoneIMRmed}[1]{\IfEqCase{#1}{{GW190930A}{0.55}{GW190929A}{-0.08}{GW190924A}{0.32}{GW190915A}{0.06}{GW190910A}{0.04}{GW190909A}{-0.32}{GW190828B}{0.11}{GW190828A}{0.46}{GW190814A}{0.01}{GW190803A}{-0.05}{GW190731A}{0.20}{GW190728A}{0.52}{GW190727A}{0.30}{GW190720A}{0.55}{GW190719A}{0.65}{GW190708A}{0.09}{GW190707A}{-0.21}{GW190706A}{0.73}{GW190701A}{-0.23}{GW190630A}{0.24}{GW190620A}{0.65}{GW190602A}{0.23}{GW190527A}{0.18}{GW190521B}{0.17}{GW190521A}{0.11}{GW190519A}{0.70}{GW190517A}{0.79}{GW190514A}{-0.34}{GW190513A}{0.22}{GW190512A}{0.001}{GW190503A}{-0.12}{GW190426A}{-1.00}{GW190425A}{0.26}{GW190424A}{0.32}{GW190421A}{-0.20}{GW190413B}{-0.10}{GW190413A}{-0.13}{GW190412A}{0.61}{GW190408A}{-0.18}}}
\newcommand{\costiltoneIMRplus}[1]{\IfEqCase{#1}{{GW190930A}{0.42}{GW190929A}{0.68}{GW190924A}{0.64}{GW190915A}{0.72}{GW190910A}{0.81}{GW190909A}{1.04}{GW190828B}{0.78}{GW190828A}{0.48}{GW190814A}{0.90}{GW190803A}{0.89}{GW190731A}{0.69}{GW190728A}{0.44}{GW190727A}{0.61}{GW190720A}{0.42}{GW190719A}{0.32}{GW190708A}{0.77}{GW190707A}{0.99}{GW190706A}{0.24}{GW190701A}{1.00}{GW190630A}{0.64}{GW190620A}{0.32}{GW190602A}{0.67}{GW190527A}{0.70}{GW190521B}{0.68}{GW190521A}{0.72}{GW190519A}{0.27}{GW190517A}{0.19}{GW190514A}{1.03}{GW190513A}{0.69}{GW190512A}{0.87}{GW190503A}{0.95}{GW190426A}{2.00}{GW190425A}{0.61}{GW190424A}{0.59}{GW190421A}{0.92}{GW190413B}{0.90}{GW190413A}{0.97}{GW190412A}{0.30}{GW190408A}{0.98}}}
\newcommand{\costilttwoIMRminus}[1]{\IfEqCase{#1}{{GW190930A}{1.02}{GW190929A}{0.91}{GW190924A}{0.90}{GW190915A}{0.90}{GW190910A}{0.77}{GW190909A}{0.67}{GW190828B}{0.85}{GW190828A}{1.11}{GW190814A}{0.79}{GW190803A}{0.83}{GW190731A}{0.95}{GW190728A}{1.11}{GW190727A}{0.99}{GW190720A}{0.96}{GW190719A}{1.14}{GW190708A}{0.92}{GW190707A}{0.65}{GW190706A}{1.18}{GW190701A}{0.71}{GW190630A}{0.94}{GW190620A}{1.13}{GW190602A}{1.04}{GW190527A}{0.99}{GW190521B}{0.88}{GW190521A}{0.89}{GW190519A}{1.15}{GW190517A}{1.08}{GW190514A}{0.61}{GW190513A}{1.03}{GW190512A}{0.90}{GW190503A}{0.78}{GW190426A}{0.00}{GW190425A}{0.87}{GW190424A}{1.01}{GW190421A}{0.67}{GW190413B}{0.84}{GW190413A}{0.79}{GW190412A}{1.15}{GW190408A}{0.72}}}
\newcommand{\costilttwoIMRmed}[1]{\IfEqCase{#1}{{GW190930A}{0.19}{GW190929A}{0.02}{GW190924A}{0.04}{GW190915A}{0.03}{GW190910A}{-0.12}{GW190909A}{-0.26}{GW190828B}{-0.03}{GW190828A}{0.31}{GW190814A}{-0.11}{GW190803A}{-0.07}{GW190731A}{0.08}{GW190728A}{0.29}{GW190727A}{0.14}{GW190720A}{0.11}{GW190719A}{0.36}{GW190708A}{0.08}{GW190707A}{-0.26}{GW190706A}{0.46}{GW190701A}{-0.21}{GW190630A}{0.09}{GW190620A}{0.40}{GW190602A}{0.21}{GW190527A}{0.11}{GW190521B}{0.05}{GW190521A}{0.0006}{GW190519A}{0.42}{GW190517A}{0.63}{GW190514A}{-0.33}{GW190513A}{0.18}{GW190512A}{0.04}{GW190503A}{-0.14}{GW190426A}{-1.00}{GW190425A}{0.16}{GW190424A}{0.20}{GW190421A}{-0.25}{GW190413B}{-0.03}{GW190413A}{-0.14}{GW190412A}{0.33}{GW190408A}{-0.18}}}
\newcommand{\costilttwoIMRplus}[1]{\IfEqCase{#1}{{GW190930A}{0.73}{GW190929A}{0.89}{GW190924A}{0.85}{GW190915A}{0.83}{GW190910A}{0.95}{GW190909A}{1.08}{GW190828B}{0.91}{GW190828A}{0.63}{GW190814A}{0.93}{GW190803A}{0.93}{GW190731A}{0.81}{GW190728A}{0.64}{GW190727A}{0.77}{GW190720A}{0.80}{GW190719A}{0.58}{GW190708A}{0.81}{GW190707A}{1.06}{GW190706A}{0.51}{GW190701A}{1.03}{GW190630A}{0.79}{GW190620A}{0.55}{GW190602A}{0.72}{GW190527A}{0.80}{GW190521B}{0.82}{GW190521A}{0.85}{GW190519A}{0.53}{GW190517A}{0.34}{GW190514A}{1.07}{GW190513A}{0.75}{GW190512A}{0.84}{GW190503A}{0.96}{GW190426A}{2.00}{GW190425A}{0.70}{GW190424A}{0.70}{GW190421A}{1.03}{GW190413B}{0.89}{GW190413A}{0.99}{GW190412A}{0.61}{GW190408A}{0.96}}}
\newcommand{\comovingdistIMRminus}[1]{\IfEqCase{#1}{{GW190930A}{256}{GW190929A}{521}{GW190924A}{174}{GW190915A}{390}{GW190910A}{537}{GW190909A}{1072}{GW190828B}{458}{GW190828A}{533}{GW190814A}{39}{GW190803A}{773}{GW190731A}{856}{GW190728A}{293}{GW190727A}{634}{GW190720A}{234}{GW190719A}{1002}{GW190708A}{301}{GW190707A}{274}{GW190706A}{940}{GW190701A}{455}{GW190630A}{310}{GW190620A}{734}{GW190602A}{697}{GW190527A}{679}{GW190521B}{359}{GW190521A}{815}{GW190519A}{714}{GW190517A}{523}{GW190514A}{997}{GW190513A}{511}{GW190512A}{414}{GW190503A}{420}{GW190426A}{143}{GW190425A}{67}{GW190424A}{706}{GW190421A}{658}{GW190413B}{905}{GW190413A}{668}{GW190412A}{164}{GW190408A}{387}}}
\newcommand{\comovingdistIMRmed}[1]{\IfEqCase{#1}{{GW190930A}{659}{GW190929A}{1382}{GW190924A}{500}{GW190915A}{1233}{GW190910A}{1351}{GW190909A}{2254}{GW190828B}{1274}{GW190828A}{1485}{GW190814A}{235}{GW190803A}{1967}{GW190731A}{2038}{GW190728A}{749}{GW190727A}{1861}{GW190720A}{643}{GW190719A}{2173}{GW190708A}{722}{GW190707A}{634}{GW190706A}{2646}{GW190701A}{1394}{GW190630A}{782}{GW190620A}{1805}{GW190602A}{1646}{GW190527A}{1598}{GW190521B}{928}{GW190521A}{2526}{GW190519A}{2189}{GW190517A}{1368}{GW190514A}{2289}{GW190513A}{1421}{GW190512A}{1120}{GW190503A}{1145}{GW190426A}{344}{GW190425A}{151}{GW190424A}{1442}{GW190421A}{1698}{GW190413B}{2293}{GW190413A}{2056}{GW190412A}{643}{GW190408A}{1104}}}
\newcommand{\comovingdistIMRplus}[1]{\IfEqCase{#1}{{GW190930A}{249}{GW190929A}{1394}{GW190924A}{165}{GW190915A}{412}{GW190910A}{475}{GW190909A}{1119}{GW190828B}{368}{GW190828A}{329}{GW190814A}{35}{GW190803A}{737}{GW190731A}{916}{GW190728A}{174}{GW190727A}{572}{GW190720A}{391}{GW190719A}{1070}{GW190708A}{240}{GW190707A}{269}{GW190706A}{787}{GW190701A}{422}{GW190630A}{331}{GW190620A}{730}{GW190602A}{811}{GW190527A}{1040}{GW190521B}{258}{GW190521A}{584}{GW190519A}{673}{GW190517A}{777}{GW190514A}{964}{GW190513A}{391}{GW190512A}{327}{GW190503A}{362}{GW190426A}{154}{GW190425A}{64}{GW190424A}{750}{GW190421A}{658}{GW190413B}{884}{GW190413A}{782}{GW190412A}{113}{GW190408A}{271}}}
\newcommand{\cosiotaIMRminus}[1]{\IfEqCase{#1}{{GW190930A}{1.53}{GW190929A}{0.69}{GW190924A}{1.73}{GW190915A}{0.63}{GW190910A}{0.84}{GW190909A}{0.97}{GW190828B}{0.84}{GW190828A}{0.29}{GW190814A}{1.45}{GW190803A}{1.18}{GW190731A}{1.09}{GW190728A}{1.31}{GW190727A}{1.39}{GW190720A}{0.15}{GW190719A}{0.95}{GW190708A}{1.18}{GW190707A}{0.39}{GW190706A}{1.01}{GW190701A}{0.58}{GW190630A}{1.50}{GW190620A}{0.45}{GW190602A}{0.81}{GW190527A}{1.26}{GW190521B}{1.37}{GW190521A}{1.43}{GW190519A}{0.81}{GW190517A}{0.39}{GW190514A}{0.96}{GW190513A}{1.65}{GW190512A}{0.96}{GW190503A}{0.20}{GW190426A}{0.84}{GW190425A}{1.42}{GW190424A}{0.99}{GW190421A}{0.83}{GW190413B}{0.81}{GW190413A}{1.61}{GW190412A}{1.27}{GW190408A}{0.94}}}
\newcommand{\cosiotaIMRmed}[1]{\IfEqCase{#1}{{GW190930A}{0.57}{GW190929A}{-0.23}{GW190924A}{0.81}{GW190915A}{-0.32}{GW190910A}{-0.13}{GW190909A}{0.05}{GW190828B}{-0.13}{GW190828A}{-0.69}{GW190814A}{0.69}{GW190803A}{0.24}{GW190731A}{0.14}{GW190728A}{0.34}{GW190727A}{0.45}{GW190720A}{-0.84}{GW190719A}{0.00}{GW190708A}{0.20}{GW190707A}{-0.59}{GW190706A}{0.08}{GW190701A}{0.71}{GW190630A}{0.53}{GW190620A}{-0.51}{GW190602A}{-0.12}{GW190527A}{0.35}{GW190521B}{0.46}{GW190521A}{0.48}{GW190519A}{-0.12}{GW190517A}{-0.55}{GW190514A}{0.01}{GW190513A}{0.76}{GW190512A}{-0.01}{GW190503A}{-0.78}{GW190426A}{-0.13}{GW190425A}{0.46}{GW190424A}{0.04}{GW190421A}{-0.10}{GW190413B}{-0.11}{GW190413A}{0.67}{GW190412A}{0.74}{GW190408A}{-0.03}}}
\newcommand{\cosiotaIMRplus}[1]{\IfEqCase{#1}{{GW190930A}{0.41}{GW190929A}{1.13}{GW190924A}{0.17}{GW190915A}{1.25}{GW190910A}{1.09}{GW190909A}{0.87}{GW190828B}{1.10}{GW190828A}{1.64}{GW190814A}{0.16}{GW190803A}{0.72}{GW190731A}{0.82}{GW190728A}{0.63}{GW190727A}{0.51}{GW190720A}{1.61}{GW190719A}{0.95}{GW190708A}{0.78}{GW190707A}{1.56}{GW190706A}{0.86}{GW190701A}{0.26}{GW190630A}{0.45}{GW190620A}{1.43}{GW190602A}{1.05}{GW190527A}{0.61}{GW190521B}{0.47}{GW190521A}{0.48}{GW190519A}{1.04}{GW190517A}{1.25}{GW190514A}{0.92}{GW190513A}{0.22}{GW190512A}{0.98}{GW190503A}{0.55}{GW190426A}{1.09}{GW190425A}{0.51}{GW190424A}{0.90}{GW190421A}{1.02}{GW190413B}{1.00}{GW190413A}{0.31}{GW190412A}{0.17}{GW190408A}{1.00}}}
\newcommand{\finalspinIMRminus}[1]{\IfEqCase{#1}{{GW190930A}{0.07}{GW190929A}{0.30}{GW190924A}{0.05}{GW190915A}{0.12}{GW190910A}{0.08}{GW190909A}{0.19}{GW190828A}{0.07}{GW190814A}{0.03}{GW190803A}{0.13}{GW190731A}{0.14}{GW190728A}{0.04}{GW190727A}{0.12}{GW190720A}{0.05}{GW190719A}{0.20}{GW190708A}{0.05}{GW190707A}{0.04}{GW190706A}{0.13}{GW190701A}{0.14}{GW190630A}{0.08}{GW190620A}{0.17}{GW190602A}{0.20}{GW190527A}{0.22}{GW190521B}{0.07}{GW190521A}{0.16}{GW190519A}{0.10}{GW190517A}{0.09}{GW190514A}{0.14}{GW190513A}{0.12}{GW190512A}{0.10}{GW190424A}{0.10}{GW190421A}{0.13}{GW190413B}{0.22}{GW190413A}{0.13}{GW190412A}{0.07}}}
\newcommand{\finalspinIMRmed}[1]{\IfEqCase{#1}{{GW190930A}{0.72}{GW190929A}{0.63}{GW190924A}{0.67}{GW190915A}{0.70}{GW190910A}{0.69}{GW190909A}{0.63}{GW190828A}{0.74}{GW190814A}{0.28}{GW190803A}{0.68}{GW190731A}{0.71}{GW190728A}{0.71}{GW190727A}{0.73}{GW190720A}{0.72}{GW190719A}{0.77}{GW190708A}{0.68}{GW190707A}{0.66}{GW190706A}{0.81}{GW190701A}{0.66}{GW190630A}{0.69}{GW190620A}{0.78}{GW190602A}{0.69}{GW190527A}{0.69}{GW190521B}{0.70}{GW190521A}{0.71}{GW190519A}{0.82}{GW190517A}{0.87}{GW190514A}{0.64}{GW190513A}{0.66}{GW190512A}{0.65}{GW190424A}{0.74}{GW190421A}{0.66}{GW190413B}{0.67}{GW190413A}{0.67}{GW190412A}{0.67}}}
\newcommand{\finalspinIMRplus}[1]{\IfEqCase{#1}{{GW190930A}{0.08}{GW190929A}{0.21}{GW190924A}{0.06}{GW190915A}{0.10}{GW190910A}{0.08}{GW190909A}{0.13}{GW190828A}{0.07}{GW190814A}{0.02}{GW190803A}{0.11}{GW190731A}{0.11}{GW190728A}{0.05}{GW190727A}{0.10}{GW190720A}{0.07}{GW190719A}{0.13}{GW190708A}{0.05}{GW190707A}{0.04}{GW190706A}{0.08}{GW190701A}{0.10}{GW190630A}{0.08}{GW190620A}{0.09}{GW190602A}{0.12}{GW190527A}{0.14}{GW190521B}{0.07}{GW190521A}{0.12}{GW190519A}{0.06}{GW190517A}{0.04}{GW190514A}{0.12}{GW190513A}{0.13}{GW190512A}{0.07}{GW190424A}{0.09}{GW190421A}{0.11}{GW190413B}{0.12}{GW190413A}{0.11}{GW190412A}{0.07}}}
\newcommand{\finalmassdetIMRminus}[1]{\IfEqCase{#1}{{GW190930A}{1.0}{GW190929A}{23.7}{GW190924A}{1.1}{GW190915A}{6.0}{GW190910A}{7.1}{GW190909A}{17.8}{GW190828A}{4.7}{GW190814A}{1.1}{GW190803A}{11.0}{GW190731A}{12.6}{GW190728A}{0.7}{GW190727A}{9.8}{GW190720A}{1.3}{GW190719A}{14.0}{GW190708A}{1.0}{GW190707A}{0.5}{GW190706A}{19.0}{GW190701A}{13.3}{GW190630A}{3.8}{GW190620A}{16.2}{GW190602A}{21.0}{GW190527A}{10.0}{GW190521B}{4.1}{GW190521A}{29.7}{GW190519A}{13.3}{GW190517A}{5.8}{GW190514A}{13.3}{GW190513A}{6.4}{GW190512A}{2.8}{GW190424A}{10.3}{GW190421A}{11.4}{GW190413B}{16.6}{GW190413A}{12.3}{GW190412A}{3.9}}}
\newcommand{\finalmassdetIMRmed}[1]{\IfEqCase{#1}{{GW190930A}{22.2}{GW190929A}{142.1}{GW190924A}{15.2}{GW190915A}{75.3}{GW190910A}{96.9}{GW190909A}{109.4}{GW190828A}{74.8}{GW190814A}{26.9}{GW190803A}{96.3}{GW190731A}{105.6}{GW190728A}{22.7}{GW190727A}{99.8}{GW190720A}{23.8}{GW190719A}{90.3}{GW190708A}{34.6}{GW190707A}{22.0}{GW190706A}{175.2}{GW190701A}{124.3}{GW190630A}{66.6}{GW190620A}{130.1}{GW190602A}{161.9}{GW190527A}{81.4}{GW190521B}{87.0}{GW190521A}{251.6}{GW190519A}{151.3}{GW190517A}{79.8}{GW190514A}{109.6}{GW190513A}{69.3}{GW190512A}{43.0}{GW190424A}{95.8}{GW190421A}{103.1}{GW190413B}{130.3}{GW190413A}{88.0}{GW190412A}{41.2}}}
\newcommand{\finalmassdetIMRplus}[1]{\IfEqCase{#1}{{GW190930A}{13.4}{GW190929A}{32.3}{GW190924A}{8.7}{GW190915A}{7.5}{GW190910A}{8.2}{GW190909A}{34.0}{GW190828A}{5.8}{GW190814A}{1.3}{GW190803A}{12.8}{GW190731A}{14.5}{GW190728A}{6.9}{GW190727A}{10.9}{GW190720A}{5.9}{GW190719A}{86.5}{GW190708A}{4.0}{GW190707A}{2.1}{GW190706A}{17.6}{GW190701A}{14.9}{GW190630A}{5.1}{GW190620A}{16.4}{GW190602A}{22.2}{GW190527A}{61.3}{GW190521B}{4.8}{GW190521A}{32.9}{GW190519A}{14.3}{GW190517A}{8.6}{GW190514A}{15.6}{GW190513A}{9.2}{GW190512A}{5.3}{GW190424A}{11.6}{GW190421A}{12.8}{GW190413B}{19.4}{GW190413A}{13.9}{GW190412A}{4.5}}}
\newcommand{\redshiftIMRminus}[1]{\IfEqCase{#1}{{GW190930A}{0.06}{GW190929A}{0.14}{GW190924A}{0.04}{GW190915A}{0.10}{GW190910A}{0.14}{GW190909A}{0.31}{GW190828B}{0.12}{GW190828A}{0.14}{GW190814A}{0.009}{GW190803A}{0.22}{GW190731A}{0.24}{GW190728A}{0.07}{GW190727A}{0.18}{GW190720A}{0.06}{GW190719A}{0.29}{GW190708A}{0.07}{GW190707A}{0.07}{GW190706A}{0.29}{GW190701A}{0.12}{GW190630A}{0.08}{GW190620A}{0.20}{GW190602A}{0.19}{GW190527A}{0.18}{GW190521B}{0.09}{GW190521A}{0.25}{GW190519A}{0.21}{GW190517A}{0.14}{GW190514A}{0.29}{GW190513A}{0.13}{GW190512A}{0.10}{GW190503A}{0.11}{GW190426A}{0.03}{GW190425A}{0.02}{GW190424A}{0.18}{GW190421A}{0.18}{GW190413B}{0.27}{GW190413A}{0.19}{GW190412A}{0.04}{GW190408A}{0.10}}}
\newcommand{\redshiftIMRmed}[1]{\IfEqCase{#1}{{GW190930A}{0.16}{GW190929A}{0.34}{GW190924A}{0.12}{GW190915A}{0.30}{GW190910A}{0.33}{GW190909A}{0.60}{GW190828B}{0.31}{GW190828A}{0.37}{GW190814A}{0.05}{GW190803A}{0.51}{GW190731A}{0.53}{GW190728A}{0.18}{GW190727A}{0.48}{GW190720A}{0.15}{GW190719A}{0.57}{GW190708A}{0.17}{GW190707A}{0.15}{GW190706A}{0.72}{GW190701A}{0.34}{GW190630A}{0.19}{GW190620A}{0.46}{GW190602A}{0.41}{GW190527A}{0.40}{GW190521B}{0.22}{GW190521A}{0.68}{GW190519A}{0.58}{GW190517A}{0.34}{GW190514A}{0.61}{GW190513A}{0.35}{GW190512A}{0.27}{GW190503A}{0.28}{GW190426A}{0.08}{GW190425A}{0.03}{GW190424A}{0.36}{GW190421A}{0.43}{GW190413B}{0.61}{GW190413A}{0.53}{GW190412A}{0.15}{GW190408A}{0.27}}}
\newcommand{\redshiftIMRplus}[1]{\IfEqCase{#1}{{GW190930A}{0.06}{GW190929A}{0.43}{GW190924A}{0.04}{GW190915A}{0.11}{GW190910A}{0.13}{GW190909A}{0.40}{GW190828B}{0.10}{GW190828A}{0.09}{GW190814A}{0.008}{GW190803A}{0.24}{GW190731A}{0.30}{GW190728A}{0.04}{GW190727A}{0.18}{GW190720A}{0.10}{GW190719A}{0.37}{GW190708A}{0.06}{GW190707A}{0.07}{GW190706A}{0.29}{GW190701A}{0.12}{GW190630A}{0.08}{GW190620A}{0.23}{GW190602A}{0.25}{GW190527A}{0.32}{GW190521B}{0.07}{GW190521A}{0.21}{GW190519A}{0.22}{GW190517A}{0.22}{GW190514A}{0.34}{GW190513A}{0.11}{GW190512A}{0.09}{GW190503A}{0.10}{GW190426A}{0.04}{GW190425A}{0.01}{GW190424A}{0.22}{GW190421A}{0.20}{GW190413B}{0.31}{GW190413A}{0.26}{GW190412A}{0.03}{GW190408A}{0.07}}}
\newcommand{\massonesourceIMRminus}[1]{\IfEqCase{#1}{{GW190930A}{2.6}{GW190929A}{34.7}{GW190924A}{2.5}{GW190915A}{7.0}{GW190910A}{6.3}{GW190909A}{11.3}{GW190828B}{6.4}{GW190828A}{4.4}{GW190814A}{1.1}{GW190803A}{7.9}{GW190731A}{9.5}{GW190728A}{2.3}{GW190727A}{7.2}{GW190720A}{3.3}{GW190719A}{13.8}{GW190708A}{3.0}{GW190707A}{1.8}{GW190706A}{15.9}{GW190701A}{9.6}{GW190630A}{7.5}{GW190620A}{14.2}{GW190602A}{18.6}{GW190527A}{11.8}{GW190521B}{5.9}{GW190521A}{16.8}{GW190519A}{13.6}{GW190517A}{7.6}{GW190514A}{8.9}{GW190513A}{9.9}{GW190512A}{5.3}{GW190503A}{7.8}{GW190426A}{2.3}{GW190425A}{0.3}{GW190424A}{7.7}{GW190421A}{8.3}{GW190413B}{12.6}{GW190413A}{6.6}{GW190412A}{4.2}{GW190408A}{4.0}}}
\newcommand{\massonesourceIMRmed}[1]{\IfEqCase{#1}{{GW190930A}{12.5}{GW190929A}{84.9}{GW190924A}{9.4}{GW190915A}{36.6}{GW190910A}{42.8}{GW190909A}{43.4}{GW190828B}{22.5}{GW190828A}{32.5}{GW190814A}{23.2}{GW190803A}{38.9}{GW190731A}{42.7}{GW190728A}{12.3}{GW190727A}{41.1}{GW190720A}{13.8}{GW190719A}{41.1}{GW190708A}{18.2}{GW190707A}{11.7}{GW190706A}{66.2}{GW190701A}{56.6}{GW190630A}{36.6}{GW190620A}{59.7}{GW190602A}{77.1}{GW190527A}{40.4}{GW190521B}{43.2}{GW190521A}{92.7}{GW190519A}{62.1}{GW190517A}{37.8}{GW190514A}{40.7}{GW190513A}{36.1}{GW190512A}{22.6}{GW190503A}{42.8}{GW190426A}{5.7}{GW190425A}{2.0}{GW190424A}{42.0}{GW190421A}{43.9}{GW190413B}{51.5}{GW190413A}{34.3}{GW190412A}{28.2}{GW190408A}{25.4}}}
\newcommand{\massonesourceIMRplus}[1]{\IfEqCase{#1}{{GW190930A}{14.9}{GW190929A}{29.2}{GW190924A}{9.6}{GW190915A}{9.3}{GW190910A}{8.9}{GW190909A}{23.3}{GW190828B}{7.9}{GW190828A}{7.0}{GW190814A}{1.3}{GW190803A}{12.9}{GW190731A}{14.0}{GW190728A}{8.7}{GW190727A}{12.9}{GW190720A}{7.4}{GW190719A}{36.8}{GW190708A}{6.5}{GW190707A}{3.7}{GW190706A}{24.2}{GW190701A}{14.0}{GW190630A}{7.6}{GW190620A}{17.4}{GW190602A}{23.0}{GW190527A}{26.2}{GW190521B}{7.5}{GW190521A}{22.5}{GW190519A}{19.2}{GW190517A}{14.9}{GW190514A}{17.2}{GW190513A}{10.6}{GW190512A}{7.5}{GW190503A}{11.3}{GW190426A}{3.9}{GW190425A}{0.6}{GW190424A}{12.5}{GW190421A}{13.8}{GW190413B}{20.2}{GW190413A}{10.8}{GW190412A}{4.6}{GW190408A}{6.3}}}
\newcommand{\masstwosourceIMRminus}[1]{\IfEqCase{#1}{{GW190930A}{3.5}{GW190929A}{9.8}{GW190924A}{2.0}{GW190915A}{6.0}{GW190910A}{7.2}{GW190909A}{11.4}{GW190828B}{2.5}{GW190828A}{5.2}{GW190814A}{0.10}{GW190803A}{8.0}{GW190731A}{10.2}{GW190728A}{2.8}{GW190727A}{10.0}{GW190720A}{2.3}{GW190719A}{6.8}{GW190708A}{3.1}{GW190707A}{1.8}{GW190706A}{15.8}{GW190701A}{12.4}{GW190630A}{4.4}{GW190620A}{11.4}{GW190602A}{17.9}{GW190527A}{8.8}{GW190521B}{6.5}{GW190521A}{19.9}{GW190519A}{13.3}{GW190517A}{9.0}{GW190514A}{9.4}{GW190513A}{3.9}{GW190512A}{3.3}{GW190503A}{9.2}{GW190426A}{0.5}{GW190425A}{0.3}{GW190424A}{8.3}{GW190421A}{9.6}{GW190413B}{13.0}{GW190413A}{6.7}{GW190412A}{1.1}{GW190408A}{4.1}}}
\newcommand{\masstwosourceIMRmed}[1]{\IfEqCase{#1}{{GW190930A}{7.7}{GW190929A}{22.6}{GW190924A}{4.8}{GW190915A}{23.8}{GW190910A}{33.6}{GW190909A}{27.8}{GW190828B}{10.5}{GW190828A}{25.7}{GW190814A}{2.58}{GW190803A}{27.6}{GW190731A}{29.4}{GW190728A}{8.1}{GW190727A}{29.9}{GW190720A}{7.8}{GW190719A}{20.1}{GW190708A}{12.8}{GW190707A}{8.4}{GW190706A}{40.6}{GW190701A}{40.0}{GW190630A}{21.9}{GW190620A}{33.5}{GW190602A}{41.2}{GW190527A}{21.0}{GW190521B}{31.9}{GW190521A}{65.3}{GW190519A}{39.3}{GW190517A}{25.1}{GW190514A}{29.6}{GW190513A}{17.2}{GW190512A}{12.8}{GW190503A}{28.5}{GW190426A}{1.5}{GW190425A}{1.4}{GW190424A}{31.8}{GW190421A}{31.0}{GW190413B}{32.3}{GW190413A}{24.9}{GW190412A}{8.8}{GW190408A}{18.4}}}
\newcommand{\masstwosourceIMRplus}[1]{\IfEqCase{#1}{{GW190930A}{1.9}{GW190929A}{16.8}{GW190924A}{1.6}{GW190915A}{6.0}{GW190910A}{6.8}{GW190909A}{12.2}{GW190828B}{3.8}{GW190828A}{4.5}{GW190814A}{0.09}{GW190803A}{8.0}{GW190731A}{9.6}{GW190728A}{1.8}{GW190727A}{7.1}{GW190720A}{2.4}{GW190719A}{11.0}{GW190708A}{2.3}{GW190707A}{1.5}{GW190706A}{15.4}{GW190701A}{10.2}{GW190630A}{6.1}{GW190620A}{13.2}{GW190602A}{19.0}{GW190527A}{13.1}{GW190521B}{5.9}{GW190521A}{18.4}{GW190519A}{12.0}{GW190517A}{6.9}{GW190514A}{9.7}{GW190513A}{7.4}{GW190512A}{3.7}{GW190503A}{7.8}{GW190426A}{0.8}{GW190425A}{0.3}{GW190424A}{7.8}{GW190421A}{8.6}{GW190413B}{12.2}{GW190413A}{6.4}{GW190412A}{1.6}{GW190408A}{3.5}}}
\newcommand{\totalmasssourceIMRminus}[1]{\IfEqCase{#1}{{GW190930A}{1.6}{GW190929A}{26.1}{GW190924A}{1.2}{GW190915A}{6.1}{GW190910A}{7.6}{GW190909A}{14.4}{GW190828B}{3.9}{GW190828A}{4.5}{GW190814A}{1.0}{GW190803A}{9.5}{GW190731A}{12.1}{GW190728A}{1.4}{GW190727A}{8.0}{GW190720A}{1.9}{GW190719A}{12.7}{GW190708A}{2.1}{GW190707A}{1.4}{GW190706A}{15.4}{GW190701A}{10.0}{GW190630A}{4.7}{GW190620A}{13.0}{GW190602A}{16.0}{GW190527A}{10.7}{GW190521B}{5.0}{GW190521A}{16.8}{GW190519A}{12.4}{GW190517A}{8.8}{GW190514A}{12.0}{GW190513A}{6.3}{GW190512A}{3.4}{GW190503A}{7.5}{GW190426A}{1.5}{GW190425A}{0.1}{GW190424A}{11.1}{GW190421A}{9.9}{GW190413B}{13.7}{GW190413A}{8.2}{GW190412A}{2.8}{GW190408A}{3.3}}}
\newcommand{\totalmasssourceIMRmed}[1]{\IfEqCase{#1}{{GW190930A}{20.4}{GW190929A}{106.6}{GW190924A}{14.2}{GW190915A}{60.5}{GW190910A}{76.2}{GW190909A}{71.6}{GW190828B}{33.2}{GW190828A}{57.9}{GW190814A}{25.8}{GW190803A}{66.7}{GW190731A}{72.1}{GW190728A}{20.6}{GW190727A}{70.8}{GW190720A}{21.7}{GW190719A}{61.2}{GW190708A}{31.3}{GW190707A}{20.3}{GW190706A}{106.8}{GW190701A}{96.4}{GW190630A}{58.8}{GW190620A}{93.5}{GW190602A}{118.6}{GW190527A}{61.6}{GW190521B}{74.9}{GW190521A}{156.3}{GW190519A}{101.3}{GW190517A}{63.7}{GW190514A}{70.6}{GW190513A}{53.7}{GW190512A}{35.6}{GW190503A}{71.3}{GW190426A}{7.2}{GW190425A}{3.4}{GW190424A}{74.2}{GW190421A}{75.0}{GW190413B}{84.2}{GW190413A}{59.5}{GW190412A}{37.0}{GW190408A}{43.9}}}
\newcommand{\totalmasssourceIMRplus}[1]{\IfEqCase{#1}{{GW190930A}{11.2}{GW190929A}{30.0}{GW190924A}{7.5}{GW190915A}{7.5}{GW190910A}{10.5}{GW190909A}{24.7}{GW190828B}{5.8}{GW190828A}{7.0}{GW190814A}{1.2}{GW190803A}{13.1}{GW190731A}{15.9}{GW190728A}{5.7}{GW190727A}{11.7}{GW190720A}{4.9}{GW190719A}{40.5}{GW190708A}{3.6}{GW190707A}{2.0}{GW190706A}{22.6}{GW190701A}{13.2}{GW190630A}{5.5}{GW190620A}{16.3}{GW190602A}{20.0}{GW190527A}{29.3}{GW190521B}{6.8}{GW190521A}{30.7}{GW190519A}{17.1}{GW190517A}{9.2}{GW190514A}{18.4}{GW190513A}{8.8}{GW190512A}{4.9}{GW190503A}{9.4}{GW190426A}{3.5}{GW190425A}{0.3}{GW190424A}{13.1}{GW190421A}{13.9}{GW190413B}{18.2}{GW190413A}{10.9}{GW190412A}{3.6}{GW190408A}{4.6}}}
\newcommand{\chirpmasssourceIMRminus}[1]{\IfEqCase{#1}{{GW190930A}{0.5}{GW190929A}{8.1}{GW190924A}{0.2}{GW190915A}{2.6}{GW190910A}{3.4}{GW190909A}{6.3}{GW190828B}{0.9}{GW190828A}{1.9}{GW190814A}{0.05}{GW190803A}{4.0}{GW190731A}{5.3}{GW190728A}{0.3}{GW190727A}{4.0}{GW190720A}{0.7}{GW190719A}{4.4}{GW190708A}{0.7}{GW190707A}{0.5}{GW190706A}{6.9}{GW190701A}{5.2}{GW190630A}{1.8}{GW190620A}{6.0}{GW190602A}{9.9}{GW190527A}{4.0}{GW190521B}{2.4}{GW190521A}{8.4}{GW190519A}{5.8}{GW190517A}{3.9}{GW190514A}{5.2}{GW190513A}{1.7}{GW190512A}{1.0}{GW190503A}{4.1}{GW190426A}{0.08}{GW190425A}{0.02}{GW190424A}{4.7}{GW190421A}{4.3}{GW190413B}{6.1}{GW190413A}{3.4}{GW190412A}{0.4}{GW190408A}{1.4}}}
\newcommand{\chirpmasssourceIMRmed}[1]{\IfEqCase{#1}{{GW190930A}{8.5}{GW190929A}{35.5}{GW190924A}{5.8}{GW190915A}{25.3}{GW190910A}{32.7}{GW190909A}{29.6}{GW190828B}{13.1}{GW190828A}{24.8}{GW190814A}{6.08}{GW190803A}{28.1}{GW190731A}{30.2}{GW190728A}{8.6}{GW190727A}{30.0}{GW190720A}{9.0}{GW190719A}{24.2}{GW190708A}{13.2}{GW190707A}{8.6}{GW190706A}{43.6}{GW190701A}{40.8}{GW190630A}{24.4}{GW190620A}{38.1}{GW190602A}{47.7}{GW190527A}{24.3}{GW190521B}{32.0}{GW190521A}{66.1}{GW190519A}{42.0}{GW190517A}{26.6}{GW190514A}{29.9}{GW190513A}{21.2}{GW190512A}{14.6}{GW190503A}{29.9}{GW190426A}{2.41}{GW190425A}{1.44}{GW190424A}{31.6}{GW190421A}{31.6}{GW190413B}{34.6}{GW190413A}{25.2}{GW190412A}{13.2}{GW190408A}{18.6}}}
\newcommand{\chirpmasssourceIMRplus}[1]{\IfEqCase{#1}{{GW190930A}{0.5}{GW190929A}{12.1}{GW190924A}{0.2}{GW190915A}{3.2}{GW190910A}{4.6}{GW190909A}{9.9}{GW190828B}{1.4}{GW190828A}{3.1}{GW190814A}{0.06}{GW190803A}{5.6}{GW190731A}{7.1}{GW190728A}{0.6}{GW190727A}{5.0}{GW190720A}{0.5}{GW190719A}{11.9}{GW190708A}{0.9}{GW190707A}{0.5}{GW190706A}{10.4}{GW190701A}{5.9}{GW190630A}{2.2}{GW190620A}{7.6}{GW190602A}{10.2}{GW190527A}{12.2}{GW190521B}{2.9}{GW190521A}{13.7}{GW190519A}{7.4}{GW190517A}{3.6}{GW190514A}{7.6}{GW190513A}{3.2}{GW190512A}{1.4}{GW190503A}{4.2}{GW190426A}{0.08}{GW190425A}{0.02}{GW190424A}{5.6}{GW190421A}{5.9}{GW190413B}{8.2}{GW190413A}{4.6}{GW190412A}{0.5}{GW190408A}{1.9}}}
\newcommand{\finalmasssourceIMRminus}[1]{\IfEqCase{#1}{{GW190930A}{1.5}{GW190929A}{26.6}{GW190924A}{1.2}{GW190915A}{5.8}{GW190910A}{7.2}{GW190909A}{13.9}{GW190828A}{4.2}{GW190814A}{1.0}{GW190803A}{9.0}{GW190731A}{11.6}{GW190728A}{1.3}{GW190727A}{7.6}{GW190720A}{1.8}{GW190719A}{12.2}{GW190708A}{2.0}{GW190707A}{1.3}{GW190706A}{14.8}{GW190701A}{9.3}{GW190630A}{4.6}{GW190620A}{12.4}{GW190602A}{15.5}{GW190527A}{10.5}{GW190521B}{4.6}{GW190521A}{15.5}{GW190519A}{11.9}{GW190517A}{8.4}{GW190514A}{11.4}{GW190513A}{6.3}{GW190512A}{3.4}{GW190424A}{10.5}{GW190421A}{9.4}{GW190413B}{13.5}{GW190413A}{7.7}{GW190412A}{2.9}}}
\newcommand{\finalmasssourceIMRmed}[1]{\IfEqCase{#1}{{GW190930A}{19.5}{GW190929A}{104.2}{GW190924A}{13.6}{GW190915A}{57.8}{GW190910A}{72.7}{GW190909A}{68.8}{GW190828A}{54.9}{GW190814A}{25.5}{GW190803A}{63.7}{GW190731A}{68.8}{GW190728A}{19.6}{GW190727A}{67.4}{GW190720A}{20.7}{GW190719A}{58.4}{GW190708A}{29.8}{GW190707A}{19.4}{GW190706A}{101.1}{GW190701A}{92.3}{GW190630A}{56.3}{GW190620A}{88.9}{GW190602A}{113.9}{GW190527A}{59.1}{GW190521B}{71.3}{GW190521A}{149.1}{GW190519A}{95.7}{GW190517A}{59.6}{GW190514A}{67.6}{GW190513A}{51.6}{GW190512A}{34.1}{GW190424A}{70.5}{GW190421A}{71.8}{GW190413B}{80.8}{GW190413A}{56.9}{GW190412A}{35.8}}}
\newcommand{\finalmasssourceIMRplus}[1]{\IfEqCase{#1}{{GW190930A}{11.5}{GW190929A}{29.0}{GW190924A}{7.7}{GW190915A}{7.1}{GW190910A}{9.9}{GW190909A}{23.7}{GW190828A}{6.6}{GW190814A}{1.2}{GW190803A}{12.4}{GW190731A}{15.0}{GW190728A}{5.9}{GW190727A}{11.0}{GW190720A}{5.1}{GW190719A}{38.9}{GW190708A}{3.7}{GW190707A}{2.1}{GW190706A}{21.4}{GW190701A}{12.4}{GW190630A}{5.3}{GW190620A}{15.3}{GW190602A}{18.7}{GW190527A}{28.2}{GW190521B}{6.4}{GW190521A}{28.4}{GW190519A}{16.3}{GW190517A}{9.3}{GW190514A}{17.7}{GW190513A}{8.7}{GW190512A}{5.1}{GW190424A}{12.4}{GW190421A}{13.4}{GW190413B}{17.7}{GW190413A}{10.3}{GW190412A}{3.7}}}
\newcommand{\radiatedenergyIMRminus}[1]{\IfEqCase{#1}{{GW190930A}{0.3}{GW190929A}{1.2}{GW190924A}{0.2}{GW190915A}{0.8}{GW190910A}{0.7}{GW190909A}{1.2}{GW190828A}{0.6}{GW190814A}{0.007}{GW190803A}{0.9}{GW190731A}{1.2}{GW190728A}{0.2}{GW190727A}{1.0}{GW190720A}{0.2}{GW190719A}{1.2}{GW190708A}{0.2}{GW190707A}{0.10}{GW190706A}{2.3}{GW190701A}{1.3}{GW190630A}{0.5}{GW190620A}{1.9}{GW190602A}{2.4}{GW190527A}{1.2}{GW190521B}{0.6}{GW190521A}{2.4}{GW190519A}{2.1}{GW190517A}{1.7}{GW190514A}{0.9}{GW190513A}{0.5}{GW190512A}{0.3}{GW190424A}{1.0}{GW190421A}{1.0}{GW190413B}{1.5}{GW190413A}{0.7}{GW190412A}{0.1}}}
\newcommand{\radiatedenergyIMRmed}[1]{\IfEqCase{#1}{{GW190930A}{0.9}{GW190929A}{2.5}{GW190924A}{0.6}{GW190915A}{2.7}{GW190910A}{3.5}{GW190909A}{2.8}{GW190828A}{3.0}{GW190814A}{0.2}{GW190803A}{3.0}{GW190731A}{3.3}{GW190728A}{1.0}{GW190727A}{3.4}{GW190720A}{1.0}{GW190719A}{2.8}{GW190708A}{1.4}{GW190707A}{0.9}{GW190706A}{5.8}{GW190701A}{4.1}{GW190630A}{2.6}{GW190620A}{4.7}{GW190602A}{5.0}{GW190527A}{2.5}{GW190521B}{3.5}{GW190521A}{7.2}{GW190519A}{5.7}{GW190517A}{4.1}{GW190514A}{2.9}{GW190513A}{2.1}{GW190512A}{1.5}{GW190424A}{3.7}{GW190421A}{3.2}{GW190413B}{3.5}{GW190413A}{2.6}{GW190412A}{1.2}}}
\newcommand{\radiatedenergyIMRplus}[1]{\IfEqCase{#1}{{GW190930A}{0.1}{GW190929A}{2.4}{GW190924A}{0.07}{GW190915A}{0.8}{GW190910A}{0.8}{GW190909A}{1.4}{GW190828A}{0.6}{GW190814A}{0.006}{GW190803A}{1.0}{GW190731A}{1.3}{GW190728A}{0.09}{GW190727A}{1.1}{GW190720A}{0.1}{GW190719A}{2.1}{GW190708A}{0.1}{GW190707A}{0.08}{GW190706A}{2.2}{GW190701A}{1.2}{GW190630A}{0.6}{GW190620A}{2.0}{GW190602A}{2.2}{GW190527A}{1.8}{GW190521B}{0.6}{GW190521A}{3.0}{GW190519A}{1.6}{GW190517A}{1.1}{GW190514A}{1.2}{GW190513A}{0.9}{GW190512A}{0.3}{GW190424A}{1.2}{GW190421A}{1.0}{GW190413B}{1.4}{GW190413A}{0.9}{GW190412A}{0.2}}}
\newcommand{\PEpercentBNSIMR}[1]{\IfEqCase{#1}{{GW190930A}{0}{GW190929A}{0}{GW190924A}{0}{GW190915A}{0}{GW190910A}{0}{GW190909A}{0}{GW190828B}{0}{GW190828A}{0}{GW190814A}{0}{GW190803A}{0}{GW190731A}{0}{GW190728A}{0}{GW190727A}{0}{GW190720A}{0}{GW190719A}{0}{GW190708A}{0}{GW190707A}{0}{GW190706A}{0}{GW190701A}{0}{GW190630A}{0}{GW190620A}{0}{GW190602A}{0}{GW190527A}{0}{GW190521B}{0}{GW190521A}{0}{GW190519A}{0}{GW190517A}{0}{GW190514A}{0}{GW190513A}{0}{GW190512A}{0}{GW190503A}{0}{GW190426A}{1}{GW190425A}{100}{GW190424A}{0}{GW190421A}{0}{GW190413B}{0}{GW190413A}{0}{GW190412A}{0}{GW190408A}{0}}}
\newcommand{\PEpercentNSBHIMR}[1]{\IfEqCase{#1}{{GW190930A}{0}{GW190929A}{0}{GW190924A}{8}{GW190915A}{0}{GW190910A}{0}{GW190909A}{0}{GW190828B}{0}{GW190828A}{0}{GW190814A}{100}{GW190803A}{0}{GW190731A}{0}{GW190728A}{0}{GW190727A}{0}{GW190720A}{0}{GW190719A}{0}{GW190708A}{0}{GW190707A}{0}{GW190706A}{0}{GW190701A}{0}{GW190630A}{0}{GW190620A}{0}{GW190602A}{0}{GW190527A}{0}{GW190521B}{0}{GW190521A}{0}{GW190519A}{0}{GW190517A}{0}{GW190514A}{0}{GW190513A}{0}{GW190512A}{0}{GW190503A}{0}{GW190426A}{99}{GW190425A}{0}{GW190424A}{0}{GW190421A}{0}{GW190413B}{0}{GW190413A}{0}{GW190412A}{0}{GW190408A}{0}}}
\newcommand{\PEpercentBBHIMR}[1]{\IfEqCase{#1}{{GW190930A}{100}{GW190929A}{100}{GW190924A}{92}{GW190915A}{100}{GW190910A}{100}{GW190909A}{100}{GW190828B}{100}{GW190828A}{100}{GW190814A}{0}{GW190803A}{100}{GW190731A}{100}{GW190728A}{100}{GW190727A}{100}{GW190720A}{100}{GW190719A}{100}{GW190708A}{100}{GW190707A}{100}{GW190706A}{100}{GW190701A}{100}{GW190630A}{100}{GW190620A}{100}{GW190602A}{100}{GW190527A}{100}{GW190521B}{100}{GW190521A}{100}{GW190519A}{100}{GW190517A}{100}{GW190514A}{100}{GW190513A}{100}{GW190512A}{100}{GW190503A}{100}{GW190426A}{0}{GW190425A}{0}{GW190424A}{100}{GW190421A}{100}{GW190413B}{100}{GW190413A}{100}{GW190412A}{100}{GW190408A}{100}}}
\newcommand{\PEpercentMassGapIMR}[1]{\IfEqCase{#1}{{GW190930A}{0}{GW190929A}{0}{GW190924A}{0}{GW190915A}{0}{GW190910A}{0}{GW190909A}{0}{GW190828B}{0}{GW190828A}{0}{GW190814A}{0}{GW190803A}{0}{GW190731A}{0}{GW190728A}{0}{GW190727A}{0}{GW190720A}{0}{GW190719A}{0}{GW190708A}{0}{GW190707A}{0}{GW190706A}{0}{GW190701A}{0}{GW190630A}{0}{GW190620A}{0}{GW190602A}{0}{GW190527A}{0}{GW190521B}{0}{GW190521A}{0}{GW190519A}{0}{GW190517A}{0}{GW190514A}{0}{GW190513A}{0}{GW190512A}{0}{GW190503A}{0}{GW190426A}{0}{GW190425A}{0}{GW190424A}{0}{GW190421A}{0}{GW190413B}{0}{GW190413A}{0}{GW190412A}{0}{GW190408A}{0}}}

\newcommand{\TIME}[1]{\IfEqCase{#1}{{GW190413A}{05:29:54}{GW190719A}{21:55:14}{GW190620A}{03:04:21}{GW190514A}{06:54:16}{GW190731A}{14:09:36}{GW190503A}{18:54:04}{GW190602A}{17:59:27}{GW190929A}{01:21:49}{GW190517A}{05:51:01}{GW190915A}{23:57:02}{GW190425A}{08:18:05}{GW190512A}{18:07:14}{GW190630A}{18:52:05}{GW190521A}{03:02:29}{GW190413B}{13:43:08}{GW190924A}{02:18:46}{GW190930A}{13:35:41}{GW190706A}{22:26:41}{GW190408A}{18:18:02}{GW190909A}{11:41:49}{GW190728A}{06:45:10}{GW190426A}{15:21:55}{GW190412A}{05:30:44}{GW190720A}{00:08:36}{GW190521B}{07:43:59}{GW190910A}{11:28:07}{GW190803A}{02:27:01}{GW190519A}{15:35:44}{GW190708A}{23:24:57}{GW190527A}{09:20:55}{GW190513A}{20:54:28}{GW190424A}{18:06:48}{GW190727A}{06:03:33}{GW190814A}{21:10:39}{GW190707A}{09:33:26}{GW190828A}{06:34:05}{GW190828B}{06:55:09}{GW190701A}{20:33:06}{GW190421A}{21:38:56}}}

\newcommand{\DATE}[1]{\IfEqCase{#1}{{GW190413A}{2019-04-13 05:29:54}{GW190719A}{2019-07-19 21:55:14}{GW190620A}{2019-06-20 03:04:21}{GW190514A}{2019-05-14 06:54:16}{GW190731A}{2019-07-31 14:09:36}{GW190503A}{2019-05-03 18:54:04}{GW190602A}{2019-06-02 17:59:27}{GW190929A}{2019-09-29 01:21:49}{GW190517A}{2019-05-17 05:51:01}{GW190915A}{2019-09-15 23:57:02}{GW190425A}{2019-04-25 08:18:05}{GW190512A}{2019-05-12 18:07:14}{GW190630A}{2019-06-30 18:52:05}{GW190521A}{2019-05-21 03:02:29}{GW190413B}{2019-04-13 13:43:08}{GW190924A}{2019-09-24 02:18:46}{GW190930A}{2019-09-30 13:35:41}{GW190706A}{2019-07-06 22:26:41}{GW190408A}{2019-04-08 18:18:02}{GW190909A}{2019-09-09 11:41:49}{GW190728A}{2019-07-28 06:45:10}{GW190426A}{2019-04-26 15:21:55}{GW190412A}{2019-04-12 05:30:44}{GW190720A}{2019-07-20 00:08:36}{GW190521B}{2019-05-21 07:43:59}{GW190910A}{2019-09-10 11:28:07}{GW190803A}{2019-08-03 02:27:01}{GW190519A}{2019-05-19 15:35:44}{GW190708A}{2019-07-08 23:24:57}{GW190527A}{2019-05-27 09:20:55}{GW190513A}{2019-05-13 20:54:28}{GW190424A}{2019-04-24 18:06:48}{GW190727A}{2019-07-27 06:03:33}{GW190814A}{2019-08-14 21:10:39}{GW190707A}{2019-07-07 09:33:26}{GW190828A}{2019-08-28 06:34:05}{GW190828B}{2019-08-28 06:55:09}{GW190701A}{2019-07-01 20:33:06}{GW190421A}{2019-04-21 21:38:56}}}

\newcommand{\NAME}[1]{\IfEqCase{#1}{{GW190413A}{GW190413\_052954}{GW190719A}{GW190719\_215514}{GW190620A}{GW190620\_030421}{GW190514A}{GW190514\_065416}{GW190731A}{GW190731\_140936}{GW190503A}{GW190503\_185404}{GW190602A}{GW190602\_175927}{GW190929A}{GW190929\_012149}{GW190517A}{GW190517\_055101}{GW190915A}{GW190915\_235702}{GW190425A}{GW190425}{GW190512A}{GW190512\_180714}{GW190630A}{GW190630\_185205}{GW190521A}{GW190521}{GW190413B}{GW190413\_134308}{GW190924A}{GW190924\_021846}{GW190930A}{GW190930\_133541}{GW190706A}{GW190706\_222641}{GW190408A}{GW190408\_181802}{GW190909A}{GW190909\_114149}{GW190728A}{GW190728\_064510}{GW190426A}{GW190426\_152155}{GW190412A}{GW190412}{GW190720A}{GW190720\_000836}{GW190521B}{GW190521\_074359}{GW190910A}{GW190910\_112807}{GW190803A}{GW190803\_022701}{GW190519A}{GW190519\_153544}{GW190708A}{GW190708\_232457}{GW190527A}{GW190527\_092055}{GW190513A}{GW190513\_205428}{GW190424A}{GW190424\_180648}{GW190727A}{GW190727\_060333}{GW190814A}{GW190814}{GW190707A}{GW190707\_093326}{GW190828A}{GW190828\_063405}{GW190828B}{GW190828\_065509}{GW190701A}{GW190701\_203306}{GW190421A}{GW190421\_213856}}}

\newcommand{\SID}[1]{\IfEqCase{#1}{{GW190413A}{S190413i}{GW190719A}{S190719an}{GW190620A}{S190620e}{GW190514A}{S190514n}{GW190731A}{S190731aa}{GW190503A}{S190503bf}{GW190602A}{S190602aq}{GW190929A}{S190929d}{GW190517A}{S190517h}{GW190915A}{S190915ak}{GW190425A}{S190425z}{GW190512A}{S190512at}{GW190630A}{S190630ag}{GW190521A}{S190521g}{GW190413B}{S190413ac}{GW190924A}{S190924h}{GW190930A}{S190930s}{GW190706A}{S190706ai}{GW190408A}{S190408an}{GW190909A}{S190909w}{GW190728A}{S190728q}{GW190426A}{S190426c}{GW190412A}{S190412m}{GW190720A}{S190720a}{GW190521B}{S190521r}{GW190910A}{S190910s}{GW190803A}{S190803e}{GW190519A}{S190519bj}{GW190708A}{S190708ap}{GW190527A}{S190527w}{GW190513A}{S190513bm}{GW190424A}{S190424ao}{GW190727A}{S190727h}{GW190814A}{S190814bv}{GW190707A}{S190707q}{GW190828A}{S190828j}{GW190828B}{S190828l}{GW190701A}{S190701ah}{GW190421A}{S190421ar}}}

\newcommand{\PUBLIC}[1]{\IfEqCase{#1}{{GW190413A}{\bf}{GW190719A}{\bf}{GW190620A}{\bf}{GW190514A}{\bf}{GW190731A}{\bf}{GW190503A}{}{GW190602A}{}{GW190929A}{\bf}{GW190517A}{}{GW190915A}{}{GW190425A}{}{GW190512A}{}{GW190630A}{}{GW190521A}{}{GW190413B}{\bf}{GW190924A}{}{GW190930A}{}{GW190706A}{}{GW190408A}{}{GW190909A}{\bf}{GW190728A}{}{GW190426A}{}{GW190412A}{}{GW190720A}{}{GW190521B}{}{GW190910A}{\bf}{GW190803A}{\bf}{GW190519A}{}{GW190708A}{\bf}{GW190527A}{\bf}{GW190513A}{}{GW190424A}{\bf}{GW190727A}{}{GW190814A}{}{GW190707A}{}{GW190828A}{}{GW190828B}{}{GW190701A}{}{GW190421A}{}}}

\newcommand{\INSTRUMENTS}[1]{\IfEqCase{#1}{{GW190413A}{HL}{GW190719A}{HL}{GW190620A}{LV}{GW190514A}{HL}{GW190731A}{HL}{GW190503A}{HLV}{GW190602A}{HLV}{GW190929A}{HLV}{GW190517A}{HLV}{GW190915A}{HLV}{GW190425A}{LV}{GW190512A}{HLV}{GW190630A}{LV}{GW190521A}{HLV}{GW190413B}{HLV}{GW190924A}{HLV}{GW190930A}{HL}{GW190706A}{HLV}{GW190408A}{HLV}{GW190909A}{HL}{GW190728A}{HLV}{GW190426A}{HLV}{GW190412A}{HLV}{GW190720A}{HLV}{GW190521B}{HL}{GW190910A}{LV}{GW190803A}{HLV}{GW190519A}{HLV}{GW190708A}{LV}{GW190527A}{HL}{GW190513A}{HLV}{GW190424A}{L}{GW190727A}{HLV}{GW190814A}{LV}{GW190707A}{HL}{GW190828A}{HLV}{GW190828B}{HLV}{GW190701A}{HLV}{GW190421A}{HL}}}

\newcommand{\PARTINSTRUMENTS}[1]{\IfEqCase{#1}{{GW190413A}{HL}{GW190719A}{HL}{GW190620A}{L}{GW190514A}{HL}{GW190731A}{HL}{GW190503A}{HL}{GW190602A}{HL}{GW190929A}{HL}{GW190517A}{HL}{GW190915A}{HL}{GW190425A}{L}{GW190512A}{HL}{GW190630A}{LV}{GW190521A}{HL}{GW190413B}{HL}{GW190924A}{HL}{GW190930A}{HL}{GW190706A}{HL}{GW190408A}{HL}{GW190909A}{HL}{GW190728A}{HL}{GW190426A}{HL}{GW190412A}{HL}{GW190720A}{HLV}{GW190521B}{HL}{GW190910A}{L}{GW190803A}{HL}{GW190519A}{HL}{GW190708A}{L}{GW190527A}{HL}{GW190513A}{HL}{GW190424A}{L}{GW190727A}{HL}{GW190814A}{LV}{GW190707A}{HL}{GW190828A}{HL}{GW190828B}{HL}{GW190701A}{HLV}{GW190421A}{HL}}}

\newcommand{\CWBALLSKYPTERRES}[1]{\IfEqCase{#1}{{GW190413A}{--}{GW190719A}{--}{GW190620A}{--}{GW190514A}{--}{GW190731A}{--}{GW190503A}{--}{GW190602A}{--}{GW190929A}{--}{GW190517A}{--}{GW190915A}{--}{GW190425A}{--}{GW190512A}{--}{GW190630A}{--}{GW190521A}{--}{GW190413B}{--}{GW190924A}{--}{GW190930A}{--}{GW190706A}{--}{GW190408A}{--}{GW190909A}{--}{GW190728A}{--}{GW190426A}{--}{GW190412A}{--}{GW190720A}{--}{GW190521B}{--}{GW190910A}{--}{GW190803A}{--}{GW190519A}{--}{GW190708A}{--}{GW190527A}{--}{GW190513A}{--}{GW190424A}{--}{GW190727A}{--}{GW190814A}{--}{GW190707A}{--}{GW190828A}{--}{GW190828B}{--}{GW190701A}{--}{GW190421A}{--}}}

\newcommand{\CWBALLSKYPASTRO}[1]{\IfEqCase{#1}{{GW190413A}{--}{GW190719A}{--}{GW190620A}{~}{GW190514A}{--}{GW190731A}{--}{GW190503A}{--}{GW190602A}{--}{GW190929A}{--}{GW190517A}{--}{GW190915A}{--}{GW190425A}{~}{GW190512A}{--}{GW190630A}{~}{GW190521A}{--}{GW190413B}{--}{GW190924A}{--}{GW190930A}{--}{GW190706A}{--}{GW190408A}{--}{GW190909A}{--}{GW190728A}{--}{GW190426A}{--}{GW190412A}{--}{GW190720A}{--}{GW190521B}{--}{GW190910A}{~}{GW190803A}{--}{GW190519A}{--}{GW190708A}{~}{GW190527A}{--}{GW190513A}{--}{GW190424A}{~}{GW190727A}{--}{GW190814A}{~}{GW190707A}{--}{GW190828A}{--}{GW190828B}{--}{GW190701A}{--}{GW190421A}{--}}}

\newcommand{\PYCBCHIGHMASSPTERRES}[1]{\IfEqCase{#1}{{GW190413A}{$0.02$}{GW190719A}{$0.18$}{GW190620A}{--}{GW190514A}{$0.04$}{GW190731A}{$0.04$}{GW190503A}{$0.00$}{GW190602A}{$0.00$}{GW190929A}{--}{GW190517A}{$0.00$}{GW190915A}{$0.00$}{GW190425A}{--}{GW190512A}{$0.00$}{GW190630A}{--}{GW190521A}{--}{GW190413B}{$0.02$}{GW190924A}{$0.00$}{GW190930A}{$0.01$}{GW190706A}{$0.00$}{GW190408A}{$0.00$}{GW190909A}{--}{GW190728A}{$0.00$}{GW190426A}{--}{GW190412A}{$0.00$}{GW190720A}{$0.00$}{GW190521B}{$0.00$}{GW190910A}{--}{GW190803A}{$0.01$}{GW190519A}{$0.00$}{GW190708A}{--}{GW190527A}{--}{GW190513A}{$0.00$}{GW190424A}{--}{GW190727A}{$0.00$}{GW190814A}{--}{GW190707A}{$0.00$}{GW190828A}{$0.00$}{GW190828B}{$0.00$}{GW190701A}{--}{GW190421A}{$0.00$}}}

\newcommand{\PYCBCHIGHMASSPASTRO}[1]{\IfEqCase{#1}{{GW190413A}{$0.98$}{GW190719A}{$0.82$}{GW190620A}{~}{GW190514A}{$0.96$}{GW190731A}{$0.96$}{GW190503A}{$1.00$}{GW190602A}{$1.00$}{GW190929A}{--}{GW190517A}{$1.00$}{GW190915A}{$1.00$}{GW190425A}{~}{GW190512A}{$1.00$}{GW190630A}{~}{GW190521A}{--}{GW190413B}{$0.98$}{GW190924A}{$1.00$}{GW190930A}{$0.99$}{GW190706A}{$1.00$}{GW190408A}{$1.00$}{GW190909A}{--}{GW190728A}{$1.00$}{GW190426A}{--}{GW190412A}{$1.00$}{GW190720A}{$1.00$}{GW190521B}{$1.00$}{GW190910A}{~}{GW190803A}{$0.99$}{GW190519A}{$1.00$}{GW190708A}{~}{GW190527A}{--}{GW190513A}{$1.00$}{GW190424A}{~}{GW190727A}{$1.00$}{GW190814A}{~}{GW190707A}{$1.00$}{GW190828A}{$1.00$}{GW190828B}{$1.00$}{GW190701A}{--}{GW190421A}{$1.00$}}}

\newcommand{\PYCBCALLSKYPTERRES}[1]{\IfEqCase{#1}{{GW190413A}{--}{GW190719A}{--}{GW190620A}{--}{GW190514A}{--}{GW190731A}{--}{GW190503A}{$0.00$}{GW190602A}{--}{GW190929A}{--}{GW190517A}{$0.00$}{GW190915A}{$0.00$}{GW190425A}{--}{GW190512A}{$0.00$}{GW190630A}{--}{GW190521A}{$0.07$}{GW190413B}{--}{GW190924A}{$0.00$}{GW190930A}{$0.00$}{GW190706A}{$0.00$}{GW190408A}{$0.00$}{GW190909A}{--}{GW190728A}{$0.00$}{GW190426A}{--}{GW190412A}{$0.00$}{GW190720A}{$0.00$}{GW190521B}{$0.00$}{GW190910A}{--}{GW190803A}{--}{GW190519A}{$0.00$}{GW190708A}{--}{GW190527A}{--}{GW190513A}{$0.00$}{GW190424A}{--}{GW190727A}{$0.00$}{GW190814A}{--}{GW190707A}{$0.00$}{GW190828A}{$0.00$}{GW190828B}{$0.00$}{GW190701A}{--}{GW190421A}{$0.11$}}}

\newcommand{\PYCBCALLSKYPASTRO}[1]{\IfEqCase{#1}{{GW190413A}{--}{GW190719A}{--}{GW190620A}{~}{GW190514A}{--}{GW190731A}{--}{GW190503A}{$1.00$}{GW190602A}{--}{GW190929A}{--}{GW190517A}{$1.00$}{GW190915A}{$1.00$}{GW190425A}{~}{GW190512A}{$1.00$}{GW190630A}{~}{GW190521A}{$0.93$}{GW190413B}{--}{GW190924A}{$1.00$}{GW190930A}{$1.00$}{GW190706A}{$1.00$}{GW190408A}{$1.00$}{GW190909A}{--}{GW190728A}{$1.00$}{GW190426A}{--}{GW190412A}{$1.00$}{GW190720A}{$1.00$}{GW190521B}{$1.00$}{GW190910A}{~}{GW190803A}{--}{GW190519A}{$1.00$}{GW190708A}{~}{GW190527A}{--}{GW190513A}{$1.00$}{GW190424A}{~}{GW190727A}{$1.00$}{GW190814A}{~}{GW190707A}{$1.00$}{GW190828A}{$1.00$}{GW190828B}{$1.00$}{GW190701A}{--}{GW190421A}{$0.89$}}}

\newcommand{\GSTLALALLSKYPTERRES}[1]{\IfEqCase{#1}{{GW190413A}{--}{GW190719A}{--}{GW190620A}{$0.00$}{GW190514A}{--}{GW190731A}{$0.03$}{GW190503A}{$0.00$}{GW190602A}{$0.00$}{GW190929A}{$0.00$}{GW190517A}{$0.00$}{GW190915A}{$0.00$}{GW190425A}{--}{GW190512A}{$0.00$}{GW190630A}{$0.00$}{GW190521A}{$0.00$}{GW190413B}{$0.05$}{GW190924A}{$0.00$}{GW190930A}{$0.08$}{GW190706A}{$0.00$}{GW190408A}{$0.00$}{GW190909A}{$0.11$}{GW190728A}{$0.00$}{GW190426A}{--}{GW190412A}{$0.00$}{GW190720A}{$0.00$}{GW190521B}{$0.00$}{GW190910A}{$0.00$}{GW190803A}{$0.01$}{GW190519A}{$0.00$}{GW190708A}{$0.00$}{GW190527A}{$0.01$}{GW190513A}{$0.00$}{GW190424A}{$0.09$}{GW190727A}{$0.00$}{GW190814A}{$0.00$}{GW190707A}{$0.00$}{GW190828A}{$0.00$}{GW190828B}{$0.00$}{GW190701A}{$0.00$}{GW190421A}{$0.00$}}}

\newcommand{\GSTLALALLSKYPASTRO}[1]{\IfEqCase{#1}{{GW190413A}{--}{GW190719A}{--}{GW190620A}{$1.00$}{GW190514A}{--}{GW190731A}{$0.97$}{GW190503A}{$1.00$}{GW190602A}{$1.00$}{GW190929A}{$1.00$}{GW190517A}{$1.00$}{GW190915A}{$1.00$}{GW190425A}{--}{GW190512A}{$1.00$}{GW190630A}{$1.00$}{GW190521A}{$1.00$}{GW190413B}{$0.95$}{GW190924A}{$1.00$}{GW190930A}{$0.92$}{GW190706A}{$1.00$}{GW190408A}{$1.00$}{GW190909A}{$0.89$}{GW190728A}{$1.00$}{GW190426A}{--}{GW190412A}{$1.00$}{GW190720A}{$1.00$}{GW190521B}{$1.00$}{GW190910A}{$1.00$}{GW190803A}{$0.99$}{GW190519A}{$1.00$}{GW190708A}{$1.00$}{GW190527A}{$0.99$}{GW190513A}{$1.00$}{GW190424A}{$0.91$}{GW190727A}{$1.00$}{GW190814A}{$1.00$}{GW190707A}{$1.00$}{GW190828A}{$1.00$}{GW190828B}{$1.00$}{GW190701A}{$1.00$}{GW190421A}{$1.00$}}}

\newcommand{\MINFAR}[1]{\IfEqCase{#1}{{GW190413A}{$7.2 \times 10^{-2}$}{GW190719A}{$1.6 \times 10^{0}$}{GW190620A}{$2.9 \times 10^{-3}$}{GW190514A}{$5.3 \times 10^{-1}$}{GW190731A}{$2.1 \times 10^{-1}$}{GW190503A}{$1.0 \times 10^{-5}$}{GW190602A}{$1.1 \times 10^{-5}$}{GW190929A}{$2.0 \times 10^{-2}$}{GW190517A}{$5.7 \times 10^{-5}$}{GW190915A}{$1.0 \times 10^{-5}$}{GW190425A}{$7.5 \times 10^{-4}$}{GW190512A}{$1.0 \times 10^{-5}$}{GW190630A}{$1.0 \times 10^{-5}$}{GW190521A}{$2.0 \times 10^{-4}$}{GW190413B}{$4.4 \times 10^{-2}$}{GW190924A}{$1.0 \times 10^{-5}$}{GW190930A}{$3.3 \times 10^{-2}$}{GW190706A}{$1.0 \times 10^{-5}$}{GW190408A}{$1.0 \times 10^{-5}$}{GW190909A}{$1.1 \times 10^{0}$}{GW190728A}{$1.0 \times 10^{-5}$}{GW190426A}{$1.4 \times 10^{0}$}{GW190412A}{$1.0 \times 10^{-5}$}{GW190720A}{$1.0 \times 10^{-5}$}{GW190521B}{$1.0 \times 10^{-5}$}{GW190910A}{$1.9 \times 10^{-5}$}{GW190803A}{$2.7 \times 10^{-2}$}{GW190519A}{$1.0 \times 10^{-5}$}{GW190708A}{$2.8 \times 10^{-5}$}{GW190527A}{$6.2 \times 10^{-2}$}{GW190513A}{$1.0 \times 10^{-5}$}{GW190424A}{$7.8 \times 10^{-1}$}{GW190727A}{$1.0 \times 10^{-5}$}{GW190814A}{$1.0 \times 10^{-5}$}{GW190707A}{$1.0 \times 10^{-5}$}{GW190828A}{$1.0 \times 10^{-5}$}{GW190828B}{$1.0 \times 10^{-5}$}{GW190701A}{$1.1 \times 10^{-2}$}{GW190421A}{$7.7 \times 10^{-4}$}}}

\newcommand{\CWBALLSKYFAR}[1]{\IfEqCase{#1}{{GW190413A}{--}{GW190719A}{--}{GW190620A}{~}{GW190514A}{--}{GW190731A}{--}{GW190503A}{$1.8 \times 10^{-3}$}{GW190602A}{$1.5 \times 10^{-2}$}{GW190929A}{--}{GW190517A}{$6.5 \times 10^{-3}$}{GW190915A}{$<$ $1.0 \times 10^{-3}$}{GW190425A}{~}{GW190512A}{$8.8 \times 10^{-1}$}{GW190630A}{~}{GW190521A}{$2.0 \times 10^{-4}$}{GW190413B}{--}{GW190924A}{--}{GW190930A}{--}{GW190706A}{$<$ $1.0 \times 10^{-3}$}{GW190408A}{$<$ $9.5 \times 10^{-4}$}{GW190909A}{--}{GW190728A}{--}{GW190426A}{--}{GW190412A}{$<$ $9.5 \times 10^{-4}$}{GW190720A}{--}{GW190521B}{$<$ $1.0 \times 10^{-4}$}{GW190910A}{~}{GW190803A}{--}{GW190519A}{$3.1 \times 10^{-4}$}{GW190708A}{~}{GW190527A}{--}{GW190513A}{--}{GW190424A}{~}{GW190727A}{$8.8 \times 10^{-2}$}{GW190814A}{~}{GW190707A}{--}{GW190828A}{$<$ $9.6 \times 10^{-4}$}{GW190828B}{--}{GW190701A}{$5.5 \times 10^{-1}$}{GW190421A}{$3.0 \times 10^{-1}$}}}

\newcommand{\CWBALLSKYIFAR}[1]{\IfEqCase{#1}{{GW190413A}{--}{GW190719A}{--}{GW190620A}{~}{GW190514A}{--}{GW190731A}{--}{GW190503A}{2.7}{GW190602A}{1.8}{GW190929A}{--}{GW190517A}{2.2}{GW190915A}{3.0}{GW190425A}{~}{GW190512A}{0.1}{GW190630A}{~}{GW190521A}{3.7}{GW190413B}{--}{GW190924A}{--}{GW190930A}{--}{GW190706A}{3.0}{GW190408A}{3.0}{GW190909A}{--}{GW190728A}{--}{GW190426A}{--}{GW190412A}{3.0}{GW190720A}{--}{GW190521B}{4.0}{GW190910A}{~}{GW190803A}{--}{GW190519A}{3.5}{GW190708A}{~}{GW190527A}{--}{GW190513A}{--}{GW190424A}{~}{GW190727A}{1.1}{GW190814A}{~}{GW190707A}{--}{GW190828A}{3.0}{GW190828B}{--}{GW190701A}{0.3}{GW190421A}{0.5}}}

\newcommand{\CWBALLSKYSNR}[1]{\IfEqCase{#1}{{GW190413A}{--}{GW190719A}{--}{GW190620A}{~}{GW190514A}{--}{GW190731A}{--}{GW190503A}{11.5}{GW190602A}{11.1}{GW190929A}{--}{GW190517A}{10.7}{GW190915A}{12.3}{GW190425A}{~}{GW190512A}{10.7}{GW190630A}{~}{GW190521A}{14.4}{GW190413B}{--}{GW190924A}{--}{GW190930A}{--}{GW190706A}{12.7}{GW190408A}{14.8}{GW190909A}{--}{GW190728A}{--}{GW190426A}{--}{GW190412A}{19.7}{GW190720A}{--}{GW190521B}{24.7}{GW190910A}{~}{GW190803A}{--}{GW190519A}{14.0}{GW190708A}{~}{GW190527A}{--}{GW190513A}{--}{GW190424A}{~}{GW190727A}{11.4}{GW190814A}{~}{GW190707A}{--}{GW190828A}{16.6}{GW190828B}{--}{GW190701A}{10.2}{GW190421A}{9.3}}}

\newcommand{\PYCBCHIGHMASSFAR}[1]{\IfEqCase{#1}{{GW190413A}{$7.2 \times 10^{-2}$}{GW190719A}{$1.6 \times 10^{0}$}{GW190620A}{~}{GW190514A}{$5.3 \times 10^{-1}$}{GW190731A}{$2.8 \times 10^{-1}$}{GW190503A}{$<$ $7.9 \times 10^{-5}$}{GW190602A}{$1.5 \times 10^{-2}$}{GW190929A}{--}{GW190517A}{$<$ $5.7 \times 10^{-5}$}{GW190915A}{$<$ $3.3 \times 10^{-5}$}{GW190425A}{~}{GW190512A}{$<$ $5.7 \times 10^{-5}$}{GW190630A}{~}{GW190521A}{--}{GW190413B}{$4.4 \times 10^{-2}$}{GW190924A}{$<$ $3.3 \times 10^{-5}$}{GW190930A}{$3.3 \times 10^{-2}$}{GW190706A}{$<$ $4.6 \times 10^{-5}$}{GW190408A}{$<$ $7.9 \times 10^{-5}$}{GW190909A}{--}{GW190728A}{$<$ $3.7 \times 10^{-5}$}{GW190426A}{--}{GW190412A}{$<$ $7.9 \times 10^{-5}$}{GW190720A}{$<$ $3.7 \times 10^{-5}$}{GW190521B}{$<$ $5.7 \times 10^{-5}$}{GW190910A}{~}{GW190803A}{$2.7 \times 10^{-2}$}{GW190519A}{$<$ $5.7 \times 10^{-5}$}{GW190708A}{~}{GW190527A}{--}{GW190513A}{$<$ $5.7 \times 10^{-5}$}{GW190424A}{~}{GW190727A}{$<$ $3.7 \times 10^{-5}$}{GW190814A}{~}{GW190707A}{$<$ $4.6 \times 10^{-5}$}{GW190828A}{$<$ $3.3 \times 10^{-5}$}{GW190828B}{$<$ $3.3 \times 10^{-5}$}{GW190701A}{--}{GW190421A}{$6.6 \times 10^{-3}$}}}

\newcommand{\PYCBCHIGHMASSIFAR}[1]{\IfEqCase{#1}{{GW190413A}{1.1}{GW190719A}{-0.2}{GW190620A}{~}{GW190514A}{0.3}{GW190731A}{0.6}{GW190503A}{4.1}{GW190602A}{1.8}{GW190929A}{--}{GW190517A}{4.2}{GW190915A}{4.5}{GW190425A}{~}{GW190512A}{4.2}{GW190630A}{~}{GW190521A}{--}{GW190413B}{1.4}{GW190924A}{4.5}{GW190930A}{1.5}{GW190706A}{4.3}{GW190408A}{4.1}{GW190909A}{--}{GW190728A}{4.4}{GW190426A}{--}{GW190412A}{4.1}{GW190720A}{4.4}{GW190521B}{4.2}{GW190910A}{~}{GW190803A}{1.6}{GW190519A}{4.2}{GW190708A}{~}{GW190527A}{--}{GW190513A}{4.2}{GW190424A}{~}{GW190727A}{4.4}{GW190814A}{~}{GW190707A}{4.3}{GW190828A}{4.5}{GW190828B}{4.5}{GW190701A}{--}{GW190421A}{2.2}}}

\newcommand{\PYCBCHIGHMASSSNR}[1]{\IfEqCase{#1}{{GW190413A}{8.6}{GW190719A}{8.0}{GW190620A}{~}{GW190514A}{8.3}{GW190731A}{8.2}{GW190503A}{12.2}{GW190602A}{11.4}{GW190929A}{--}{GW190517A}{10.2}{GW190915A}{12.7}{GW190425A}{~}{GW190512A}{12.2}{GW190630A}{~}{GW190521A}{--}{GW190413B}{9.0}{GW190924A}{12.4}{GW190930A}{9.8}{GW190706A}{12.3}{GW190408A}{13.6}{GW190909A}{--}{GW190728A}{13.4}{GW190426A}{--}{GW190412A}{17.8}{GW190720A}{10.5}{GW190521B}{24.0}{GW190910A}{~}{GW190803A}{8.6}{GW190519A}{13.0}{GW190708A}{~}{GW190527A}{--}{GW190513A}{11.9}{GW190424A}{~}{GW190727A}{11.8}{GW190814A}{~}{GW190707A}{12.8}{GW190828A}{15.3}{GW190828B}{10.8}{GW190701A}{--}{GW190421A}{10.2}}}

\newcommand{\PYCBCALLSKYFAR}[1]{\IfEqCase{#1}{{GW190413A}{--}{GW190719A}{--}{GW190620A}{~}{GW190514A}{--}{GW190731A}{--}{GW190503A}{$3.7 \times 10^{-2}$}{GW190602A}{--}{GW190929A}{--}{GW190517A}{$1.8 \times 10^{-2}$}{GW190915A}{$8.6 \times 10^{-4}$}{GW190425A}{~}{GW190512A}{$3.8 \times 10^{-5}$}{GW190630A}{~}{GW190521A}{$1.1 \times 10^{0}$}{GW190413B}{--}{GW190924A}{$<$ $6.3 \times 10^{-5}$}{GW190930A}{$3.4 \times 10^{-2}$}{GW190706A}{$6.7 \times 10^{-5}$}{GW190408A}{$<$ $2.5 \times 10^{-5}$}{GW190909A}{--}{GW190728A}{$<$ $1.6 \times 10^{-5}$}{GW190426A}{--}{GW190412A}{$<$ $3.1 \times 10^{-5}$}{GW190720A}{$<$ $2.0 \times 10^{-5}$}{GW190521B}{$<$ $1.8 \times 10^{-5}$}{GW190910A}{~}{GW190803A}{--}{GW190519A}{$<$ $1.8 \times 10^{-5}$}{GW190708A}{~}{GW190527A}{--}{GW190513A}{$3.7 \times 10^{-4}$}{GW190424A}{~}{GW190727A}{$3.5 \times 10^{-3}$}{GW190814A}{~}{GW190707A}{$<$ $1.0 \times 10^{-5}$}{GW190828A}{$<$ $1.5 \times 10^{-5}$}{GW190828B}{$5.8 \times 10^{-5}$}{GW190701A}{--}{GW190421A}{$1.9 \times 10^{0}$}}}

\newcommand{\PYCBCALLSKYIFAR}[1]{\IfEqCase{#1}{{GW190413A}{--}{GW190719A}{--}{GW190620A}{~}{GW190514A}{--}{GW190731A}{--}{GW190503A}{1.4}{GW190602A}{--}{GW190929A}{--}{GW190517A}{1.8}{GW190915A}{3.1}{GW190425A}{~}{GW190512A}{4.4}{GW190630A}{~}{GW190521A}{-0.0}{GW190413B}{--}{GW190924A}{4.2}{GW190930A}{1.5}{GW190706A}{4.2}{GW190408A}{4.6}{GW190909A}{--}{GW190728A}{4.8}{GW190426A}{--}{GW190412A}{4.5}{GW190720A}{4.7}{GW190521B}{4.8}{GW190910A}{~}{GW190803A}{--}{GW190519A}{4.8}{GW190708A}{~}{GW190527A}{--}{GW190513A}{3.4}{GW190424A}{~}{GW190727A}{2.5}{GW190814A}{~}{GW190707A}{5.0}{GW190828A}{4.8}{GW190828B}{4.2}{GW190701A}{--}{GW190421A}{-0.3}}}

\newcommand{\PYCBCALLSKYSNR}[1]{\IfEqCase{#1}{{GW190413A}{--}{GW190719A}{--}{GW190620A}{~}{GW190514A}{--}{GW190731A}{--}{GW190503A}{12.2}{GW190602A}{--}{GW190929A}{--}{GW190517A}{10.4}{GW190915A}{13.0}{GW190425A}{~}{GW190512A}{12.2}{GW190630A}{~}{GW190521A}{12.6}{GW190413B}{--}{GW190924A}{12.5}{GW190930A}{9.7}{GW190706A}{11.7}{GW190408A}{13.5}{GW190909A}{--}{GW190728A}{13.4}{GW190426A}{--}{GW190412A}{17.9}{GW190720A}{10.6}{GW190521B}{24.0}{GW190910A}{~}{GW190803A}{--}{GW190519A}{13.0}{GW190708A}{~}{GW190527A}{--}{GW190513A}{11.8}{GW190424A}{~}{GW190727A}{11.5}{GW190814A}{~}{GW190707A}{12.8}{GW190828A}{15.3}{GW190828B}{10.8}{GW190701A}{--}{GW190421A}{10.2}}}

\newcommand{\GSTLALALLSKYFAR}[1]{\IfEqCase{#1}{{GW190413A}{--}{GW190719A}{--}{GW190620A}{$2.9 \times 10^{-3}$$^\dagger$}{GW190514A}{--}{GW190731A}{$2.1 \times 10^{-1}$}{GW190503A}{$<$ $1.0 \times 10^{-5}$}{GW190602A}{$1.1 \times 10^{-5}$}{GW190929A}{$2.0 \times 10^{-2}$}{GW190517A}{$9.6 \times 10^{-4}$}{GW190915A}{$<$ $1.0 \times 10^{-5}$}{GW190425A}{$7.5 \times 10^{-4}$$^\dagger$}{GW190512A}{$<$ $1.0 \times 10^{-5}$}{GW190630A}{$<$ $1.0 \times 10^{-5}$}{GW190521A}{$1.2 \times 10^{-3}$}{GW190413B}{$3.8 \times 10^{-1}$}{GW190924A}{$<$ $1.0 \times 10^{-5}$}{GW190930A}{$5.8 \times 10^{-1}$}{GW190706A}{$<$ $1.0 \times 10^{-5}$}{GW190408A}{$<$ $1.0 \times 10^{-5}$}{GW190909A}{$1.1 \times 10^{0}$}{GW190728A}{$<$ $1.0 \times 10^{-5}$}{GW190426A}{$1.4 \times 10^{0}$}{GW190412A}{$<$ $1.0 \times 10^{-5}$}{GW190720A}{$<$ $1.0 \times 10^{-5}$}{GW190521B}{$<$ $1.0 \times 10^{-5}$}{GW190910A}{$1.9 \times 10^{-5}$$^\dagger$}{GW190803A}{$3.2 \times 10^{-2}$}{GW190519A}{$<$ $1.0 \times 10^{-5}$}{GW190708A}{$2.8 \times 10^{-5}$$^\dagger$}{GW190527A}{$6.2 \times 10^{-2}$}{GW190513A}{$<$ $1.0 \times 10^{-5}$}{GW190424A}{$7.8 \times 10^{-1}$$^\dagger$}{GW190727A}{$<$ $1.0 \times 10^{-5}$}{GW190814A}{$<$ $1.0 \times 10^{-5}$}{GW190707A}{$<$ $1.0 \times 10^{-5}$}{GW190828A}{$<$ $1.0 \times 10^{-5}$}{GW190828B}{$<$ $1.0 \times 10^{-5}$}{GW190701A}{$1.1 \times 10^{-2}$}{GW190421A}{$7.7 \times 10^{-4}$}}}

\newcommand{\GSTLALALLSKYIFAR}[1]{\IfEqCase{#1}{{GW190413A}{--}{GW190719A}{--}{GW190620A}{2.5$^\dagger$}{GW190514A}{--}{GW190731A}{0.7}{GW190503A}{5.0}{GW190602A}{4.9}{GW190929A}{1.7}{GW190517A}{3.0}{GW190915A}{5.0}{GW190425A}{3.1$^\dagger$}{GW190512A}{5.0}{GW190630A}{5.0}{GW190521A}{2.9}{GW190413B}{0.4}{GW190924A}{5.0}{GW190930A}{0.2}{GW190706A}{5.0}{GW190408A}{5.0}{GW190909A}{-0.0}{GW190728A}{5.0}{GW190426A}{-0.2}{GW190412A}{5.0}{GW190720A}{5.0}{GW190521B}{5.0}{GW190910A}{4.7$^\dagger$}{GW190803A}{1.5}{GW190519A}{5.0}{GW190708A}{4.5$^\dagger$}{GW190527A}{1.2}{GW190513A}{5.0}{GW190424A}{0.1$^\dagger$}{GW190727A}{5.0}{GW190814A}{5.0}{GW190707A}{5.0}{GW190828A}{5.0}{GW190828B}{5.0}{GW190701A}{2.0}{GW190421A}{3.1}}}

\newcommand{\GSTLALALLSKYSNR}[1]{\IfEqCase{#1}{{GW190413A}{--}{GW190719A}{--}{GW190620A}{10.9}{GW190514A}{--}{GW190731A}{8.5}{GW190503A}{12.1}{GW190602A}{12.1}{GW190929A}{9.9}{GW190517A}{10.6}{GW190915A}{13.1}{GW190425A}{13.0}{GW190512A}{12.3}{GW190630A}{15.6}{GW190521A}{15.0}{GW190413B}{10.0}{GW190924A}{13.2}{GW190930A}{10.0}{GW190706A}{12.3}{GW190408A}{14.7}{GW190909A}{8.5}{GW190728A}{13.6}{GW190426A}{10.1}{GW190412A}{18.9}{GW190720A}{11.7}{GW190521B}{24.4}{GW190910A}{13.4}{GW190803A}{9.0}{GW190519A}{12.0}{GW190708A}{13.1}{GW190527A}{8.9}{GW190513A}{12.3}{GW190424A}{10.0}{GW190727A}{12.3}{GW190814A}{22.2}{GW190707A}{13.0}{GW190828A}{16.0}{GW190828B}{11.1}{GW190701A}{11.6}{GW190421A}{10.6}}}

\newcommand{\OBSERVINGINSTRUMENTS}[1]{\IfEqCase{#1}{{GW190413A}{HLV}{GW190719A}{HL}{GW190620A}{LV}{GW190514A}{HL}{GW190731A}{HL}{GW190503A}{HLV}{GW190602A}{HLV}{GW190929A}{HLV}{GW190517A}{HLV}{GW190915A}{HLV}{GW190425A}{LV}{GW190512A}{HLV}{GW190630A}{LV}{GW190521A}{HLV}{GW190413B}{HLV}{GW190924A}{HLV}{GW190930A}{HL}{GW190706A}{HLV}{GW190408A}{HLV}{GW190909A}{HL}{GW190728A}{HLV}{GW190426A}{HLV}{GW190412A}{HLV}{GW190720A}{HLV}{GW190521B}{HL}{GW190910A}{LV}{GW190803A}{HLV}{GW190519A}{HLV}{GW190708A}{LV}{GW190527A}{HL}{GW190513A}{HLV}{GW190424A}{L}{GW190727A}{HLV}{GW190814A}{LV}{GW190707A}{HL}{GW190828A}{HLV}{GW190828B}{HLV}{GW190701A}{HLV}{GW190421A}{HL}}}

\newcommand{\PEINSTRUMENTS}[1]{\IfEqCase{#1}{{GW190413A}{HLV}{GW190719A}{HL}{GW190620A}{LV}{GW190514A}{HL}{GW190731A}{HL}{GW190503A}{HLV}{GW190602A}{HLV}{GW190929A}{HLV}{GW190517A}{HLV}{GW190915A}{HLV}{GW190425A}{LV}{GW190512A}{HLV}{GW190630A}{LV}{GW190521A}{HLV}{GW190413B}{HLV}{GW190924A}{HLV}{GW190930A}{HL}{GW190706A}{HLV}{GW190408A}{HLV}{GW190909A}{HL}{GW190728A}{HLV}{GW190426A}{HLV}{GW190412A}{HLV}{GW190720A}{HLV}{GW190521B}{HL}{GW190910A}{(H)LV}{GW190803A}{HLV}{GW190519A}{HLV}{GW190708A}{LV}{GW190527A}{HL}{GW190513A}{HLV}{GW190424A}{L}{GW190727A}{HLV}{GW190814A}{(H)LV}{GW190707A}{HL}{GW190828A}{HLV}{GW190828B}{HLV}{GW190701A}{HLV}{GW190421A}{HL}}}

\newcommand{\MITIGATIONMETHOD}[1]{\IfEqCase{#1}{{GW190413A}{None}{GW190719A}{None}{GW190620A}{None}{GW190514A}{L1 glitch subtraction, glitch-only model}{GW190731A}{None}{GW190503A}{L1 glitch subtraction, glitch-only model}{GW190602A}{None}{GW190929A}{None}{GW190517A}{None}{GW190915A}{None}{GW190425A}{L1 glitch subtraction, glitch-only model}{GW190512A}{None}{GW190630A}{None}{GW190521A}{None}{GW190413B}{L1 glitch subtraction, glitch-only model}{GW190924A}{L1 glitch subtraction, glitch-only model}{GW190930A}{None}{GW190706A}{None}{GW190408A}{None}{GW190909A}{None}{GW190728A}{None}{GW190426A}{None}{GW190412A}{None}{GW190720A}{None}{GW190521B}{None}{GW190910A}{None}{GW190803A}{None}{GW190519A}{None}{GW190708A}{None}{GW190527A}{None}{GW190513A}{L1 glitch subtraction, glitch-only model}{GW190424A}{L1 glitch subtraction, glitch-only model}{GW190727A}{\fixme{L1 $f_\text{min}$: 50 Hz}}{GW190814A}{L1 $f_\text{min}$: 30 Hz; H1 non-observing data used}{GW190707A}{None}{GW190828A}{None}{GW190828B}{None}{GW190701A}{L1 glitch subtraction, glitch+signal model}{GW190421A}{None}}}

\newcommand{\OTHERDQ}[1]{\IfEqCase{#1}{{GW190413A}{--}{GW190719A}{--}{GW190620A}{--}{GW190514A}{--}{GW190731A}{--}{GW190503A}{--}{GW190602A}{--}{GW190929A}{--}{GW190517A}{--}{GW190915A}{--}{GW190425A}{--}{GW190512A}{--}{GW190630A}{--}{GW190521A}{--}{GW190413B}{--}{GW190924A}{--}{GW190930A}{--}{GW190706A}{--}{GW190408A}{--}{GW190909A}{--}{GW190728A}{--}{GW190426A}{--}{GW190412A}{--}{GW190720A}{--}{GW190521B}{--}{GW190910A}{--}{GW190803A}{--}{GW190519A}{--}{GW190708A}{--}{GW190527A}{--}{GW190513A}{--}{GW190424A}{--}{GW190727A}{--}{GW190814A}{--}{GW190707A}{--}{GW190828A}{--}{GW190828B}{--}{GW190701A}{--}{GW190421A}{--}}}

\newtoggle{cleancomments}
\togglefalse{cleancomments}

\iftoggle{cleancomments}{

}{

}

\newcommand{\IMRP}{\textsc{IMRPhenomPv2}}
\newcommand{\IMRPHM}{\textsc{IMRPhenomPv3HM}}
\newcommand{\IMRPNRT}{\textsc{IMRPhenomPv2\_NRTidal}}
\newcommand{\SEOBNR}{\textsc{SEOBNRv4}}
\newcommand{\SEOBNRHM}{\textsc{SEOBNRv4HM}}
\newcommand{\SEOBROM}{\textsc{SEOBNRv4\_ROM}}
\newcommand{\SEOBHM}{\textsc{SEOBNRv4PHM}}
\newcommand{\SEOBHMROM}{\textsc{SEOBNRv4HM\_ROM}}
\newcommand{\NRSur}{\textsc{NRSur7dq4}}

\newcommand{\cmark}{\ding{51}}

\newcommand{\linf}{\textsc{LALInference}}
\newcommand{\lal}{\textsc{LALSuite}}

\newcommand{\bw}{\textsc{BayesWave}}
\newcommand{\gstlal}{\textsc{GstLAL}}
\newcommand{\pycbc}{\textsc{PyCBC}}
\newcommand{\cwb}{\textsc{cWB}}

\newcommand{\msun}{\ensuremath{\mathrm{M}_\odot}}

\newcommand{\Nevents}{24}

\def\dk{\delta\kappa}

\def\dk1{\delta\kappa_1}
\def\dk2{\delta\kappa_2}

\newcommand{\SimBestSymEvents}{GW151226, \NAME{GW190412A}, \NAME{GW190720A}, and \NAME{GW190728A}}
\newcommand{\SimStdCut}{150}
\newcommand{\SimBestPosUL}[1]{\IfEqCase{#1}{{NAME0}{GW151226}{VALUE0}{11.33}{NAME1}{\NAME{GW190412A}}{VALUE1}{110.89}}}
\newcommand{\SimCombinedCI}[1]{\IfEqCase{#1}{{HIER_POP}{\ensuremath{-23.2^{+52.2}_{-62.4}}}{SIMPLE_POP}{\ensuremath{-15.2^{+15.9}_{-19.0}}}{HIER_POP_NEG}{72.54}{HIER_POP_POS}{59.97}{SIMPLE_POP_NEG}{29.96}{SIMPLE_POP_POS}{9.01}{HIER_MU}{\ensuremath{-24.6^{+30.7}_{-35.3}}}{HIER_SIGMA}{52.7}}}
\newcommand{\SimBF}[1]{\IfEqCase{#1}{{SYM}{1.1}{POS}{2.0}}}
\newcommand{\LivMgUL}{\ensuremath{3.09 \times 10^{-23}}}
\newcommand{\LivMgSolarUL}{\ensuremath{3.16\times 10^{-23}}}
\newcommand{\LivImprov}[1]{\IfEqCase{#1}{{AMP}{2.6}{MG}{1.5}{EXP}{2.1}{MG_SOLAR}{1.0}}}
\newcommand{\LivEvents}[1]{\IfEqCase{#1}{{GWTC-2}{31}{GWTC-1}{7}}}
\newcommand{\LivBootstrapNumFracWidthRatioGreaterThanPointFive}{\ensuremath{9}}
\newcommand{\LivBootstrapFracWidthRatioAvg}{\ensuremath{0.12}}
\newcommand{\LivBootstrapFracWidthRatioMax}{\ensuremath{1.7}}
\newcommand{\LivBootstrapFracWidthRatioMaxName}{\NAME{GW190828B}}
\newcommand{\LivBootstrapFracWidthRatioMaxSign}{\ensuremath{ < 0}}
\newcommand{\LivBootstrapFracWidthRatioMaxAlpha}{\ensuremath{4}}

\newcommand{\ParBestSimplePopP}[1]{\IfEqCase{#1}{{NAME}{\ensuremath{\delta\hat{\varphi}_{-2}}}{VALUE}{\ensuremath{-1.10^{+1.20}_{-1.20}\times 10^{-3}}}{QGR}{93\%}}}
\newcommand{\ParWorstSimplePopP}[1]{\IfEqCase{#1}{{NAME}{\ensuremath{\delta\hat{\varphi}_{6l}}}{VALUE}{\ensuremath{-0.80^{+1.32}_{-1.29}}}{QGR}{84\%}}}
\newcommand{\ParBestSimplePopS}[1]{\IfEqCase{#1}{{NAME}{\ensuremath{\delta\hat{\varphi}_{-2}}}{VALUE}{\ensuremath{-0.70^{+1.00}_{-1.00}\times 10^{-3}}}{QGR}{86\%}}}
\newcommand{\ParWorstSimplePopS}[1]{\IfEqCase{#1}{{NAME}{\ensuremath{\delta\hat{\varphi}_{6l}}}{VALUE}{\ensuremath{-0.47^{+1.17}_{-1.17}}}{QGR}{74\%}}}
\newcommand{\ParBestHierPopP}[1]{\IfEqCase{#1}{{NAME}{\ensuremath{\delta\hat{\varphi}_{-2}}}{VALUE}{\ensuremath{-0.97^{+4.62}_{-4.07}\times 10^{-3}}}{QGR}{68\%}}}
\newcommand{\ParWorstHierPopP}[1]{\IfEqCase{#1}{{NAME}{\ensuremath{\delta\hat{\varphi}_{6l}}}{VALUE}{\ensuremath{-0.42^{+1.67}_{-1.50}}}{QGR}{69\%}}}
\newcommand{\ParBestHierZgrP}[1]{\IfEqCase{#1}{{NAME}{\ensuremath{\delta\hat{\varphi}_{7}}}{VALUE}{\ensuremath{0.03}}}}
\newcommand{\ParWorstHierZgrP}[1]{\IfEqCase{#1}{{NAME}{\ensuremath{\delta\hat{\beta}_{2}}}{VALUE}{\ensuremath{0.81}}}}
\newcommand{\ParBestHierPopS}[1]{\IfEqCase{#1}{{NAME}{\ensuremath{\delta\hat{\varphi}_{-2}}}{VALUE}{\ensuremath{-0.72^{+3.41}_{-3.76}\times 10^{-3}}}{QGR}{69\%}}}
\newcommand{\ParWorstHierPopS}[1]{\IfEqCase{#1}{{NAME}{\ensuremath{\delta\hat{\varphi}_{6l}}}{VALUE}{\ensuremath{-0.22^{+1.52}_{-1.56}}}{QGR}{60\%}}}
\newcommand{\ParBestHierZgrS}[1]{\IfEqCase{#1}{{NAME}{\ensuremath{\delta\hat{\varphi}_{4}}}{VALUE}{\ensuremath{0.08}}}}
\newcommand{\ParWorstHierZgrS}[1]{\IfEqCase{#1}{{NAME}{\ensuremath{\delta\hat{\varphi}_{3}}}{VALUE}{\ensuremath{0.71}}}}
\newcommand{\ParAlphaBetaBest}[1]{\IfEqCase{#1}{{Zh_NAME}{\ensuremath{\delta\hat{\alpha}_{2}}}{Zh_VALUE}{0.3}{Qh_NAME}{\ensuremath{\delta\hatdalpha2}}{Qh_VALUE}{21.4}{Zs_NAME}{\ensuremath{\delta\hat{\alpha}_{2}}}{Zs_VALUE}{0.6}}}

\newcommand{\pyRingTruncationTimeForH}[1]{\IfEqCase{#1}{ {GW150914}{ 1126259462.42322 } } }

\newcommand{\pyRingHighestTIGERLogOddsRatio}{ 0.07 }
\newcommand{\pyRingHighestTIGERLogOddsRatioEventName}{ GW170823 }
\newcommand{\pyRingCombinedTIGERLogOddsRatio}{ -0.61 }

\newcommand{\pyRingFrequencyDeviationPop}{$ \delta \hat{f}_{221} = 0.02^{+ 0.29 }_{- 0.33 } $ } 

\newcommand{\pSEOBFrequencyDeviationPop}{$\delta \hat{f}_{220} = 0.03^{+ 0.38 }_{- 0.35 } $ }
\newcommand{\pSEOBDampingTimeDeviationPop}{$\delta \hat{\tau}_{220} = 0.16^{+ 0.98 }_{- 0.98 } $}
\newcommand{\pSEOBFrequencyDeviationMean}{$\mu = 0.03^{+ 0.17 }_{- 0.18 }  $}
\newcommand{\pSEOBFrequencyDeviationStd}{$\sigma < 0.37 $}
\newcommand{\pSEOBDampingTimeDeviationMean}{$\mu = 0.16^{+ 0.47 }_{- 0.46 }  $}
\newcommand{\pSEOBDampingTimeDeviationStd}{$\sigma < 0.88 $}

\newcommand{\twoc}[1]{\multicolumn{2}{c}{#1}}

\newcommand{\threec}[1]{\multicolumn{3}{c}{#1}}

\begin{document}

\title{Tests of General Relativity with Binary Black Holes from the second LIGO–Virgo Gravitational-Wave Transient Catalog}

\iftoggle{endauthorlist}{
 \let\mymaketitle\maketitle
 \let\myauthor\author
 \let\myaffiliation\affiliation
 \author{The LIGO Scientific Collaboration and the Virgo Collaboration}
}{
 \iftoggle{fullauthorlist}{

\author{R.~Abbott}
\affiliation{LIGO, California Institute of Technology, Pasadena, CA 91125, USA}
\author{T.~D.~Abbott}
\affiliation{Louisiana State University, Baton Rouge, LA 70803, USA}
\author{S.~Abraham}
\affiliation{Inter-University Centre for Astronomy and Astrophysics, Pune 411007, India}
\author{F.~Acernese}
\affiliation{Dipartimento di Farmacia, Universit\`a di Salerno, I-84084 Fisciano, Salerno, Italy  }
\affiliation{INFN, Sezione di Napoli, Complesso Universitario di Monte S.Angelo, I-80126 Napoli, Italy  }
\author{K.~Ackley}
\affiliation{OzGrav, School of Physics \& Astronomy, Monash University, Clayton 3800, Victoria, Australia}
\author{A.~Adams}
\affiliation{Christopher Newport University, Newport News, VA 23606, USA}
\author{C.~Adams}
\affiliation{LIGO Livingston Observatory, Livingston, LA 70754, USA}
\author{R.~X.~Adhikari}
\affiliation{LIGO, California Institute of Technology, Pasadena, CA 91125, USA}
\author{V.~B.~Adya}
\affiliation{OzGrav, Australian National University, Canberra, Australian Capital Territory 0200, Australia}
\author{C.~Affeldt}
\affiliation{Max Planck Institute for Gravitational Physics (Albert Einstein Institute), D-30167 Hannover, Germany}
\affiliation{Leibniz Universit\"at Hannover, D-30167 Hannover, Germany}
% ...

\collaboration{The LIGO Scientific Collaboration and the Virgo Collaboration}

 }{
 \author{The LIGO Scientific Collaboration and the Virgo Collaboration}
}
}

\date[\relax]{compiled \today}

\begin{abstract}
Gravitational waves enable tests of general relativity in the highly dynamical and strong-field regime.
Using events detected by LIGO--Virgo up to 1 October 2019, we evaluate the consistency of the data with predictions from the theory.
We first establish that residuals from the best-fit waveform are consistent with detector noise, and that the low- and high-frequency parts of the signals are in agreement.
We then consider parametrized modifications to the waveform by varying post-Newtonian and phenomenological coefficients, improving past constraints by factors of ${\sim}2$; we also find consistency with Kerr black holes when we specifically target signatures of the spin-induced quadrupole moment.
Looking for gravitational-wave dispersion, we tighten constraints on Lorentz-violating coefficients by a factor of ${\sim}\LivImprov{AMP}$ and bound the mass of the graviton to $m_g \leq \LivMgUL \text{ eV}/c^2$ with 90\% credibility.
We also analyze the properties of the merger remnants by measuring ringdown frequencies and damping times, constraining fractional deviations away from the Kerr frequency to \pSEOBFrequencyDeviationPop for the fundamental quadrupolar mode, and \pyRingFrequencyDeviationPop for the first overtone; additionally, we find no evidence for postmerger echoes.
Finally, we determine that our data are consistent with tensorial polarizations through a template-independent method.
When possible, we assess the validity of general relativity based on collections of events analyzed jointly.
We find no evidence for new physics beyond general relativity, for black hole mimickers, or for any unaccounted systematics.

\end{abstract}

\maketitle

\section{Introduction}
\label{sec:intro}
General relativity (GR) remains our most accurate theory of gravity, having withstood many experimental tests in the Solar System \cite{Will:2014kxa} as well as binary pulsar \cite{Wex:2014nva,Will:2014kxa}, cosmological \cite{Clifton:2011jh,Ferreira:2019xrr} and gravitational-wave (GW) observations \cite{TheLIGOScientific:2016src,TheLIGOScientific:2016pea,Yunes:2016jcc,Abbott:2017vtc,Ezquiaga:2017ekz,Sakstein:2017xjx,Creminelli:2017sry,Baker:2017hug,Abbott:2017oio,Abbott:2018lct,LIGOScientific:2019fpa}. Many of these tests probe regimes where gravitational fields are weak, spacetime curvature is small, and characteristic velocities are not comparable to the speed of light. Observations of compact binary coalescences enable us to test GR in extreme environments of strong gravitational fields, large spacetime curvature, and velocities comparable to the speed of light; high post-Newtonian (PN) order calculations and numerical relativity (NR) simulations are required to accurately model the emitted GW signal \cite{TheLIGOScientific:2016src,TheLIGOScientific:2016pea,Abbott:2018lct,LIGOScientific:2019fpa}. 

We report results from tests of GR on binary black hole (BBH) signals using the second Gravitational-wave Transient catalog (GWTC-2) \cite{GWTC2}. The GWTC-2 catalog includes all observations reported in the first catalog (GWTC-1) \cite{GWTC1}, covering the first (O1) and second (O2) observing runs, as well as new events identified in the first half of the third observing run (O3a) of the Advanced LIGO and Advanced Virgo detectors \cite{GWTC2}. We focus on the most significant signals, requiring them to have been detected with a false-alarm rate (FAR) $< 10^{-3}~\mathrm{yr}^{-1}$.

A current limitation on tests of beyond-GR physics with compact binary coalescences is the lack of understanding of the strong-field merger regime in nearly all modified theories of gravity. This restricts our analysis to testing the null hypothesis, taken to be GR, using model-independent or parametrized tests of GR \cite{Blanchet:1994ez,Blanchet:1994ex,Arun:2006hn,Arun:2006yw,Yunes:2009ke,Mishra:2010tp,Li:2011cg,Li:2011vx,Agathos:2013upa,TheLIGOScientific:2016src,Meidam:2017dgf,Abbott:2018lct,LIGOScientific:2019fpa,Islam:2019dmk,Capano:2020dix}. An important goal in constraining beyond-GR theories is the development of model-dependent tests, requiring analytical waveforms and NR simulations in alternative theories of gravity across the binary parameter space. Unfortunately, there is still a lack of alternative theories of gravity that are mathematically well-posed, physically viable, and provide sufficiently well-defined alternative predictions for the GW signal emitted by two coalescing compact objects. Recent NR studies have begun to model astrophysically relevant binary black hole mergers in beyond-GR theories \cite{Okounkova:2017yby,Hirschmann:2017psw,Witek:2018dmd,Okounkova:2019zjf,Okounkova:2020rqw} and numerous advances have been made deriving the analytical equations of motion and gravitational waveforms in such theories \cite{Sotiriou:2006pq,Yagi:2011xp,Mirshekari:2013vb,Lang:2013fna,Lang:2014osa,Sennett:2016klh,Sennett:2016rwa,Julie:2017rpw,Khalil:2018aaj,Julie:2018lfp,Bernard:2018hta,Bernard:2018ivi,Julie:2019sab,Sennett:2019bpc}. However, it is often unknown whether the full theories are well-posed and a significant amount of work is required before the results can be used in the context of GW data analysis. 

The approach taken here is therefore to (i) check the consistency of GR predictions with the data, and (ii) introduce parametrized modifications to GR waveforms in order to constrain the degree to which the deviations from the GR predictions agree with the data. As in \cite{LIGOScientific:2019fpa}, the results in this paper should be treated as observational constraints on deviations from GR. Such limits are a quantitative indication of the degree to which the data are described by GR but can also be reinterpreted in the context of a given modified theory of gravity to produce constraints, subject to a number of assumptions \cite{Yunes:2016jcc,Nair:2019iur}. Our analyses do not reveal any inconsistency with GR and the results improve on the previous tests of GR using the BBHs observed in O1 and O2 \cite{TheLIGOScientific:2016src,TheLIGOScientific:2016pea,Abbott:2017vtc,Abbott:2017oio,Abbott:2018lct,LIGOScientific:2019fpa}.

The analyses performed in this paper can be broken down into four broad categories. In order to test the consistency of the GR predictions in a generic way, we look for residual power after subtracting the best-fit GR waveform from the data. We also separately study the low-frequency and high-frequency portions of an observed signal, and evaluate the agreement of the inferred parameters. To constrain specific deviations from GR, we perform parametrized tests targeting the generation of GWs and the propagation of the GW signal. %
All these approaches were already implemented in \cite{LIGOScientific:2019fpa} for GWTC-1 signals.
In addition, we introduce a new suite of analyses: 
an extension of the parametrized test considering terms from the spin-induced quadrupole moment of the binary components, dedicated studies of the remnant properties (ringdown and echoes), and a new method for probing the geometry of GW polarizations. 

The tests considered here are not all independent, and will have some degree of overlap or redundancy. Whilst a detailed discussion and study of the complex relationships between the tests is beyond the scope of this paper, it is important to highlight potential complementarity between the analyses. For example, any physics that modifies the generation of GWs would also likely lead to modifications to their propagation. Similarly, physics that modifies the nature of the remnant object might also predict modifications to the earlier inspiral dynamics.
Furthermore, several types of deviations from GR may be picked up simultaneously by multiple analyses.

The rapid increase in the number of observed binary coalescences has driven interest in how we can best combine information from a set of measurements. In order to address this question, we employ hierarchical inference on a subset of our analyses to parametrize and constrain the distribution of observed beyond-GR parameters for different sources \cite{Zimmerman:2019wzo,Isi:2019asy}. This allows us to make quantitative statements about the overall agreement of our observations with the null hypothesis that GR is correct and that no strong systematics are present. Such measurements are qualitatively more general than combined constraints previously presented in \cite{LIGOScientific:2019fpa}. In Sec.~\ref{sec:inference} we discuss parameter inference for individual events and 
detail how the hierarchical analysis is performed on the full set of measurements.

Our constraints on deviations from GR are currently dominated by statistical uncertainty induced by detector noise \cite{TheLIGOScientific:2016src,Abbott:2016wiq,LIGOScientific:2019fpa}. Yet, the statistical uncertainty can be reduced by combining the results from multiple events. Additional uncertainty will arise from systematic error in the calibration of the detectors and power spectral density (PSD) estimation, as well as errors in the modeling of GW waveforms in GR; unlike uncertainty induced by detector noise, such errors do not improve when combining multiple events and therefore will dominate the uncertainty budget for sufficiently large catalogs of merger events.
Most of the tests in this paper are sensitive to such systematics, which could mimic a deviation from GR.
However, we do not find any evidence of GR violations that cannot be accounted for by possible systematics.

This paper is organized as follows. Section~\ref{sec:events} provides an overview of the data used in the analysis. It also defines the event selection criteria and discusses which GW events are used to produce the individual and combined results presented in this paper. We provide details about gravitational waveforms and data analysis methods in Sec.~\ref{sec:inference}. In Sec.~\ref{sec:con} we present the residuals test, and the inspiral--merger--ringdown (IMR) consistency test. In Sec.~\ref{sec:gen} we outline tests of GW generation, including generic parametrized modifications and a test of the spin-induced quadrupole moment. In Sec.~\ref{sec:liv} we describe tests of GW propagation using a modified dispersion relation. We present tests of the remnant properties in Sec.~\ref{sec:rem} and study GW polarizations in Sec.~\ref{sec:pol}. Finally, we conclude with Sec.~\ref{sec:conclusion}. 

Data products associated with the results of analyses in this paper can be found in \cite{GWTC2:TGR:release}. The GW strain data for all events are available at the Gravitational Wave Open Science Center \cite{Abbott:2019ebz,GWOSC}.

\section{Data, events, and significance}
\label{sec:events}
The analyses presented here use data taken during O3a by Advanced LIGO 
\cite{TheLIGOScientific:2014jea} and Advanced Virgo \cite{TheVirgo:2014hva}. 
O3a extended from 1 April 2019 to 1 October 2019. All three detectors achieved 
sensitivities significantly better than those in the previous observing run \cite{GWTC1}.  
Calibration \cite{Karki:2016pht,Cahillane:2017vkb,Viets:2017yvy,Sun:2020wke} accuracy of a few percent in 
amplitude and a few degrees in phase was achieved at all sites. To improve the precision of parameter estimation, 
various noise subtraction methods \cite{Driggers:2018gii,Davis:2018yrz,Vajente:2019ycy,Cornish:2014kda} were applied to some of the
events used here (see {Table~V} in \cite{GWTC2} for the list of events requiring such mitigation). 
See \cite{GWTC2} for detailed discussion of instrument performance and data quality for O3a.

We present results for the detections of possible BBH events in O3a with FAR $< 10^{-3}$ per year, as reported by any of the pipelines featured in \cite{GWTC2}.
This threshold is stricter than the one in \cite{LIGOScientific:2019fpa} to accommodate the increased number of events within computational constraints.
The {\Nevents} selected events, and some of their key properties, are listed in Table~\ref{tab:events}.
Out of those, \NAME{GW190814A}{} is the only one to have been identified as a possible neutron star--black hole (NSBH) system based on the inferred component masses, although the true nature of the secondary object remains unknown \cite{GW190814}.
In this paper, we start from the null hypothesis that all signals analyzed (including \NAME{GW190814A}) correspond to BBHs as described by GR, and proceed to seek evidence in the data to challenge this (we find none).
We do not study the likely binary neutron star signal \NAME{GW190425A}{} \cite{GW190425}.

Detection significance is provided by two pipelines that rely on GR templates (\pycbc{}~\cite{pycbc-github, Canton:2014ena, Usman:2015kfa} and
\gstlal{}~\cite{Sachdev:2019vvd,Messick:2016aqy}, both relying on the waveform models described in \cite{Sathyaprakash:1991mt, Blanchet:1995ez, Blanchet:2005tk,
Buonanno:2009zt} and \cite{Bohe:2016gbl}), and by one pipeline that does not (\textsc{coherent WaveBurst}, henceforth \cwb{} \cite{Klimenko:2008fu, Klimenko:2015ypf, TheLIGOScientific:2016uux}).
Making use of a measure of significance that assumes the validity of GR could
potentially lead to biases in the selection of events to be tested, systematically
disfavoring signals in which a GR violation would be most evident (e.g., \cite{Chia:2020psj}).
\cwb{} would detect at least some of the conceivable chirp-like signals with sufficient departures from GR that they would be missed by the templated searches.
Nonetheless, we cannot fully discard the existence of a hidden population of signals exhibiting large deviations from GR, which could escape both modeled and unmodeled searches.

Out of all the events reported in \cite{GWTC2}, only the massive event \NAME{GW190521A}{} was identified with greater significance by the unmodeled search.
This can be explained as a consequence of the system's high mass, which led to a short signal with only ${\sim}4$ cycles visible in our detectors \cite{GW190521g,GW190521g:imp}.
This fact makes it more difficult to evaluate consistency with GR for this event than for other (less massive) systems which remain in the sensitive band of our detectors for a longer period.
This is especially true for tests targeting the inspiral, since there is little signal-to-noise ratio (SNR) before the merger ($\mathrm{SNR} \approx 4.7$, computed as in Sec.~\ref{sec:imr});
on the other hand, this signal is highly suitable for studies of black hole (BH) ringdown \cite{GW190521g:imp}.

We consider each of the GW events individually, carrying out different analyses depending on the properties of each signal.
Some of the tests presented here, such as the IMR consistency test in Sec.~\ref{sec:imr} and the parametrized tests in Sec.~\ref{sec:gen}, distinguish between the inspiral and the postinspiral regimes of the signal.
The remnant-focused analyses of Sec.~\ref{sec:rem} are only meaningful for systems massive enough for the postinspiral signal to be detectable by LIGO--Virgo.
Finally, studies of polarization content are only feasible for detections involving the full three-detector network.
We choose which analyses to apply in each case following pre-established selection criteria based on the signal power recovered in different frequency regimes, or the number of involved detectors.
Table \ref{tab:events} indicates which events have met the selection criteria for each analysis; further details are provided in the sections below.

Having a large number of detections also allows us to make statements about the validity of GR from the set of measurements as a whole.
Ideally, we would like to constrain the properties of the true population of signals that exist in Nature---for example, if GR is correct, the population distribution of parametrized deviations from GR would be a $\delta$ function at the point corresponding to no deviation.
However, this would require an understanding of our detection efficiency as a function of these deviations \cite{Loredo:2004nn,Mandel:2018mve}, as well as a joint model for the distribution of individual event properties and deviations from GR \cite{GWTC2:rates}.
Because no such comprehensive modeling is available, we do not attempt to make any statements about possible intrinsic populations, but rather measure the distribution of deviations from GR across \emph{observed} signals.
Our strategies for doing so are outlined in Sec.~\ref{sec:inference:populations}.

Given the increased significance threshold for inclusion in this paper, we dispense with the two-tiered selection criterion applied in \cite{LIGOScientific:2019fpa}. 
Instead, we make combined statements using all events in our selection.
When possible, we also combine our results for O3a with those from preceding observation runs that satisfy our selection criterion.
That includes all events analyzed in \cite{LIGOScientific:2019fpa} except GW151012 and GW170729; that is: GW150914, GW151226, GW170104, GW170608, GW170809, GW170814, GW170818, and GW170823.%
\footnote{Unlike in this paper, combined results in \cite{LIGOScientific:2019fpa} did not include GW170818 because it was only detected by a single pipeline.}
This is done for tests already presented in \cite{LIGOScientific:2019fpa} (residuals test, IMR consistency, parametrized tests, and modified dispersion relations), as well as for new analyses for which pre-O3a results are presented here for the first time (spin-induced moments, ringdown, and polarizations).

In some cases we perform tests on events that yield uninformative results, so that the posterior distribution extends across the full extent of the prior.
This means that upper limits in such cases are determined by the prior, and thus are arbitrary.
However, this is not a problem when considering the set of measurements as a whole using the techniques described in Sec.~\ref{sec:inference:populations}.
\begin{table*}
\caption{\label{tab:events}
List of O3a events considered in this paper.
The first block of columns gives the names of the events and lists the instruments involved in each detection, as well as some relevant properties obtained assuming GR: luminosity distance $D_\text{L}$, redshifted total mass $(1+z)M$, redshifted chirp mass $(1+z)\mathcal{M}$, redshifted final mass $(1+z)M_\text{f}$, dimensionless final spin $\chi_\text{f} = c |\vec{S}_{\rm f}| / (G M^2_{\rm f})$, and signal-to-noise ratio SNR.
Reported quantities correspond to the median and 90\% symmetric credible intervals, as computed in Table VI in \cite{GWTC2}.
The last block of columns indicates which analyses are performed on a given event according to the selection criteria in Sec.~\ref{sec:events}: RT = residuals test (Sec.~\ref{sec:res}); IMR = inspiral-merger-ringdown consistency test (Sec.~\ref{sec:imr}); PAR = parametrized tests of GW generation (Sec.~\ref{sec:par}); SIM = spin-induced moments (Sec.~\ref{sec:sim}); MDR = modified GW dispersion relation (Sec.~\ref{sec:liv}); RD = ringdown (Sec.~\ref{sec:rin}); ECH = echoes searches (Sec.~\ref{sec:ech}); POL = polarization content (Sec.~\ref{sec:pol}).
}
\begin{tabular}{ l l l l l l@{\;} l@{\;} l@{\;} l l c l c c c c c c c c}
\toprule
\multirow{2}{*}{Event} & \hphantom{X} & \multirow{2}{*}{Inst.} & \hphantom{X} & \multicolumn{5}{c}{Properties} & \hphantom{} & \multirow{2}{*}{SNR} & \hphantom{X} & \multicolumn{8}{c}{Tests performed}\\
\cline{5-9}
\cline{13-20}
& & & & $D_\text{L}$ & {\scriptsize $(1+z)$}$M$ & {\scriptsize $(1+z)$}$\mathcal{M}$ & {\scriptsize $(1+z)$}$M_\text{f}$ & $\chi_\text{f}$ & & & & RT & IMR & PAR \ & SIM & MDR & RD & ECH & POL \\
& & & & [Gpc] & [$M_\odot$] & [$M_\odot$] & [$M_\odot$] & & & & & & & & & & &\\
\midrule
\NAME{GW190408A} & &\OBSERVINGINSTRUMENTS{GW190408A} & & $\luminositydistancemed{GW190408A}^{+\luminositydistanceplus{GW190408A}}_{-\luminositydistanceminus{GW190408A}}$ & $\totalmassdetmed{GW190408A}^{+\totalmassdetplus{GW190408A}}_{-\totalmassdetminus{GW190408A}}$ & $\chirpmassdetmed{GW190408A}^{+\chirpmassdetplus{GW190408A}}_{-\chirpmassdetminus{GW190408A}}$ & $\finalmassdetmed{GW190408A}^{+\finalmassdetplus{GW190408A}}_{-\finalmassdetminus{GW190408A}}$ & $\finalspinmed{GW190408A}^{+\finalspinplus{GW190408A}}_{-\finalspinminus{GW190408A}}$ & & $\networkmatchedfiltersnrIMRmed{GW190408A}^{+\networkmatchedfiltersnrIMRplus{GW190408A}}_{-\networkmatchedfiltersnrIMRminus{GW190408A}}$ & & \cmark & \cmark & \cmark & \cmark & \cmark & \cmark & \cmark & \cmark \\[0.075cm]
\NAME{GW190412A} & &\OBSERVINGINSTRUMENTS{GW190412A} & & $\luminositydistancemed{GW190412A}^{+\luminositydistanceplus{GW190412A}}_{-\luminositydistanceminus{GW190412A}}$ & $\totalmassdetmed{GW190412A}^{+\totalmassdetplus{GW190412A}}_{-\totalmassdetminus{GW190412A}}$ & $\chirpmassdetmed{GW190412A}^{+\chirpmassdetplus{GW190412A}}_{-\chirpmassdetminus{GW190412A}}$ & $\finalmassdetmed{GW190412A}^{+\finalmassdetplus{GW190412A}}_{-\finalmassdetminus{GW190412A}}$ & $\finalspinmed{GW190412A}^{+\finalspinplus{GW190412A}}_{-\finalspinminus{GW190412A}}$ & & $\networkmatchedfiltersnrIMRmed{GW190412A}^{+\networkmatchedfiltersnrIMRplus{GW190412A}}_{-\networkmatchedfiltersnrIMRminus{GW190412A}}$ & & \cmark &   --   & \cmark & \cmark & \cmark & -- &   \cmark   & \cmark \\[0.075cm]
\NAME{GW190421A} & &\OBSERVINGINSTRUMENTS{GW190421A} & & $\luminositydistancemed{GW190421A}^{+\luminositydistanceplus{GW190421A}}_{-\luminositydistanceminus{GW190421A}}$ & $\totalmassdetmed{GW190421A}^{+\totalmassdetplus{GW190421A}}_{-\totalmassdetminus{GW190421A}}$ & $\chirpmassdetmed{GW190421A}^{+\chirpmassdetplus{GW190421A}}_{-\chirpmassdetminus{GW190421A}}$ & $\finalmassdetmed{GW190421A}^{+\finalmassdetplus{GW190421A}}_{-\finalmassdetminus{GW190421A}}$ & $\finalspinmed{GW190421A}^{+\finalspinplus{GW190421A}}_{-\finalspinminus{GW190421A}}$ & & $\networkmatchedfiltersnrIMRmed{GW190421A}^{+\networkmatchedfiltersnrIMRplus{GW190421A}}_{-\networkmatchedfiltersnrIMRminus{GW190421A}}$ & & \cmark & \cmark & \cmark &   --   & \cmark & \cmark & \cmark &   --   \\[0.075cm]
\NAME{GW190503A} & &\OBSERVINGINSTRUMENTS{GW190503A} & & $\luminositydistancemed{GW190503A}^{+\luminositydistanceplus{GW190503A}}_{-\luminositydistanceminus{GW190503A}}$ & $\totalmassdetmed{GW190503A}^{+\totalmassdetplus{GW190503A}}_{-\totalmassdetminus{GW190503A}}$ & $\chirpmassdetmed{GW190503A}^{+\chirpmassdetplus{GW190503A}}_{-\chirpmassdetminus{GW190503A}}$ & $\finalmassdetmed{GW190503A}^{+\finalmassdetplus{GW190503A}}_{-\finalmassdetminus{GW190503A}}$ & $\finalspinmed{GW190503A}^{+\finalspinplus{GW190503A}}_{-\finalspinminus{GW190503A}}$ & & $\networkmatchedfiltersnrIMRmed{GW190503A}^{+\networkmatchedfiltersnrIMRplus{GW190503A}}_{-\networkmatchedfiltersnrIMRminus{GW190503A}}$ & & \cmark & \cmark & \cmark &   --   & \cmark & \cmark & \cmark & \cmark \\[0.075cm]
\NAME{GW190512A} & &\OBSERVINGINSTRUMENTS{GW190512A} & & $\luminositydistancemed{GW190512A}^{+\luminositydistanceplus{GW190512A}}_{-\luminositydistanceminus{GW190512A}}$ & $\totalmassdetmed{GW190512A}^{+\totalmassdetplus{GW190512A}}_{-\totalmassdetminus{GW190512A}}$ & $\chirpmassdetmed{GW190512A}^{+\chirpmassdetplus{GW190512A}}_{-\chirpmassdetminus{GW190512A}}$ & $\finalmassdetmed{GW190512A}^{+\finalmassdetplus{GW190512A}}_{-\finalmassdetminus{GW190512A}}$ & $\finalspinmed{GW190512A}^{+\finalspinplus{GW190512A}}_{-\finalspinminus{GW190512A}}$ & & $\networkmatchedfiltersnrIMRmed{GW190512A}^{+\networkmatchedfiltersnrIMRplus{GW190512A}}_{-\networkmatchedfiltersnrIMRminus{GW190512A}}$ & & \cmark &   --   & \cmark & \cmark & \cmark & \cmark & \cmark &   \cmark   \\[0.075cm]
\NAME{GW190513A} & &\OBSERVINGINSTRUMENTS{GW190513A} & & $\luminositydistancemed{GW190513A}^{+\luminositydistanceplus{GW190513A}}_{-\luminositydistanceminus{GW190513A}}$ & $\totalmassdetmed{GW190513A}^{+\totalmassdetplus{GW190513A}}_{-\totalmassdetminus{GW190513A}}$ & $\chirpmassdetmed{GW190513A}^{+\chirpmassdetplus{GW190513A}}_{-\chirpmassdetminus{GW190513A}}$ & $\finalmassdetmed{GW190513A}^{+\finalmassdetplus{GW190513A}}_{-\finalmassdetminus{GW190513A}}$ & $\finalspinmed{GW190513A}^{+\finalspinplus{GW190513A}}_{-\finalspinminus{GW190513A}}$ & & $\networkmatchedfiltersnrIMRmed{GW190513A}^{+\networkmatchedfiltersnrIMRplus{GW190513A}}_{-\networkmatchedfiltersnrIMRminus{GW190513A}}$ & & \cmark & \cmark & \cmark &   --   & \cmark & \cmark & \cmark &   \cmark   \\[0.075cm]
\NAME{GW190517A} & &\OBSERVINGINSTRUMENTS{GW190517A} & & $\luminositydistancemed{GW190517A}^{+\luminositydistanceplus{GW190517A}}_{-\luminositydistanceminus{GW190517A}}$ & $\totalmassdetmed{GW190517A}^{+\totalmassdetplus{GW190517A}}_{-\totalmassdetminus{GW190517A}}$ & $\chirpmassdetmed{GW190517A}^{+\chirpmassdetplus{GW190517A}}_{-\chirpmassdetminus{GW190517A}}$ & $\finalmassdetmed{GW190517A}^{+\finalmassdetplus{GW190517A}}_{-\finalmassdetminus{GW190517A}}$ & $\finalspinmed{GW190517A}^{+\finalspinplus{GW190517A}}_{-\finalspinminus{GW190517A}}$ & & $\networkmatchedfiltersnrIMRmed{GW190517A}^{+\networkmatchedfiltersnrIMRplus{GW190517A}}_{-\networkmatchedfiltersnrIMRminus{GW190517A}}$ & & \cmark &   --   & \cmark &   --   & \cmark & -- & \cmark & \cmark \\[0.075cm]
\NAME{GW190519A} & &\OBSERVINGINSTRUMENTS{GW190519A} & & $\luminositydistancemed{GW190519A}^{+\luminositydistanceplus{GW190519A}}_{-\luminositydistanceminus{GW190519A}}$ & $\totalmassdetmed{GW190519A}^{+\totalmassdetplus{GW190519A}}_{-\totalmassdetminus{GW190519A}}$ & $\chirpmassdetmed{GW190519A}^{+\chirpmassdetplus{GW190519A}}_{-\chirpmassdetminus{GW190519A}}$ & $\finalmassdetmed{GW190519A}^{+\finalmassdetplus{GW190519A}}_{-\finalmassdetminus{GW190519A}}$ & $\finalspinmed{GW190519A}^{+\finalspinplus{GW190519A}}_{-\finalspinminus{GW190519A}}$ & & $\networkmatchedfiltersnrIMRmed{GW190519A}^{+\networkmatchedfiltersnrIMRplus{GW190519A}}_{-\networkmatchedfiltersnrIMRminus{GW190519A}}$ & & \cmark & \cmark & \cmark &   --   & \cmark & \cmark & \cmark & \cmark \\[0.075cm]
\NAME{GW190521A} & &\OBSERVINGINSTRUMENTS{GW190521A} & & $\luminositydistancemed{GW190521A}^{+\luminositydistanceplus{GW190521A}}_{-\luminositydistanceminus{GW190521A}}$ & $\totalmassdetmed{GW190521A}^{+\totalmassdetplus{GW190521A}}_{-\totalmassdetminus{GW190521A}}$ & $\chirpmassdetmed{GW190521A}^{+\chirpmassdetplus{GW190521A}}_{-\chirpmassdetminus{GW190521A}}$ & $\finalmassdetmed{GW190521A}^{+\finalmassdetplus{GW190521A}}_{-\finalmassdetminus{GW190521A}}$ & $\finalspinmed{GW190521A}^{+\finalspinplus{GW190521A}}_{-\finalspinminus{GW190521A}}$ & & $\networkmatchedfiltersnrIMRmed{GW190521A}^{+\networkmatchedfiltersnrIMRplus{GW190521A}}_{-\networkmatchedfiltersnrIMRminus{GW190521A}}$ & & \cmark &   --   & \cmark &   --   & -- & \cmark & \cmark & \cmark \\[0.075cm]
\NAME{GW190521B} & &\OBSERVINGINSTRUMENTS{GW190521B} & & $\luminositydistancemed{GW190521B}^{+\luminositydistanceplus{GW190521B}}_{-\luminositydistanceminus{GW190521B}}$ & $\totalmassdetmed{GW190521B}^{+\totalmassdetplus{GW190521B}}_{-\totalmassdetminus{GW190521B}}$ & $\chirpmassdetmed{GW190521B}^{+\chirpmassdetplus{GW190521B}}_{-\chirpmassdetminus{GW190521B}}$ & $\finalmassdetmed{GW190521B}^{+\finalmassdetplus{GW190521B}}_{-\finalmassdetminus{GW190521B}}$ & $\finalspinmed{GW190521B}^{+\finalspinplus{GW190521B}}_{-\finalspinminus{GW190521B}}$ & & $\networkmatchedfiltersnrIMRmed{GW190521B}^{+\networkmatchedfiltersnrIMRplus{GW190521B}}_{-\networkmatchedfiltersnrIMRminus{GW190521B}}$ & & \cmark & \cmark & \cmark & \cmark & \cmark & \cmark & \cmark &   --   \\[0.075cm]
\NAME{GW190602A} & &\OBSERVINGINSTRUMENTS{GW190602A} & & $\luminositydistancemed{GW190602A}^{+\luminositydistanceplus{GW190602A}}_{-\luminositydistanceminus{GW190602A}}$ & $\totalmassdetmed{GW190602A}^{+\totalmassdetplus{GW190602A}}_{-\totalmassdetminus{GW190602A}}$ & $\chirpmassdetmed{GW190602A}^{+\chirpmassdetplus{GW190602A}}_{-\chirpmassdetminus{GW190602A}}$ & $\finalmassdetmed{GW190602A}^{+\finalmassdetplus{GW190602A}}_{-\finalmassdetminus{GW190602A}}$ & $\finalspinmed{GW190602A}^{+\finalspinplus{GW190602A}}_{-\finalspinminus{GW190602A}}$ & & $\networkmatchedfiltersnrIMRmed{GW190602A}^{+\networkmatchedfiltersnrIMRplus{GW190602A}}_{-\networkmatchedfiltersnrIMRminus{GW190602A}}$ & & \cmark &   --   & \cmark &   --   & \cmark & \cmark & \cmark & \cmark \\[0.075cm]
\NAME{GW190630A} & &\OBSERVINGINSTRUMENTS{GW190630A} & & $\luminositydistancemed{GW190630A}^{+\luminositydistanceplus{GW190630A}}_{-\luminositydistanceminus{GW190630A}}$ & $\totalmassdetmed{GW190630A}^{+\totalmassdetplus{GW190630A}}_{-\totalmassdetminus{GW190630A}}$ & $\chirpmassdetmed{GW190630A}^{+\chirpmassdetplus{GW190630A}}_{-\chirpmassdetminus{GW190630A}}$ & $\finalmassdetmed{GW190630A}^{+\finalmassdetplus{GW190630A}}_{-\finalmassdetminus{GW190630A}}$ & $\finalspinmed{GW190630A}^{+\finalspinplus{GW190630A}}_{-\finalspinminus{GW190630A}}$ & & $\networkmatchedfiltersnrIMRmed{GW190630A}^{+\networkmatchedfiltersnrIMRplus{GW190630A}}_{-\networkmatchedfiltersnrIMRminus{GW190630A}}$ & & \cmark & \cmark & \cmark &   \cmark   & \cmark & -- & \cmark &   --   \\[0.075cm]
\NAME{GW190706A} & &\OBSERVINGINSTRUMENTS{GW190706A} & & $\luminositydistancemed{GW190706A}^{+\luminositydistanceplus{GW190706A}}_{-\luminositydistanceminus{GW190706A}}$ & $\totalmassdetmed{GW190706A}^{+\totalmassdetplus{GW190706A}}_{-\totalmassdetminus{GW190706A}}$ & $\chirpmassdetmed{GW190706A}^{+\chirpmassdetplus{GW190706A}}_{-\chirpmassdetminus{GW190706A}}$ & $\finalmassdetmed{GW190706A}^{+\finalmassdetplus{GW190706A}}_{-\finalmassdetminus{GW190706A}}$ & $\finalspinmed{GW190706A}^{+\finalspinplus{GW190706A}}_{-\finalspinminus{GW190706A}}$ & & $\networkmatchedfiltersnrIMRmed{GW190706A}^{+\networkmatchedfiltersnrIMRplus{GW190706A}}_{-\networkmatchedfiltersnrIMRminus{GW190706A}}$ & & \cmark & \cmark & \cmark &   --   & \cmark & \cmark & \cmark & \cmark \\[0.075cm]
\NAME{GW190707A} & &\OBSERVINGINSTRUMENTS{GW190707A} & & $\luminositydistancemed{GW190707A}^{+\luminositydistanceplus{GW190707A}}_{-\luminositydistanceminus{GW190707A}}$ & $\totalmassdetmed{GW190707A}^{+\totalmassdetplus{GW190707A}}_{-\totalmassdetminus{GW190707A}}$ & $\chirpmassdetmed{GW190707A}^{+\chirpmassdetplus{GW190707A}}_{-\chirpmassdetminus{GW190707A}}$ & $\finalmassdetmed{GW190707A}^{+\finalmassdetplus{GW190707A}}_{-\finalmassdetminus{GW190707A}}$ & $\finalspinmed{GW190707A}^{+\finalspinplus{GW190707A}}_{-\finalspinminus{GW190707A}}$ & & $\networkmatchedfiltersnrIMRmed{GW190707A}^{+\networkmatchedfiltersnrIMRplus{GW190707A}}_{-\networkmatchedfiltersnrIMRminus{GW190707A}}$ & & \cmark &   --   & \cmark &   \cmark   & \cmark & -- & \cmark &   --   \\[0.075cm]
\NAME{GW190708A} & &\OBSERVINGINSTRUMENTS{GW190708A} & & $\luminositydistancemed{GW190708A}^{+\luminositydistanceplus{GW190708A}}_{-\luminositydistanceminus{GW190708A}}$ & $\totalmassdetmed{GW190708A}^{+\totalmassdetplus{GW190708A}}_{-\totalmassdetminus{GW190708A}}$ & $\chirpmassdetmed{GW190708A}^{+\chirpmassdetplus{GW190708A}}_{-\chirpmassdetminus{GW190708A}}$ & $\finalmassdetmed{GW190708A}^{+\finalmassdetplus{GW190708A}}_{-\finalmassdetminus{GW190708A}}$ & $\finalspinmed{GW190708A}^{+\finalspinplus{GW190708A}}_{-\finalspinminus{GW190708A}}$ & & $\networkmatchedfiltersnrIMRmed{GW190708A}^{+\networkmatchedfiltersnrIMRplus{GW190708A}}_{-\networkmatchedfiltersnrIMRminus{GW190708A}}$ & & \cmark &   --   & \cmark & \cmark & \cmark & \cmark & \cmark &   --   \\[0.075cm]
\NAME{GW190720A} & &\OBSERVINGINSTRUMENTS{GW190720A} & & $\luminositydistancemed{GW190720A}^{+\luminositydistanceplus{GW190720A}}_{-\luminositydistanceminus{GW190720A}}$ & $\totalmassdetmed{GW190720A}^{+\totalmassdetplus{GW190720A}}_{-\totalmassdetminus{GW190720A}}$ & $\chirpmassdetmed{GW190720A}^{+\chirpmassdetplus{GW190720A}}_{-\chirpmassdetminus{GW190720A}}$ & $\finalmassdetmed{GW190720A}^{+\finalmassdetplus{GW190720A}}_{-\finalmassdetminus{GW190720A}}$ & $\finalspinmed{GW190720A}^{+\finalspinplus{GW190720A}}_{-\finalspinminus{GW190720A}}$ & & $\networkmatchedfiltersnrIMRmed{GW190720A}^{+\networkmatchedfiltersnrIMRplus{GW190720A}}_{-\networkmatchedfiltersnrIMRminus{GW190720A}}$ & & \cmark &   --   & \cmark & \cmark & \cmark & -- & \cmark & \cmark \\[0.075cm]
\NAME{GW190727A} & &\OBSERVINGINSTRUMENTS{GW190727A} & & $\luminositydistancemed{GW190727A}^{+\luminositydistanceplus{GW190727A}}_{-\luminositydistanceminus{GW190727A}}$ & $\totalmassdetmed{GW190727A}^{+\totalmassdetplus{GW190727A}}_{-\totalmassdetminus{GW190727A}}$ & $\chirpmassdetmed{GW190727A}^{+\chirpmassdetplus{GW190727A}}_{-\chirpmassdetminus{GW190727A}}$ & $\finalmassdetmed{GW190727A}^{+\finalmassdetplus{GW190727A}}_{-\finalmassdetminus{GW190727A}}$ & $\finalspinmed{GW190727A}^{+\finalspinplus{GW190727A}}_{-\finalspinminus{GW190727A}}$ & & $\networkmatchedfiltersnrIMRmed{GW190727A}^{+\networkmatchedfiltersnrIMRplus{GW190727A}}_{-\networkmatchedfiltersnrIMRminus{GW190727A}}$ & & \cmark & \cmark & \cmark &   --   & \cmark & \cmark & \cmark & \cmark \\[0.075cm]
\NAME{GW190728A} & &\OBSERVINGINSTRUMENTS{GW190728A} & & $\luminositydistancemed{GW190728A}^{+\luminositydistanceplus{GW190728A}}_{-\luminositydistanceminus{GW190728A}}$ & $\totalmassdetmed{GW190728A}^{+\totalmassdetplus{GW190728A}}_{-\totalmassdetminus{GW190728A}}$ & $\chirpmassdetmed{GW190728A}^{+\chirpmassdetplus{GW190728A}}_{-\chirpmassdetminus{GW190728A}}$ & $\finalmassdetmed{GW190728A}^{+\finalmassdetplus{GW190728A}}_{-\finalmassdetminus{GW190728A}}$ & $\finalspinmed{GW190728A}^{+\finalspinplus{GW190728A}}_{-\finalspinminus{GW190728A}}$ & & $\networkmatchedfiltersnrIMRmed{GW190728A}^{+\networkmatchedfiltersnrIMRplus{GW190728A}}_{-\networkmatchedfiltersnrIMRminus{GW190728A}}$ & & \cmark &   --   & \cmark &   \cmark   & \cmark & -- & \cmark & \cmark \\[0.075cm]
\NAME{GW190814A} & &\OBSERVINGINSTRUMENTS{GW190814A}\footnote{Parameter estimation for \NAME{GW190814A}{} made use of data from the three instruments, HLV, although search pipelines only considered LV \cite{GW190814}.} & & $\luminositydistancemed{GW190814A}^{+\luminositydistanceplus{GW190814A}}_{-\luminositydistanceminus{GW190814A}}$ & $\totalmassdetmed{GW190814A}^{+\totalmassdetplus{GW190814A}}_{-\totalmassdetminus{GW190814A}}$ & $\chirpmassdetmed{GW190814A}^{+\chirpmassdetplus{GW190814A}}_{-\chirpmassdetminus{GW190814A}}$ & $\finalmassdetmed{GW190814A}^{+\finalmassdetplus{GW190814A}}_{-\finalmassdetminus{GW190814A}}$ & $\finalspinmed{GW190814A}^{+\finalspinplus{GW190814A}}_{-\finalspinminus{GW190814A}}$ & & $\networkmatchedfiltersnrIMRmed{GW190814A}^{+\networkmatchedfiltersnrIMRplus{GW190814A}}_{-\networkmatchedfiltersnrIMRminus{GW190814A}}$ & & \cmark &   \cmark   & \cmark & -- & \cmark & -- &   --   &   --   \\[0.075cm]
\NAME{GW190828A} & &\OBSERVINGINSTRUMENTS{GW190828A} & & $\luminositydistancemed{GW190828A}^{+\luminositydistanceplus{GW190828A}}_{-\luminositydistanceminus{GW190828A}}$ & $\totalmassdetmed{GW190828A}^{+\totalmassdetplus{GW190828A}}_{-\totalmassdetminus{GW190828A}}$ & $\chirpmassdetmed{GW190828A}^{+\chirpmassdetplus{GW190828A}}_{-\chirpmassdetminus{GW190828A}}$ & $\finalmassdetmed{GW190828A}^{+\finalmassdetplus{GW190828A}}_{-\finalmassdetminus{GW190828A}}$ & $\finalspinmed{GW190828A}^{+\finalspinplus{GW190828A}}_{-\finalspinminus{GW190828A}}$ & & $\networkmatchedfiltersnrIMRmed{GW190828A}^{+\networkmatchedfiltersnrIMRplus{GW190828A}}_{-\networkmatchedfiltersnrIMRminus{GW190828A}}$ & & \cmark & \cmark & \cmark & \cmark & \cmark & \cmark & \cmark & \cmark \\[0.075cm]
\NAME{GW190828B} & &\OBSERVINGINSTRUMENTS{GW190828B} & & $\luminositydistancemed{GW190828B}^{+\luminositydistanceplus{GW190828B}}_{-\luminositydistanceminus{GW190828B}}$ & $\totalmassdetmed{GW190828B}^{+\totalmassdetplus{GW190828B}}_{-\totalmassdetminus{GW190828B}}$ & $\chirpmassdetmed{GW190828B}^{+\chirpmassdetplus{GW190828B}}_{-\chirpmassdetminus{GW190828B}}$ & $\finalmassdetmed{GW190828B}^{+\finalmassdetplus{GW190828B}}_{-\finalmassdetminus{GW190828B}}$ & $\finalspinmed{GW190828B}^{+\finalspinplus{GW190828B}}_{-\finalspinminus{GW190828B}}$ & & $\networkmatchedfiltersnrIMRmed{GW190828B}^{+\networkmatchedfiltersnrIMRplus{GW190828B}}_{-\networkmatchedfiltersnrIMRminus{GW190828B}}$ & & \cmark &   --   & \cmark & \cmark & \cmark & -- & \cmark & \cmark \\[0.075cm]
\NAME{GW190910A} & &\OBSERVINGINSTRUMENTS{GW190910A} & & $\luminositydistancemed{GW190910A}^{+\luminositydistanceplus{GW190910A}}_{-\luminositydistanceminus{GW190910A}}$ & $\totalmassdetmed{GW190910A}^{+\totalmassdetplus{GW190910A}}_{-\totalmassdetminus{GW190910A}}$ & $\chirpmassdetmed{GW190910A}^{+\chirpmassdetplus{GW190910A}}_{-\chirpmassdetminus{GW190910A}}$ & $\finalmassdetmed{GW190910A}^{+\finalmassdetplus{GW190910A}}_{-\finalmassdetminus{GW190910A}}$ & $\finalspinmed{GW190910A}^{+\finalspinplus{GW190910A}}_{-\finalspinminus{GW190910A}}$ & & $\networkmatchedfiltersnrIMRmed{GW190910A}^{+\networkmatchedfiltersnrIMRplus{GW190910A}}_{-\networkmatchedfiltersnrIMRminus{GW190910A}}$ & & \cmark & \cmark & \cmark &   --   & \cmark & \cmark & \cmark &   --   \\[0.075cm]
\NAME{GW190915A} & &\OBSERVINGINSTRUMENTS{GW190915A} & & $\luminositydistancemed{GW190915A}^{+\luminositydistanceplus{GW190915A}}_{-\luminositydistanceminus{GW190915A}}$ & $\totalmassdetmed{GW190915A}^{+\totalmassdetplus{GW190915A}}_{-\totalmassdetminus{GW190915A}}$ & $\chirpmassdetmed{GW190915A}^{+\chirpmassdetplus{GW190915A}}_{-\chirpmassdetminus{GW190915A}}$ & $\finalmassdetmed{GW190915A}^{+\finalmassdetplus{GW190915A}}_{-\finalmassdetminus{GW190915A}}$ & $\finalspinmed{GW190915A}^{+\finalspinplus{GW190915A}}_{-\finalspinminus{GW190915A}}$ & & $\networkmatchedfiltersnrIMRmed{GW190915A}^{+\networkmatchedfiltersnrIMRplus{GW190915A}}_{-\networkmatchedfiltersnrIMRminus{GW190915A}}$ & & \cmark &   --   & \cmark &   --   & \cmark & \cmark & \cmark & \cmark \\[0.075cm]
\NAME{GW190924A} & &\OBSERVINGINSTRUMENTS{GW190924A} & & $\luminositydistancemed{GW190924A}^{+\luminositydistanceplus{GW190924A}}_{-\luminositydistanceminus{GW190924A}}$ & $\totalmassdetmed{GW190924A}^{+\totalmassdetplus{GW190924A}}_{-\totalmassdetminus{GW190924A}}$ & $\chirpmassdetmed{GW190924A}^{+\chirpmassdetplus{GW190924A}}_{-\chirpmassdetminus{GW190924A}}$ & $\finalmassdetmed{GW190924A}^{+\finalmassdetplus{GW190924A}}_{-\finalmassdetminus{GW190924A}}$ & $\finalspinmed{GW190924A}^{+\finalspinplus{GW190924A}}_{-\finalspinminus{GW190924A}}$ & & $\networkmatchedfiltersnrIMRmed{GW190924A}^{+\networkmatchedfiltersnrIMRplus{GW190924A}}_{-\networkmatchedfiltersnrIMRminus{GW190924A}}$ & & \cmark &   --   & \cmark & \cmark & \cmark & -- & \cmark & \cmark \\[0.075cm]
\bottomrule
\end{tabular}
\end{table*}

\section{Parameter inference}
\label{sec:inference}

\subsection{Individual events}
\label{sec:inference:individual}

The foundation for almost all of the tests presented in this paper
are the waveform models that describe the GW signal emitted from a coalescing 
compact binary.
The only exception is 
the polarization analysis (Sec.~\ref{sec:pol}), which relies on null-stream projections of the data \cite{Sutton:2009gi,Pang:2020pfz}.
In GR, the GW signal from a BBH on a quasicircular orbit
is fully characterized by $15$ parameters \cite{Romero-Shaw:2020owr}.
These include the intrinsic parameters (the masses $m_{1,2}$ and spin angular momenta $\vec S_{1,2}$   
of the binary components), and extrinsic ones (the luminosity distance, 
the location of the binary in the sky, the orientation of its orbit with 
respect to observer's line of sight, its polarization angle, and the reference time and 
orbital phase). The dominant effects of the BHs' spin 
angular momenta on the waveform comes from the spin components along the orbital axis. 
However, the other components of the spins lead to precession of 
the spin vectors and the binary’s orbital plane, introducing modulations 
into the GW amplitude and phase \cite{Apostolatos:1994mx,Kidder:1995zr}. We find that aligned-spin waveform models 
are sufficient for many events in this paper, but we analyze all events with 
at least one precessing waveform model, to take these effects into account.

The working null hypothesis throughout the paper is that all events are quasicircular BBHs in GR, with no measurable systematics.
In principle, a BBH waveform could be affected by the presence of eccentricity, which is not included in any of the waveform models we use.
The presence of significant eccentricity could result in systematic errors mimicking a deviation from GR \cite{Ramos-Buades:2019uvh,Moore:2019vjj,Romero-Shaw:2019itr}.
If evidence for such a deviation was found, extra work would be required to discard eccentricity, matter effects (for less massive systems), or other systematics.

For a majority of the tests we employ two waveform families to model signals from BBHs in GR. 
One is the non-precessing effective-one-body (EOB) waveform family \SEOBNR ~\cite{Bohe:2016gbl}, 
an analytical model that takes inputs from post-Newtonian theory, 
BH perturbation theory, the gravitational self-force formalism,  
and NR simulations. For computational efficiency in the analyses, we 
use a frequency-domain reduced-order model for \SEOBNR~known as \SEOBROM~\cite{Bohe:2016gbl}.
There exists a precessing EOB waveform model \textsc{SEOBNRv4P} \cite{Pan:2013rra,Babak:2016tgq,Ossokine:2020kjp}, which has been employed in \cite{GWTC2}, but we do not use it here due to its high computational cost.
The other waveform family is the precessing 
phenomenological waveform family \IMRP{} \cite{Husa:2015iqa,Khan:2015jqa,Hannam:2013oca}, 
a frequency-domain model that describes the spin precession effects in terms of two effective parameters by twisting up the underlying aligned-spin model \cite{Schmidt:2010it,Schmidt:2012rh,Schmidt:2014iyl}. 
The aligned-spin model is itself calibrated to hybrid waveforms, which are constructed by stitching 
together waveforms from the inspiral part (modeled using the SEOBNRv2 \cite{Taracchini:2013} model without calibration from NR) 
and the merger--ringdown part (modeled using NR simulations) of the coalescence.
The two waveform models, \IMRP{} and \SEOBROM, are employed to help gauge systematics, 
as discussed in detail in Sec.~\ref{sec:par}.
Although a detailed study of waveform systematics is beyond the scope of this paper, 
relevant studies can be found in \cite{Bohe:2016gbl,Khan:2015jqa,Cotesta:2018fcv,Varma:2018mmi,Cotesta:2020qhw,Ramos-Buades:2020noq,Pratten:2020fqn,Pratten:2020ceb,Ossokine:2020kjp}.

During O3a, we observed a number of events for which higher-order (non-quadrupole) multipole moments of the radiation were shown to affect parameter estimation;
this includes \NAME{GW190412A}{} \cite{GW190412}, \NAME{GW190521A}{} \cite{GW190521g, GW190521g:imp}, and \NAME{GW190814A}{}  \cite{GW190814}.
Where possible and appropriate, we employ one of three waveform models incorporating higher moments (HMs): \IMRPHM{} \cite{Khan:2018fmp, Khan:2019kot}, \SEOBHMROM{} \cite{Cotesta:2018fcv,Cotesta:2020qhw}, or \NRSur{} \cite{Varma:2019csw}.
\IMRPHM{} is a successor of \IMRP{} that includes two-spin precession \cite{Chatziioannou:2017tdw} and the $(\ell, |m|)=(2, 2), (2, 1), (3, 3), (3, 2), (4, 4), (4, 3)$ multipoles; \SEOBHMROM{} is built upon \SEOBNRHM{} which incorporates $(\ell, |m|) = (2, 2), (2,1), (3, 3), (4, 4), (5, 5)$; finally, \NRSur{} is a surrogate model that is built by directly interpolating NR simulations, accounting for all spin degrees of freedom and all multipoles with $\ell \leq 4$, in the coprecessing frame.
When we use \IMRP{}, ~\IMRPHM{}, and \NRSur{}, we impose a prior $m_2/m_1 \geq 1/18, 1/18, 1/6$, respectively, on the mass ratio, as these waveform families are not known to be valid for lower $m_2/m_1$.
Whenever we make use of a waveform other than \IMRP{} or \SEOBROM{}, we state so explicitly in the text.

A majority of the tests presented in this paper are performed using the \linf{} 
code~\cite{Veitch:2014wba} in the LIGO Scientific Collaboration Algorithm Library 
Suite (\lal{})~\cite{lalsuite}. 
This code is designed to carry out Bayesian inference using two possible sampling algorithms: Markov-chain Monte Carlo (MCMC), and nested sampling. 
More detail on how the binary parameters are estimated can be found in Sec.~V of \cite{GWTC2}. 
In \linf{} analyses, the PSD used was either estimated at the time of each event using the 
\bw{} code ~\cite{Cornish:2014kda, Littenberg:2014oda} or estimated near the time of an event using Welch's method \cite{Welch_psd}.
Unless otherwise specified, the prior distributions of various GR parameters (intrinsic and extrinsic) for each event 
are the same as in \cite{GWTC2}.
The priors on non-GR parameters specific to each test
are discussed in their respective sections below.  
Other quantities such as the frequency range (over which the matched-filter output is computed)  
for each event is kept the same as in \cite{GWTC2}, unless otherwise specified.

Exceptions to the use of \linf{} include the residuals test of Sec.~\ref{sec:res}, the IMR consistency test of Sec.~\ref{sec:imr}, one of the ringdown studies in Sec.~\ref{sec:rin}, and the polarization analysis of Sec.~\ref{sec:pol}.
The residuals test uses \bw{} directly to carry out inference on the residual data.
Additional to \linf{}, the IMR consistency test also employs a parallelized nested sampling pipeline \textsc{pBilby} \cite{Ashton:2018jfp,Smith:2019ucc,Romero-Shaw:2020owr}.
The damped-sinusoid ringdown analysis is carried out with the \textsc{pyRing} pipeline \cite{Carullo:2019flw,Isi:2019aib}.
The polarization analysis is carried out with the \textsc{BANTAM} pipeline \cite{Pang:2020pfz}.

Finally, we assumed the same cosmology for all the events in this paper to infer their unredshifted masses and the proper distances (as required in Sec.~\ref{sec:liv}).
Specifically, we take $H_0 = 67.90 \text{ km s}^{-1} \text{ Mpc}^{-1}$ for the Hubble constant, and $\Omega_\text{m} = 0.3065$ and $\Omega_\Lambda = 0.6935$ for the matter and dark energy density parameters (``TT+lowP+lensing+ext'' values from~\cite{Ade:2015xua}).

\subsection{Sets of measurements}
\label{sec:inference:populations}

\newcommand{\p}{\hat{p}}
\newcommand{\deltap}{\delta \p}
\newcommand{\popmu}{\mu}
\newcommand{\popsig}{\sigma}
\newcommand{\eventmu}{\tilde{\mu}}
\newcommand{\eventsig}{\tilde{\sigma}}

\newcommand{\ppe}{\delta\hat{\varphi}}
\newcommand{\nevent}{N}
\newcommand{\nparam}{P}
\newcommand{\data}{\vec{d}}

There are multiple statistical strategies for drawing inferences from a set of events, each carrying its own set of assumptions about the nature of potential deviations from GR and how they may manifest in our signals.
For simplicity, \cite{LIGOScientific:2019fpa} reported constraints assuming that deviations from GR would manifest equally across events, independent of source properties.
This is only strictly justifiable when the deviation parameters are known by construction to be the same for all detected events (or some known function of the source properties).
This is the case for probes of the propagation of GWs (e.g., dispersion), where the propagation effects can reasonably be assumed to affect all sources equally (barring a known dependence on the luminosity distance, which is explicitly factored out of the analysis).
However, it is generally not the case for parametrized tests of GW generation, wherein waveforms are allowed to deviate in arbitrary (albeit controlled) ways from the GR prediction.

To relax the assumption of shared deviations across events, in this paper we apply the hierarchical inference technique proposed and implemented for GWTC-1 events in \cite{Zimmerman:2019wzo,Isi:2019asy}.
We apply this procedure to the IMR consistency test (Sec.~\ref{sec:imr}), the waveform generation tests (Sec.~\ref{sec:gen}), and the ringdown analyses (Sec.~\ref{sec:rin}).
The strategy consists of modeling non-GR parameters for each event in our pool as drawn from a common underlying distribution, whose properties we infer coherently from the data for all events as whole \cite{Bovy:2009,Loredo:2004nn}.
The nature of such unknown distribution would be determined by the true theory of gravity and the population of sources (e.g.,~the magnitude of the departure from GR could be a function of the total mass of the binary), convolved with any biases affecting our selection of events.
By comparing the inferred distribution to the GR prediction (no deviation for any of the events), we obtain a null test of GR from our whole set of observations.

Unlike other contexts in which hierarchical techniques are used (notably, the study of astrophysical populations \cite{LIGOScientific:2018jsj,GWTC2:rates}), the goal here is always to characterize the distribution of measured quantities for the events in our set, not to make inferences about underlying astrophysical distributions that are not directly accessible (as discussed in Sec.~\ref{sec:events}).
This simplifies our hierarchical model, which does not attempt to deconvolve selection biases. 
However, it limits the kinds of conclusions we may draw from our observations, since they will necessarily pertain strictly to the signals that we have detected and analyzed.

Although the true nature of the hyperdistribution could be arbitrarily complex, we may always capture its essential features by means of a moment expansion.
To achieve this, we model the true values of each beyond-GR parameter in our pool of events as drawn from a Gaussian of unknown mean $\popmu$ and standard deviation $\popsig$ \cite{Isi:2019asy}.
This is a suitable choice because the Gaussian is the least informative distribution (i.e.,~it has maximum entropy conditional on the first two moments) \cite{Shannon:1948zz}.
GR is recovered for $\sigma=0$ and $\mu=x_{\rm GR}$, where $x_{\rm GR}$ is the GR prediction for the parameter at hand (e.g.,~$x_{\rm GR}=0$ for parameters defined as a fractional deviation away from GR).
As the number of detections increases in the future, we may enhance flexibility by including additional moments in our model (akin to adding further terms in a series expansion).
In spite of its simplicity, the Gaussian parametrization has been shown to work effectively even when the true distribution presents highly nontrivial features, like correlations across the beyond-GR parameters \cite{Isi:2019asy}.
A set of measurements not conforming to GR would be identified through posteriors on $\mu$ and $\sigma$ that are inconsistent with the GR values, at the 90\% credible level.

We obtain posteriors on the hyperparameters $\mu$ and $\sigma$ through a joint analysis of the set of detections, using the \textsc{Stan}-based \cite{JSSv076i01} infrastructure developed in \cite{Isi:2019asy}.
We summarize the results from that hierarchical analysis through the population-marginalized distribution for the beyond-GR parameters, also known as the \emph{observed population predictive distribution} \cite{GWTC2:rates}.
For a given beyond-GR parameter $x$, this distribution $p(x \mid d)$ is the expectation for $x$ after marginalizing over the hyperparameters $\mu$ and $\sigma$, 
\begin{equation} \label{eq:inf:hier_dist}
p(x \mid d) =
\int p(x \mid \mu, \sigma)\, 
p(\mu,\sigma \mid d)\, {\rm d} \mu\, {\rm d}\sigma\, ,
\end{equation}
where $d$ represents the data for \emph{all} detected events, and $p(x \mid \mu, \sigma) \sim {\cal N}(\mu, \sigma)$ by construction \cite{Isi:2019asy}.
Since we are characterizing a group of observations, not an astrophysical distribution, there is no factor in Eq.~\eqref{eq:inf:hier_dist} accounting for selection biases.
A posterior expectation $p(x \mid d)$ that supports $x=x_{\rm GR}$ is a necessary, but insufficient, condition for establishing agreement with GR---since we must also have $\sigma$ consistent with zero.
If GR is correct and in the absence of systematics, $p(x \mid d)$ should approach a Dirac $\delta$ function at $x_{\rm GR}$ with increasing number of observations.
Assuming $x_{\rm GR}$ is supported by $p(x \mid d)$, the width of this distribution is a measure of our uncertainty about deviations from GR in this parameter after combining all events.

Requiring that all events share the same value of the beyond-GR parameter is equivalent to demanding $\sigma=0$.
Fixing $\sigma=0$, the hierarchical method reduces to the approach of multiplying likelihoods from individual events \cite{Zimmerman:2019wzo}, as done in \cite{LIGOScientific:2019fpa}.
Equation \eqref{eq:inf:hier_dist} may then be interpreted as a posterior on the value of $x$, and is identical to the combined posteriors as computed in \cite{LIGOScientific:2019fpa}.
In the sections below, we present both types of combined results (inferred $\sigma$, and fixed $\sigma=0$), facilitating comparisons to previously reported constraints.
For a concrete demonstration of the usefulness of the hierarchical approach see Sec.~\ref{sec:imr} (and the related Appendix \ref{app:imr}), where we show how this technique succesfully identifies a subset of signals not conforming to the null hypothesis (due to known systematics, in this case), while the multiplied-likelihood approach does not.

Finally, under certain circumstances, statements from the set of measurements may be obtained by studying the empirical distribution of some detection statistic for a frequentist null test of the hypothesis that GR is a good description of the data.
As for the residuals test (Sec.~\ref{sec:res}), this may be done if the analysis yields a distribution of $p$-values, obtained by comparing some detection statistic against an empirical background distribution for each event.
If the null hypothesis holds, we expect the resulting $p$-values to be uniformly distributed in the interval $[0,\,1]$.
Agreement with this expectation can be quantified through a meta $p$-value obtained through Fisher's method \cite{Fisher1948}.
It can also be represented visually through a probability--probability (PP) plot, displaying the fraction of events yielding $p$-values smaller than or equal to any given number: under the null hypothesis, the PP plot should be diagonal (see also Appendix~\ref{app:res}).

\section{Consistency tests}
\label{sec:con}

    \subsection{Residuals test}
    \label{sec:res}
    \newcommand{\ResWorstP}[1]{\IfEqCase{#1}{{NAME}{\NAME{GW190421A}}{FF90}{0.81}{N}{193.0}{SNR90}{7.52}{SNRGR}{10.47}{p_SNR90}{0.07}{run}{O3a}{Events}{\NAME{GW190421A}}}}
\newcommand{\ResBestP}[1]{\IfEqCase{#1}{{NAME}{\NAME{GW190727A}}{FF90}{0.92}{N}{193.0}{SNR90}{4.88}{SNRGR}{11.62}{p_SNR90}{0.97}{run}{O3a}{Events}{\NAME{GW190727A}}}}
\newcommand{\ResWorstSNR}[1]{\IfEqCase{#1}{{NAME}{\NAME{GW190408A}}{FF90}{0.88}{N}{193.0}{SNR90}{8.48}{SNRGR}{16.06}{p_SNR90}{0.15}{run}{O3a}{Events}{\NAME{GW190408A}}}}
\newcommand{\ResMetaPvalue}{0.39}
\newcommand{\ResNevents}{34}

\newcommand{\bsn}{{\cal B}^{\rm S}_{\rm N}}
\newcommand{\bsg}{{\cal B}^{\rm S}_{\rm G}}
\newcommand{\rhores}{\mathrm{SNR}_{90}}%
\newcommand{\rhosig}{\mathrm{SNR}_\mathrm{GR}}
\newcommand{\rhodiff}{\mathrm{SNR}_\mathrm{res}}
\newcommand{\rhonoise}{\rhores^\mathrm{n}}
\newcommand{\ff}{\mathrm{FF}_{90}}
\newcommand{\resN}{193}

\begin{table}
\caption{Waveforms subtracted to study residuals in Sec.~\ref{sec:res}.}
\label{tab:res:wfs}
\begin{tabular}{l@{\quad}l@{\quad}l@{\quad}l}
\toprule
Event            & Ref.                           & Approximant & Ref.                                            \\
\midrule                                          
\NAME{GW190412A} & \cite{GW190412}                & \IMRPHM{}   & \cite{Khan:2018fmp, Khan:2019kot}               \\
\NAME{GW190521A} & \cite{GW190521g,GW190521g:imp} & \NRSur{}    & \cite{Varma:2018mmi}                            \\
\NAME{GW190814A} & \cite{GW190814}                & \IMRPHM{}   & \cite{Khan:2018fmp, Khan:2019kot}               \\
\midrule                                          
All others       & \cite{GWTC2}                   & \IMRP{}     & \cite{Husa:2015iqa,Khan:2015jqa,Hannam:2013oca} \\
\bottomrule
\end{tabular}
\end{table}

A generic way of quantifying the success of our GR waveforms in describing the data is to study the residual strain after subtracting the best-fit template for each event \cite{LIGOScientific:2019hgc}.
Residual analyses are sensitive to any sort of modeling systematics, whether they arise from a deviation from GR or more prosaic reasons.
Results from similar studies were previously presented in \cite{TheLIGOScientific:2016src,LIGOScientific:2019fpa,GW190814,GW190521g:imp}.

We follow the procedure described in \cite{LIGOScientific:2019fpa}.
For each event in our set, we subtract the maximum likelihood (best-fit) GR-based waveform from the data to obtain residuals for a $1~\mathrm{s}$ window centered on the trigger time reported in \cite{GWTC2}.
Except for the three events detailed in Table~\ref{tab:res:wfs}, we obtain the GR prediction using the \IMRP{} waveform family.%
\footnote{For \NAME{GW190814A}{}, we also used \SEOBHM{}, which yielded results consistent with \IMRPHM{} \cite{GW190814}.}
We then use \bw{} to place a 90\%-credible upper-limit on the leftover coherent signal-to-noise ratio (SNR).
To evaluate whether this value, $\rhores$, is consistent with instrumental noise fluctuations, we measure the coherent power in \resN{} sets of noise-only detector data around each event.
This yields a $p$-value for noise-producing coherent power with $\rhonoise$ greater than or equal to the residual value $\rhores$, i.e., $p = P(\rhonoise \ge \rhores \mid \mathrm{noise})$.

Our results for O3a events are summarized in Table \ref{tab:residuals} (see Table~II in \cite{LIGOScientific:2019fpa} for O1 and O2 events).
For each event, we present the values of the residual $\rhores$, as well as the corresponding fitting factor $\ff = \rhosig\, / (\rhodiff^2 + \rhosig^2)^{1/2}$, where $\rhodiff$ is the coherent residual SNR and $\rhosig$ is the SNR of the best-fit template.
This quantifies agreement between the best-fit template and the data as being better than $\ff\times100\%$ \cite{LIGOScientific:2019fpa,TheLIGOScientific:2016src}.
Table \ref{tab:residuals} also shows the $\rhores$ $p$-values.

\begin{figure}
	\centering
	\includegraphics[width=\columnwidth]{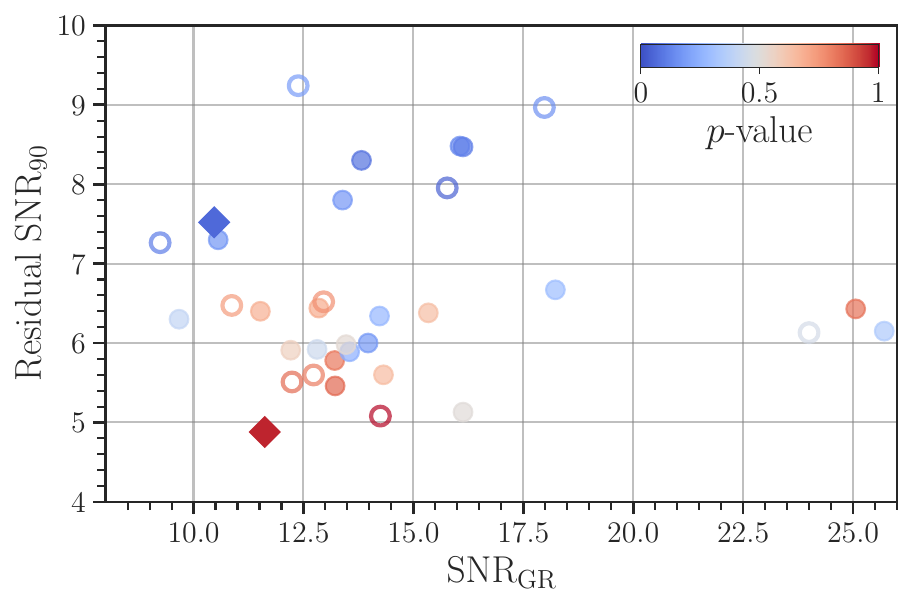}
	\caption{Upper limit on the residual network SNR ($\rhores$) for each event, as a function of SNR recovered by the maximum-likelihood template ($\mathrm{SNR}_\mathrm{GR}$), with the corresponding $p$-value shown in color (see Table~\ref{tab:residuals}).
Solid (empty) markers indicate events detected in O3a (O1 or O2).
Diamonds highlight the O3a events yielding the highest ({\protect\ResBestP{NAME}}) and lowest ({\protect\ResWorstP{NAME}}) $p$-values, $p=\protect\ResBestP{p_SNR90}$ and $p=\protect\ResWorstP{p_SNR90}$ respectively.
}
	\label{fig:res:snr}
\end{figure}

Figure \ref{fig:res:snr} displays the $\rhores$ values reported in Table \ref{tab:residuals} as a function of the SNR of the best-fit template, with $\rhores$ $p$-values encoded in the marker colors;
events preceding O3 are identified by an empty marker (see Table II in \cite{LIGOScientific:2019fpa}).
If the GR model is a good fit for the data, the magnitude of $\rhores$ should depend only on the state of the instruments at the time of each event, not on the amplitude of the subtracted template.
This is consistent with Fig.~\ref{fig:res:snr}, which reveals no sign of such a trend.

The variation in $\rhores$ is linked to the distribution of the corresponding $p$-values, as suggested by Fig.~\ref{fig:res:snr}.
The O3a event yielding the highest (lowest) $p$-value is \ResBestP{NAME}{} (\ResWorstP{NAME}) with $\rhores=\ResBestP{SNR90}$ and $p=\ResBestP{p_SNR90}$ ($\rhores=\ResWorstP{SNR90}$ and $p=\ResWorstP{p_SNR90}$), and is highlighted in Fig.~\ref{fig:res:snr} by a red (blue) diamond.
Although \ResWorstSNR{NAME}{} is the O3a event with the highest residual power ($\rhores = \ResWorstSNR{SNR90}$), the $p$-value of \ResWorstSNR{p_SNR90}{} indicates that this is not inconsistent with the background distribution.
Two pre-O3a events, GW170814 and GW170818, yielded higher $\rhores$ than \ResWorstSNR{NAME}{}~\cite{LIGOScientific:2019fpa}, as seen in Fig.~\ref{fig:res:snr}.

\begin{figure}
	\centering
	\includegraphics[width=\columnwidth]{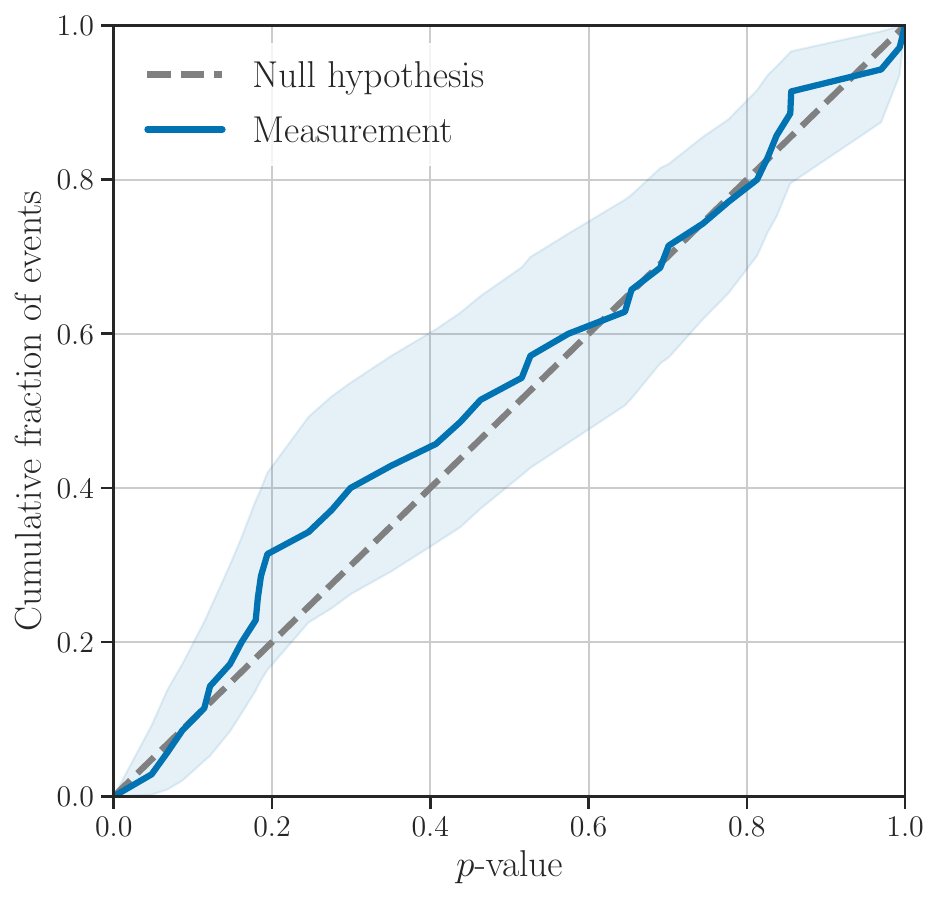}
	\caption{Fraction of events yielding a residuals-test $p$-value less than or equal to the abscissa. The light-blue band marks the 90\%-credible region for our measurement, factoring in the uncertainty due to a finite number of both events and background instantiations (Appendix~\ref{app:res}). The meta $p$-value for a uniform distribution is \ResMetaPvalue.
}
	\label{fig:res:pp}
\end{figure}

The set of $p$-values shown in Table \ref{tab:residuals} is consistent with all coherent residual power being due to instrumental noise.
Assuming that this is indeed the case, we expect the $p$-values to be uniformly distributed over $[0, 1]$.
Agreement with a uniform distribution is represented via the PP plot in Fig.~\ref{fig:res:pp}, which shows that the measurement agrees with the null hypothesis (diagonal line) within 90\% credibility (computed as detailed in Appendix~\ref{app:res}).
We also compute a meta $p$-value for a uniform distribution of \ResMetaPvalue{} (see Sec.~\ref{sec:inference:populations}).
This demonstrates no statistically significant deviations between the observed residual power and the detector noise around the set of events.

\begin{table}
\caption{Results of the residuals analysis (Sec.~\ref{sec:res}).
For each event, we present the SNR of the subtracted GR waveform ($\mathrm{SNR}_{\rm GR}$), the 90\%-credible upper limit on the residual network SNR ($\rhores$), a corresponding lower limit on the fitting factor ($\ff$), and the $p$-value.
}
\label{tab:residuals}
\centering
\begin{tabular}{lrccc}
\toprule
           Events & ${\rm SNR}_{\rm GR}$ & Residual ${\rm SNR}_{90}$ & ${\rm FF}_{90}$ & $p$-value \\
\midrule
 \NAME{GW190408A} &                16.06 &                      8.48 &            0.88 &      0.15 \\
 \NAME{GW190412A} &                18.23 &                      6.67 &            0.94 &      0.30 \\
 \NAME{GW190421A} &                10.47 &                      7.52 &            0.81 &      0.07 \\
 \NAME{GW190503A} &                13.21 &                      5.78 &            0.92 &      0.83 \\
 \NAME{GW190512A} &                12.81 &                      5.92 &            0.91 &      0.44 \\
 \NAME{GW190513A} &                12.85 &                      6.44 &            0.89 &      0.70 \\
 \NAME{GW190517A} &                11.52 &                      6.40 &            0.87 &      0.69 \\
 \NAME{GW190519A} &                15.34 &                      6.38 &            0.92 &      0.65 \\
 \NAME{GW190521A} &                14.23 &                      6.34 &            0.91 &      0.28 \\
 \NAME{GW190521B} &                25.71 &                      6.15 &            0.97 &      0.35 \\
 \NAME{GW190602A} &                13.22 &                      5.46 &            0.92 &      0.86 \\
 \NAME{GW190630A} &                16.13 &                      5.13 &            0.95 &      0.52 \\
 \NAME{GW190706A} &                13.39 &                      7.80 &            0.86 &      0.18 \\
 \NAME{GW190707A} &                13.55 &                      5.89 &            0.92 &      0.25 \\
 \NAME{GW190708A} &                13.97 &                      6.00 &            0.92 &      0.19 \\
 \NAME{GW190720A} &                10.56 &                      7.30 &            0.82 &      0.18 \\
 \NAME{GW190727A} &                11.62 &                      4.88 &            0.92 &      0.97 \\
 \NAME{GW190728A} &                13.47 &                      5.98 &            0.91 &      0.53 \\
 \NAME{GW190814A} &                25.06 &                      6.43 &            0.97 &      0.84 \\
 \NAME{GW190828A} &                16.13 &                      8.47 &            0.89 &      0.12 \\
 \NAME{GW190828B} &                 9.67 &                      6.30 &            0.84 &      0.41 \\
 \NAME{GW190910A} &                14.32 &                      5.60 &            0.93 &      0.65 \\
 \NAME{GW190915A} &                13.82 &                      8.30 &            0.86 &      0.09 \\
 \NAME{GW190924A} &                12.21 &                      5.91 &            0.90 &      0.57 \\
\bottomrule
\end{tabular}

\end{table}

    \subsection{Inspiral--merger--ringdown consistency test}
    \label{sec:imr}
    \newcommand{\ImrMfHierMu}[1]{\IfEqCase{#1}{{LOWM}{\ensuremath{0.02^{+0.11}_{-0.09}}}{GWTC2}{\ensuremath{0.13^{+0.18}_{-0.12}}}{GWTC1}{\ensuremath{0.15^{+0.33}_{-0.22}}}}}
\newcommand{\ImrChifHierMu}[1]{\IfEqCase{#1}{{LOWM}{\ensuremath{-0.06^{+0.15}_{-0.16}}}{GWTC2}{\ensuremath{-0.01^{+0.12}_{-0.12}}}{GWTC1}{\ensuremath{-0.03^{+0.21}_{-0.21}}}}}
\newcommand{\ImrMfHierSigma}[1]{\IfEqCase{#1}{{LOWM}{\ensuremath{0.17}}{GWTC2}{\ensuremath{0.32}}{GWTC1}{\ensuremath{0.55}}}}
\newcommand{\ImrChifHierSigma}[1]{\IfEqCase{#1}{{LOWM}{\ensuremath{0.34}}{GWTC2}{\ensuremath{0.20}}{GWTC1}{\ensuremath{0.32}}}}
\newcommand{\ImrMfHierPop}[1]{\IfEqCase{#1}{{LOWM}{\ensuremath{0.02^{+0.20}_{-0.17}}}{GWTC2}{\ensuremath{0.12^{+0.42}_{-0.32}}}{GWTC1}{\ensuremath{0.13^{+0.68}_{-0.49}}}}}
\newcommand{\ImrChifHierPop}[1]{\IfEqCase{#1}{{LOWM}{\ensuremath{-0.05^{+0.36}_{-0.41}}}{GWTC2}{\ensuremath{-0.01^{+0.21}_{-0.24}}}{GWTC1}{\ensuremath{-0.03^{+0.37}_{-0.35}}}}}
\newcommand{\ImrGWTCTWO}[1]{\IfEqCase{#1}{{DMFGWTC2PHENOM}{\ensuremath{-0.04^{+0.08}_{-0.06}}}{DMFGWTC2SEOB}{\ensuremath{0.01^{+0.09}_{-0.08}}}{DCHIFGWTC2PHENOM}{\ensuremath{-0.09^{+0.11}_{-0.08}}}{DCHIFGWTC2SEOB}{\ensuremath{-0.05^{+0.11}_{-0.09}}}}}
\newcommand{\ImrEVENTSTATS}[1]{\IfEqCase{#1}{{GW170814DMFGWTC2PHENOM}{\ensuremath{0.63^{+0.55}_{-0.81}}}{GW170814DCHIFGWTC2PHENOM}{\ensuremath{-0.03^{+0.54}_{-0.26}}}{GW170814GRQUANTGWTC2}{\ensuremath{22.9}}{GW170818DMFGWTC2PHENOM}{\ensuremath{0.16^{+0.41}_{-0.30}}}{GW170818DCHIFGWTC2PHENOM}{\ensuremath{-0.13^{+0.53}_{-0.52}}}{GW170818GRQUANTGWTC2}{\ensuremath{26.8}}{S190412mDMFGWTC2PHENOM}{\ensuremath{-0.56^{+0.63}_{-0.88}}}{S190412mDCHIFGWTC2PHENOM}{\ensuremath{-0.11^{+0.98}_{-0.31}}}{S190412mGRQUANTGWTC2}{\ensuremath{69.0}}{GW170823DMFGWTC2PHENOM}{\ensuremath{0.79^{+0.21}_{-0.46}}}{GW170823DCHIFGWTC2PHENOM}{\ensuremath{0.16^{+0.44}_{-0.55}}}{GW170823GRQUANTGWTC2}{\ensuremath{93.3}}{S190828jDMFGWTC2PHENOM}{\ensuremath{0.03^{+0.34}_{-0.25}}}{S190828jDCHIFGWTC2PHENOM}{\ensuremath{-0.08^{+0.39}_{-0.34}}}{S190828jGRQUANTGWTC2}{\ensuremath{21.5}}{S190408anDMFGWTC2PHENOM}{\ensuremath{0.02^{+0.40}_{-0.27}}}{S190408anDCHIFGWTC2PHENOM}{\ensuremath{-0.01^{+0.47}_{-0.44}}}{S190408anGRQUANTGWTC2}{\ensuremath{11.4}}{S190503bfDMFGWTC2PHENOM}{\ensuremath{0.62^{+0.40}_{-0.71}}}{S190503bfDCHIFGWTC2PHENOM}{\ensuremath{0.33^{+0.44}_{-0.50}}}{S190503bfGRQUANTGWTC2}{\ensuremath{53.2}}{S190727hDMFGWTC2PHENOM}{\ensuremath{0.55^{+0.24}_{-0.28}}}{S190727hDCHIFGWTC2PHENOM}{\ensuremath{-0.06^{+0.47}_{-0.48}}}{S190727hGRQUANTGWTC2}{\ensuremath{98.7}}{GW150914DMFGWTC2PHENOM}{\ensuremath{0.21^{+0.31}_{-0.22}}}{GW150914DCHIFGWTC2PHENOM}{\ensuremath{0.15^{+0.47}_{-0.40}}}{GW150914GRQUANTGWTC2}{\ensuremath{55.7}}{S190521rDMFGWTC2PHENOM}{\ensuremath{0.10^{+0.36}_{-0.22}}}{S190521rDCHIFGWTC2PHENOM}{\ensuremath{0.07^{+0.30}_{-0.28}}}{S190521rGRQUANTGWTC2}{\ensuremath{0.0}}{S190630agDMFGWTC2PHENOM}{\ensuremath{-0.13^{+0.26}_{-0.20}}}{S190630agDCHIFGWTC2PHENOM}{\ensuremath{-0.15^{+0.43}_{-0.28}}}{S190630agGRQUANTGWTC2}{\ensuremath{58.8}}{GW170104DMFGWTC2PHENOM}{\ensuremath{-0.07^{+0.33}_{-0.23}}}{GW170104DCHIFGWTC2PHENOM}{\ensuremath{-0.10^{+0.47}_{-0.37}}}{GW170104GRQUANTGWTC2}{\ensuremath{29.0}}{S190513bmDMFGWTC2PHENOM}{\ensuremath{-0.03^{+0.33}_{-0.27}}}{S190513bmDCHIFGWTC2PHENOM}{\ensuremath{-0.21^{+0.55}_{-0.40}}}{S190513bmGRQUANTGWTC2}{\ensuremath{35.0}}{GW170809DMFGWTC2PHENOM}{\ensuremath{-0.03^{+0.34}_{-0.28}}}{GW170809DCHIFGWTC2PHENOM}{\ensuremath{-0.22^{+0.50}_{-0.45}}}{GW170809GRQUANTGWTC2}{\ensuremath{26.6}}{S190814bvDMFGWTC2PHENOM}{\ensuremath{-0.29^{+0.46}_{-0.87}}}{S190814bvDCHIFGWTC2PHENOM}{\ensuremath{-0.87^{+0.26}_{-0.11}}}{S190814bvGRQUANTGWTC2}{\ensuremath{99.9}}{S190421arDMFGWTC2PHENOM}{\ensuremath{0.95^{+0.23}_{-0.89}}}{S190421arDCHIFGWTC2PHENOM}{\ensuremath{0.08^{+0.44}_{-0.50}}}{S190421arGRQUANTGWTC2}{\ensuremath{78.7}}{S190706aiDMFGWTC2PHENOM}{\ensuremath{0.45^{+0.28}_{-0.34}}}{S190706aiDCHIFGWTC2PHENOM}{\ensuremath{0.09^{+0.66}_{-0.51}}}{S190706aiGRQUANTGWTC2}{\ensuremath{96.5}}{S190910sDMFGWTC2PHENOM}{\ensuremath{0.30^{+0.59}_{-0.45}}}{S190910sDCHIFGWTC2PHENOM}{\ensuremath{0.02^{+0.51}_{-0.61}}}{S190910sGRQUANTGWTC2}{\ensuremath{29.3}}{S190519bjDMFGWTC2PHENOM}{\ensuremath{0.47^{+0.26}_{-0.45}}}{S190519bjDCHIFGWTC2PHENOM}{\ensuremath{0.08^{+0.44}_{-0.48}}}{S190519bjGRQUANTGWTC2}{\ensuremath{85.6}}}}
\newcommand{\EVENTSELECTION}[1]{\IfEqCase{#1}{{S190408anFCIMR}{\ensuremath{164}}{S190408anFCTIGER}{\ensuremath{68}}{S190408anOPTSNRPREIMR}{\ensuremath{13.6}}{S190408anOPTSNRPOSTIMR}{\ensuremath{6.4}}{S190408anOPTSNRPOSTTIGER}{\ensuremath{12.5}}{S190408anOPTSNRPRETIGER}{\ensuremath{8.3}}{S190408anOPTSNR}{\ensuremath{15.0}}{S190412mFCIMR}{\ensuremath{213}}{S190412mFCTIGER}{\ensuremath{83}}{S190412mOPTSNRPREIMR}{\ensuremath{18.2}}{S190412mOPTSNRPOSTIMR}{\ensuremath{5.9}}{S190412mOPTSNRPOSTTIGER}{\ensuremath{11.8}}{S190412mOPTSNRPRETIGER}{\ensuremath{15.1}}{S190412mOPTSNR}{\ensuremath{19.1}}{S190421arFCIMR}{\ensuremath{82}}{S190421arFCTIGER}{\ensuremath{36}}{S190421arOPTSNRPREIMR}{\ensuremath{8.1}}{S190421arOPTSNRPOSTIMR}{\ensuremath{6.6}}{S190421arOPTSNRPOSTTIGER}{\ensuremath{10.0}}{S190421arOPTSNRPRETIGER}{\ensuremath{2.9}}{S190421arOPTSNR}{\ensuremath{10.4}}{S190425zFCIMR}{\ensuremath{2725}}{S190425zFCTIGER}{\ensuremath{1001}}{S190425zOPTSNRPREIMR}{\ensuremath{0.0}}{S190425zOPTSNRPOSTIMR}{\ensuremath{0.0}}{S190425zOPTSNRPOSTTIGER}{\ensuremath{0.1}}{S190425zOPTSNRPRETIGER}{\ensuremath{12.6}}{S190425zOPTSNR}{\ensuremath{12.6}}{S190503bfFCIMR}{\ensuremath{99}}{S190503bfFCTIGER}{\ensuremath{39}}{S190503bfOPTSNRPREIMR}{\ensuremath{11.5}}{S190503bfOPTSNRPOSTIMR}{\ensuremath{7.5}}{S190503bfOPTSNRPOSTTIGER}{\ensuremath{13.0}}{S190503bfOPTSNRPRETIGER}{\ensuremath{4.3}}{S190503bfOPTSNR}{\ensuremath{13.7}}{S190512atFCIMR}{\ensuremath{199}}{S190512atFCTIGER}{\ensuremath{87}}{S190512atOPTSNRPREIMR}{\ensuremath{12.4}}{S190512atOPTSNRPOSTIMR}{\ensuremath{3.1}}{S190512atOPTSNRPOSTTIGER}{\ensuremath{7.4}}{S190512atOPTSNRPRETIGER}{\ensuremath{10.5}}{S190512atOPTSNR}{\ensuremath{12.8}}{S190513bmFCIMR}{\ensuremath{125}}{S190513bmFCTIGER}{\ensuremath{48}}{S190513bmOPTSNRPREIMR}{\ensuremath{11.2}}{S190513bmOPTSNRPOSTIMR}{\ensuremath{7.2}}{S190513bmOPTSNRPOSTTIGER}{\ensuremath{12.2}}{S190513bmOPTSNRPRETIGER}{\ensuremath{5.1}}{S190513bmOPTSNR}{\ensuremath{13.3}}{S190517hFCIMR}{\ensuremath{167}}{S190517hFCTIGER}{\ensuremath{41}}{S190517hOPTSNRPREIMR}{\ensuremath{9.7}}{S190517hOPTSNRPOSTIMR}{\ensuremath{5.4}}{S190517hOPTSNRPOSTTIGER}{\ensuremath{10.5}}{S190517hOPTSNRPRETIGER}{\ensuremath{3.4}}{S190517hOPTSNR}{\ensuremath{11.1}}{S190519bjFCIMR}{\ensuremath{78}}{S190519bjFCTIGER}{\ensuremath{23}}{S190519bjOPTSNRPREIMR}{\ensuremath{10.0}}{S190519bjOPTSNRPOSTIMR}{\ensuremath{11.2}}{S190519bjOPTSNRPOSTTIGER}{\ensuremath{15.0}}{S190519bjOPTSNRPRETIGER}{\ensuremath{0.0}}{S190519bjOPTSNR}{\ensuremath{15.0}}{S190521gFCIMR}{\ensuremath{37}}{S190521gFCTIGER}{\ensuremath{14}}{S190521gOPTSNRPREIMR}{\ensuremath{4.6}}{S190521gOPTSNRPOSTIMR}{\ensuremath{13.1}}{S190521gOPTSNRPOSTTIGER}{\ensuremath{13.9}}{S190521gOPTSNRPRETIGER}{\ensuremath{0.0}}{S190521gOPTSNR}{\ensuremath{13.9}}{S190521rFCIMR}{\ensuremath{105}}{S190521rFCTIGER}{\ensuremath{40}}{S190521rOPTSNRPREIMR}{\ensuremath{23.4}}{S190521rOPTSNRPOSTIMR}{\ensuremath{9.9}}{S190521rOPTSNRPOSTTIGER}{\ensuremath{23.5}}{S190521rOPTSNRPRETIGER}{\ensuremath{9.7}}{S190521rOPTSNR}{\ensuremath{25.4}}{S190602aqFCIMR}{\ensuremath{56}}{S190602aqFCTIGER}{\ensuremath{22}}{S190602aqOPTSNRPREIMR}{\ensuremath{5.6}}{S190602aqOPTSNRPOSTIMR}{\ensuremath{11.8}}{S190602aqOPTSNRPOSTTIGER}{\ensuremath{13.1}}{S190602aqOPTSNRPRETIGER}{\ensuremath{0.0}}{S190602aqOPTSNR}{\ensuremath{13.1}}{S190630agFCIMR}{\ensuremath{135}}{S190630agFCTIGER}{\ensuremath{50}}{S190630agOPTSNRPREIMR}{\ensuremath{14.0}}{S190630agOPTSNRPOSTIMR}{\ensuremath{8.2}}{S190630agOPTSNRPOSTTIGER}{\ensuremath{14.1}}{S190630agOPTSNRPRETIGER}{\ensuremath{8.1}}{S190630agOPTSNR}{\ensuremath{16.3}}{S190706aiFCIMR}{\ensuremath{67}}{S190706aiFCTIGER}{\ensuremath{19}}{S190706aiOPTSNRPREIMR}{\ensuremath{7.8}}{S190706aiOPTSNRPOSTIMR}{\ensuremath{10.1}}{S190706aiOPTSNRPOSTTIGER}{\ensuremath{12.7}}{S190706aiOPTSNRPRETIGER}{\ensuremath{0.0}}{S190706aiOPTSNR}{\ensuremath{12.7}}{S190707qFCIMR}{\ensuremath{391}}{S190707qFCTIGER}{\ensuremath{161}}{S190707qOPTSNRPREIMR}{\ensuremath{13.2}}{S190707qOPTSNRPOSTIMR}{\ensuremath{2.0}}{S190707qOPTSNRPOSTTIGER}{\ensuremath{5.5}}{S190707qOPTSNRPRETIGER}{\ensuremath{12.2}}{S190707qOPTSNR}{\ensuremath{13.4}}{S190708apFCIMR}{\ensuremath{261}}{S190708apFCTIGER}{\ensuremath{103}}{S190708apOPTSNRPREIMR}{\ensuremath{13.0}}{S190708apOPTSNRPOSTIMR}{\ensuremath{4.2}}{S190708apOPTSNRPOSTTIGER}{\ensuremath{8.0}}{S190708apOPTSNRPRETIGER}{\ensuremath{11.1}}{S190708apOPTSNR}{\ensuremath{13.7}}{S190720aFCIMR}{\ensuremath{405}}{S190720aFCTIGER}{\ensuremath{126}}{S190720aOPTSNRPREIMR}{\ensuremath{10.5}}{S190720aOPTSNRPOSTIMR}{\ensuremath{1.4}}{S190720aOPTSNRPOSTTIGER}{\ensuremath{5.2}}{S190720aOPTSNRPRETIGER}{\ensuremath{9.2}}{S190720aOPTSNR}{\ensuremath{10.5}}{S190727hFCIMR}{\ensuremath{96}}{S190727hFCTIGER}{\ensuremath{35}}{S190727hOPTSNRPREIMR}{\ensuremath{10.0}}{S190727hOPTSNRPOSTIMR}{\ensuremath{7.2}}{S190727hOPTSNRPOSTTIGER}{\ensuremath{12.2}}{S190727hOPTSNRPRETIGER}{\ensuremath{2.0}}{S190727hOPTSNR}{\ensuremath{12.3}}{S190728qFCIMR}{\ensuremath{426}}{S190728qFCTIGER}{\ensuremath{157}}{S190728qOPTSNRPREIMR}{\ensuremath{12.5}}{S190728qOPTSNRPOSTIMR}{\ensuremath{1.4}}{S190728qOPTSNRPOSTTIGER}{\ensuremath{5.3}}{S190728qOPTSNRPRETIGER}{\ensuremath{11.4}}{S190728qOPTSNR}{\ensuremath{12.6}}{S190814bvFCIMR}{\ensuremath{207}}{S190814bvFCTIGER}{\ensuremath{137}}{S190814bvOPTSNRPREIMR}{\ensuremath{23.9}}{S190814bvOPTSNRPOSTIMR}{\ensuremath{6.9}}{S190814bvOPTSNRPOSTTIGER}{\ensuremath{10.9}}{S190814bvOPTSNRPRETIGER}{\ensuremath{22.3}}{S190814bvOPTSNR}{\ensuremath{24.8}}{S190828jFCIMR}{\ensuremath{132}}{S190828jFCTIGER}{\ensuremath{45}}{S190828jOPTSNRPREIMR}{\ensuremath{13.8}}{S190828jOPTSNRPOSTIMR}{\ensuremath{8.5}}{S190828jOPTSNRPOSTTIGER}{\ensuremath{15.1}}{S190828jOPTSNRPRETIGER}{\ensuremath{6.0}}{S190828jOPTSNR}{\ensuremath{16.2}}{S190828lFCIMR}{\ensuremath{201}}{S190828lFCTIGER}{\ensuremath{80}}{S190828lOPTSNRPREIMR}{\ensuremath{9.4}}{S190828lOPTSNRPOSTIMR}{\ensuremath{3.0}}{S190828lOPTSNRPOSTTIGER}{\ensuremath{7.6}}{S190828lOPTSNRPRETIGER}{\ensuremath{6.3}}{S190828lOPTSNR}{\ensuremath{9.9}}{S190910sFCIMR}{\ensuremath{92}}{S190910sFCTIGER}{\ensuremath{35}}{S190910sOPTSNRPREIMR}{\ensuremath{9.6}}{S190910sOPTSNRPOSTIMR}{\ensuremath{10.7}}{S190910sOPTSNRPOSTTIGER}{\ensuremath{14.0}}{S190910sOPTSNRPRETIGER}{\ensuremath{3.3}}{S190910sOPTSNR}{\ensuremath{14.4}}{S190915akFCIMR}{\ensuremath{122}}{S190915akFCTIGER}{\ensuremath{46}}{S190915akOPTSNRPREIMR}{\ensuremath{12.8}}{S190915akOPTSNRPOSTIMR}{\ensuremath{3.1}}{S190915akOPTSNRPOSTTIGER}{\ensuremath{12.6}}{S190915akOPTSNRPRETIGER}{\ensuremath{3.7}}{S190915akOPTSNR}{\ensuremath{13.1}}{S190924hFCIMR}{\ensuremath{580}}{S190924hFCTIGER}{\ensuremath{239}}{S190924hOPTSNRPREIMR}{\ensuremath{12.2}}{S190924hOPTSNRPOSTIMR}{\ensuremath{1.3}}{S190924hOPTSNRPOSTTIGER}{\ensuremath{3.4}}{S190924hOPTSNRPRETIGER}{\ensuremath{11.8}}{S190924hOPTSNR}{\ensuremath{12.2}}}}
\newcommand{\QGR}{\ensuremath{\mathcal{Q}_{\rm GR}}}
\newcommand{\dMf}{\ensuremath{\Delta M_{\rm f} / \bar{M}_{\rm f}}}
\newcommand{\dchif}{\ensuremath{\Delta \chi_{\rm f} / \bar{\chi}_{\rm f}}}

GR predicts that the final state of the coalescence of two BHs will be a single perturbed Kerr BH \cite{Campanelli:2008dv,Owen:2009sb,Owen:2010vw,Bhagwat:2017tkm}. Assuming that GR is valid, the mass and spin of the remnant BH inferred from the low-frequency portion of the signal should be consistent with those measured from the high-frequency part \cite{Hughes:2004vw,Ghosh:2015jra,Ghosh:2017gfp}, where the low- and high-frequency regimes roughly correspond to the inspiral and postinspiral, respectively, when considering the dominant mode \cite{Ghosh:2017gfp}. This provides a consistency test for GR, related to the remnant-focused studies we present in Sec.~\ref{sec:rem} and the postinspiral coefficients in Sec.~\ref{sec:par}.

\begin{figure}
	\begin{center}
	\includegraphics[width=3.5in]{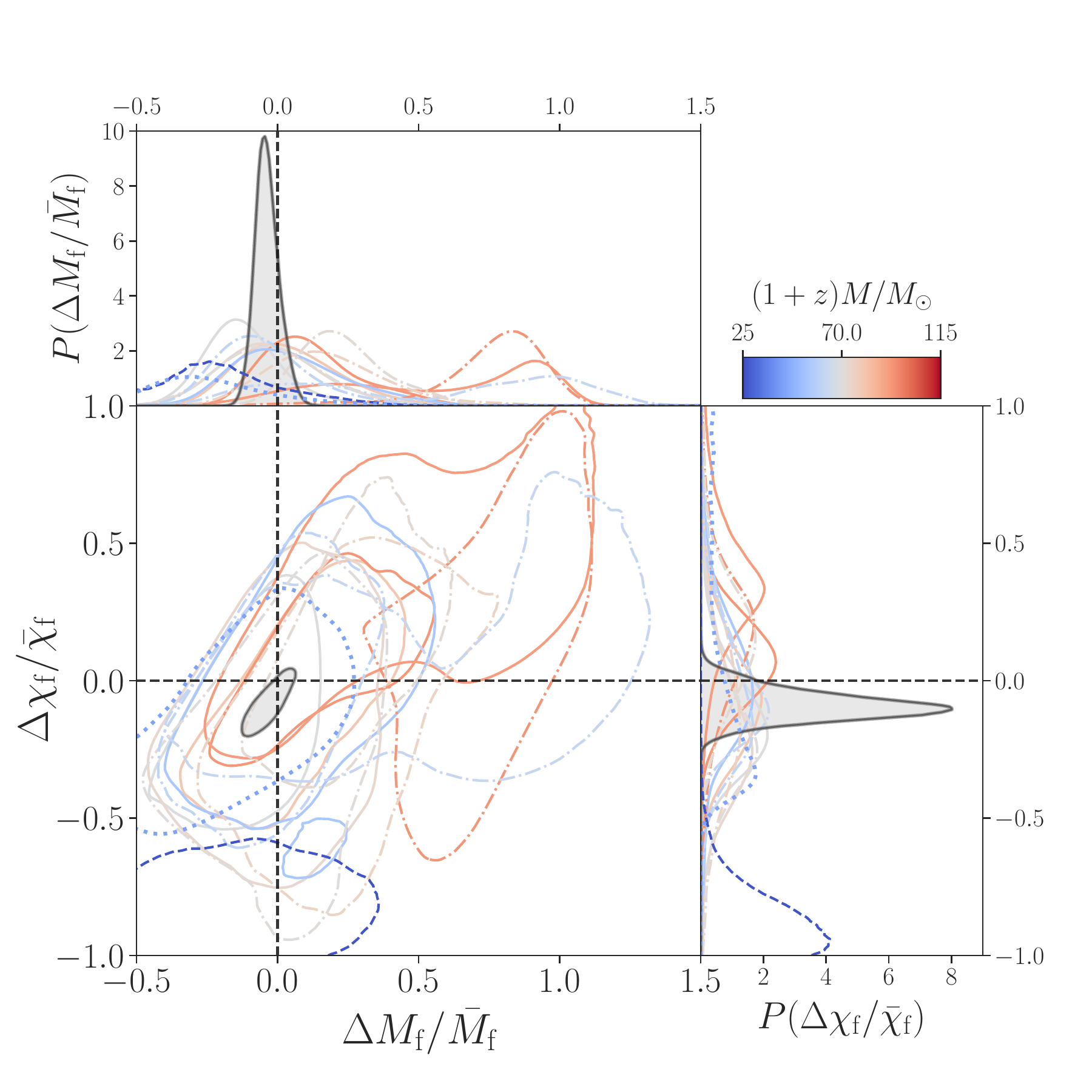}
	\end{center}
	\caption{Results of the IMR consistency test for the selected BBH events with median $(1+z) M < 100 M_{\odot}$ (see Table~\ref{tab:imr_test_params}). The main panel shows the 90\% credible regions of the posteriors for $(\dMf, \dchif)$ assuming a uniform prior, with the cross marking the expected value for GR. The side panels show the marginalized posterior for $\dMf$ and $\dchif$. The gray distribution correspond to the product of all the individual posteriors. O3a (pre-O3a) events are plotted with solid (dot--dashed) traces. Color encodes the redshifted total mass in solar masses, with a turnover between blue and red around the median of the $(1+z) M/ M_{\odot}$ distribution for the plotted events.
  The results for \protect\NAME{GW190412A}{} and \protect\NAME{GW190814A}{} are identified by dotted and dashed contours, respectively. The two events with contours that do not enclose the origin are GW170823 (dot--dashed) and \protect\NAME{GW190814A}{} (dashed). \protect\NAME{GW190408A}{} has a multimodal posterior that results in the small contour (blue) away from zero.   
  }
	\label{fig:imr_test_posteriors}
\end{figure}

We take the cutoff frequency $f_\text{c}^{\rm IMR}$ between the inspiral and postinspiral regimes to be the $m = 2$ mode GW frequency of the innermost stable circular orbit of a Kerr BH, with mass $M_{\rm f}$ and dimensionless spin magnitude $\chi_{\rm f}$ estimated from the full BBH signal assuming GR.
The final mass and spin are calculated by averaging NR-calibrated final-state fits \cite{Healy:2016lce,Hofmann:2016yih,Jimenez-Forteza:2016oae}, where the aligned-spin final spin fits are augmented by a contribution from the in-plane spins \cite{Bohe:PPv2,spinfit-T1600168}.
We compute $f_\text{c}^{\rm IMR}$ from augmented NR-calibrated fits applied to the posterior median values for the masses and spins of the binary components. We then independently estimate the binary's parameters from the low- (high-) frequency portion of the signal, restricting the Fourier-domain likelihood calculation to frequencies below (above) the cutoff frequency $f_\text{c}^{\rm IMR}$. The two independent estimates of the source parameters are used to infer the posterior distributions of $M_{\rm f}$ and  $\chi_{\rm f}$ using the augmented NR-calibrated final-state fits. 
For the signal to be consistent with GR, the two estimates must be consistent with each other.

For this test, we require the inspiral and postinspiral portions of the signal to be informative. As a proxy for the amount of information that can be extracted from each part of the signal, we calculate the SNR of the inspiral and postinspiral part of the signal using the preferred waveform model for each event (Table \ref{tab:res:wfs}), evaluated at the maximum a posteriori parameters for the complete IMR posterior distributions \cite{GWTC2}. As in \cite{LIGOScientific:2019fpa}, we only apply the IMR consistency test to events that have $\textrm{SNR} > 6$ in both regions. When studying the set of measurements as a whole (cf.~Sec~\ref{sec:inference:populations}), we impose an additional criterion on the median redshifted total mass such that $(1+z) M < 100\, M_{\odot}$. This additional cut further ensures that the binary contains sufficient information in the inspiral regime because the test would be strongly biased for heavy BBHs.
A criterion based on mass was not applied in \cite{LIGOScientific:2019fpa} because most GWTC-1 events automatically satisfied it.
The cutoff frequency and SNRs for all events used in this analysis are detailed in Table~\ref{tab:imr_test_params}.\footnote{The frequency $f_{\rm c}^{\rm IMR}$ was determined using preliminary parameter inference results and the values in Table~\ref{tab:imr_test_params} may slightly differ to those obtained using the posterior samples in GWTC-2. However, the test is robust against small changes to the cutoff frequency \cite{Ghosh:2017gfp}.}

\begin{table}
\caption{Results from the IMR consistency test (Sec.~\ref{sec:imr}). $f_\text{c}^{\rm IMR}$ denotes the cutoff frequency between the inspiral and postinspiral regimes; $\rho_\mathrm{IMR}$, $\rho_\mathrm{insp}$, and $\rho_\mathrm{postinsp}$ are the SNR in the full signal, the inspiral part, and the postinspiral part respectively; and the GR quantile $\QGR$ denotes the fraction of the likelihood enclosed by the isoprobability contour that passes through the GR value, with smaller values indicating better consistency with GR. For lower SNRs, the likelihood is typically broader and $\QGR$ is generally higher. An asterisk denotes events with median $(1+z) M > 100 M_{\odot}$, for which we expect strong systematics. We highlight \protect\NAME{GW190412A}{} with a dagger as we show results for comparison to \cite{GW190412}, but the event is not used in the joint likelihood as the postinspiral SNR is below the threshold for inclusion. The difference in the results for GWTC-1 events compared to \cite{LIGOScientific:2019fpa} is due to the change in priors. 
}
\label{tab:imr_test_params}
\centering
\begin{tabular}{l@{~} c@{\quad}r@{\quad}r@{\quad} r@{\quad} r}
\toprule
Event 		& $f_\text{c}^{\rm IMR}$ [Hz]  & $\rho_\mathrm{IMR}$ & $\rho_\mathrm{insp}$ & $\rho_\mathrm{postinsp}$ & $\QGR$ [\%] \\
\midrule
GW150914	& 132	  & 25.3	& 19.4	& 16.1	& \ImrEVENTSTATS{GW150914GRQUANTGWTC2}\phantom{${}^*$} \\	
GW170104	& 143	  & 13.7	& 10.9	& 8.5	& \ImrEVENTSTATS{GW170104GRQUANTGWTC2}\phantom{${}^*$} \\	
GW170809	& 136	  & 12.7	& 10.6	& 7.1	& \ImrEVENTSTATS{GW170809GRQUANTGWTC2}\phantom{${}^*$} \\
GW170814	& 161	  & 16.8	& 15.3	& 7.2	& \ImrEVENTSTATS{GW170814GRQUANTGWTC2}\phantom{${}^*$} \\
GW170818	& 128	  & 12.0	& 9.3	& 7.2	& \ImrEVENTSTATS{GW170818GRQUANTGWTC2}\phantom{${}^*$} \\
GW170823	& 102	  & 11.9	& 7.9	& 8.5	& \ImrEVENTSTATS{GW170823GRQUANTGWTC2}\phantom{${}^*$} \\ 
\midrule
\NAME{GW190408A}	& \EVENTSELECTION{S190408anFCIMR}	  &  \EVENTSELECTION{S190408anOPTSNR} 	&  \EVENTSELECTION{S190408anOPTSNRPREIMR} & \EVENTSELECTION{S190408anOPTSNRPOSTIMR} 	& \ImrEVENTSTATS{S190408anGRQUANTGWTC2}\phantom{${}^*$} \\
\NAME{GW190412A}	&  \EVENTSELECTION{S190412mFCIMR}	  &  \EVENTSELECTION{S190412mOPTSNR} 	&  \EVENTSELECTION{S190412mOPTSNRPREIMR} & \EVENTSELECTION{S190412mOPTSNRPOSTIMR} 	& \ImrEVENTSTATS{S190412mGRQUANTGWTC2}{${}^{\dagger}$} \\
\NAME{GW190421A}	& \EVENTSELECTION{S190421arFCIMR}	  &  \EVENTSELECTION{S190421arOPTSNR} 	&  \EVENTSELECTION{S190421arOPTSNRPREIMR} & \EVENTSELECTION{S190421arOPTSNRPOSTIMR} 	& \ImrEVENTSTATS{S190421arGRQUANTGWTC2}{${}^*$} \\
\NAME{GW190503A}	& \EVENTSELECTION{S190503bfFCIMR}	  &  \EVENTSELECTION{S190503bfOPTSNR} 	&  \EVENTSELECTION{S190503bfOPTSNRPREIMR} & \EVENTSELECTION{S190503bfOPTSNRPOSTIMR} 	& \ImrEVENTSTATS{S190503bfGRQUANTGWTC2}\phantom{${}^*$} \\
\NAME{GW190513A}	&  $\EVENTSELECTION{S190513bmFCIMR}$ %
&  \EVENTSELECTION{S190513bmOPTSNR} 	&  \EVENTSELECTION{S190513bmOPTSNRPREIMR} & \EVENTSELECTION{S190513bmOPTSNRPOSTIMR} 	& \ImrEVENTSTATS{S190513bmGRQUANTGWTC2}\phantom{${}^*$} \\
\NAME{GW190519A}	&  \EVENTSELECTION{S190519bjFCIMR}	  &  \EVENTSELECTION{S190519bjOPTSNR} 	&  \EVENTSELECTION{S190519bjOPTSNRPREIMR} & \EVENTSELECTION{S190519bjOPTSNRPOSTIMR} 	& \ImrEVENTSTATS{S190519bjGRQUANTGWTC2}{${}^*$} \\
\NAME{GW190521B}	&  \EVENTSELECTION{S190521rFCIMR}	  &  \EVENTSELECTION{S190521rOPTSNR} 	&  \EVENTSELECTION{S190521rOPTSNRPREIMR} & \EVENTSELECTION{S190521rOPTSNRPOSTIMR} 	& \ImrEVENTSTATS{S190521rGRQUANTGWTC2}\phantom{${}^*$} \\
\NAME{GW190630A}	&   \EVENTSELECTION{S190630agFCIMR}	  &  \EVENTSELECTION{S190630agOPTSNR} 	&  \EVENTSELECTION{S190630agOPTSNRPREIMR} & \EVENTSELECTION{S190630agOPTSNRPOSTIMR} 	& \ImrEVENTSTATS{S190630agGRQUANTGWTC2}\phantom{${}^*$} \\
\NAME{GW190706A}	&   \EVENTSELECTION{S190706aiFCIMR}	  &  \EVENTSELECTION{S190706aiOPTSNR} 	&  \EVENTSELECTION{S190706aiOPTSNRPREIMR} & \EVENTSELECTION{S190706aiOPTSNRPOSTIMR} 	& \ImrEVENTSTATS{S190706aiGRQUANTGWTC2}{${}^*$} \\
\NAME{GW190727A} 	& \EVENTSELECTION{S190727hFCIMR}	  &  \EVENTSELECTION{S190727hOPTSNR} 	&  \EVENTSELECTION{S190727hOPTSNRPREIMR} & \EVENTSELECTION{S190727hOPTSNRPOSTIMR} 	& \ImrEVENTSTATS{S190727hGRQUANTGWTC2}{${}^*$} \\
\NAME{GW190814A}  &  \EVENTSELECTION{S190814bvFCIMR}	  &  \EVENTSELECTION{S190814bvOPTSNR} 	&  \EVENTSELECTION{S190814bvOPTSNRPREIMR} & \EVENTSELECTION{S190814bvOPTSNRPOSTIMR} 	& \ImrEVENTSTATS{S190814bvGRQUANTGWTC2}\phantom{${}^*$} \\
\NAME{GW190828A} 	&  $\EVENTSELECTION{S190828jFCIMR}$ %
&  \EVENTSELECTION{S190828jOPTSNR} 	&  \EVENTSELECTION{S190828jOPTSNRPREIMR} & \EVENTSELECTION{S190828jOPTSNRPOSTIMR} 	& \ImrEVENTSTATS{S190828jGRQUANTGWTC2}\phantom{${}^*$} \\
\NAME{GW190910A} 	&  \EVENTSELECTION{S190910sFCIMR}	  &  \EVENTSELECTION{S190910sOPTSNR} 	&  \EVENTSELECTION{S190910sOPTSNRPREIMR} & \EVENTSELECTION{S190910sOPTSNRPOSTIMR} 	& \ImrEVENTSTATS{S190910sGRQUANTGWTC2}{${}^*$} \\
\bottomrule
\end{tabular}
\end{table}
In order to constrain possible departures from GR, we introduce two dimensionless parameters that quantify the fractional difference between the two estimates
\begin{align}
\frac{\Delta M_{\rm f}}{ \bar{M}_{\rm f}} &= 2 \frac{ M_{\rm f}^{\rm{insp}} - M_{\rm f}^{\rm{postinsp}} }{ M_{\rm f}^{\rm{insp}} + M_{\rm f}^{\rm{postinsp}} },  \\
\frac{\Delta \chi_{\rm f}}{ \bar{\chi}_{\rm f}}&= 2 \frac{ \chi_{\rm f}^{\rm{insp}} - \chi_{\rm f}^{\rm{postinsp}} }{ \chi_{\rm f}^{\rm{insp}} + \chi_{\rm f}^{\rm{postinsp}} } , 
\end{align}
\newline
where the superscripts denote the estimate of the mass or the spin from the inspiral and postinspiral portions of the signal \cite{Ghosh:2015jra}.
As in \cite{LIGOScientific:2019fpa}, we perform parameter estimation using uniform priors for the component masses and spin magnitudes and an isotropic prior on the spin orientations; this choice induces a highly non-uniform effective prior in $\dMf$ and $\dchif$.
In order to alleviate this, and in contrast with \cite{LIGOScientific:2019fpa}, we re-weight the posteriors to work with a uniform prior for the deviation parameters.
This eliminates confounding factors and has the advantage of more clearly conveying the information gained from the data.
For example, binary configurations with comparable mass ratios and $\chi_{\rm eff} \sim 0$ will lead to a remnant spin ${\sim 0.7}$ \cite{Healy:2016lce,Hofmann:2016yih,Jimenez-Forteza:2016oae}, which means that the $\chi_{\rm f}$ prior is concentrated around this value and that, consequently, the $\Delta \chi_{\rm f}$ is concentrated around 0; this leads to artificially narrow $\Delta \chi_{\rm f}$ posteriors that should not be interpreted as a strong constraint from the data on deviations from GR.

We summarize our results in Fig.~\ref{fig:imr_test_posteriors}, where we represent the two-dimensional posteriors for all GWTC-2 events analyzed by means of their 90\% credible level.
The contours are colored as a function of the median redshifted total binary mass $(1+z)M$, as inferred from the full waveform assuming GR, and we only include events with $(1+z)M < 100\, M_\odot$.
Events preceding O3a are identified with a dot--dashed trace and were already analyzed in \cite{LIGOScientific:2019fpa}.
However, distributions in Fig.~\ref{fig:imr_test_posteriors} here are generally broader than Fig.~2 of that paper because our results represent posteriors using a uniform prior. Although \NAME{GW190412A}{} does not meet the SNR threshold for this test, we highlight the posteriors for this event in Fig.~\ref{fig:imr_test_posteriors} for comparison to previously published results \cite{GW190412}.

We find that the \NAME{GW190412A}{} and \NAME{GW190814A}{} postinspiral distance posteriors are cut off by the upper prior bounds on the distance, $3~\mathrm{Gpc}$ and $2~\mathrm{Gpc}$, respectively. Due to the low SNR in the postinspiral, the distance posterior is cut off by the prior even when increasing the upper bound on the volumetric distance prior $p(D_\text{L}) \propto D_\text{L}^2$. The IMR consistency results for these events are therefore unavoidably dependent on the choice of priors. To mitigate such issues, we have chosen upper bounds that lead to a small probability density near the cutoff. For future applications of the test we will consider ways to impose \emph{a priori} selection cuts to exclude such cases from consideration. 

The fraction of the posterior enclosed by the isoprobability contours that pass through the GR value, i.e., the two-dimensional GR quantile $\QGR$, for each event is given in Table~\ref{tab:imr_test_params}, where smaller values indicate better consistency with GR. For low (high) SNRs, the posteriors will be broader (narrower) and $\QGR$ will be higher (lower) if GR is the correct hypothesis. The binary with the smallest $\QGR$ is \protect\NAME{GW190521B}{}, which has a small but non-zero quantile that is rounded to zero in Table~\ref{tab:imr_test_params}.
For binaries with masses $(1+z) M > 100 M_{\odot}$ we typically observe $\QGR > 50\%$, which can be explained by the known systematics mentioned above. 
See Appendix~\ref{app:imr} for a more detailed exposition of mass-related systematics.
Of the binaries below the mass threshold, \protect\NAME{GW190814A}{} has the highest quantile, $\QGR = \ImrEVENTSTATS{S190814bvGRQUANTGWTC2} \%$, but has a relatively low SNR in the postinspiral regime and a relatively low redshifted mass; the other notable outlier is GW170823,  $\QGR = \ImrEVENTSTATS{GW170823GRQUANTGWTC2} \%$, which has the lowest SNR and a relatively high redshifted mass, $(1+z) M \approx 93 \, M_\odot$.
For \protect\NAME{GW190814A}{}, the likelihood for the final spin fractional deviation shows a notable departure from the GR value.%
\footnote{The \protect\NAME{GW190814A}{} posterior was truncated at $\Delta\chi_f/\bar{\chi}_f = -1$ in this analysis, but we have confirmed this has no effect on $\QGR$.}
However, \protect\NAME{GW190814A}{} was a higher mass ratio event with very small spins, resulting in an inferred final spin of $\chi_{\rm f} \sim 0.28$ \cite{GW190814}. As a consequence of the low SNR, the postinspiral regime is uninformative and the posterior is dominated by the prior which peaks at $\chi_{\rm f} \sim 0.7$. In contrast, the masses and spins are very accurately measured in the inspiral regime and a final spin of $\chi_{\rm f} \sim 0.28$ is recovered. The apparent departure from GR can be explained by the mismatch in the information recovered between the two regimes.

\begin{figure}[tb] 
	\begin{center}
	\includegraphics[width=\columnwidth]{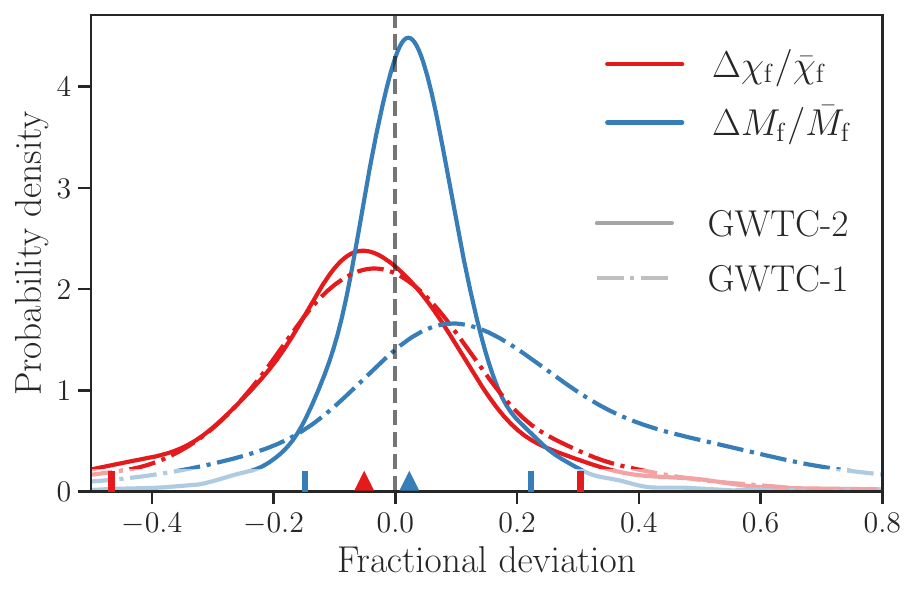}
	\end{center}
	\caption{Distributions for the remnant mass (blue) and spin (red) fractional deviations, as obtained by hierarchically combining the results in Fig.~\ref{fig:imr_test_posteriors} (solid trace). For comparison, we also show the result obtained using only GWTC-1 events (dot dashed trace).
  The probability densities summarize our expectation for the fraction of observed events with a given value of $\dMf$ and $\dchif$, as defined in Eq.~\eqref{eq:inf:hier_dist}. GR predicts no deviation on either parameter (vertical dashed line). Triangles mark the GWTC-2 medians, and vertical bars the symmetric 90\%-credible intervals.}
	\label{fig:imr_hier}
\end{figure}

We may interpret results from our set of observations collectively through hierarchical models for the mass and spin deviations, as described in Sec.~\ref{sec:inference:populations}. Here we treat $\dMf$ and $\dchif$ as independent parameters; future implementations may consider them jointly.
With 90\% credibility, we constrain the population hyperparameters $(\mu,\, \sigma)$ to be $(\ImrMfHierMu{LOWM},\, < \ImrMfHierSigma{LOWM})$ and $(\ImrChifHierMu{LOWM},\, < \ImrChifHierSigma{LOWM})$ for $\dMf$ and $\dchif$ respectively, consistent with GR  ($\mu=\sigma=0$) for both parameters (posteriors provided in Appendix \ref{app:imr}).
In Fig.~\ref{fig:imr_hier}, we represent the result through the population-marginalized expectation for $\dMf$ (blue) and $\dchif$ (red), as defined in Eq.~\eqref{eq:inf:hier_dist}.
This measurement constrains $\dMf = \ImrMfHierPop{LOWM}$ and $\dchif = \ImrChifHierPop{LOWM}$, quite consistent with the expectation from GR.

If we assume that the fractional deviations take the same value for all events, then we obtain the less-conservative combined posterior shown in gray in Fig.~\ref{fig:imr_test_posteriors}.
We find $\dMf = \ImrGWTCTWO{DMFGWTC2PHENOM}$ and $\dchif = \ImrGWTCTWO{DCHIFGWTC2PHENOM}$, also consistent with the GR values.

Had we included the high-mass events discussed above in the analysis, for which IMR tests are known to exhibit systematic offsets, the hierarchical method would have resulted in modest tension with GR, as discussed more fully in Appendix~\ref{app:imr}.  The hierarchical method with $\sigma = 0$ (assuming all events have the same deviation parameters) does not find any inconsistency when high-mass events are included, so we conclude that in this case the full hierarchical method is more sensitive to these (systematics-induced) deviations from GR.

This analysis used \IMRP{} or \IMRPHM{} waveforms for the same events for which they were used for the residuals analysis, given in Table~\ref{tab:res:wfs}.
In order to gauge systematic errors arising from imperfect waveform modeling, we also produce results using the non-precessing \SEOBROM{} model, but these results exclude \NAME{GW190412A}{} and \NAME{GW190814A}{} due to the relative importance of HMs. Despite the differences between the two waveform approximants, the posteriors are in broad agreement and we find no qualitative difference in the results (see Appendix~\ref{app:imr}). This is in agreement with the expectation that systematic errors will be subdominant to statistical errors for the typical SNRs reported in GWTC-2 \cite{Ghosh:2017gfp}.

\section{Tests of gravitational wave generation}
\label{sec:gen}

    \subsection{Generic modifications}
    \label{sec:par}
    \begin{table}
\caption{Parametrized test event selection for all binaries meeting the FAR $< 10^{-3}~\mathrm{yr}^{-1}$ threshold. Here $f_{\text{c}}^{\rm PAR}$ denotes the cutoff frequency used to demarcate the division between the inspiral, and postinspiral regimes; $\rho_{\rm IMR}$, $\rho_{\rm insp}$, and $\rho_{\rm postinsp}$ are the \textit{optimal} SNRs of the full signal, the inspiral, and postinspiral regions respectively. The last two columns denote if the event is included in parametrized tests on the inspiral (PI) and postinspiral (PPI) respectively.  GW190814 is excluded due to the impact of HMs, see Appendix~\ref{app:par}. 
}
\label{tab:par_events}
\centering
\begin{tabular}{lc@{\quad}rrr@{\quad}rr}
\toprule
\rm{Event} &  $f_{\text c}^{\rm PAR} \; [ \rm{Hz} ]$ & $\rho_{\rm IMR}$ &  $\rho_{\rm insp}$ &   $\rho_{\rm postinsp}$ & PI & PPI  \\
\midrule
GW150914 &    50 &    24.7  &           9.6 &            22.8& \cmark & \cmark \\
GW151226 &    153 &    12.3  &           11.1 &            5.3 & \cmark & $-$ \\
GW170104 &    60 &    13.4  &           7.9 &            11.3 & \cmark & \cmark \\
GW170608 &    179 &    15.8  &           14.8 &            6.3 & \cmark & \cmark \\
GW170809 &    54 &    12.0  &           5.8 &            10.9 & $-$ & \cmark \\
GW170814 &    58 &    16.3  &           9.1 &            13.6 & \cmark & \cmark \\
GW170818 &    48 &    10.8  &           4.5 &            10.1 & $-$ & \cmark \\
GW170823 &    40 &    11.5  &           4.2 &            11.1 & $-$ & \cmark \\
\hline
\NAME{GW190408A} &    \EVENTSELECTION{S190408anFCTIGER} &    \EVENTSELECTION{S190408anOPTSNR} &               \EVENTSELECTION{S190408anOPTSNRPRETIGER} &               \EVENTSELECTION{S190408anOPTSNRPOSTTIGER} & \cmark & \cmark \\
\NAME{GW190412A} &     \EVENTSELECTION{S190412mFCTIGER} &    \EVENTSELECTION{S190412mOPTSNR} &               \EVENTSELECTION{S190412mOPTSNRPRETIGER} &               \EVENTSELECTION{S190412mOPTSNRPOSTTIGER} & \cmark & \cmark \\
\NAME{GW190421A} &     \EVENTSELECTION{S190421arFCTIGER} &    \EVENTSELECTION{S190421arOPTSNR} &               \EVENTSELECTION{S190421arOPTSNRPRETIGER} &               \EVENTSELECTION{S190421arOPTSNRPOSTTIGER} & $-$ & \cmark \\
\NAME{GW190503A} &     \EVENTSELECTION{S190503bfFCTIGER} &    \EVENTSELECTION{S190503bfOPTSNR} &               \EVENTSELECTION{S190503bfOPTSNRPRETIGER} &               \EVENTSELECTION{S190503bfOPTSNRPOSTTIGER} & $-$ & \cmark \\
\NAME{GW190512A} &     \EVENTSELECTION{S190512atFCTIGER} &    \EVENTSELECTION{S190512atOPTSNR} &               \EVENTSELECTION{S190512atOPTSNRPRETIGER} &               \EVENTSELECTION{S190512atOPTSNRPOSTTIGER} & \cmark & \cmark \\
\NAME{GW190513A} &     \EVENTSELECTION{S190513bmFCTIGER} &    \EVENTSELECTION{S190513bmOPTSNR} &               \EVENTSELECTION{S190513bmOPTSNRPRETIGER} &               \EVENTSELECTION{S190513bmOPTSNRPOSTTIGER} & $-$ & \cmark \\
\NAME{GW190517A} &     \EVENTSELECTION{S190517hFCTIGER} &    \EVENTSELECTION{S190517hOPTSNR} &               \EVENTSELECTION{S190517hOPTSNRPRETIGER} &               \EVENTSELECTION{S190517hOPTSNRPOSTTIGER} & $-$ & \cmark \\
\NAME{GW190519A} &     \EVENTSELECTION{S190519bjFCTIGER} &    \EVENTSELECTION{S190519bjOPTSNR} &               \EVENTSELECTION{S190519bjOPTSNRPRETIGER} &               \EVENTSELECTION{S190519bjOPTSNRPOSTTIGER} & $-$ & \cmark \\
\NAME{GW190521A} &     \EVENTSELECTION{S190521gFCTIGER} &    \EVENTSELECTION{S190521gOPTSNR} &               \EVENTSELECTION{S190521gOPTSNRPRETIGER} &               \EVENTSELECTION{S190521gOPTSNRPOSTTIGER} & $-$ & \cmark \\
\NAME{GW190521B} &     \EVENTSELECTION{S190521rFCTIGER} &    \EVENTSELECTION{S190521rOPTSNR} &               \EVENTSELECTION{S190521rOPTSNRPRETIGER} &               \EVENTSELECTION{S190521rOPTSNRPOSTTIGER} & \cmark & \cmark \\
\NAME{GW190602A} &    \EVENTSELECTION{S190602aqFCTIGER} &    \EVENTSELECTION{S190602aqOPTSNR} &               \EVENTSELECTION{S190602aqOPTSNRPRETIGER} &               \EVENTSELECTION{S190602aqOPTSNRPOSTTIGER} & $-$ & \cmark \\
\NAME{GW190630A} &    \EVENTSELECTION{S190630agFCTIGER} &    \EVENTSELECTION{S190630agOPTSNR} &               \EVENTSELECTION{S190630agOPTSNRPRETIGER} &               \EVENTSELECTION{S190630agOPTSNRPOSTTIGER} & \cmark & \cmark \\
\NAME{GW190706A} &     \EVENTSELECTION{S190706aiFCTIGER} &    \EVENTSELECTION{S190706aiOPTSNR} &               \EVENTSELECTION{S190706aiOPTSNRPRETIGER} &               \EVENTSELECTION{S190706aiOPTSNRPOSTTIGER} & $-$ & \cmark \\
\NAME{GW190707A} &    \EVENTSELECTION{S190707qFCTIGER} &    \EVENTSELECTION{S190707qOPTSNR} &               \EVENTSELECTION{S190707qOPTSNRPRETIGER} &               \EVENTSELECTION{S190707qOPTSNRPOSTTIGER} & \cmark & $-$ \\
\NAME{GW190708A} &    \EVENTSELECTION{S190708apFCTIGER} &    \EVENTSELECTION{S190708apOPTSNR} &               \EVENTSELECTION{S190708apOPTSNRPRETIGER} &               \EVENTSELECTION{S190708apOPTSNRPOSTTIGER} & \cmark & \cmark \\
\NAME{GW190720A} &    \EVENTSELECTION{S190720aFCTIGER} &    \EVENTSELECTION{S190720aOPTSNR} &               \EVENTSELECTION{S190720aOPTSNRPRETIGER} &               \EVENTSELECTION{S190720aOPTSNRPOSTTIGER} & \cmark & $-$ \\
\NAME{GW190727A} &     \EVENTSELECTION{S190727hFCTIGER} &    \EVENTSELECTION{S190727hOPTSNR} &               \EVENTSELECTION{S190727hOPTSNRPRETIGER} &               \EVENTSELECTION{S190727hOPTSNRPOSTTIGER} & $-$ & \cmark \\
\NAME{GW190728A} &    \EVENTSELECTION{S190728qFCTIGER} &    \EVENTSELECTION{S190728qOPTSNR} &               \EVENTSELECTION{S190728qOPTSNRPRETIGER} &               \EVENTSELECTION{S190728qOPTSNRPOSTTIGER} & \cmark & $-$ \\
\NAME{GW190814A} &    \EVENTSELECTION{S190814bvFCTIGER} &    \EVENTSELECTION{S190814bvOPTSNR} &               \EVENTSELECTION{S190814bvOPTSNRPRETIGER} &               \EVENTSELECTION{S190814bvOPTSNRPOSTTIGER} & \cmark & \cmark \\
\NAME{GW190828A} &     \EVENTSELECTION{S190828jFCTIGER} &    \EVENTSELECTION{S190828jOPTSNR} &               \EVENTSELECTION{S190828jOPTSNRPRETIGER} &               \EVENTSELECTION{S190828jOPTSNRPOSTTIGER} & \cmark & \cmark \\
\NAME{GW190828B} &     \EVENTSELECTION{S190828lFCTIGER} &    \EVENTSELECTION{S190828lOPTSNR} &               \EVENTSELECTION{S190828lOPTSNRPRETIGER} &               \EVENTSELECTION{S190828lOPTSNRPOSTTIGER} & \cmark & \cmark \\
\NAME{GW190910A} &    \EVENTSELECTION{S190910sFCTIGER} &    \EVENTSELECTION{S190910sOPTSNR} &               \EVENTSELECTION{S190910sOPTSNRPRETIGER} &               \EVENTSELECTION{S190910sOPTSNRPOSTTIGER} & $-$ & \cmark \\
\NAME{GW190915A} &     \EVENTSELECTION{S190915akFCTIGER} &    \EVENTSELECTION{S190915akOPTSNR} &               \EVENTSELECTION{S190915akOPTSNRPRETIGER} &               \EVENTSELECTION{S190915akOPTSNRPOSTTIGER} & $-$ & \cmark \\
\NAME{GW190924A} &    \EVENTSELECTION{S190924hFCTIGER} &    \EVENTSELECTION{S190924hOPTSNR} &               \EVENTSELECTION{S190924hOPTSNRPRETIGER} &               \EVENTSELECTION{S190924hOPTSNRPOSTTIGER} & \cmark & $-$ \\
\bottomrule
\end{tabular}

\end{table}

\begin{figure*}%
  \centering
  \includegraphics[width=\textwidth]{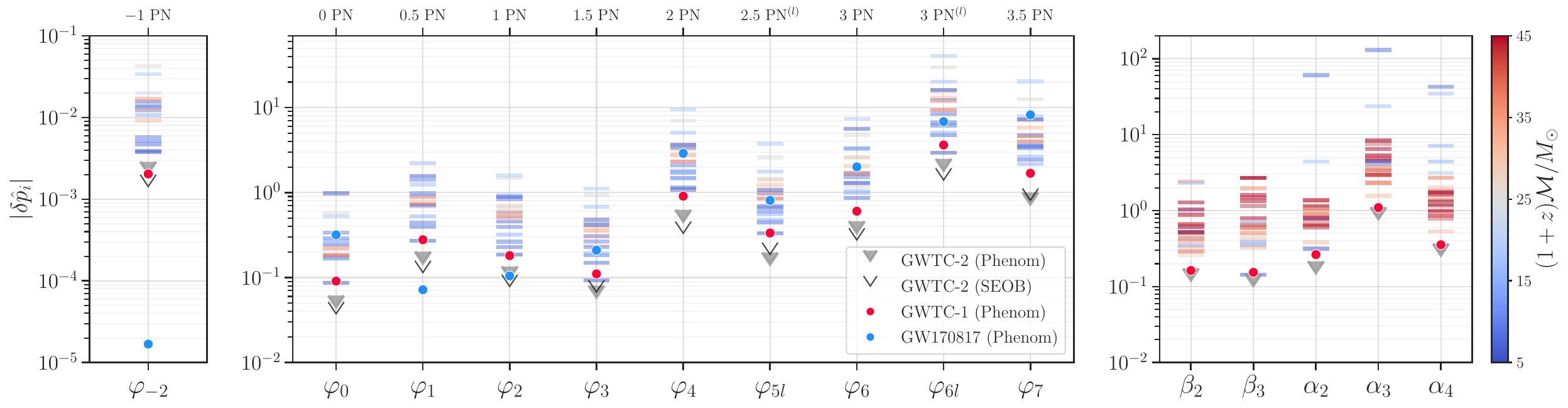}
  \caption{$90\%$ upper bounds on the absolute magnitude of the GR violating parameters $\delta \hat{p}_i$. The left and middle panels show the $-1$PN through $3.5$PN inspiral coefficients, while the right panel shows the postinspiral coefficients $\{\delta\hat{\beta}_i,\, \delta\hat{\alpha}_i\}$. Constraints obtained from individual events with \IMRP{} are represented by horizontal stripes, colored by the median redshifted chirp mass $(1+z)\mathcal{M}$, inferred assuming GR. Gray triangles (black wedges) mark the constraints obtained with \IMRP{} (\SEOBROM{}) when all GWTC-2 events are combined assuming a shared deviation from GR. For reference, we show the equivalent results for GWTC-1 (\IMRP{}) and the individual constraints from GW170817 (\IMRPNRT{}), as red and blue circles respectively. }
  \label{fig:tig:pn_bounds}
\end{figure*}

Parametrized tests of GW generation allow us to quantify generic deviations from GR predictions. Such corrections could arise as modifications to the binding energy and angular momentum of the source, or as modifications to the energy and angular momentum flux, both leading to modified equations of motion. In this section, we focus on constraining deviations from GR by introducing parametric deformations to an underlying GR waveform model. 

The early inspiral of compact binaries is well described by the PN approximation \cite{Blanchet:1995ez,Kidder:1995zr,Damour:2001bu,Blanchet:2005tk,Blanchet:2006gy,Arun:2008kb,Bohe:2012mr,Marsat:2012fn,Bohe:2013cla,Blanchet:2013haa,Damour:2014jta,Bohe:2015ana}, a perturbative approach to solving the Einstein field equations in which we perform an expansion in terms of a small velocity parameter $v/c$. Once the intrinsic parameters of the binary are fixed, the coefficients at different orders of $v/c$ in the PN series are uniquely determined. A consistency test of GR using the PN phase coefficients was first proposed in \cite{Blanchet:1994ez,Blanchet:1994ex,Arun:2006hn,Arun:2006yw,Mishra:2010tp}, and a general model independent parametrization was introduced in \cite{Yunes:2009ke}. A Bayesian framework based on the general parametrization was introduced in \cite{Li:2011cg,Li:2011vx,Agathos:2013upa}, with subsequent extensions to the late-inspiral and postinspiral coefficients being introduced in \cite{Meidam:2017dgf}.  

In order to constrain GR violations, we adopt two approaches. In the first approach, we directly constrain the analytical coefficients that describe the phase evolution of the \IMRP{} waveform model \cite{Hannam:2013oca,Husa:2015iqa,Khan:2015jqa}. The frequency-domain GW phase $\varphi (f)$ of \IMRP{} can be broken down into three key regions: inspiral, intermediate, and merger--ringdown. The inspiral in \IMRP{} is described by a PN expansion augmented with higher order pseudo-PN coefficients calibrated against EOB--NR hybrid waveforms. The PN phase evolution is written as a closed-form frequency domain expression by employing the stationary phase approximation. The intermediate and merger--ringdown regimes are described by analytical phenomenological expressions. The cutoff frequency $f_{\text{c}}^{\rm PAR}$ between the inspiral and intermediate region in \IMRP{} is defined to be $G M (1+z) f_{\text{c}}^{\rm PAR} / c^3 = 0.018$, where $z$ is the redshift and $f_{\text{c}}^{\rm PAR}$ is independent of the intrinsic parameters of the binary. We use $p_i$ to collectively denote all of the inspiral $\lbrace \varphi_i \rbrace$ and postinspiral $\lbrace \alpha_i, \beta_i \rbrace$ parameters.  The deviations from GR are expressed in terms of relative shifts $\delta \hat{p}_i$ in the waveform coefficients $p_i \rightarrow (1 + \delta \hat{p}_i )p_i$, which are introduced as additional free parameters to be constrained by the data. 

The second approach \cite{Abbott:2018lct} can apply modifications to the inspiral of any underlying waveform model, analytical or non-analytical, by adding corrections that correspond to deformations of a given inspiral coefficient $\delta \hat{\varphi}_i$ at low frequencies and tapering the corrections to $0$ at the cutoff frequency $f_{\text{c}}^{\rm PAR}$. The second approach is applied to the non-analytical model \SEOBROM{} \cite{Purrer:2015tud}, a frequency-domain reduced-order model for the \SEOBNR{} waveform approximant \cite{Bohe:2016gbl}. There is a subtle difference in the way in which deviations from GR are introduced and parametrized in the two approaches. In the first approach, we directly constrain the fractional deviations in the non-spinning portion of the phase whereas in the second approach the fractional deviations are also applied to the spin sector. As in \cite{LIGOScientific:2019fpa}, the posteriors in the second approach are mapped post-hoc to the parametrization used in the first approach, consistent with previously presented results. See Sec.~\ref{sec:rin} for an SEOB-based analysis of the postmerger signal, interpreted in the context of studies of the remnant properties.

We constrain deviations from the PN phase coefficients predicted by GR using deviation parameters $\delta \hat{\varphi}_i$. Here $i$ denotes the power of $v/c$ beyond the leading order Newtonian contribution to the phase $\varphi(f)$. The frequency dependence of the phase coefficients is given by $f^{(i-5)/3}$, so that $\delta \hat{\varphi}_i$ quantifies deviations to the $i/2$ PN order. We constrain coefficients up to $3.5$PN $(i=7)$, including terms that have a logarithmic dependence occurring at $2.5$ and $3$PN order. The non-logarithmic term at $2.5$PN $(i=5)$ cannot be constrained as it is degenerate with the coalescence phase. The coefficients describing deviations from GR were introduced in Eq.~(19) of \cite{Li:2011cg}. In addition, we include a coefficient at $i=-2$ corresponding to an effective $-1$PN term that, in some circumstances, can be interpreted as arising from the emission of dipolar radiation. The full set of inspiral parameters that we constrain is therefore
\begin{align}
\left\lbrace \delta \hat{\varphi}_{-2} , \delta \hat{\varphi}_{0} , \delta \hat{\varphi}_{1}, \delta \hat{\varphi}_{2}, \delta \hat{\varphi}_{3}, \delta \hat{\varphi}_{4}, \delta \hat{\varphi}_{5l}, \delta \hat{\varphi}_{6},  \delta \hat{\varphi}_{6l}, \delta \hat{\varphi}_{7} \right\rbrace .
\end{align}
\newline 
The inspiral deviations are expressed as shifts to the part of the PN coefficients with no spin dependence, $\varphi_i^{\rm NS}$, i.e., $\varphi_i \rightarrow (1 + \delta\hat{\varphi}_i ) \varphi_i^{\rm NS} + \varphi_i^{\rm S}$, where $\varphi_i^{S}$ denotes the spin-dependent part of the aligned-spin PN coefficients. This is the same parametrization that has been previously used \cite{TheLIGOScientific:2016src,TheLIGOScientific:2016pea,Abbott:2017oio,Abbott:2018lct,LIGOScientific:2019fpa} and circumvents the potential singular behavior observed when the spin-dependent terms cancel with the non-spinning term. In GR, the coefficients occurring at $-1$PN and $0.5$PN vanish, so we parametrize $\delta \hat{\varphi}_{-2}$ and $\delta \hat{\varphi}_{1}$ as \textit{absolute} deviations, with a prefactor equal to the $0$PN coefficient; all other coefficients represent \emph{fractional} deviations around the GR value. We derive constraints on the inspiral coefficients using the \IMRP{} and \SEOBROM{} analyses. 

\begin{figure*}[h!tp]
  \centering
\includegraphics[keepaspectratio, width=\textwidth]{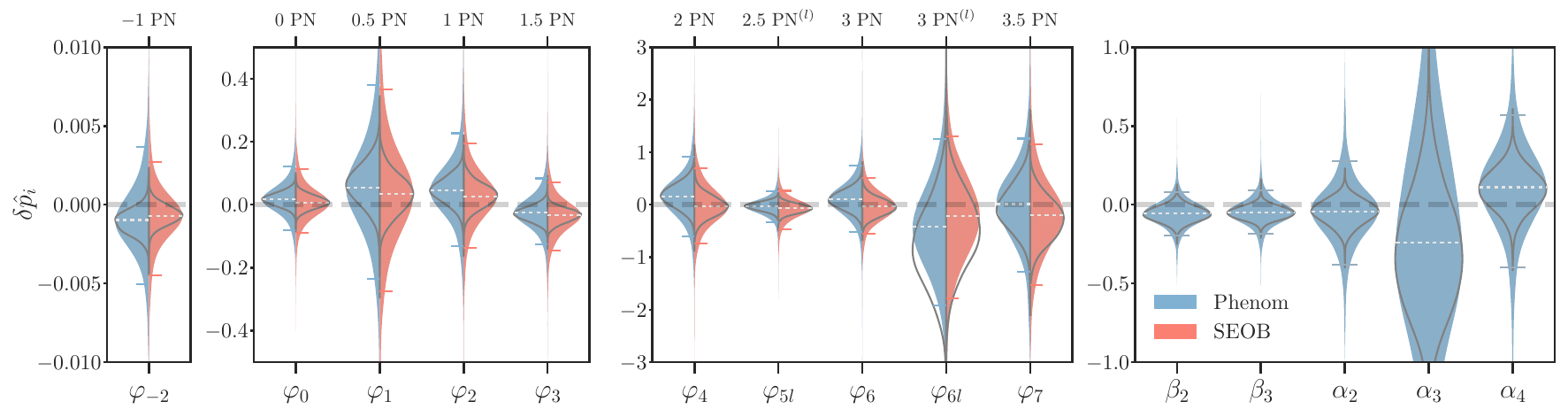}
\caption{Combined GWTC-2 BBH results for parametrized violations of GR obtained from the designated events in Table~\ref{tab:par_events}, for each deviation parameter $\delta\hat{p}_i$ (abscissa). The probability densities shown in color represent the population-marginalized expectation, Eq.~\eqref{eq:inf:hier_dist}, obtained from a hierarchical analysis allowing independent GR deviations for each event.
In contrast, the unfilled black distributions result from restricting all events to share a common value of each parameter.
Phenom (SEOB) results were obtained with \IMRP{} (\SEOBROM) and are shown in blue (red); the $\{\beta_i,\,\alpha_i\}$ coefficients are not probed with SEOB, as they are intrinsic to Phenom waveforms.
For the hierarchical results, error bars denote symmetric 90\%-credible intervals and a white dashed line marks the median.
The dashed horizontal line at $\delta\hat{p}_i= 0$ highlights the expected GR value.
}
\label{fig:par:both}
\end{figure*}

Besides the inspiral, the intermediate and merger--ringdown model in \IMRP{} is analytical and allows for parametrized deviations of the phenomenological coefficients that describe these regimes, denoted by $\lbrace \delta \hat{\beta}_2 , \delta \hat{\beta}_3 \rbrace$ and $\lbrace \delta \hat{\alpha}_2 , \delta \hat{\alpha}_3, \delta \hat{\alpha}_4 \rbrace$ respectively. The parameters $\delta \hat{\beta}_i$ explicitly capture deformations in the NR calibrated coefficients $\beta_i$ in the intermediate regime, whereas the parameters $\delta \hat{\alpha}_i$ describe deformations of the merger--ringdown coefficients $\alpha_i$ obtained from a mix of BH perturbation theory and calibration to NR \cite{Husa:2015iqa,Khan:2015jqa}. We omit $\delta\hat \alpha_5$ as this occurs in the same term as $\delta\hat \alpha_4$, see Eq.~(13) of \cite{Khan:2015jqa}, meaning that there will be a degree of degeneracy between the two coefficients. 

As detailed in Sec.~\ref{sec:intro}, we consider all binaries that meet the significance threshold of FAR $< 10^{-3}~\mathrm{yr}^{-1}$ and impose the additional requirement that the $\rm{SNR} > 6$ in the inspiral regime ($\delta \hat{\varphi}_i$) or postinspiral regime ($\delta \hat{\beta}_i$ and $\delta \hat{\alpha}_i$) respectively for an event to be included in the analyses, as data below these SNR thresholds fails to provide meaningful constraints. In contrast to the selection criteria used in \cite{LIGOScientific:2019fpa}, GW170818 meets the FAR threshold applied in this analysis and is included in the joint constraints. The SNRs and cutoff frequencies for all events are detailed in Table~\ref{tab:par_events}. 

For three of the events considered in this analysis, HMs have a non-trivial impact on parameter estimation and must be taken into account.
This is the case for \NAME{GW190412A}{} and \NAME{GW190814A}{}, which show evidence of detectable HM power \cite{GW190412,GW190814}, and for \NAME{GW190521A}{}, which does not \cite{GW190521g,GW190521g:imp}.
We perform the parametrized tests using \IMRPHM{} and, for \NAME{GW190412A}{} and \NAME{GW190814A}{}, \SEOBHMROM{}.
By construction, parametrized deformations in \IMRPHM{} are propagated to the HMs through approximate rescalings of the $(2,2)$ mode with no new coefficients being introduced. The framework used for the \SEOBHMROM{} analysis is extended to HMs in an analogous way. We show the posterior distributions for \NAME{GW190412A}{} and \NAME{GW190814A}{}, the two events that show measurable HM power, in Appendix~\ref{app:par}. 

We use \linf{} to calculate the posterior probability distributions of the parameters characterizing the waveform \cite{Veitch:2014wba}. The parametrization used here recovers GR in the limit $\delta \hat{p}_i \rightarrow 0$, enabling us to verify consistency with GR if the posteriors of $\delta \hat{p}_i$ have support at $0$. As in previous analyses, we only allow the coefficients $\delta \hat{p}_i$ to vary one at a time. Despite the lack of generality, this approach is effective at detecting deviations from GR that do not just modify a single coefficient \cite{Sampson:2013lpa,Meidam:2017dgf,Haster:2020nxf}. In particular, the coefficients will be sensitive to corrections that occur at generic PN orders even when varying a coefficient that corresponds to some fixed PN order \cite{Meidam:2017dgf,Haster:2020nxf}. Allowing the test to vary multiple coefficients simultaneously can often lead to posteriors that are less informative, with the single-coefficient templates often being preferred to the templates with multiple parameters in the context of Bayesian model selection \cite{Sampson:2013lpa}. Varying multiple coefficients simultaneously would therefore not improve the efficiency of detecting violations of GR \cite{Sampson:2013lpa}.
On the other hand, nontrivial multicoefficient deviations may be detected even when only one $\delta \hat{p}_i$ is allowed to vary at a time \cite{Isi:2019asy}.
We adopt uniform priors on $\delta \hat{p}_i$ that are symmetric about zero. 
Due to the way in which parametrized deformations are implemented, evaluating a model in certain regions of the parameter space can lead to pathologies and unphysical effects. This can result in multimodal posterior distributions or other systematic errors, see the discussion in Appendix~\ref{app:par}. 

In Fig.~\ref{fig:tig:pn_bounds} we show the $90\%$ upper bounds on the absolute magnitude of the GR violating coefficients, $| \delta\hat p_i |$. The individual bounds are colored by the mean redshifted chirp mass, $(1+z)\mathcal{M}$, as inferred assuming GR (Table~\ref{tab:events}). The results for GWTC-2 include all new BBHs reported in \cite{GWTC2} plus the BBHs reported in GWTC-1 \cite{GWTC1}, combined by assuming a shared value of the coefficient across events (i.e.,~by multiplying the individual likelihoods). Whilst the combined results for GWTC-1 and GWTC-2 do not include the two BNS events, GW170817 and GW190425, in Fig.~\ref{fig:tig:pn_bounds} we show the results for GW170817 separately for comparison to previously published results \cite{Abbott:2018lct}.

We broadly see that lighter binaries contribute prominently to our constraint on the inspiral coefficients and heavier binaries drive the constraints on the postinspiral coefficients. This is to be expected as more (less) of the inspiral moves into the sensitivity of the detectors as we decrease (increase) the mass and we suppress (enhance) the SNR in the postinspiral. For all coefficients, bar the $-1$PN and $0.5$PN terms, the joint-likelihood bounds determined using GWTC-1 and GWTC-2 BBHs improve on all previous constraints \cite{Abbott:2018lct,LIGOScientific:2019fpa}. The tightest bounds on the $-1$PN and $0.5$PN coefficients come from GW170817, which improves on the GWTC-2 BBH constraints by a factor of $120$ and $2.2$ respectively. We find that the combined GWTC-2 results improve on the GWTC-1 constraints by a factor ${\sim} 1.9$ for the inspiral coefficients and ${\sim} 1.4$ for the postinspiral coefficients respectively. This improvement is broadly consistent with the factor expected from the increased number of events, $\sqrt{17/5} \approx 1.8$ for the inspiral and $\sqrt{26/7} \approx 1.9$ for the postinspiral respectively. 
Neglecting the $-1$PN coefficient, we find that the $0$PN term is the best constrained parameter, $| \delta\hat \varphi_{0}| \lesssim 4.4 \times 10^{-2}$. However, this bound is weaker than the $90\%$ upper bound inferred from the orbital-period derivative $\dot{P}_{\rm orb}$ of the double pulsar J0737$-$3039 by a factor ${\sim}3$ \cite{Yunes:2010qb,Wex:2014nva}. 

Although all results from individual events offer support for the GR value, a small fraction of them contain $\delta\hat{p}_i = 0$ only in the tails. 
This is the case for some of the coefficients for \NAME{GW190519A}, \NAME{GW190521B}, \NAME{GW190814A}, \NAME{GW190828B}, and \NAME{GW190924A}.
Yet, given the large number of events and coefficients analyzed, this is not surprising: for GR signals in Gaussian noise, we would expect on average approximately $1$ out of $10$ independent trials to return $\delta\hat{p}_i = 0$ outside the 90\%-credible level just from statistical fluctuations.

To evaluate the set of measurements holistically, we produce the population-marginalized distributions for each parameter $\delta\hat{p}_i$ following the method described in Sec.~\ref{sec:inference:populations}; the result is the filled distributions in Fig.~\ref{fig:par:both}.
These distributions represent our best knowledge of the possible values of the $\delta\hat{p}_i$'s from all LIGO--Virgo BBHs with $\mathrm{FAR} < 10^{-3}~\mathrm{yr}^{-1}$ to date.
For comparison, Fig.~\ref{fig:par:both} also shows the joint likelihoods obtained by restricting the deviation to be the same for all events (unfilled black distributions), which were used to derive the combined GWTC-2 constraints in Fig.~\ref{fig:tig:pn_bounds}.

\begin{figure}
\raggedright
\includegraphics[width=0.95\columnwidth]{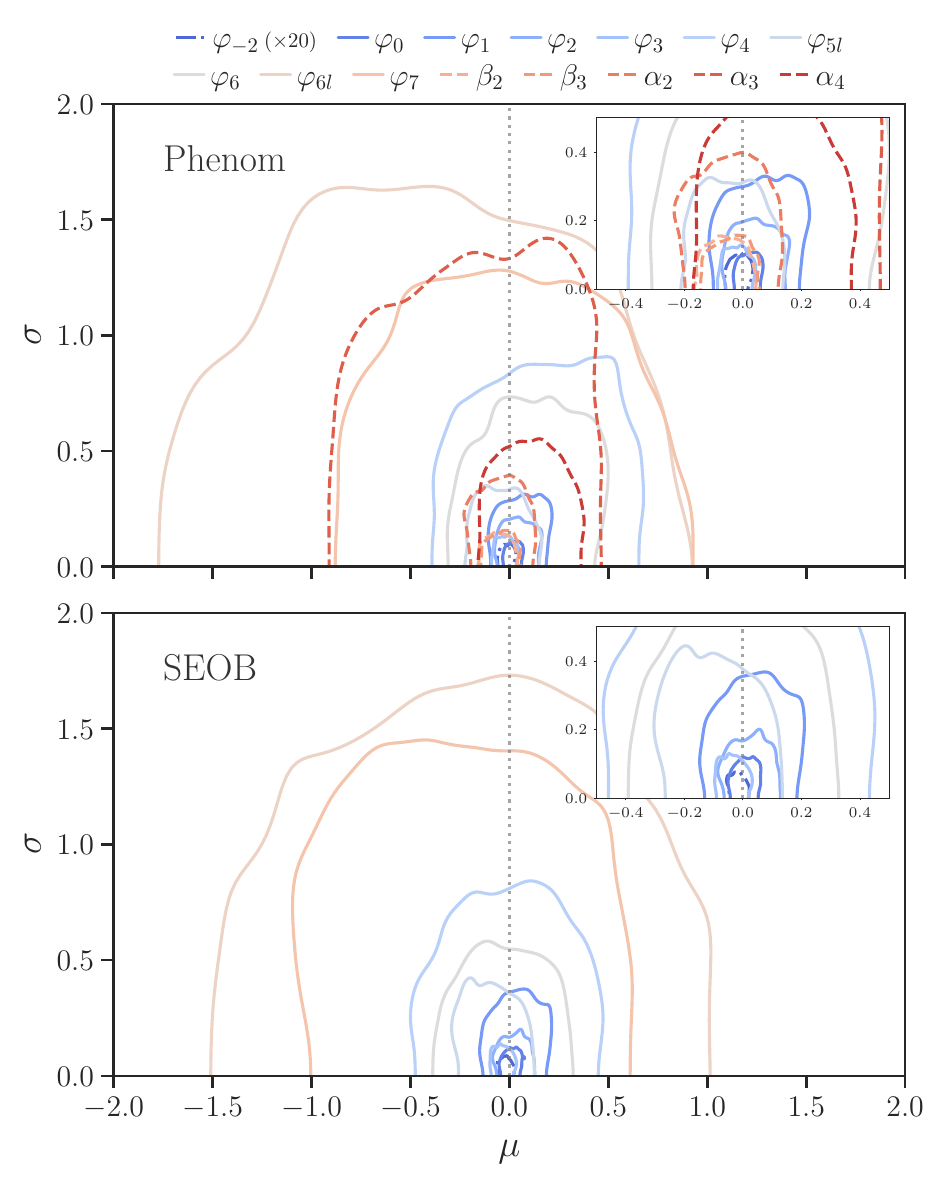}
\caption{Hyperparameter measurements for the parametrized-deviation coefficients.
Contours enclose 90\% of the posterior probability for the $\mu$ and $\sigma$ hyperparameters corresponding to each of the $\delta\hat{p}_i$ coefficients, as indicated by the legend.
The top (bottom) panel shows \IMRP{} (\SEOBROM{}) results, corresponding to the blue (red) distributions in Fig.~\ref{fig:par:both}.
The insets provides a closer look around $\mu=\sigma=0$, our baseline expectation in the absence of GR violations or measurement systematics; all contours enclose this point.
As in Table~\ref{tab:par}, the values for $\varphi_{-2}$ have been rescaled by a factor of $20$ for ease of display.}
\label{fig:par:contour}
\end{figure}

All population-marginalized distributions are consistent with GR, with $\delta \hat{p}_i=0$ lying close to the median for most parameters, and always within the 90\% credible symmetric interval.
The medians, 90\% credible intervals, and GR quantiles $Q_{\rm GR} = P(\delta\hat{p}_i < 0)$ of these distributions are presented in Table~\ref{tab:par}, together with equivalent quantities for the joint-likelihood approach.
A value of $Q_{\rm GR}$ significantly different from 50\% indicates that the null hypothesis falls in the tails of the distribution.
The quantiles may also be directly translated into $z$-scores defined by $z_{\rm GR} = \Phi^{-1}(Q_{\rm GR})$, where $\Phi^{-1}$ is the inverse cumulative distribution function for a standard normal random variable.
The $z$-score encodes the distance of the posterior mean away from zero in units of standard deviation (discussed below).

In terms of the overall magnitude of the allowed fractional deviations, the parameter constrained most tightly by the hierarchical analysis is $\ParBestHierPopP{NAME} = \ParBestHierPopP{VALUE}$, within 90\% credibility.
On the other hand, the loosest constraint comes from $\ParWorstHierPopP{NAME} = \ParWorstHierPopP{VALUE}$, also within 90\% credibility.
In both cases, however, the null-hypothesis lies close to the median, with $Q_{\rm GR} = \ParBestHierPopP{QGR}$ and $Q_{\rm GR} = \ParWorstHierPopP{QGR}$ respectively.
The magnitude of the constraint, however, is parametrization-dependent and may not be meaningful outside the context of a specific theory \cite{Yunes:2009ke,Yunes:2016jcc,Chua:2020oxn}.

Agreement with GR requires not only that the distributions in Fig.~\ref{fig:par:both} support $\delta\hat{p}_i = 0$, but also that the measured hyperparameters be consistent with $\mu=\sigma=0$ (see Sec.~\ref{sec:inference:populations}).
This is indeed the case, as can be inferred from the 90\% credible measurements shown in Fig.~\ref{fig:par:contour}, and summarized in the third and fourth columns of Table \ref{tab:par}.
The implications of the hyperparameter measurement are concisely captured by the two-dimensional GR quantile $\mathcal{Q}_{\rm GR}$, defined as the isoprobability contour passing through $\mu=\sigma=0$:
a posterior with $\mathcal{Q}_{\rm GR} = 0$ peaks at the GR expectation, with larger values indicating reduced support.

\begin{figure}
\raggedright
\includegraphics[width=0.95\columnwidth]{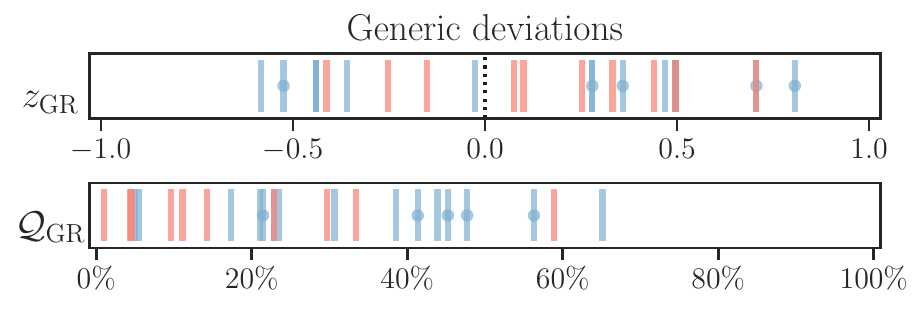}\\
\hspace{1.5pt}\includegraphics[width=0.935\columnwidth]{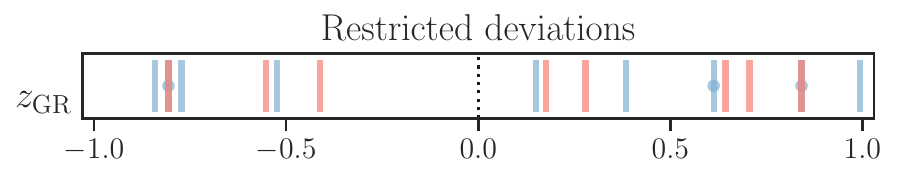}
\caption{Figures of merit for the GWTC-2 parametrized tests results. Each vertical stripe corresponds to a given $\delta\hat{p}_i$ as estimated using \IMRP{} (blue) or \SEOBROM{} (red); circles identify the postmerger coefficients $\{\delta\hat{\alpha}_i,\,\delta\hat{\beta}_i\}$. The top two strips summarize the hierarchical results for generic deviations across events: the $z$-score for $\delta\hat{p}_i = 0$, $z_{\rm GR}$, and the two-dimensional quantile for the hyperparameters $\mu=\sigma=0$, $\mathcal{Q}_{\rm GR}$.
The bottom strip shows an equivalent $z$-score obtained by restricting to identical deviations across events.
The generic (restricted) $z$-scores correspond to the filled (unfilled) distributions in Fig.~\ref{fig:par:both}, and $\mathcal{Q}_{\rm GR}$ to those in Fig.~\ref{fig:par:contour}.}
\label{fig:par:summary}
\end{figure}

Figure~\ref{fig:par:summary} summarizes the main conclusions from this section through a visualization of $z_{\rm GR}$ and $\mathcal{Q}_{\rm GR}$ from the hierarchical analysis (top and middle), and of $z_{\rm GR}$ from the joint-likelihood analysis (bottom).
Each $\delta\hat{p}_i$ is represented by a vertical stripe, with the postmerger $\{\delta\hat{\beta}_i,\, \delta\hat{\alpha}_i\}$ coefficients identified by an additional circle.
The figure suggests that the postmerger parameters may behave distinctly from the rest, tending to show more pronounced excursions away from the baseline expectation ($z_{\rm GR} \approx 0$).
In any case, because 1$\sigma$ outliers are not unlikely and the null hypothesis lies well within the 90\% credible regions for all coefficients (Table~\ref{tab:par}), we conclude that there is no statistically significant evidence for GR violations.

\begin{table}
\caption{Results from parametrized tests of GW generation (Sec.~\ref{sec:par}).
Combined constraints on each deviation parameter $\delta\hat{p}_i$ from the full set of GWTC-2 BBH measurements using the \IMRP{} or \SEOBROM{} waveforms, as indicated by P or S respectively in the second column.
The general constraints do not assume the deviation takes the same value for all events, and are summarized by the hyperdistribution mean $\mu$ and standard deviation $\sigma$, as well as the inferred direct constraint on $\delta\hat{p}_i$ (defined in Sec.~\ref{sec:inference:populations}).
The restricted constraints assume a common value of the parameter shared by all events, and are summarized by the constraint on $\delta\hat{p}_i$.
All quantities represent the median and 90\%-credible intervals excepting $\sigma$, for which we provide an upper limit.
For both general and restricted results, $Q_{\rm GR}$ is the GR quantile associated with Fig.~\ref{fig:par:both}.
}

\label{tab:par}
\centering
\begin{tabular}{ccrrrrrrrr}
\toprule
              $\hat{p}_i$ & WF & \phantom{X} & \multicolumn{4}{c}{General} & \phantom{X} & \multicolumn{2}{c}{Restricted} \\
\cline{4-7} \cline{9-10}
               &   & \phantom{X} &                       $\mu$ &    $\sigma$ &                    $\delta\hat{p}_i$ &       $Q_{\rm GR}$ & \phantom{X} &                    $\delta\hat{p}_i$ &       $Q_{\rm GR}$ \\
\midrule
${\varphi}_{-2}$ & P &             &  $-0.02^{+0.04}_{-0.03}$ &  $<0.08$ &  $-0.02^{+0.09}_{-0.08}$ &  $68\%$ &             &  $-0.02^{+0.02}_{-0.02}$ &  $93\%$ \\
         ${}^{[\times 20]}$ & S &             &  $-0.01^{+0.03}_{-0.03}$ &  $<0.07$ &  $-0.01^{+0.07}_{-0.08}$ &  $69\%$ &             &  $-0.01^{+0.02}_{-0.02}$ &  $86\%$ \\[4pt]
${\varphi}_{0}$ & P &             &   $0.02^{+0.05}_{-0.04}$ &  $<0.09$ &   $0.02^{+0.10}_{-0.10}$ &  $33\%$ &             &   $0.02^{+0.04}_{-0.03}$ &  $20\%$ \\
               & S &             &   $0.01^{+0.05}_{-0.04}$ &  $<0.10$ &   $0.01^{+0.11}_{-0.09}$ &  $44\%$ &             &   $0.01^{+0.03}_{-0.04}$ &  $34\%$ \\[4pt]
${\varphi}_{1}$ & P &             &   $0.06^{+0.14}_{-0.13}$ &  $<0.27$ &   $0.05^{+0.32}_{-0.29}$ &  $33\%$ &             &   $0.07^{+0.10}_{-0.11}$ &  $15\%$ \\
               & S &             &   $0.03^{+0.14}_{-0.14}$ &  $<0.29$ &   $0.03^{+0.33}_{-0.31}$ &  $40\%$ &             &   $0.03^{+0.11}_{-0.10}$ &  $29\%$ \\[4pt]
${\varphi}_{2}$ & P &             &   $0.05^{+0.09}_{-0.09}$ &  $<0.17$ &   $0.04^{+0.18}_{-0.18}$ &  $28\%$ &             &   $0.04^{+0.07}_{-0.07}$ &  $14\%$ \\
               & S &             &   $0.03^{+0.08}_{-0.08}$ &  $<0.15$ &   $0.03^{+0.17}_{-0.16}$ &  $34\%$ &             &   $0.03^{+0.06}_{-0.06}$ &  $21\%$ \\[4pt]
${\varphi}_{3}$ & P &             &  $-0.02^{+0.05}_{-0.05}$ &  $<0.10$ &  $-0.02^{+0.11}_{-0.10}$ &  $69\%$ &             &  $-0.03^{+0.04}_{-0.04}$ &  $90\%$ \\
               & S &             &  $-0.03^{+0.05}_{-0.05}$ &  $<0.10$ &  $-0.03^{+0.11}_{-0.11}$ &  $76\%$ &             &  $-0.03^{+0.02}_{-0.04}$ &  $96\%$ \\[4pt]
${\varphi}_{4}$ & P &             &   $0.14^{+0.44}_{-0.41}$ &  $<0.72$ &   $0.16^{+0.76}_{-0.77}$ &  $33\%$ &             &   $0.17^{+0.36}_{-0.36}$ &  $22\%$ \\
               & S &             &  $-0.01^{+0.40}_{-0.36}$ &  $<0.65$ &  $-0.03^{+0.72}_{-0.70}$ &  $53\%$ &             &  $-0.05^{+0.33}_{-0.30}$ &  $57\%$ \\[4pt]
${\varphi}_{5l}$ & P &             &  $-0.03^{+0.15}_{-0.15}$ &  $<0.27$ &  $-0.04^{+0.29}_{-0.30}$ &  $61\%$ &             &  $-0.02^{+0.12}_{-0.15}$ &  $65\%$ \\
               & S &             &  $-0.08^{+0.16}_{-0.17}$ &  $<0.33$ &  $-0.07^{+0.34}_{-0.39}$ &  $67\%$ &             &  $-0.08^{+0.15}_{-0.15}$ &  $80\%$ \\[4pt]
${\varphi}_{6}$ & P &             &   $0.10^{+0.32}_{-0.32}$ &  $<0.56$ &   $0.10^{+0.64}_{-0.62}$ &  $36\%$ &             &   $0.08^{+0.30}_{-0.27}$ &  $30\%$ \\
               & S &             &  $-0.04^{+0.30}_{-0.27}$ &  $<0.47$ &  $-0.03^{+0.54}_{-0.52}$ &  $54\%$ &             &  $-0.05^{+0.27}_{-0.27}$ &  $61\%$ \\[4pt]
${\varphi}_{6l}$ & P &             &  $-0.41^{+1.07}_{-1.01}$ &  $<1.27$ &  $-0.42^{+1.67}_{-1.50}$ &  $69\%$ &             &  $-0.80^{+1.32}_{-1.29}$ &  $84\%$ \\
               & S &             &  $-0.22^{+0.96}_{-1.01}$ &  $<1.31$ &  $-0.22^{+1.52}_{-1.56}$ &  $60\%$ &             &  $-0.47^{+1.17}_{-1.17}$ &  $74\%$ \\[4pt]
${\varphi}_{7}$ & P &             &   $0.02^{+0.70}_{-0.75}$ &  $<1.09$ &   $0.01^{+1.25}_{-1.29}$ &  $49\%$ &             &  $-0.08^{+0.75}_{-0.66}$ &  $56\%$ \\
               & S &             &  $-0.23^{+0.70}_{-0.68}$ &  $<1.16$ &  $-0.20^{+1.36}_{-1.32}$ &  $63\%$ &             &  $-0.29^{+0.63}_{-0.66}$ &  $76\%$ \\[4pt]
\midrule
${\beta}_{2}$ & P &             &  $-0.06^{+0.07}_{-0.08}$ &  $<0.12$ &  $-0.06^{+0.14}_{-0.14}$ &  $79\%$ &             &  $-0.07^{+0.08}_{-0.07}$ &  $90\%$ \\[4pt]
${\beta}_{3}$ & P &             &  $-0.05^{+0.08}_{-0.08}$ &  $<0.12$ &  $-0.05^{+0.14}_{-0.14}$ &  $76\%$ &             &  $-0.05^{+0.07}_{-0.06}$ &  $90\%$ \\[4pt]
${\alpha}_{2}$ & P &             &  $-0.04^{+0.13}_{-0.15}$ &  $<0.30$ &  $-0.04^{+0.32}_{-0.33}$ &  $61\%$ &             &  $-0.04^{+0.11}_{-0.13}$ &  $73\%$ \\[4pt]
${\alpha}_{3}$ & P &             &  $-0.23^{+0.65}_{-0.56}$ &  $<1.10$ &  $-0.24^{+1.36}_{-1.19}$ &  $64\%$ &             &  $-0.32^{+0.62}_{-0.55}$ &  $80\%$ \\[4pt]
${\alpha}_{4}$ & P &             &   $0.11^{+0.22}_{-0.23}$ &  $<0.44$ &   $0.11^{+0.46}_{-0.51}$ &  $30\%$ &             &   $0.10^{+0.19}_{-0.22}$ &  $21\%$ \\[4pt]
\bottomrule
\end{tabular}
\end{table}

The results from this section can be used to place constraints on individual theories by reinterpreting the coefficients $\delta\hat \varphi_i$ within the parametrized post-Einstein (ppE) framework given a theory-dependent mapping \cite{Yunes:2009ke,Yunes:2016jcc}. Recently, \cite{Nair:2019iur} used the coefficients $\delta\hat \varphi_i$ to place constraints on higher-curvature theories in the small-coupling approximation, focusing on two specific examples: Einstein-dilaton-Gauss--Bonnet and dynamical Chern--Simons gravity. The improved constraints on the coefficients $\delta\hat \varphi_i$ provided here will allow for tighter constraints on the coupling constants in such theories under similar (nontrivial) assumptions.

    \subsection{Spin-induced quadrupole moment}
    \label{sec:sim}
    
The leading order spin-induced multipole
moment, the spin-induced quadrupole moment, is a measure of the degree of an object's 
oblateness due to its spin, specifically of its effect on the surrounding gravitational field \cite{Poisson:1997ha,Laarakkers:1997hb,Ryan:1995wh}.
If the object is in an inspiraling binary, this effect will become imprinted in the GW waveform 
at specific PN orders, helping us identify the object's nature and composition  \cite{Krishnendu:2017shb}.
For a compact object with mass $m$ and spin $\chi$, the spin-induced quadrupole moment is given by 
\begin{equation}\label{eq:sim} 
	Q= -\, \kappa\, \chi^2 m^3,
\end{equation}
where $\kappa$ is the spin-induced quadrupole moment coefficient,
which depends on the equation of state, mass, and spin of the compact object. 
Due to the no-hair conjecture \cite{Hansen:1974zz,Carter,Gurlebeck:2015xpa},  
$\kappa$ is unity for BHs in GR, while it may take a range of values for neutron stars or BH mimickers \cite{Poisson:1997ha,Laarakkers:1997hb,Ryan:1995wh,Uchikata:2015yma}. 
For example, depending upon the equation of state, the value of $\kappa$ can vary 
between ${\sim}2$ and ${\sim}14$ for a spinning neutron star \cite{Pappas:2012qg,Pappas:2012ns,Harry:2018hke}, and between ${\sim}10$ 
and ${\sim}150$ for slowly spinning boson stars \cite{FDRyan1997,Herdeiro:2014goa,Baumann:2018vus,Chia:2020psj}. 
The spin-induced quadrupole moments first appear along with the self-spin terms in the GW phasing formula as a 2PN leading-order effect \cite{Poisson:1997ha}.   
In this paper, we also incorporate 3PN corrections to the GW phase due to the spin-induced quadrupole moment of binary components \cite{Arun:2008kb,Mishra:2016whh}.  
As shown in \cite{Krishnendu:2019tjp}, the measurement accuracy of these parameters is largely correlated 
with masses and spins of the binary system.  Despite the degeneracy, the presence of spin terms at other PN orders 
as well as the non-spinning PN coefficients help to break the correlations of $\kappa$ with 
spins and mass parameters, permitting its measurement for spinning binary systems.  
It has been demonstrated in the past that it is possible to measure spin-induced 
multipole moments for intermediate mass-ratio \cite{Brown:2006pj,Rodriguez:2012} and extreme mass-ratio inspirals \cite{Barack:2006pq,Babak:2017tow}. 
This parameter can also be constrained through electromagnetic observations of active galactic nuclei (see \cite{Laine:2020dnr} 
for a recent measurement) and supermassive BHs \cite{Akiyama:2019cqa}.

\begin{figure}
  \centering
  \includegraphics[width=\columnwidth]{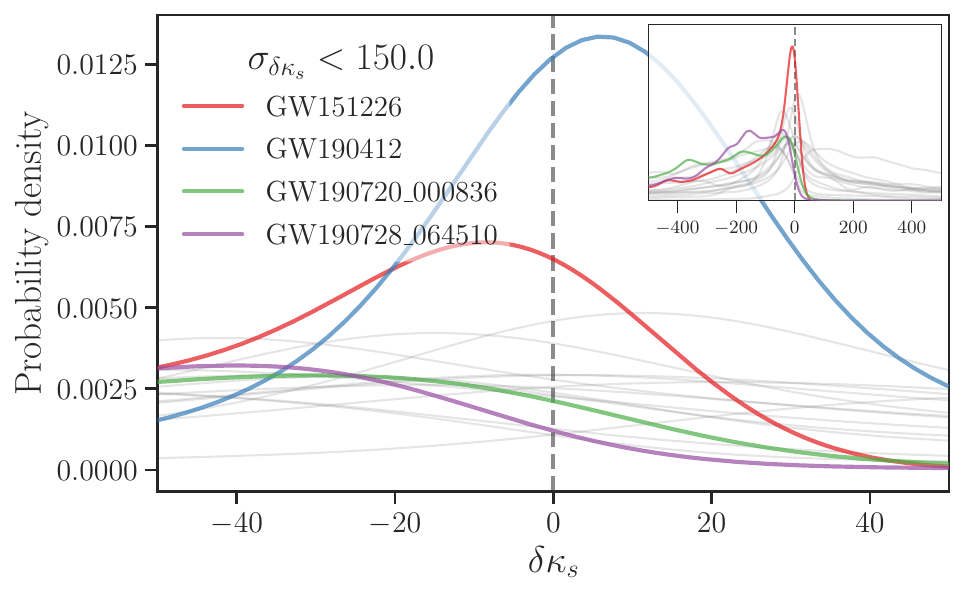}
  \caption{
  Posterior probability distribution on the spin-induced quadrupole moment parameter $\delta\kappa_s$ from the GWTC-2 events listed in the SIM column of Table~\ref{tab:events}.
  We highlight \protect\SimBestSymEvents{}, as they yield the tightest distributions (with standard deviation $\sigma_{\delta\kappa_s} < \SimStdCut$); other events are shown in gray.
  The inset expands the plot range to the full range of the prior, removing \protect\NAME{GW190412A}{} to facilitate display of the other events.
  The vertical dashed line at $\delta\kappa_s=0$ marks the Kerr BBH expectation.
  }
  \label{fig:sim}
\end{figure}

In principle, the BH nature of the binary components can be probed by measuring their individual spin-induced quadrupole moment coefficients $\kappa_1$ and $\kappa_2$, parametrized as deviations away from unity $\delta\kappa_{1}$ and $\delta\kappa_{2}$.
However, for the stellar-mass compact binaries accessible to LIGO and Virgo, it is often difficult to simultaneously constrain $\delta\kappa_{1}$ and $\delta\kappa_{2}$ due to the strong degeneracies between these and other binary parameters, like the spins and masses~\cite{Krishnendu:2017shb,Krishnendu:2018nqa}. 
We define the symmetric and anti-symmetric combinations of the individual deviation parameters as
$\delta\kappa_s=(\delta\kappa_1+\delta\kappa_2)/2$ and
$\delta\kappa_a=(\delta\kappa_1-\delta\kappa_2)/2$, but in this analysis we
restrict $\delta\kappa_a = 0$, implying $\delta\kappa_1=\delta\kappa_2=\delta\kappa_s$.
The assumption $\delta\kappa_a = 0$ also demands that the two compact objects be of the same kind 
which holds well when both the objects are BHs. For non-BH binaries, this restriction 
leads to stronger implications, requiring the two compact objects to have similar 
masses and equation of state as  $\delta\kappa_1$ and $\delta\kappa_2$ are functions of these. 
Having a non-BH compact object in the binary will violate these 
restrictions, which could lead to systematic biases in the estimation of $\delta \kappa_s$. 
For non-BBH signals, the value of $\delta\kappa_s$ would be offset 
from zero, given the definition, and it is unlikely for such offsets to be completely compensated by the aforementioned systematics. Therefore, the posteriors of $\delta\kappa_s$ 
for non-BBH signals will tend to peak away from zero, hinting at the presence of an exotic compact object.

For a  more general test of BBH nature, one might also include effects such as the tidal 
deformations that arise due to the object's binary companion \cite{Sennett:2017etc,Johnson-McDaniel:2018uvs,Pacilio:2020jza}
and tidal heating \cite{Cardoso:2017cfl,Maselli:2017cmm,Datta:2019epe,Datta:2019euh,Datta:2020gem,Datta:2020rvo} along with the spin-induced deformations. 
The present test does not consider these effects but focuses only on spin-induced deformations.

\begin{figure}
  \centering
  \includegraphics[width=\columnwidth]{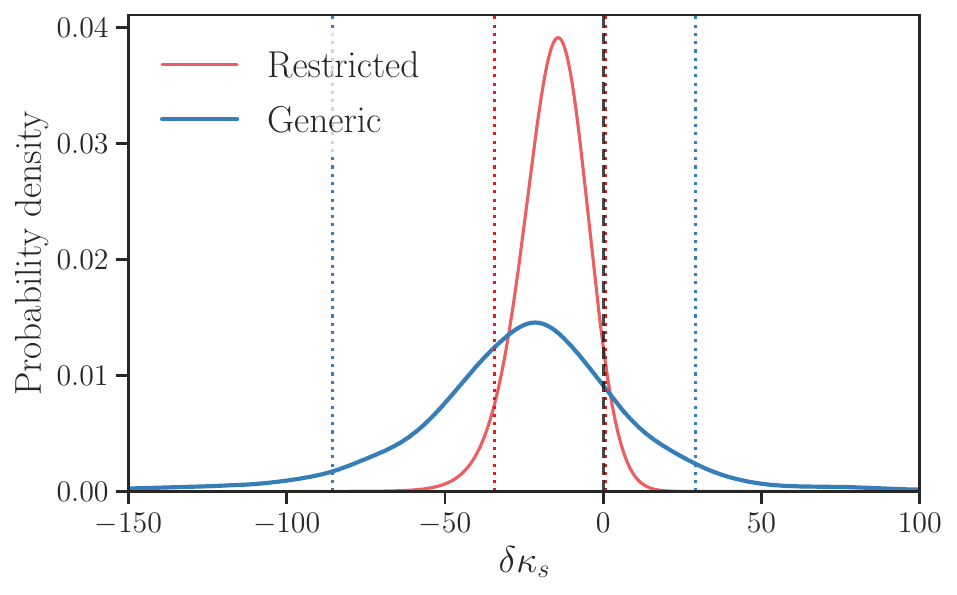}
  \caption{
  Combined measurement on the spin-induced quadrupole moment parameter $\delta\kappa_s$ from the set of all of events in Fig.~\ref{fig:sim}.
  The red curve (restricted) represents the posterior obtained assuming $\delta\kappa_s$ takes the same value for all events.
The blue histogram (generic) was obtained by hierarchically combining events without that assumption, as in Eq.~\eqref{eq:inf:hier_dist}.
Dotted lines bound symmetric 90\%-credible intervals, $\delta\kappa_s = \protect\SimCombinedCI{HIER_POP}$ ($\delta\kappa_s = \protect\SimCombinedCI{SIMPLE_POP}$) for the generic (restricted) case.
The Kerr BBH value ($\delta\kappa_s=0$) is marked by a dashed line.
}
  \label{fig:sim_combined}
\end{figure}

We perform this analysis on the compact binaries observed in O1, O2 and O3a. 
Though the spin-induced effects for non-BH compact objects are not 
modeled beyond the inspiral phase, as a null test of BBH nature, the analysis was performed by including the full inspiral, 
merger, and ringdown phases, using a waveform model built on \IMRP{}. In this model, only the inspiral phase of the waveform (defined as in Sec.~\ref{sec:par}) is modified in terms of $\delta\kappa_1$ and $\delta\kappa_2$.
For \NAME{GW190412A}, which showed evidence of HMs \cite{GW190412}, we employed a waveform model built on 
\IMRPHM{} with the same modifications in terms of $\delta\kappa_1$ and $\delta\kappa_2$ as for the model based on \IMRP{}.
We apply this test only to the events in Table~\ref{tab:events} 
that have SNR of 6 or more in the inspiral phase under the GR BBH assumption (same threshold as in Table~\ref{tab:par_events}); we apply the same criteria to the GWTC-1 events.
In this paper, we do not apply this test on GW190814 as the outcome of the test on GW190814 has already been discussed in \cite{GW190814} 
and we have not gained any new insights since then.

We employ a uniform prior on $\delta\kappa_s$ 
in the range $[-500, 500]$.
The prior limits at $\pm 500$ were chosen so they safely encompass the known models of BH mimickers, including gravastars and other exotic objects that may have  $\delta\kappa_s < 0$  \cite{Uchikata:2015yma}.
As elsewhere in this paper, the $\delta\kappa_s$ constraints apply exclusively to the set of events analyzed, and do not preclude the existence of objects with $|\delta\kappa_s|$ high enough to be missed by our search pipelines \cite{Chia:2020psj}.

Figure \ref{fig:sim} shows the measurement of $\delta\kappa_s$ from individual events.
We find that $\delta\kappa_s$ is poorly constrained for the majority of events, which can be attributed to the low spin of these events \cite{GWTC2}.
From Eq.~\eqref{eq:sim},  it is clear that the quadrupole moment vanishes when the spins are zero, irrespective of the value of $\kappa$.
Therefore, any meaningful upper limit on $\kappa$ would require the lower limit on at least 
one of the spin magnitudes to exclude zero. If this condition is not met, the posteriors of $\delta\kappa_s$ would rail against the priors in this analysis. 
The dependence of the upper limit of $\kappa$ on the spin magnitudes was studied in \cite{Krishnendu:2019tjp}.
In Fig.~\ref{fig:sim}, we highlight the events with the most concentrated $\delta\kappa_s$ posteriors, with a sample standard deviation $\sigma_{\delta\kappa_s} < \SimStdCut{}$: \SimBestSymEvents{}.
We do not quote symmetric credible intervals from individual events, since all of the posteriors present tails reaching the edge of the prior on at least one side.

We may narrow down the scope of the test by focusing on the $\delta\kappa_s >0$ region of our prior, 
which is well constrained by a subset of the events.
Doing so is well motivated in the context of neutron stars \cite{Laarakkers:1997hb,Pappas:2012qg,Pappas:2012ns} and specific BH mimickers such as boson stars \cite{FDRyan1997} 
for which  $\kappa_s>1$.
Restricting to positive $\delta\kappa_s$, the two events providing the tightest upper limits are \SimBestPosUL{NAME0}[] and \SimBestPosUL{NAME1}{}, with 90\% credible bounds of $\delta\kappa_s < \SimBestPosUL{VALUE0}{}$ and $\delta\kappa_s < \SimBestPosUL{VALUE1}{}$ respectively.

Figure~\ref{fig:sim_combined} shows the distributions on $\delta\kappa_s$ obtained by considering all the events collectively.
Though most of the individual signals yielded poor constraints, the set is not completely uninformative: as can be seen from Fig.~\ref{fig:sim}, most of the posteriors have markedly stronger support in regions close to zero, even though they extend to the edge of the prior.
This is reflected by the combined results of Fig.~\ref{fig:sim_combined}, which disfavor large values of $|\delta\kappa_s|$.
The blue histogram represents the population-marginalized posterior obtained without assuming a unique value of $\delta\kappa_s$ across events, using the hierarchical approach of Sec.~\ref{sec:inference:populations}. 
With $90\%$ credibility, this analysis determines $\delta\kappa_s = \SimCombinedCI{HIER_POP}$, which indicates that the events considered are consistent with a population 
dominated by Kerr BBHs (within the given uncertainty).
The distribution hyperparameters are also consistent with the null-hypothesis ($\mu=\sigma=0$), with $\mu = \SimCombinedCI{HIER_MU}$ and $\sigma < \SimCombinedCI{HIER_SIGMA}$.
Both $\mu$ and the population-marginalized posterior of Fig.~\ref{fig:sim_combined} inherit the asymmetry of the individual events, which tend to be skewed towards $\delta\kappa_s < 0$ (cf.~inset in Fig.~\ref{fig:sim}); this suggests that negative values of $\delta\kappa_s$ are harder to constrain.
Conditional on positive values, the generic population results constrain $\delta\kappa_s < \SimCombinedCI{HIER_POP_POS}$.

The red curve in Fig.~\ref{fig:sim_combined} represents the joint-likelihood posterior obtained by restricting $\kappa_s$ to take the same value for all the events.
Under that assumption, we find $\delta\kappa_s = \SimCombinedCI{SIMPLE_POP}$ and, conditional on positive values, $\delta\kappa_s < \SimCombinedCI{SIMPLE_POP_POS}$.
The hypothesis that all of the events considered are Kerr BBHs ($\delta\kappa_s=0$) is preferred over an alternative proposal that all of them are not with a shared $\delta\kappa_s \neq 0$, with a log Bayes factor of $\log_{10} \mathcal{B}^{\rm Kerr}_{\delta\kappa_s\,\neq\,0} = \SimBF{SYM}{}$, or log Bayes factor of $\log_{10} \mathcal{B}^{\rm Kerr}_{\delta\kappa_s\,\geq\,0} = \SimBF{POS}{}$ if only allowing $\delta\kappa_s \geq 0$.

\section{Tests of gravitational wave propagation}
\label{sec:liv}
In GR, GWs far from their source propagate along null geodesics, with energy $E$ and momentum $p$ related by the dispersion
relation $E^2 = p^2c^2$, where $c$ is the speed of light.
Extensions to GR may violate this in several ways, e.g., by endowing the graviton with a mass.
To probe generalized dispersion relations, we adopt the common phenomenological modification to GR introduced
in \cite{Mirshekari:2011yq} and applied to LIGO and Virgo data in \cite{Abbott:2017vtc,LIGOScientific:2019fpa}:
\begin{equation}
\label{eq:liv:dispersion}
E^2 = p^2c^2 + A_\alpha p^\alpha c^\alpha\, ,
\end{equation} 
where $A_\alpha$ and $\alpha$ are phenomenological parameters, and 
GR is recovered if $A_\alpha = 0$ for all $\alpha$.
To leading order, Eq.~\eqref{eq:liv:dispersion} may encompass a variety of predictions from different extensions to GR
\cite{Mirshekari:2011yq,Yunes:2016jcc,Will:1997bb,Calcagni:2009kc,AmelinoCamelia:2002wr,
Horava:2009uw,Sefiedgar:2010we,Kostelecky:2016kfm};
this includes massive gravity for $\alpha = 0$ and $A_\alpha > 0$, with a graviton mass $m_g = A_0^{1/2} c^{-2}$
\cite{Will:1997bb}.
As in \cite{LIGOScientific:2019fpa}, we consider $\alpha$ values from $0$ to $4$ in steps of $0.5$,
excluding $\alpha = 2$, which is degenerate with an overall time delay.
A nonzero $A_\alpha$ manifests itself in the data as a frequency-dependent dephasing of the GW signal, which builds up as the wave propagates towards Earth and hence increases with the source comoving distance, potentially enhancing weak GR deviations.

\begin{table*}
\caption{\label{tab:liv:results_summary}
Results for the modified dispersion analysis (Sec.~\ref{sec:liv}).
The table shows 90\%-credible upper bounds on the graviton mass $m_g$ and the absolute value of the modified dispersion relation parameter $A_\alpha$, as well as the GR quantiles $Q_\text{GR}$.
The $<$ and $>$ labels denote the upper bound on $|A_\alpha|$ when assuming $A_\alpha < 0$ and $>0$, respectively, and $\bar{A}_\alpha = A_\alpha/\mathrm{eV}^{2-\alpha}$ is dimensionless. 
Rows compare the GWTC-1 results from \cite{LIGOScientific:2019fpa} to the GWTC-2 results.
}
\centering
\resizebox{\textwidth}{!}{\begin{tabular}{ccc*{7}{cccc}ccc}
 \toprule
 & $m_g$ & & \threec{$|\bar{A}_0|$} & & \threec{$|\bar{A}_{0.5}|$} & & \threec{$|\bar{A}_1|$} & & \threec{$|\bar{A}_{1.5}|$} & & \threec{$|\bar{A}_{2.5}|$} & & \threec{$|\bar{A}_3|$} & & \threec{$|\bar{A}_{3.5}|$} & & \threec{$|\bar{A}_4|$}\\
 \cline{4-6}
 \cline{8-10}
 \cline{12-14}
 \cline{16-18}
 \cline{20-22}
 \cline{24-26}
 \cline{28-30}
 \cline{32-34}
  & [$10^{-23}$ & & $<$ & $>$ & $Q_\text{GR}$ & & $<$ & $>$ & $Q_\text{GR}$ & & $<$ & $>$ & $Q_\text{GR}$ & & $<$ & $>$ & $Q_\text{GR}$ & & $<$ & $>$ & $Q_\text{GR}$ & & $<$ & $>$ & $Q_\text{GR}$ & & $<$ & $>$ & $Q_\text{GR}$ & & $<$ & $>$ & $Q_\text{GR}$\\
 &  eV$/c^2$] & & \twoc{[$10^{-45}$]} & [\%] & & \twoc{[$10^{-38}$]} & [\%] & & \twoc{[$10^{-32}$]} & [\%] & & \twoc{[$10^{-26}$]} & [\%] & & \twoc{[$10^{-14}$]} & [\%] & & \twoc{[$10^{-8}$]} & [\%] & & \twoc{[$10^{-2}$]} & [\%] & & \twoc{[$10^{4}$]} & [\%]\\
 \midrule
GWTC-1 & 4.70 &   & 7.99 & 3.39 &  79 &   & 1.17 & 0.70 &  73 &   & 2.51 & 1.21 &  70 &   & 6.96 & 3.70 &  86 &   & 5.05 & 8.01 &  28 &   & 2.94 & 3.66 &  25 &   & 2.01 & 3.73 &  35 &   & 1.44 & 2.34 &  34 \\
{\bf GWTC-2} & 3.09 &   & 1.75 & 1.37 &  66 &   & 0.46 & 0.28 &  66 &   & 1.00 & 0.52 &  79 &   & 3.35 & 1.47 &  83 &   & 1.74 & 2.43 &  31 &   & 1.08 & 2.17 &  17 &   & 0.76 & 1.57 &  12 &   & 0.64 & 0.88 &  25 \\
\bottomrule
\end{tabular}
}
\end{table*}

The analysis makes use of a modified version of the \IMRP{} waveform (checks for systematics using \SEOBHMROM{} were presented in \cite{LIGOScientific:2019fpa}).
We use Eq.~(3) of \cite{LIGOScientific:2019fpa} to compute the dephasing for a given $A_\alpha$.%
\footnote{There was a typographic error in Eq.~(4) of Ref.~\cite{LIGOScientific:2019fpa}: the $1/(\alpha - 2)$ exponent should instead be $1/(2 - \alpha)$.}
This expression was derived in \cite{Mirshekari:2011yq} by treating waves emitted at a given time as particles that travel at the particle velocity $v_p = pc^2/E$ associated with the wave's instantaneous frequency.
Different dephasings can arise from different prescriptions, e.g., using the group velocity instead, but the corresponding bound on $A_\alpha$ can be obtained by rescaling with an appropriate factor in most cases.
See discussions after Eq.~(5) in \cite{LIGOScientific:2019fpa} for details.

\begin{figure}
  \centering
  \includegraphics[width=\columnwidth]{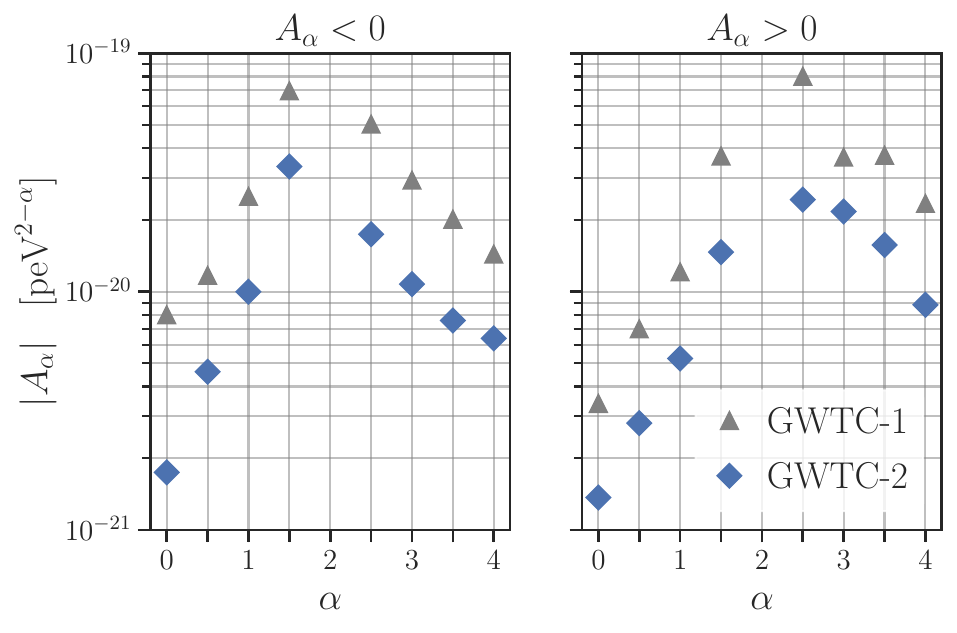}
\caption{90\% credible upper bounds on the absolute value of the modified dispersion relation parameter $A_\alpha$. 
The upper limits are derived from the distributions in Fig.~\ref{fig:liv:results_violin}, treating the positive and negative values of $A_\alpha$ separately. Picoelectronvolts provide a convenient scale 
because $1~\mathrm{peV} \simeq h\times 250~\mathrm{Hz}$, where $250~\mathrm{Hz}$ is close to the most sensitive frequencies of the LIGO and Virgo instruments.
Marker style distinguishes the new GWTC-2 results from the previous GWTC-1 results in \cite{LIGOScientific:2019fpa}.}
  \label{fig:liv:results_summary}
\end{figure}

We assume priors flat in $A_\alpha$ except when reporting the mass of the graviton, where we use a prior flat in that quantity.
We analyze \LivEvents{GWTC-2}{} events from GWTC-2 satisfying our FAR threshold (see Sec.~\ref{sec:events} and Table \ref{tab:events}).%
\footnote{We do not consider \NAME{GW190521A}{} because we were unable to obtain well-converged results for that event without using HMs, which are not yet implemented for this test. We have analyzed \NAME{GW190412A}{} and \NAME{GW190814A}{} without HMs, despite evidence that HMs contribute to those signals. However, we have checked that this does not bias the results through an injection study for \NAME{GW190412A}{} and $\alpha = 0$. We have also confirmed that excluding \NAME{GW190412A}{} and \NAME{GW190814A}{} would affect the combined results in Table \ref{tab:events} by only ${\sim}5\%$ on average (12\% in the worst case).}
Since we can take $A_\alpha$ and $m_g$ to be universal parameters, results from different events can be easily combined by multiplying the individual likelihoods.
Although we only discuss the overall combined results here, individual-event posteriors are available in \cite{GWTC2:TGR:release}, as for other tests.

\begin{figure}
\includegraphics[width=\columnwidth]{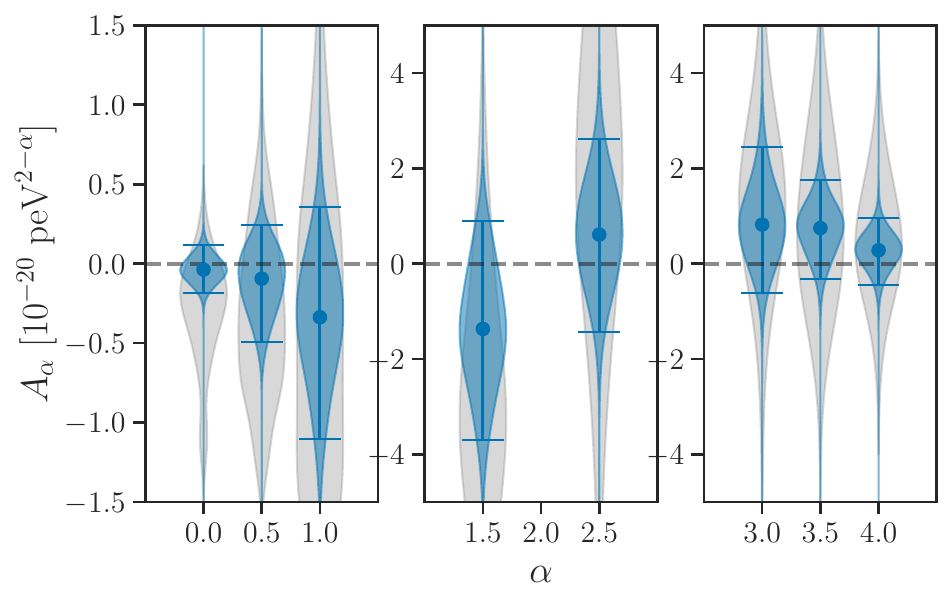}
\caption{Violin plots of the full posteriors on the modified dispersion relation parameter $A_\alpha$ calculated from the GWTC-2 events (blue), with the $90\%$ credible interval around the median indicated. For comparison, we also show the GWTC-1 previous measurement (gray), reported in \cite{LIGOScientific:2019fpa}.}
\label{fig:liv:results_violin}
\end{figure}

We show our results in Table~\ref{tab:liv:results_summary} and Figs.~\ref{fig:liv:results_summary} and \ref{fig:liv:results_violin}.
Table~\ref{tab:liv:results_summary} and Fig.~\ref{fig:liv:results_summary} present constraints on the allowed amount of dispersion through the 90\%-credible upper limits on $|A_\alpha|$, computed separately for $A_\alpha >0$ and $A_\alpha < 0$.
There is noticeable improvement when combining GWTC-2 results with respect to the previous result in \cite{LIGOScientific:2019fpa}.
This is the case for both positive and negative amplitudes, meaning that we are more tightly constraining these quantities closer to the nondispersive, GR prediction ($A_\alpha = 0$).
The average improvement in the $|A_\alpha|$ upper limits relative to \cite{LIGOScientific:2019fpa} is a factor $\LivImprov{AMP}$, although this fluctuates slightly across choices of $\alpha$.
Overall, this is consistent with the factor of $\sqrt{\LivEvents{GWTC-2}/\LivEvents{GWTC-1}} \approx \LivImprov{EXP}$ naively expected from the increase in the number of events analyzed.%
\footnote{We have analyzed 8 events from GWTC-1, one more than for the combined results in \cite{LIGOScientific:2019fpa} because those excluded GW170818.}%

Upper limits on the $A_\alpha$ parameters can be uncertain due to the difficulty in accurately sampling the long tails of the posteriors.
To quantify this uncertainty, we follow a Bayesian bootstrapping procedure~\cite{rubin1981}, as done previously in \cite{Abbott:2017vtc,LIGOScientific:2019fpa},
with $2000$ bootstrap realizations for each value of $\alpha$ and sign of $A_\alpha$.
We find that the average width of the 90\%-credible interval of the individual-event upper limits is a factor of \LivBootstrapFracWidthRatioAvg{} of the reported upper limit itself, i.e., the average uncertainty in the upper limit is \LivBootstrapFracWidthRatioAvg{}.
Out of all upper limits, \LivBootstrapNumFracWidthRatioGreaterThanPointFive{} carry fractional uncertainties larger than $0.5$.
The most uncertain upper limit is that for \LivBootstrapFracWidthRatioMaxName{} and $A_{\LivBootstrapFracWidthRatioMaxAlpha{}} \LivBootstrapFracWidthRatioMaxSign{}$, with a fractional uncertainty of \LivBootstrapFracWidthRatioMax{}.

Figure~\ref{fig:liv:results_violin} shows the overall posterior obtained for negative and positive values of $A_\alpha$.
The enhanced stringency of our measurements relative to our previous GWTC-1 results is also visible here, as seen in the smaller size of the blue violins with respect to the gray, and the fact that the medians (blue circles) are generally closer to the GR value.
The latter is also manifested in the GR quantiles $Q_{\rm GR} = P(A_\alpha < 0)$ in Table~\ref{tab:liv:results_summary}, which tend to be closer to 50\% ($Q_{\rm GR}=50\%$ implies the distribution is centered on the GR value).

From our combined GWTC-2 data, we bound the graviton mass to be $m_g \leq \LivMgUL \mathrm{eV}/c^2$, with 90\% credibility (Table~\ref{tab:liv:results_summary}).
This represents an improvement of a factor of $\LivImprov{MG}{}$ relative to \cite{LIGOScientific:2019fpa}.
The new measurement is comparable to the most recent Solar System bound of $ \LivMgSolarUL \text{ eV}/c^2$, also with 90\% credibility  \cite{Bernus:2020szc}.

\section{Remnant properties}
\label{sec:rem}

    \subsection{Ringdown}
    \label{sec:rin}
In GR, the remnant object resulting from the coalescence of two astrophysical BHs is a perturbed Kerr BH. This remnant BH will gradually relax to its Kerr stationary state by emitting GWs corresponding to a specific set of characteristic quasi-normal modes (QNMs), whose frequency $f$ and damping time $\tau$ depend solely on the BH mass $M_\mathrm{f}$ and the dimensionless spin $\chi_\mathrm{f}$. This last stage of the coalescence is known as \emph{ringdown}.
The description of the ringdown stage is based on the final state conjecture \cite{Penrose:1969pc,Penrose:2002col,2002nmgm.meet...28K, Chrusciel:2012jk} stating that the physical spectrum of QNMs is exclusively determined by the final BH mass and spin (the no-hair conjecture \cite{Zeldovich2,Israel,Carter,Hawking1972,Robinson,Mazur,Bunting, Dafermos:2008en}) and that the Kerr solution is an attractor of BH spacetimes in astrophysical scenarios.\footnote{In principle such frequencies and damping times would also depend on the electric charge of the remnant BH. However, for astrophysically relevant scenarios the BH charge is expected to be negligible \cite{Gibbons:1975kk,Blandford:1977ds,1982PhRvD..25.2509H}.}

By analyzing the postmerger signal from a BBH coalescence independently of the preceding inspiral, we can verify the final state conjecture, test the nature of the remnant object (complementary to the searches for GW echoes discussed in Sec.~\ref{sec:ech}), and estimate directly the remnant mass and spin assuming it is a Kerr BH---which, in turn, allows us to test GR's prediction for the energy and angular momentum emitted during the coalescence (complementary to the IMR consistency test discussed in Sec.~\ref{sec:imr}, and the postinspiral parameters in Sec.~\ref{sec:par}). This set of analyses is referred to as \emph{BH spectroscopy} \cite{Detweiler:1980gk, Dreyer:2003bv, LISA_spectroscopy, Gossan:2011ha, PhysRevD.90.064009, Carullo:2018sfu, Brito:2018rfr, Carullo:2019flw, Isi:2019aib, Bhagwat:2019bwv, Bhagwat:2019dtm,Cabero:2019zyt}. Unlike the IMR consistency test, a ringdown-only analysis is not contaminated by frequency mixing with other phases of the signal and it does not require a large amount of SNR in the inspiral regime (the lack of such SNR is why the IMR consistency test was unable to be applied to \NAME{GW190521A}{} \cite{GW190521g,GW190521g:imp}, for instance).

The complex-valued GW waveform during ringdown can be expressed as a superposition of damped sinusoids:
\begin{widetext}
\begin{equation}\label{eq:ring_wf}
\begin{aligned}
	h_{+}(t) - i h_{\times}(t) = \sum_{\ell = 2}^{+\infty} \sum_{m = - \ell}^{\ell} \sum_{n = 0}^{+\infty} \; & \; \mathcal{A}_{\ell m n} \; \exp \left[ -\frac{t-t_0}{(1+z)\tau_{\ell m n}} \right] \exp \left[ -\frac{2\pi i f_{\ell m n}(t-t_0)}{1+z} \right] {}_{-2}S_{\ell m n}(\theta, \phi, \chi_{\rm f}), \\
\end{aligned}
\end{equation}
\end{widetext}
where $z$ is the cosmological redshift, and the $(\ell,m,n)$ indices label the QNMs. The angular multipoles are denoted by $\ell$ and $m$, while $n$ orders modes of a given $(\ell,m)$ by decreasing damping time.
The frequency and the damping time for each ringdown mode can be computed for a perturbed isolated BH as a function of its mass $M_\mathrm{f}$ and spin $\chi_\mathrm{f}$ \cite{Vishveshwara1970b, Press1971, Teukolsky:1973ha, Chandrasekhar:1975zza}. For each $(\ell,m,n)$, there are in principle two associated frequencies and damping times: those for a \emph{prograde} mode, with $\mathrm{sgn}(f_{\ell m n}) = \mathrm{sgn}(m)$, and those for a \emph{retrograde} mode, with $\mathrm{sgn}(f_{\ell m n}) \neq \mathrm{sgn}(m)$---retrograde modes are not expected to be relevant \cite{LISA_spectroscopy}, so we do not include them in Eq.~\eqref{eq:ring_wf}. The frequency and damping time of the $+|m|$ mode are related to those of the $-|m|$ mode by $f_{\ell m n} = - f_{\ell -m n}$ and $\tau_{\ell m n} = \tau_{\ell -m n}$ for $m \neq 0$.
The complex amplitudes $\mathcal{A}_{\ell mn}$ characterize the excitation and the phase of each ringdown mode at a reference time $t_0$, which for a BBH merger can be predicted from numerical simulations \cite{Kamaretsos:2012bs, PhysRevD.90.124032, MMRDNP}. 
In general, $\mathcal{A}_{\ell mn}$ is independent of $\mathcal{A}_{\ell -m n}$.

The angular dependence of the GW waveform is contained in the spin-weighted spheroidal harmonics ${}_{-2}S_{\ell m n}(\theta, \phi, \chi_{\rm f})$, where $\theta, \phi$ are the polar and azimuthal angles in a frame centered on the remnant BH and aligned with its angular momentum.
We approximate these functions through the spin-weighted spherical harmonics ${}_{-2}Y_{\ell m}(\theta, \phi)$, which introduces mode-mixing between QNMs with the same $m$ index but different $\ell$ indices \cite{Buonanno:2006ui,Kelly:2012nd,Berti:2014fga}.
Except in one case, as indicated below, models in this section do not account for this effect.
However, mode-mixing is expected to be negligible for the modes we consider, in particular for the dominant $\ell=|m|=2$ mode \cite{Buonanno:2006ui,Kelly:2012nd,Berti:2014fga}.

We present results from two approaches: a time-domain ringdown analysis \textsc{pyRing} \cite{Carullo:2019flw,Isi:2019aib}, and a parametrized version of an aligned-spin EOB waveform model with HMs called \textsc{pSEOBNRv4HM} \cite{Brito:2018rfr, Cotesta:2018fcv}.

\subsubsection{The \textsc{pyRing} analysis}
The \textsc{pyRing} analysis infers the remnant BH parameters based on the ringdown part of a signal alone. The analysis is completely formulated in the time domain \cite{Carullo:2019flw,Isi:2019aib} for both the likelihood function and waveform templates, hence avoiding spectral leakage from previous stages of the coalescence as would arise in a frequency-domain analysis when Fourier transforming a template with an abrupt start \cite{Carullo:2019flw,Isi:2019aib,Cabero:2017avf}.
We employ four different waveform templates, each constructed with different sets of assumptions in order to obtain agnostic measurements of the QNM frequencies and damping times, and to explore the contribution of modes other than the least damped mode ($n=0$).

\begin{table*}
\caption{\label{tab:rin:final_mass_spin}The median, and symmetric $90\%$-credible intervals, of the redshifted final mass and final spin, inferred from the full IMR analysis (IMR) and the \textsc{pyRing} analysis with three different waveform models ($\mathrm{Kerr_{220}}$, $\mathrm{Kerr_{221}}$, and $\mathrm{Kerr}_\mathrm{HM}$). We quantify the contribution of the HMs using log Bayes factors $\log_{10} \mathcal{B}^{\rm HM}_{\rm 220}$, where a positive value reflects the presence of HMs in the data. Similarly, we quantify the contribution of the first overtone using log Bayes factors $\log_{10} \mathcal{B}^{\rm 221}_{\rm 220}$, where a positive value reflects the presence of the first overtone in the data. We also quantify the level of agreement with GR for each event using log odds ratios $\log_{10} \mathcal{O}^{\rm modGR}_{\rm GR}$ comparing the generic modified-GR hypothesis with GR. The catalog-combined log odds ratio is slightly negative ($\pyRingCombinedTIGERLogOddsRatio{}$), and the log odds ratios for individual events are also inconclusive, showing no evidence that the Kerr metric is insufficient.}
\scalebox{0.95}{

\begin{tabular}{lllllllllllrrrr}
\toprule
Event & \multicolumn{4}{c}{Redshifted final mass} & \hphantom{X} & \multicolumn{4}{c}{Final spin} & \hphantom{X} & \multicolumn{1}{c}{Higher} & \hphantom{X} & \multicolumn{2}{c}{Overtones} \\
& \multicolumn{4}{c}{$(1+z)M_\mathrm{f} \; [\msun{}]$} & \hphantom{X} & \multicolumn{4}{c}{$\chi_{\mathrm{f}}$} & \hphantom{X} & \multicolumn{1}{c}{modes} & \hphantom{X} &  \multicolumn{2}{c}{} \\[0.075cm]
\cline{2-5}
\cline{7-10}
\cline{12-12}
\cline{14-15}
& IMR & $\mathrm{Kerr_{220}}$ & $\mathrm{Kerr_{221}}$ & $\mathrm{Kerr_{HM}}$ & \hphantom{X} & IMR & $\mathrm{Kerr_{220}}$ & $\mathrm{Kerr_{221}}$ & $\mathrm{Kerr_{HM}}$ & \hphantom{X} &  \multicolumn{1}{c}{$\log_{10} \mathcal{B}^{\rm HM}_{\rm 220}$} & \hphantom{X} & \multicolumn{1}{c}{$\log_{10} \mathcal{B}^{\rm 221}_{\rm 220}$} & \multicolumn{1}{c}{$\log_{10} \mathcal{O}^{\rm modGR}_{\rm GR}$} \\
\midrule

GW150914 &
$ 68.8^{+ 3.6 }_{- 3.1 } $ &
$ 59.5^{+ 19.7 }_{- 10.6 } $ &
$ 67.3^{+ 15.8 }_{- 13.3 } $ &
$ 220.2^{+ 180.4 }_{- 144.5 } $ &
\hphantom{X} &
$ 0.69^{+ 0.05 }_{- 0.04 } $ &
$ 0.44^{+ 0.39 }_{- 0.39 } $ &
$ 0.61^{+ 0.24 }_{- 0.43 } $ &
$ 0.60^{+ 0.23 }_{- 0.47 } $ &
\hphantom{X} & 
$ 0.01 $ &
\hphantom{X} &
$ 0.15 $ &
$ -0.22 $ \\[0.075cm]

GW170104 &
$ 58.5^{+ 4.6 }_{- 4.1 } $ &
$ 61.8^{+ 20.8 }_{- 14.1 } $ &
$ 61.5^{+ 20.0 }_{- 12.8 } $ &
$ 214.1^{+ 179.9 }_{- 126.7 } $ &
\hphantom{X} &
$ 0.66^{+ 0.08 }_{- 0.11 } $ &
$ 0.27^{+ 0.42 }_{- 0.24 } $ &
$ 0.47^{+ 0.41 }_{- 0.41 } $ &
$ 0.58^{+ 0.25 }_{- 0.48 } $ &
\hphantom{X} & 
$ 0.05 $ &
\hphantom{X} &
$ -0.66 $ &
$ -0.11 $ \\[0.075cm]

GW170814 &
$ 59.7^{+ 3.0 }_{- 2.3 } $ &
$ 74.1^{+ 326.9 }_{- 58.1 } $ &
$ 56.2^{+ 41.2 }_{- 13.2 } $ &
$ 188.6^{+ 126.5 }_{- 120.9 } $ &
\hphantom{X} &
$ 0.72^{+ 0.07 }_{- 0.05 } $ &
$ 0.52^{+ 0.42 }_{- 0.47 } $ &
$ 0.54^{+ 0.37 }_{- 0.47 } $ &
$ 0.61^{+ 0.23 }_{- 0.53 } $ &
\hphantom{X} & 
$ 0.01 $ &
\hphantom{X} &
$ -0.69 $ &
$ -0.05 $ \\[0.075cm]

GW170823 &
$ 88.8^{+ 11.2 }_{- 10.2 } $ &
$ 73.5^{+ 29.1 }_{- 16.5 } $ &
$ 83.6^{+ 40.6 }_{- 16.1 } $ &
$ 136.5^{+ 137.9 }_{- 69.7 } $ &
\hphantom{X} &
$ 0.72^{+ 0.09 }_{- 0.12 } $ &
$ 0.47^{+ 0.39 }_{- 0.41 } $ &
$ 0.45^{+ 0.45 }_{- 0.40 } $ &
$ 0.65^{+ 0.21 }_{- 0.54 } $ &
\hphantom{X} & 
$ 0.04 $ &
\hphantom{X} &
$ -1.17 $ &
$ 0.07 $ \\[0.075cm]

GW190408\_181802 &
$ 53.0^{+ 3.2 }_{- 3.4 } $ &
$ 201.7^{+ 280.5 }_{- 188.4 } $ &
$ 51.0^{+ 26.0 }_{- 14.6 } $ &
$ 122.0^{+ 164.2 }_{- 76.1 } $ &
\hphantom{X} &
$ 0.67^{+ 0.06 }_{- 0.07 } $ &
$ 0.42^{+ 0.49 }_{- 0.38 } $ &
$ 0.51^{+ 0.39 }_{- 0.45 } $ &
$ 0.55^{+ 0.29 }_{- 0.56 } $ &
\hphantom{X} & 
$ 0.02 $ &
\hphantom{X} &
$ -1.66 $ &
$ 0.00 $ \\[0.075cm]

GW190512\_180714 &
$ 43.5^{+ 4.0 }_{- 2.8 } $ &
$ 72.5^{+ 58.4 }_{- 44.5 } $ &
$ 41.3^{+ 60.9 }_{- 13.4 } $ &
$ 108.1^{+ 107.2 }_{- 35.0 } $ &
\hphantom{X} &
$ 0.65^{+ 0.07 }_{- 0.07 } $ &
$ 0.53^{+ 0.40 }_{- 0.48 } $ &
$ 0.46^{+ 0.39 }_{- 0.41 } $ &
$ 0.70^{+ 0.16 }_{- 0.51 } $ &
\hphantom{X} & 
$ 0.01 $ &
\hphantom{X} &
$ -1.00 $ &
$ -0.10 $ \\[0.075cm]

GW190513\_205428 &
$ 70.6^{+ 11.5 }_{- 6.7 } $ &
$ 62.7^{+ 364.0 }_{- 36.3 } $ &
$ 71.4^{+ 23.5 }_{- 12.5 } $ &
$ 129.3^{+ 159.1 }_{- 75.2 } $ &
\hphantom{X} &
$ 0.68^{+ 0.14 }_{- 0.12 } $ &
$ 0.41^{+ 0.46 }_{- 0.37 } $ &
$ 0.34^{+ 0.44 }_{- 0.31 } $ &
$ 0.57^{+ 0.27 }_{- 0.55 } $ &
\hphantom{X} & 
$ 0.05 $ &
\hphantom{X} &
$ -0.89 $ &
$ -0.10 $ \\[0.075cm]

GW190519\_153544 &
$ 146.8^{+ 14.7 }_{- 15.4 } $ &
$ 124.9^{+ 37.9 }_{- 26.1 } $ &
$ 128.4^{+ 31.9 }_{- 25.5 } $ &
$ 181.8^{+ 59.2 }_{- 49.0 } $ &
\hphantom{X} &
$ 0.79^{+ 0.07 }_{- 0.13 } $ &
$ 0.52^{+ 0.33 }_{- 0.45 } $ &
$ 0.56^{+ 0.26 }_{- 0.42 } $ &
$ 0.72^{+ 0.14 }_{- 0.31 } $ &
\hphantom{X} & 
$ 0.11 $ &
\hphantom{X} &
$ -0.03 $ &
$ -0.27 $ \\[0.075cm]

GW190521 &
$ 256.6^{+ 36.6 }_{- 30.4 } $ &
$ 292.5^{+ 50.4 }_{- 58.9 } $ &
$ 296.9^{+ 41.4 }_{- 40.2 } $ &
$ 303.4^{+ 29.9 }_{- 41.6 } $ &
\hphantom{X} &
$ 0.71^{+ 0.12 }_{- 0.16 } $ &
$ 0.79^{+ 0.12 }_{- 0.31 } $ &
$ 0.82^{+ 0.09 }_{- 0.16 } $ &
$ 0.81^{+ 0.07 }_{- 0.17 } $ &
\hphantom{X} & 
$ 0.28 $ &
\hphantom{X} &
$ -0.06 $ &
$ -0.37 $ \\[0.075cm]

GW190521\_074359 &
$ 88.0^{+ 4.3 }_{- 4.8 } $ &
$ 87.7^{+ 24.4 }_{- 20.5 } $ &
$ 90.4^{+ 16.8 }_{- 16.6 } $ &
$ 101.8^{+ 147.9 }_{- 29.2 } $ &
\hphantom{X} &
$ 0.72^{+ 0.05 }_{- 0.07 } $ &
$ 0.66^{+ 0.25 }_{- 0.53 } $ &
$ 0.72^{+ 0.16 }_{- 0.33 } $ &
$ 0.77^{+ 0.11 }_{- 0.54 } $ &
\hphantom{X} & 
$ 0.04 $ &
\hphantom{X} &
$ 0.04 $ &
$ -0.35 $ \\[0.075cm]

GW190602\_175927 &
$ 163.8^{+ 20.7 }_{- 18.3 } $ &
$ 151.2^{+ 67.9 }_{- 29.0 } $ &
$ 173.2^{+ 50.1 }_{- 37.6 } $ &
$ 258.7^{+ 53.9 }_{- 70.7 } $ &
\hphantom{X} &
$ 0.70^{+ 0.10 }_{- 0.14 } $ &
$ 0.33^{+ 0.41 }_{- 0.30 } $ &
$ 0.58^{+ 0.25 }_{- 0.42 } $ &
$ 0.74^{+ 0.12 }_{- 0.28 } $ &
\hphantom{X} & 
$ 0.03 $ &
\hphantom{X} &
$ -0.87 $ &
$ -0.13 $ \\[0.075cm]

GW190706\_222641 &
$ 171.1^{+ 20.0 }_{- 23.7 } $ &
$ 142.3^{+ 53.6 }_{- 29.9 } $ &
$ 153.1^{+ 34.2 }_{- 28.6 } $ &
$ 209.9^{+ 151.1 }_{- 129.0 } $ &
\hphantom{X} &
$ 0.78^{+ 0.09 }_{- 0.18 } $ &
$ 0.46^{+ 0.40 }_{- 0.41 } $ &
$ 0.58^{+ 0.26 }_{- 0.43 } $ &
$ 0.63^{+ 0.22 }_{- 0.49 } $ &
\hphantom{X} & 
$ 0.03 $ &
\hphantom{X} &
$ 0.04 $ &
$ -0.36 $ \\[0.075cm]

GW190708\_232457 &
$ 34.4^{+ 2.7 }_{- 0.7 } $ &
$ 387.3^{+ 104.4 }_{- 370.2 } $ &
$ 35.1^{+ 30.9 }_{- 21.3 } $ &
$ 105.8^{+ 166.6 }_{- 61.0 } $ &
\hphantom{X} &
$ 0.69^{+ 0.04 }_{- 0.04 } $ &
$ 0.32^{+ 0.54 }_{- 0.29 } $ &
$ 0.37^{+ 0.47 }_{- 0.34 } $ &
$ 0.59^{+ 0.25 }_{- 0.58 } $ &
\hphantom{X} & 
$ 0.01 $ &
\hphantom{X} &
$ -1.25 $ &
$ 0.07 $ \\[0.075cm]

GW190727\_060333 &
$ 99.2^{+ 10.7 }_{- 9.8 } $ &
$ 84.8^{+ 47.8 }_{- 72.2 } $ &
$ 98.6^{+ 34.7 }_{- 22.2 } $ &
$ 108.9^{+ 120.0 }_{- 65.6 } $ &
\hphantom{X} &
$ 0.73^{+ 0.10 }_{- 0.10 } $ &
$ 0.59^{+ 0.34 }_{- 0.52 } $ &
$ 0.63^{+ 0.30 }_{- 0.54 } $ &
$ 0.60^{+ 0.24 }_{- 0.54 } $ &
\hphantom{X} & 
$ 0.02 $ &
\hphantom{X} &
$ -2.08 $ &
$ -0.20 $ \\[0.075cm]

GW190828\_063405 &
$ 75.7^{+ 6.0 }_{- 5.2 } $ &
$ 82.6^{+ 39.0 }_{- 32.4 } $ &
$ 82.3^{+ 24.3 }_{- 24.6 } $ &
$ 102.5^{+ 150.3 }_{- 45.9 } $ &
\hphantom{X} &
$ 0.75^{+ 0.06 }_{- 0.07 } $ &
$ 0.85^{+ 0.13 }_{- 0.67 } $ &
$ 0.84^{+ 0.13 }_{- 0.62 } $ &
$ 0.69^{+ 0.17 }_{- 0.60 } $ &
\hphantom{X} & 
$ 0.05 $ &
\hphantom{X} &
$ -1.51 $ &
$ -0.24 $ \\[0.075cm]

GW190910\_112807 &
$ 97.0^{+ 9.3 }_{- 7.1 } $ &
$ 118.3^{+ 28.6 }_{- 32.2 } $ &
$ 107.4^{+ 31.9 }_{- 27.0 } $ &
$ 154.0^{+ 98.9 }_{- 43.2 } $ &
\hphantom{X} &
$ 0.70^{+ 0.08 }_{- 0.07 } $ &
$ 0.80^{+ 0.15 }_{- 0.53 } $ &
$ 0.75^{+ 0.18 }_{- 0.48 } $ &
$ 0.79^{+ 0.09 }_{- 0.32 } $ &
\hphantom{X} & 
$ 0.19 $ &
\hphantom{X} &
$ -1.18 $ &
$ -0.05 $ \\[0.075cm]

GW190915\_235702 &
$ 74.8^{+ 7.9 }_{- 7.4 } $ &
$ 270.6^{+ 197.4 }_{- 257.2 } $ &
$ 72.0^{+ 246.4 }_{- 16.2 } $ &
$ 246.1^{+ 153.7 }_{- 122.6 } $ &
\hphantom{X} &
$ 0.70^{+ 0.09 }_{- 0.11 } $ &
$ 0.50^{+ 0.41 }_{- 0.45 } $ &
$ 0.42^{+ 0.46 }_{- 0.38 } $ &
$ 0.60^{+ 0.23 }_{- 0.42 } $ &
\hphantom{X} & 
$ 0.10 $ &
\hphantom{X} &
$ -1.43 $ &
$ 0.03 $ \\[0.075cm]

\bottomrule
\end{tabular}
}
\end{table*}

The $\mathrm{Kerr_{220}}$ template corresponds to the $\ell=|m|=2,\, n=0$ contribution (i.e., the $220$ mode) of Eq.~\eqref{eq:ring_wf}, where the frequencies and damping times are predicted as a function of $\left( M_\mathrm{f}, \chi_\mathrm{f} \right)$ by GR, while the complex amplitudes are kept as free parameters.
The remnant mass and spin were sampled with uniform priors.
The $\mathrm{Kerr_{221}}$ template is similar to $\mathrm{Kerr_{220}}$ but incorporates the first overtone ($n=1$) for $\ell=|m|=2$ in addition to the fundamental mode.
We do not consider a higher number of overtones since they are not expected to be relevant at current sensitivity \cite{Giesler:2019uxc,Isi:2019aib,Ota:2019bzl,Forteza:2020hbw}.
Uniform priors on the remnant mass and spin were also adopted.

The $\mathrm{Kerr}_\mathrm{HM}$ template includes all fundamental prograde modes with $\ell \leq 4$, with the angular dependence parametrized using spin-weighted spherical harmonics, taking into account mode-mixing
 \cite{MMRDNP}.
NR fits are used to compute amplitudes as a function of the initial binary parameters, and frequencies and damping times as a function of the remnant parameters where both the initial binary parameters and the remnant parameters are sampled independently with uniform priors.

We use as a reference time $t_0$, which is chosen based on an estimate of the peak of the strain $(h_+^2 + h_{\times}^2)$ from the full IMR analyses assuming GR.\footnote{For events in O1 and O2, the waveform approximant used in the full IMR analyses was \SEOBROM. As for events in O3a, the waveform approximant used in the full IMR analyses was \IMRP{}, except for \NAME{GW190521A}{}, where \NRSur{} was used instead.}
When overtones ($n>0$) are included, for instance in a $\mathrm{Kerr_{221}}$ template, we fit the data starting at $t_0$ itself \cite{Giesler:2019uxc, Isi:2019aib}, while for a $\mathrm{Kerr_{220}}$ template we start the fit $10 GM_\mathrm{f} (1+z)/c^3$ after $t_0$, which is when the least damped mode is expected to dominate the signal. Fits with $\mathrm{Kerr}_\mathrm{HM}$ templates, on the other hand, start at $15 GM_\mathrm{f} (1+z)/c^3$ after $t_0$. The sky locations and start times at each detector are released in \cite{GWTC2:TGR:release}.

We analyze all the GWTC-2 BBHs and report results for those events where the remnant parameters were constrained compared to the adopted prior bounds, and for which the Bayesian evidence favors the presence of a signal over pure Gaussian noise when using our most sensitive template ($\mathrm{Kerr_{221}}$). 
Estimates of the remnant parameters obtained through the three waveform templates ($\mathrm{Kerr_{220}}$, $\mathrm{Kerr_{221}}$, and $\mathrm{Kerr}_\mathrm{HM}$) are reported in Table \ref{tab:rin:final_mass_spin}. Fitting the data at an earlier time increases the SNR available when using this template, which is reflected in tighter constraints of the remnant parameters as shown in Table \ref{tab:rin:final_mass_spin} for the $\mathrm{Kerr_{221}}$ template. In all cases, at $90\%$ credibility, the two-dimensional $M_{\rm f}$--$\chi_{\rm f}$ posterior distributions from the three waveform templates include GR predictions coming from the full IMR analyses \cite{GWTC2}.
For \NAME{GW190521A}{}, the results reported in \cite{GW190521g, GW190521g:imp} are not identical to the ones reported here as the previous analyses did not include the negative-$m$ mode, and we have updated to use a more precise value for the reference time. The lower frequency cut-off for this event was also changed from $20~\mathrm{Hz}$ to $11~\mathrm{Hz}$. None of the conclusions previously reported for \NAME{GW190521A}{} are affected by these small changes.

We use log Bayes factors to quantify the contribution of overtones or HMs during ringdown. In Table~\ref{tab:rin:final_mass_spin}, we report the log Bayes factors $\log_{10} \mathcal{B}^{\rm HM}_{\rm 220}$ comparing a fit with all modes in $\mathrm{Kerr}_\mathrm{HM}$, versus one with only the $\ell= |m| =2,\, n=0$ mode; this computation provides no strong evidence for the presence of HMs.
We also present the log Bayes factors $\log_{10} \mathcal{B}^{\rm 221}_{\rm 220}$ comparing the results obtained when fitting the full postmerger signal using the $n=0,1$ modes against the template including the $n=0$ mode only, with both templates starting at the reference time $t_0$.
The data show evidence for the presence of overtones only for loud signals (for example \NAME{GW190521B}{} shows such evidence), although in all cases estimates of the remnant parameters tend to get closer to the full IMR waveform estimates when including overtones.

To achieve a test of the final state conjecture and quantify the level of agreement with GR, we modify the $\mathrm{Kerr_{221}}$ template to allow for fractional deviations in the frequency and damping time with respect to their GR predictions for the $221$ mode (the first overtone). Meanwhile, the frequency and the damping time of the better-measured $220$ mode remain the same as their GR predictions as functions of the remnant mass $M_{\mathrm f}$ and spin $\chi_{\mathrm f}$ to help constrain the remnant properties. This approach, compared to allowing for deviations in the fundamental mode, has the advantage of lowering the impact of priors on the remnant mass and spin recovery, as well as the impact of correlations among the deviation parameters and the remnant parameters.
We sample over the regular Kerr parameters and the fractional deviations with uniform priors in the $[-1,1]$ range for the frequency $\delta \hat{f}_{221}$ and in the $[-0.9, 1]$ range for the damping time $\delta \hat{\tau}_{221}$.\footnote{The lower prior bound on the damping time deviation is set by the discrete analysis time resolution.} The posteriors on the fractional deviations quantify the agreement of the $221$ mode with the Kerr prediction.

Additionally, we may follow \cite{PhysRevD.90.064009,Li:2011cg, Li:2011vx} to compute a log odds ratio $\log_{10} \mathcal{O}^{\rm modGR}_{\rm GR}$ for deviations from the Kerr ringdown.
We define the baseline GR hypothesis $\mathcal{H}_{\rm GR}$ to be the proposition that both the fractional deviation parameters vanish, i.e.,~$\delta \hat{f}_{221} = \delta \hat{\tau}_{221} = 0$. Similarly, we define the modified GR hypothesis $\mathcal{H}_{\rm modGR}$ to be the proposition that \emph{at least one} of the fractional deviation parameters is non-zero, with the priors above.
We may construct $\mathcal{H}_{\rm modGR}$ from three sub-hypotheses, which we label $\mathcal{H}_{\delta \hat{f}_{221}}$, $\mathcal{H}_{\delta \hat{\tau}_{221}}$, and $\mathcal{H}_{\delta \hat{f}_{221},\;\delta \hat{\tau}_{221}}$.
For $\mathcal{H}_{\delta \hat{f}_{221}}$, we write the frequencies and damping times for the $220$ and the $221$ mode as
\begin{equation}
\label{eq:rin_H_f_221}
\mathcal{H}_{\delta \hat{f}_{221}} \equiv
\begin{cases}
	\; f_{220} & = f_{220}^{\text{GR}}(M_{\mathrm f}, \chi_{\mathrm f}) \\
	\; \tau_{220} & = \tau_{220}^{\text{GR}}(M_{\mathrm f}, \chi_{\mathrm f}) \\
	\; f_{221} & = f_{221}^{\text{GR}}(M_{\mathrm f}, \chi_{\mathrm f}) (1 + \delta \hat{f}_{221}) \\
	\; \tau_{221} & = \tau_{221}^{\text{GR}}(M_{\mathrm f}, \chi_{\mathrm f}) \\	
\end{cases}\, ,
\end{equation}
where the ``GR'' superscript indicates the Kerr value corresponding to a given $M_{\rm f}$ and $\chi_{\rm f}$.
Similarly, for $\mathcal{H}_{\delta \hat{\tau}_{221}}$, we write the frequencies and damping times for the $220$ and the $221$ mode as
\begin{equation}
\label{eq:rin_H_tau_221}
\mathcal{H}_{\delta \hat{\tau}_{221}} \equiv
\begin{cases}
	\; f_{220} & = f_{220}^{\text{GR}}(M_{\mathrm f}, \chi_{\mathrm f}) \\
	\; \tau_{220} & = \tau_{220}^{\text{GR}}(M_{\mathrm f}, \chi_{\mathrm f}) \\
	\; f_{221} & = f_{221}^{\text{GR}}(M_{\mathrm f}, \chi_{\mathrm f}) \\
	\; \tau_{221} & = \tau_{221}^{\text{GR}}(M_{\mathrm f}, \chi_{\mathrm f}) (1 + \delta \hat{\tau}_{221})  \\	
\end{cases}\, .
\end{equation}
Finally, for $\mathcal{H}_{\delta \hat{f}_{221},\;\delta \hat{\tau}_{221}}$, we again write the frequencies and damping times as
\begin{equation}
\label{eq:rin_H_f_tau_221}
\mathcal{H}_{\delta \hat{f}_{221},\;\delta \hat{\tau}_{221}} \equiv
\begin{cases}
	\; f_{220} & = f_{220}^{\text{GR}}(M_{\mathrm f}, \chi_{\mathrm f}) \\
	\; \tau_{220} & = \tau_{220}^{\text{GR}}(M_{\mathrm f}, \chi_{\mathrm f}) \\
	\; f_{221} & = f_{221}^{\text{GR}}(M_{\mathrm f}, \chi_{\mathrm f})(1 + \delta \hat{f}_{221}) \\
	\; \tau_{221} & = \tau_{221}^{\text{GR}}(M_{\mathrm f}, \chi_{\mathrm f}) (1 + \delta \hat{\tau}_{221}) \\	
\end{cases}\, ,
\end{equation}
allowing deviations in both frequency and damping time of the $221$ mode simultaneously.

If we assign equal prior weight to both the GR and modified-GR hypotheses, then the odds ratio is
\begin{equation}
	\mathcal{O}^{\rm modGR}_{\rm GR} = \dfrac{1}{3} \left( \mathcal{B}^{\delta \hat{f}_{221}}_{\rm GR} + \mathcal{B}^{\delta \hat{\tau}_{221}}_{\rm GR} + \mathcal{B}^{\delta \hat{f}_{221},\delta \hat{\tau}_{221}}_{\rm GR} \right).
\end{equation}
The log odds ratios $\log_{10} \mathcal{O}^{\rm modGR}_{\rm GR}$ are reported in Table~\ref{tab:rin:final_mass_spin} for each event. Among all the events considered, \pyRingHighestTIGERLogOddsRatioEventName{} has the highest $\log_{10} \mathcal{O}^{\rm modGR}_{\rm GR}$ with a value of $\pyRingHighestTIGERLogOddsRatio{}$, which is not statistically significant. We also find a catalog-combined log odds ratio of $\pyRingCombinedTIGERLogOddsRatio{}$, in favor of the GR hypothesis that the Kerr metric is sufficient to describe the observed ringdown signals. 

\begin{figure}
\includegraphics[width=\columnwidth]{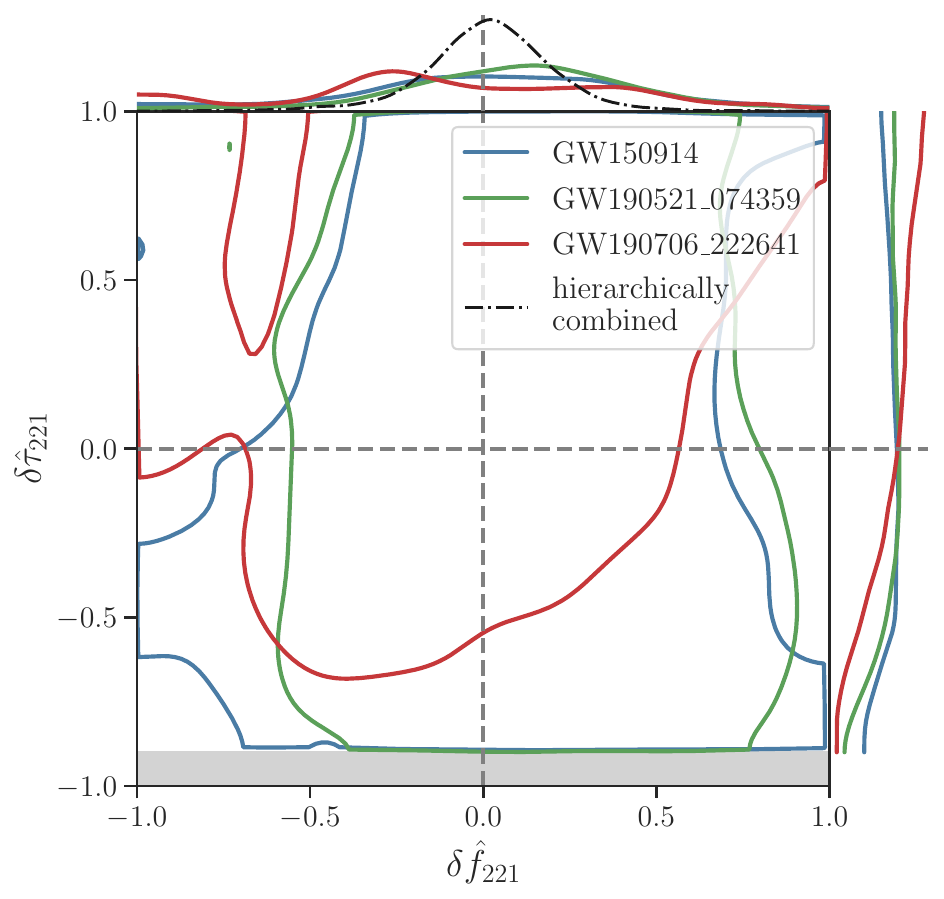}
\caption{The $90\%$ credible region of the joint posterior distribution of the fractional deviations of the frequency $\delta \hat{f}_{221}$ and the damping time $\delta \hat{\tau}_{221}$, and their marginalized posterior distributions, for the $\ell=|m|=2,n=1$ mode from the \textsc{pyRing} analysis, where we allow both the frequency and the damping time of the $221$ mode to deviate from the GR predictions. Here we show measurements from individual events where the data prefer the waveform model with both the fundamental and the first overtone ($n = 0, 1$) modes over the model with only the $n = 0$ fundamental mode. The measurements of the fractional deviation of the frequency from individual events, and as a set of measurements (using all $17$ events), both show consistency with GR. The fractional deviation of the damping time is mostly unconstrained.}
\label{fig:rin:qnm_deviation_221_pyRing}
\end{figure}

Figure \ref{fig:rin:qnm_deviation_221_pyRing} shows both the 1D marginal and the joint posterior distributions for $\delta \hat{f}_{221}$ and $\delta \hat{\tau}_{221}$ obtained from individual GW events where we allow both the frequency and the damping time of the $221$ mode to deviate from the GR predictions (i.e., the $\mathcal{H}_{\delta \hat{f}_{221},\;\delta \hat{\tau}_{221}}$ hypothesis). We only show results from GW events where the data prefer the waveform model with both the fundamental and the first overtone ($n = 0, 1$) modes over the model with only the $n = 0$ fundamental mode with $\log_{10} \mathcal{B}^{\rm 221}_{\rm 220} > 0$. The measurements show consistency with GR for the frequency. As for the damping time, it is essentially unconstrained, except for events with low SNR in the ringdown (such as \NAME{GW190727A}{}) where the posterior distribution of $\delta \hat{\tau}_{221}$ rails towards the lower prior bound $-0.9$, as the data show little evidence of the first overtone.
The results broadly agree with previous analyses for GW150914 \cite{Isi:2019aib}, although the truncation time chosen here ($t_0 = \pyRingTruncationTimeForH{GW150914}$ GPS in Hanford) is slightly later than in \cite{Isi:2019aib,TheLIGOScientific:2016src}.\footnote{To err on the conservative side, \textsc{pyRing} internally approximates the truncation time to the next element on the time axis. Peak times for all events are available under the \texttt{trigtime} entry of the released configuration files \cite{GWTC2:TGR:release}.}
A hierarchical analysis of the set of measurements using all $17$ events constrains the frequency deviations to \pyRingFrequencyDeviationPop{}, in agreement with the Kerr hypothesis.
The hierarchical analysis is uninformative for $\delta \hat{\tau}_{221}$ within the prior bounds considered.

Finally, as another test of the consistency of the ringdown signals with GR, we use a template which consists of a single damped sinusoid to fit the data, where the frequency, damping time, and complex amplitude are considered as free parameters without imposing any predictions from GR.
This means that, for this template, we assume neither that the remnant object is a Kerr BH, nor that it originated from a BBH coalescence. We place uniform priors on the frequency, damping time, log of the magnitude, and the phase of the complex amplitude.
The frequency and damping time obtained by fitting this template to the data are shown in Table \ref{tab:rin:freq_tau_results}, where we report $90\%$ credible intervals from the marginalized posteriors for each of these two parameters.
The values show good agreement with the results from full IMR analyses where GR is assumed, except for GW170814, \NAME{GW190512A}{}, \NAME{GW190828A}{}, and \NAME{GW190910A}{}, where the estimates of the damping time from the \textsc{pyRing} analysis are higher than the estimates from the full IMR analyses. Nevertheless, in all these cases the contours of the $90\%$ credible region in the frequency-damping time space from the two analyses actually do overlap.
We observed that events with low SNR in the ringdown often show overestimations of the damping time with respect to the median value obtained using the full IMR waveform. 
To assess whether the overestimation is caused by detector noise fluctuations, 
we injected simulated IMR waveforms with parameters consistent with \NAME{GW190828A}{}, close to the coalescence time of the event.
The injections show a similar behavior to what was observed in the actual event, with 3 out of 10 injections having the injected value lying outside the $90\%$ credible interval of the damping time.
The same injections performed in a zero noise configuration instead always have the posterior distributions of the damping time peaking at the injected value, suggesting that the overestimation of the damping time is associated with the detector noise fluctuations.

\subsubsection{The \textsc{pSEOBNRv4HM} analysis}
The \textsc{pSEOBNRv4HM} ringdown analysis uses a parametrized version of a spinning EOB waveform model with HMs, calibrated on non-precessing binaries \cite{Brito:2018rfr, Cotesta:2018fcv}.
The analysis uses the frequency-domain likelihood function while the waveform model is constructed in the time domain. 
In this model the effective frequency and damping time of the $220$ mode are written in terms of fractional deviations from their nominal GR values: $f_{220} = f_{220}^{\rm GR} (1+\delta \hat{f}_{220})$ and $\tau_{220} = \tau_{220}^{\rm GR} (1+\delta \hat{\tau}_{220})$ \cite{Brito:2018rfr}, where $\delta \hat{f}_{220}$ and $\delta \hat{\tau}_{220}$ are estimated directly from the data using the parameter inference techniques described in Sec.~\ref{sec:inference}, and $f_{220}^{\rm GR}$, $\tau_{220}^{\rm GR}$ are computed using the mass and spin of the BH remnant as determined by NR fits reported in \cite{Cotesta:2018fcv}.

We performed this analysis only on O3a events with a median redshifted total mass $>90M_{\odot}$ since this analysis is computationally expensive, and we expect these events to give the best measurements among all the O3a events. Table~\ref{tab:rin:freq_tau_results} shows the redshifted effective frequency $f_{220}$ and the redshifted effective damping time $\tau_{220}$ of the $220$ mode inferred from this analysis.

The frequency and the damping time inferred from the \textsc{pSEOB} analysis are also in good agreement with the full IMR measurements that assume GR, except for \NAME{GW190521A}{}, \NAME{GW190727A}{}, and \NAME{GW190910A}{} where the estimates of the damping time from the \textsc{pSEOB} analysis are higher than the estimates from the full IMR analyses. Nevertheless, in all these cases the 2D $90\%$ credible regions do overlap.
In order to better understand this issue, we investigated possible biases due to properties of the detector noise. We injected a set of simulated numerical relativity signals with parameters consistent with \NAME{GW190521A}{} into real data immediately adjacent to the event, and ran the \textsc{pSEOB} analysis on them. For $3$ out of $5$ injections around the event we recover posteriors that overestimate the damping time and for which the injected GR value lies outside the $90\%$ credible interval, suggesting that the overestimation of the damping time for \NAME{GW190521A}{} is a possible artifact of noise fluctuations. A similar study was conducted with \textsc{pyRing} using the damped sinusoid model for \NAME{GW190828A}{} and we also observed overestimations of the damping time. This suggests that the overestimation of the damping time is a common systematic error for low-SNR signals.

\begin{table}
\caption{\label{tab:rin:freq_tau_results}The median value and symmetric $90\%$ credible interval of the redshifted frequency and damping time estimated using the full IMR analysis (IMR), the \textsc{pyRing} analysis with a single damped sinusoid (DS), and the \textsc{pSEOBNRv4HM} analysis (pSEOB).}
\scalebox{0.9}{
\begin{tabular}{llllllll}
\toprule
Event & \multicolumn{3}{c}{Redshifted} & \hphantom{X} & \multicolumn{3}{c}{Redshifted} \\
& \multicolumn{3}{c}{frequency [Hz]} & \hphantom{X} & \multicolumn{3}{c}{damping time [ms]} \\[0.075cm]
\cline{2-4}
\cline{6-8}
& IMR & DS & pSEOB & \hphantom{X} & IMR & DS & pSEOB \\
\midrule

GW150914 &
$248^{+8}_{-7}$ &
$247^{+15}_{-16}$ &
$-$ &
\hphantom{X} &
$4.2^{+0.3}_{-0.2}$ &
$4.3^{+4.0}_{-1.9}$ &
$-$
\\[0.075cm]

GW170104 &
$287^{+15}_{-25}$ &
$220^{+78}_{-93}$ &
$-$ &
\hphantom{X} &
$3.5^{+0.4}_{-0.3}$ &
$2.4^{+15.7}_{-1.5}$ &
$-$
\\[0.075cm]

GW170814 &
$293^{+11}_{-14}$ &
$525^{+332}_{-331}$ &
$-$ &
\hphantom{X} &
$3.7^{+0.3}_{-0.2}$ &
$24.7^{+22.5}_{-19.3}$ &
$-$
\\[0.075cm]

GW170823 &
$197^{+17}_{-17}$ &
$216^{+558}_{-53}$ &
$-$ &
\hphantom{X} &
$5.5^{+1.0}_{-0.8}$ &
$8.4^{+34.1}_{-5.6}$ &
$-$
\\[0.075cm]

GW190408\_181802 &
$319^{+11}_{-20}$ &
$464^{+503}_{-418}$ &
$-$ &
\hphantom{X} &
$3.2^{+0.3}_{-0.3}$ &
$10.3^{+32.4}_{-9.1}$ &
$-$
\\[0.075cm]

GW190421\_213856 &
$162^{+14}_{-14}$ &
$-$ &
$171^{+50}_{-16}$ &
\hphantom{X} &
$6.3^{+1.2}_{-0.8}$ &
$-$ &
$8.5^{+5.3}_{-4.2}$
\\[0.075cm]

GW190503\_185404 &
$191^{+17}_{-15}$ &
$-$ &
$265^{+501}_{-79}$ &
\hphantom{X} &
$5.3^{+0.8}_{-0.8}$ &
$-$ &
$3.5^{+3.4}_{-1.8}$
\\[0.075cm]

GW190512\_180714 &
$381^{+33}_{-42}$ &
$214^{+525}_{-34}$ &
$-$ &
\hphantom{X} &
$2.6^{+0.2}_{-0.2}$ &
$22.9^{+23.7}_{-18.6}$ &
$-$
\\[0.075cm]

GW190513\_205428 &
$241^{+26}_{-28}$ &
$277^{+480}_{-108}$ &
$-$ &
\hphantom{X} &
$4.3^{+1.1}_{-0.4}$ &
$6.1^{+22.7}_{-4.7}$ &
$-$
\\[0.075cm]

GW190519\_153544 &
$127^{+9}_{-9}$ &
$126^{+11}_{-17}$ &
$124^{+12}_{-13}$ &
\hphantom{X} &
$9.5^{+1.7}_{-1.5}$ &
$9.7^{+9.2}_{-4.1}$ &
$10.3^{+3.6}_{-3.1}$
\\[0.075cm]

GW190521 &
$68^{+4}_{-4}$ &
$65^{+3}_{-3}$ &
$67^{+2}_{-2}$ &
\hphantom{X} &
$15.8^{+3.9}_{-2.5}$ &
$22.3^{+12.6}_{-7.5}$ &
$30.7^{+7.7}_{-7.4}$
\\[0.075cm]

GW190521\_074359 &
$198^{+7}_{-7}$ &
$195^{+15}_{-17}$ &
$205^{+15}_{-12}$ &
\hphantom{X} &
$5.4^{+0.4}_{-0.4}$ &
$7.8^{+7.3}_{-3.5}$ &
$5.3^{+1.5}_{-1.2}$
\\[0.075cm]

GW190602\_175927 &
$105^{+10}_{-9}$ &
$97^{+12}_{-23}$ &
$99^{+15}_{-15}$ &
\hphantom{X} &
$10.0^{+2.0}_{-1.4}$ &
$8.8^{+12.6}_{-3.9}$ &
$8.8^{+5.4}_{-3.6}$
\\[0.075cm]

GW190706\_222641 &
$108^{+11}_{-10}$ &
$109^{+7}_{-12}$ &
$112^{+7}_{-8}$ &
\hphantom{X} &
$10.9^{+2.4}_{-2.2}$ &
$19.6^{+25.5}_{-12.5}$ &
$19.4^{+7.2}_{-8.9}$
\\[0.075cm]

GW190708\_232457 &
$497^{+10}_{-46}$ &
$768^{+152}_{-717}$ &
$-$ &
\hphantom{X} &
$2.1^{+0.2}_{-0.1}$ &
$26.5^{+21.3}_{-24.4}$ &
$-$
\\[0.075cm]

GW190727\_060333 &
$178^{+18}_{-16}$ &
$351^{+587}_{-276}$ &
$201^{+11}_{-21}$ &
\hphantom{X} &
$6.1^{+1.1}_{-0.8}$ &
$20.5^{+26.0}_{-17.5}$ &
$15.4^{+5.3}_{-6.1}$
\\[0.075cm]

GW190828\_063405 &
$239^{+10}_{-11}$ &
$248^{+294}_{-14}$ &
$-$ &
\hphantom{X} &
$4.8^{+0.6}_{-0.5}$ &
$16.3^{+25.5}_{-9.7}$ &
$-$
\\[0.075cm]

GW190910\_112807 &
$177^{+8}_{-8}$ &
$165^{+9}_{-11}$ &
$174^{+12}_{-8}$ &
\hphantom{X} &
$5.9^{+0.8}_{-0.5}$ &
$12.8^{+16.2}_{-6.0}$ &
$9.5^{+3.1}_{-2.7}$
\\[0.075cm]

GW190915\_235702 &
$232^{+14}_{-18}$ &
$512^{+445}_{-473}$ &
$-$ &
\hphantom{X} &
$4.6^{+0.8}_{-0.6}$ &
$15.7^{+30.3}_{-14.3}$ &
$-$
\\[0.075cm]

\bottomrule
\end{tabular}}
\end{table}

\begin{figure}
  \centering
  \includegraphics[width=\columnwidth]{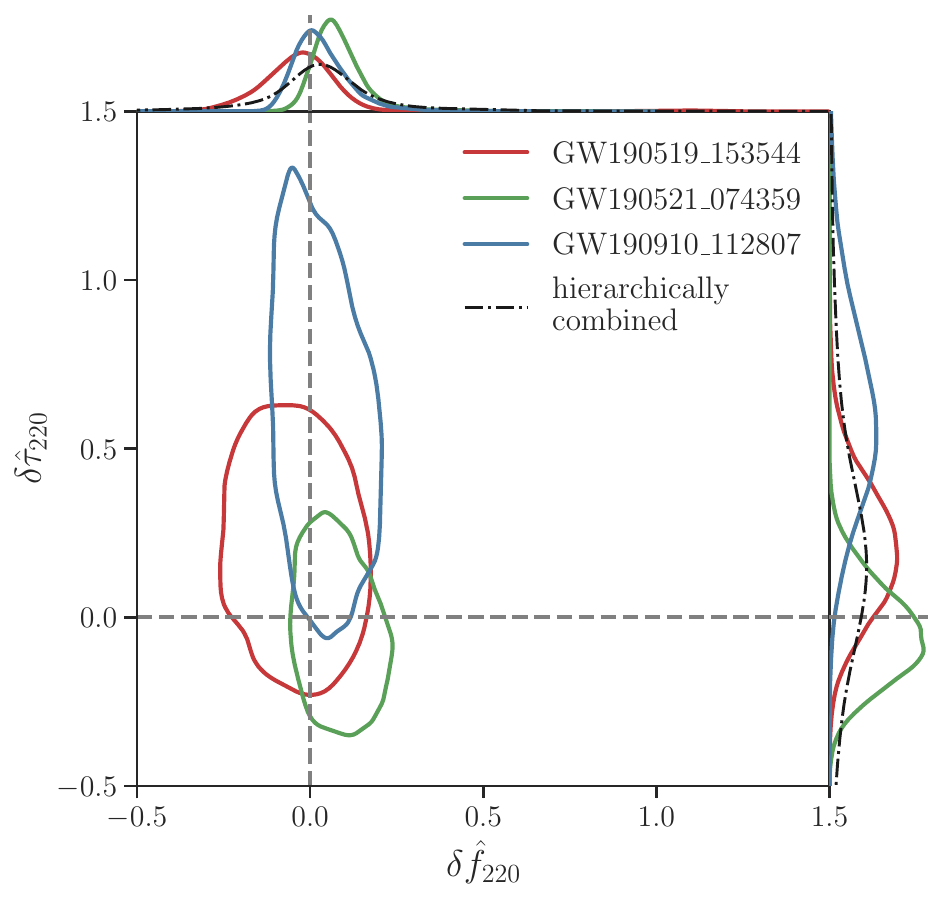}
  \caption{The $90\%$ credible region of the joint posterior distribution of the fractional deviations of the frequency $\delta \hat{f}_{220}$ and the damping time $\delta \hat{\tau}_{220}$, and their marginalized posterior distributions, for the $\ell=|m|=2,n=0$ mode from the \textsc{pSEOBNRv4HM} analysis. We only include events that have SNR $>8$ in both the inspiral and postinspiral stage in this plot where we have sufficient information to break the degeneracy between the binary total mass and the fractional deviation parameters in the absence of measurable HMs. The measurements of the fractional deviations for individual events, and as a set of measurements, both show consistency with GR.}
  \label{fig:rin:qnm_deviation_220_pSEOBNR}
\end{figure}

In Fig.~\ref{fig:rin:qnm_deviation_220_pSEOBNR}, we show the $90\%$ credible region of the joint posterior distribution of the frequency and damping time deviations, as well as their respective marginalized distributions.
We only include events that have SNR $>8$ in both the inspiral and postinspiral regimes, with cutoff frequencies as in Table \ref{tab:imr_test_params}.
This is because, in order to make meaningful inferences about $\delta \hat{f}_{220}$ and $\delta \hat{\tau}_{220}$ with \textsc{pSEOB} in the absence of measurable HMs, the signal must contain sufficient information in the inspiral and merger stages to break the degeneracy between the binary total mass and the GR deviations.
The fractional deviations obtained this way quantify the agreement between the pre- and postmerger portions of the waveform, and are thus not fully analogous to the \textsc{pyRing} quantities.

From Fig.~\ref{fig:rin:qnm_deviation_220_pSEOBNR}, the frequency and the damping time of the $220$ mode are consistent with the GR prediction ($\delta \hat{f}_{220} = \delta \hat{\tau}_{220} = 0$) for \NAME{GW190519A}{} and \NAME{GW190521B}{}, while for \NAME{GW190910A}{} it shows excellent agreement with GR for $\delta \hat{f}_{220}$ but the GR prediction has only little support in the marginalized posterior distribution of $\delta \hat{\tau}_{220}$.

In spite of the low number of events, we also apply the hierarchical framework to the marginal distributions in Fig.~\ref{fig:rin:qnm_deviation_220_pSEOBNR}.
The population-marginalized constraints are \pSEOBFrequencyDeviationPop{} and \pSEOBDampingTimeDeviationPop{}, which are consistent with GR for both parameters.
The $\delta \hat{\tau}_{220}$ measurement is uninformative, which is not surprising given the spread of the \NAME{GW190910A}{} result and the low number of events.
The hyperparameters also reflect this, since they are constrained for $\delta \hat{f}_{220}$ (\pSEOBFrequencyDeviationMean{}, \pSEOBFrequencyDeviationStd{}) but uninformative for $\delta\hat{\tau}_{220}$ (\pSEOBDampingTimeDeviationMean{}, \pSEOBDampingTimeDeviationStd{}).
The bounds for the fractional deviation in frequency for the $220$ mode, from the \textsc{pSEOB} analysis, and for the $221$ mode, from the \textsc{pyRing} analysis, can be used to cast constraints on specific theories of modified gravity that predict non-zero values of these deviations \cite{Cardoso:2019mqo,McManus:2019ulj}, as well as to bound possible deviations in the ringdown spectrum caused by a non-Kerr-BH remnant object (see, e.g., \cite{Maggio:2020jml}).

    \subsection{Echoes}
    \label{sec:ech}
\newcommand{\echoescombinedlogbayesfactor}{-5.02}
\newcommand{\echoeslowestlogbayesfactorevent}{\NAME{GW190412A}}
\newcommand{\echoeslowestlogbayesfactor}{-1.30}
\newcommand{\echoeshighestlogbayesfactorevent}{\NAME{GW190915A}}
\newcommand{\echoeshighestlogbayesfactor}{0.17}
It is hypothesized that there may be compact objects having a light ring and a reflective surface located between the light ring and the would-be event horizon. These compact objects are referred to as exotic compact objects (ECOs), for example gravastars \cite{Mazur:2001fv} and fuzzballs \cite{Lunin:2001jy,Lunin:2002qf}. When an ECO is formed as the remnant of a compact binary coalescence, a train of repeating pulses known as GW echoes are emitted from the ECO in the late postmerger stage in addition to the usual ringdown we expect from BHs. The effective potential barrier and the reflective surface act like a cavity trapping the GWs. Unlike BHs, which have a purely in-going boundary condition at the event horizon, the GWs trapped in the cavity will be reflected back and forth between the potential barrier and the surface, emitting pulses of waves towards infinity when some of the waves are transmitted through the potential barrier and escape \cite{Kokkotas:1995av, Tominaga:1999iy, Ferrari:2000sr, Volkel:2017kfj, Volkel:2018hwb, Cardoso:2019rvt}. Detecting these GW echoes would be clear evidence of the existence of these proposed ECOs \cite{Cardoso_BH_mimick, Cardoso:2016oxy, Maselli:2017tfq}, though there are still no full and viable models of ECOs that produce echoes \cite{Maggio:2018ivz, 2018CQGra..35oLT01P, Chen:2019hfg, Cardoso:2019rvt, Coates:2019bun}.

We employ a template-based approach \cite{Lo:2018sep} that uses the model proposed in \cite{Abedi:2016hgu} to search for GW echoes. The waveform model takes the ringdown part of an IMR waveform and repeats the modulated ringdown waveform according to five additional echo parameters which control the relative amplitude of the echoes, the damping factor between each echo, the start time of ringdown, the time of the first echo with respect to the merger, and the time delay between each echo. We adopt a uniform prior for each of the echo parameters. We used \IMRP{} as the IMR waveform approximant for all the events we analyzed except for \NAME{GW190521A}{} where \NRSur{} was used instead. The pipeline computes the log Bayes factor $\log_{10} \mathcal{B}^{\rm IMRE}_{\rm IMR}$ of the data being describable by an inspiral--merger--ringdown--echoes (IMRE) waveform versus an IMR waveform, and uses it as the detection statistic to identify the existence of echoes in the data.

We analyze $31$ BBH signals from GWTC-2 passing our false-alarm rate threshold (see Sec.~\ref{sec:events} and Table~\ref{tab:events}) and report the search results of GW echoes in Table~\ref{tab:ech:search_results}.\footnote{We do not analyze \NAME{GW190814A}{} because the long data segment and high sampling rate it requires makes the analysis prohibitively expensive.} No statistically significant evidence of echoes was found in the data; it was reported in \cite{Lo:2018sep} that for detector noise fluctuations typical for O1, a detection threshold for $\log_{10} \mathcal{B}^{\rm IMRE}_{\rm IMR}$ was found to be roughly $2.48$ by empirically constructing the background distribution of the Bayes factor if we require the false-alarm probability to be $\lesssim 3 \times 10^{-7}$. The event \echoeshighestlogbayesfactorevent{} has the highest $\log_{10} \mathcal{B}^{\rm IMRE}_{\rm IMR}$ of merely $\echoeshighestlogbayesfactor$, which indicates negligible support for the presence of GW echoes in the data.
While we did not present the Bayes factor for GW151012 and GW170729 here as their corresponding FARs are above the threshold, the results are consistent with no significant evidence of echoes being found in the data.
The null results for O1 and O2 events are consistent with what was reported in \cite{Ashton:2016xff, Westerweck:2017hus, Lo:2018sep, Nielsen:2018lkf, Uchikata:2019frs, Wang:2020ayy}. The posterior distributions of the extra echo parameters mostly recover their corresponding prior distributions, consistent with the fact that we did not detect any echoes in the data.

\begin{table}
\caption{\label{tab:ech:search_results} Results of search for GW echoes. A positive value of the log Bayes factor $\log_{10} \mathcal{B}^{\rm IMRE}_{\rm IMR}$ indicates a preference for the IMRE model over the IMR model, while a negative value of the log Bayes factor suggests instead a preference for the IMR model over the IMRE model.}
\scalebox{0.92}{

\begin{tabular}{lrc|clr}
\toprule
Event & $\log_{10} \mathcal{B}^{\rm IMRE}_{\rm IMR}$ & \hphantom{X} & \hphantom{X} & Event & $\log_{10} \mathcal{B}^{\rm IMRE}_{\rm IMR}$ \\
\midrule

GW150914 & $-0.57$ & \hphantom{X} & \hphantom{X} & GW170809 & $-0.22$ \\[0.075cm]

GW151226 & $-0.08$ & \hphantom{X} & \hphantom{X} & GW170814 & $-0.49$ \\[0.075cm]

GW170104 & $-0.53$ & \hphantom{X} & \hphantom{X} & GW170818 & $-0.62$ \\[0.075cm]

GW170608 & $-0.44$ & \hphantom{X} & \hphantom{X} & GW170823 & $-0.34$ \\[0.075cm]

\midrule

\NAME{GW190408A} & $-0.93$ & \hphantom{X} & \hphantom{X} & \NAME{GW190706A} & $-0.10$ \\[0.075cm]

\NAME{GW190412A} & $-1.30$ & \hphantom{X} & \hphantom{X} & \NAME{GW190707A} & $0.08$ \\[0.075cm]

\NAME{GW190421A} & $-0.11$ & \hphantom{X} & \hphantom{X} & \NAME{GW190708A} & $-0.87$ \\[0.075cm]

\NAME{GW190503A} & $-0.36$ & \hphantom{X} & \hphantom{X} & \NAME{GW190720A} & $-0.45$ \\[0.075cm]

\NAME{GW190512A} & $-0.56$ & \hphantom{X} & \hphantom{X} & \NAME{GW190727A} & $0.01$ \\[0.075cm]

\NAME{GW190513A} & $-0.03$ & \hphantom{X} & \hphantom{X} & \NAME{GW190728A} & $0.01$ \\[0.075cm]

\NAME{GW190517A} & $0.16$ & \hphantom{X} & \hphantom{X} & \NAME{GW190828A} & $0.10$ \\[0.075cm]

\NAME{GW190519A} & $-0.10$ & \hphantom{X} & \hphantom{X} & \NAME{GW190828B} & $-0.01$ \\[0.075cm]

\NAME{GW190521A} & $-1.82$ & \hphantom{X} & \hphantom{X} & \NAME{GW190910A} & $-0.22$ \\[0.075cm]

\NAME{GW190521B} & $-0.72$ & \hphantom{X} & \hphantom{X} & \NAME{GW190915A} & $0.17$ \\[0.075cm]

\NAME{GW190602A} & $0.13$ & \hphantom{X} & \hphantom{X} & \NAME{GW190924A} & $-0.03$ \\[0.075cm]

\NAME{GW190630A} & $0.08$ & \hphantom{X} & \hphantom{X} &  &  \\[0.075cm]

\bottomrule
\end{tabular}
}
\end{table}

\section{Polarizations}
\label{sec:pol}
\newcommand{\PolWorstV}[1]{\IfEqCase{#1}{{NAME}{\NAME{GW190503A}}{BF}{-0.072}}}
\newcommand{\PolWorstS}[1]{\IfEqCase{#1}{{NAME}{\NAME{GW190720A}}{BF}{0.074}}}
\newcommand{\PolBestV}[1]{\IfEqCase{#1}{{NAME}{\NAME{GW190513A}}{BF}{0.139}}}
\newcommand{\PolBestS}[1]{\IfEqCase{#1}{{NAME}{\NAME{GW190513A}}{BF}{1.380}}}

Generic metric theories of gravity may allow up to six GW polarizations \cite{Eardley:1973br,Eardley:1974nw}.
These correspond to the two tensor modes (helicity $\pm 2$) allowed in GR, plus two additional vector modes (helicity $\pm1$), and two scalar modes (helicity 0).
The polarization content of a GW is imprinted in the relative amplitudes of the outputs at different detectors, as determined by the corresponding antenna patterns \cite{Will:2014kxa,Chatziioannou:2012rf,Isi:2017equ,Callister:2017ocg,Isi:2017fbj}.
This fact can be used to reconstruct the GW polarization content from the data, although a five-detector network would be needed to do this optimally with transient signals.
The existing three-detector network may be used to distinguish between some specific subsets of all the possible polarization combinations.

We previously reported constraints on extreme polarization alternatives (full tensor versus full vector, and full tensor versus full scalar) in \cite{Abbott:2017oio,Abbott:2018lct,LIGOScientific:2019fpa}, using a simplified analysis that relied on GR templates \cite{Isi:2017fbj}.
None of the events analyzed (GW170814, GW170817, and GW170818) disfavored the tensorial hypothesis.
Because the source sky location was known from electromagnetic observations \cite{Monitor:2017mdv}, the results were strongest for GW170817, which we found to be highly inconsistent with the full-vector and full-scalar hypotheses with (base ten) log Bayes factors ${\gtrsim}20$ \cite{Abbott:2018lct}.
Although this is strong evidence against vector or scalar being the only possible GW polarization, it does not strictly preclude scenarios in which only some sources produce vector-only or scalar-only GWs.

\begin{table}
\caption{
Base-ten logarithms of Bayes factors for different polarization hypotheses: full-tensor versus full-vector ($\log_{10} \mathcal{B}^T_V$), and full-tensor versus full-scalar ($\log_{10} \mathcal{B}^T_S$). These results were obtained with the waveform independent method described in Sec.~\ref{sec:pol}.
They are less informative than those in \cite{Abbott:2017oio,Abbott:2018lct,LIGOScientific:2019fpa} because the present method does not attempt to track the signal phase across time.
}
\label{tab:pol}
\centering
\begin{tabular}{l@{\qquad}r@{\qquad}r}
\toprule
            Event & $\log_{10} {\cal B}^T_V$ & $\log_{10} {\cal B}^T_S$ \\
\midrule
         GW170809 &                  $0.078$ &                  $0.421$ \\[2pt]
         GW170814 &                 $-0.032$ &                  $0.740$ \\[2pt]
         GW170818 &                  $0.002$ &                  $0.344$ \\[2pt]
\midrule
 \NAME{GW190408A} &                  $0.076$ &                  $0.480$ \\[2pt]
 \NAME{GW190412A} &                  $0.079$ &                  $0.539$ \\[2pt]
 \NAME{GW190503A} &                 $-0.072$ &                  $1.245$ \\[2pt]
 \NAME{GW190512A} &                 $-0.024$ &                  $0.346$ \\[2pt]
 \NAME{GW190513A} &                  $0.139$ &                  $1.380$ \\[2pt]
 \NAME{GW190517A} &                  $0.008$ &                  $0.730$ \\[2pt]
 \NAME{GW190519A} &                  $0.067$ &                  $0.799$ \\[2pt]
 \NAME{GW190521A} &                  $0.093$ &                  $1.156$ \\[2pt]
 \NAME{GW190602A} &                 $-0.064$ &                  $0.373$ \\[2pt]
 \NAME{GW190706A} &                  $0.052$ &                  $0.771$ \\[2pt]
 \NAME{GW190720A} &                  $0.034$ &                  $0.074$ \\[2pt]
 \NAME{GW190727A} &                  $0.087$ &                  $1.024$ \\[2pt]
 \NAME{GW190728A} &                 $-0.024$ &                  $0.083$ \\[2pt]
 \NAME{GW190828A} &                  $0.063$ &                  $0.851$ \\[2pt]
 \NAME{GW190828B} &                 $-0.034$ &                  $0.084$ \\[2pt]
 \NAME{GW190915A} &                  $0.020$ &                  $1.238$ \\[2pt]
 \NAME{GW190924A} &                 $-0.051$ &                  $0.384$ \\[2pt]
\bottomrule
\end{tabular}
\end{table}

Here we probe the same extreme polarization hypotheses as in previous studies, but through a different technique that does not rely on specific waveform models.
This null-stream based polarization test is a Bayesian implementation of the null stream construct proposed in~\cite{GuerselTinto1989}, generalized to vector and scalar antenna patterns \cite{Chatziioannou:2012rf,Pang:2020pfz}.
A null stream is a linear combination of the data streams from different detectors that is known to be free of true GWs with a given helicity and sky location, irrespective of the GW waveform.
Any excess power remaining in the null stream must have been produced by a GW signal whose helicity or sky location is not what was assumed.
We quantify such excess power by means of the null energy, as defined in \cite{Sutton:2009gi}.
If the polarization modes and the sky location of the GW signal are correctly specified, this quantity will fluctuate solely due to instrumental noise and will follow a chi-squared distribution \cite{Sutton:2009gi}.
This provides a likelihood function for the hypothesis that the data contain a signal with a given helicity and sky location. 
By marginalizing over the source location, we may obtain the evidences of different polarization hypotheses and compute Bayes factors comparing them.
We take a uniform distribution over the celestial sphere as our sky-location prior, and compute evidences through an extended version of the \textsc{BANTAM} pipeline presented in \cite{Pang:2020pfz}.

In Table~\ref{tab:pol}, we present the resulting Bayes factors for full-tensor versus full-vector $\mathcal{B}^T_V$, and full-tensor versus full-scalar $\mathcal{B}^T_S$.
None of the signals analyzed favor either of the non-GR hypotheses (full-vector, or full-scalar) to any significant degree.
The Bayes factors in Table \ref{tab:pol} are less informative than those in \cite{Abbott:2017oio,Abbott:2018lct,LIGOScientific:2019fpa} because the present method does not attempt to track the signal phase across time, relying only on signal power added incoherently across time--frequency pixels of the null stream \cite{Sutton:2009gi}.
The events yielding the lowest Bayes factors are \PolWorstV{NAME}{} and \PolWorstS{NAME}{}, with $\log_{10} \mathcal{B}^T_V = \PolWorstV{BF}{}$ and $\log_{10} \mathcal{B}^T_S = \PolWorstS{BF}{}$ respectively;
on the other hand, the event yielding the highest Bayes factors is \PolBestV{NAME}{} for both vector and scalar, with $\log_{10} \mathcal{B}^T_V = \PolBestV{BF}{}$ and $\log_{10} \mathcal{B}^T_S = \PolBestS{BF}{}$ respectively.

\begin{figure}
	\centering
	\includegraphics[width=\columnwidth]{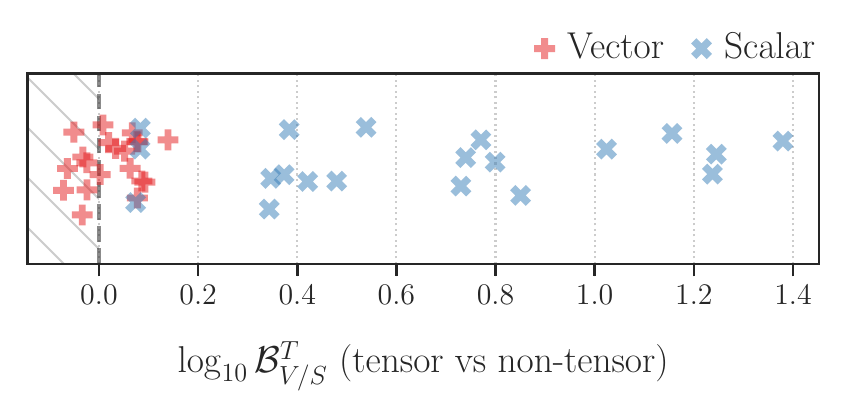}
	\caption{Distribution of $\log_{10}$ Bayes factors for different polarization hypotheses: full-tensor versus full-vector (red), and full-tensor versus full-scalar (blue).
  The horizontal axis of this strip plot represents the logarithm of $\mathcal{B}^T_{V/S}$ in Table \ref{tab:pol}, with each red/blue marker corresponding to a single event; the vertical axis carries no meaning.
  Values of $\log_{10} {\cal B}^T_{V/S} < 0$ indicate a preference for the nontensor hypothesis (hatched region).
  The different spreads of the sets of markers are as expected for GR signals and no event reaches large negative values of $\log_{10} {\cal B}^T_{V/S}$, meaning all signals are consistent with tensor polarizations.
}
	\label{fig:pol}
\end{figure}

The distributions of $\log_{10} \mathcal{B}^T_V$ and $\log_{10} \mathcal{B}^T_S$ are as expected from GR signals with the observed SNRs \cite{T2000405}.
As is clear from Fig.~\ref{fig:pol}, the scalar results more decisively favor the tensor hypothesis than the vector ones.
The asymmetry between the vector and scalar results is explained by the intrinsic geometries of the LIGO--Virgo antenna patterns, which make scalar waves easier to distinguish \cite{T2000405}.
As in previous studies, we conclude there is no evidence for pure vector or pure scalar polarizations.

\section{Conclusions and outlook}
\label{sec:conclusion}
GWs give us an opportunity to observationally probe the nature of gravity in its strong-field, dynamical regime, which is difficult to access by other means.
With an ever-growing number of detections, we are now able to put GR to the test with increasing precision and in qualitatively new ways.
In this paper, we presented eight tests of GR and the nature of BHs using signals from the latest LIGO--Virgo catalog, GWTC-2 \cite{GWTC2}. 
These tests leverage different aspects of GW physics to constrain the null hypothesis that our signals were produced by merging Kerr BHs in agreement with Einstein's theory, and that our GR-based models are sufficient to capture their behavior.
We find that all of the LIGO--Virgo detections analyzed are consistent with GR, and do not find any evidence for deviations from theoretical expectations, or unknown systematics.

We began by checking the consistency of the data with the GR prediction in a generic way through the residuals and IMR consistency tests (Sec.~\ref{sec:con}).
We found that, for all events, residual data obtained after subtracting a best-fit GR waveform are consistent with instrumental noise (Sec.~\ref{sec:res}), and confirmed that events return compatible parameter estimates when the low- and high-frequency regimes are analyzed separately (Sec.~\ref{sec:imr}).

Next we focused on controlled deviations away from the GR prediction for the GW waveform (Sec.~\ref{sec:par}).
Allowing for corrections to the GW phasing through inspiral PN parameters, as well as phenomenological merger-ringdown coefficients, we found no evidence for GR deviations, and improved previous constraints in \cite{LIGOScientific:2019fpa} by a factor of ${\sim}2$.
We also targeted specific deviations in the GW phasing due to modifications to the spin-induced quadrupole moment of the binary components, obtaining broad constraints in agreement with the Kerr hypothesis (Sec.~\ref{sec:sim}).
Through a generalized dispersion relation, we tested GR's prediction that GWs propagate without dispersion and that the graviton is massless (Sec.~\ref{sec:liv}). We found no evidence for GW dispersion, and tightened previous constraints on Lorentz-violating dispersion parameters by a factor of ${\sim}\LivImprov{AMP}$. We constrained the mass of the graviton to be $m_g \leq \LivMgUL{} \text{ eV}/c^2$ with 90\% credibility---an improvement of a factor of $\LivImprov{MG}$ over the GWTC-1 measurement \cite{LIGOScientific:2019fpa}, and comparable to the Solar System bounds \cite{Bernus:2020szc}.

The detection of relatively high-mass events, coupled with the development of novel analysis techniques, allowed us to probe the properties of the merger remnant through targeted studies of the signal after merger.
We validated the expectation that the remnants were Kerr BHs, constraining QNM frequencies and damping times (Sec.~\ref{sec:rin}).
The results show agreement with Kerr remnants: the population-marginalized constraint on the fractional deviation away from the Kerr frequency is \pSEOBFrequencyDeviationPop for the $220$ mode, and \pyRingFrequencyDeviationPop for the $221$ mode at 90\% credibility.
In addition, we considered the existence of GW echoes---repetitions of the postmerger signal that could signal the presence of some reflective structure near the presumed event horizon of the remnant object, absent for classical BHs (Sec.~\ref{sec:ech}).
A search for such excess power after the main signal using periodic templates yielded no significant evidence for echoes.

Finally, we studied the polarization content of GWs with a new approach that does not make use of templates to reconstruct the signal power (Sec.~\ref{sec:pol}). With only three active detectors, we cannot simultaneously constrain all the six possible GW polarizations allowed in generic metric theories of gravity (scalar, vector, and tensor). Instead, as in previous studies, we compared the likelihood of having purely scalar or purely vector polarizations against the pure tensor case, predicted by GR. 
We found no evidence in favor of non-tensor GWs.

Our conclusions come from the analysis of multiple BBH signals, studied individually and collectively.
To understand our measurements holistically, we made use of a variety of statistical techniques, including hierarchical Bayesian inference, to evaluate the agreement of our set of measurements with the expectation from GR.
As the number of GW detections continues to grow, these strategies will become increasingly indispensable as tools to properly interpret our data and their agreement with theory, as well as to tease out potential disagreements that would be indiscernible from individual signals.
With constantly improving detectors and analysis capabilities, we will continue to expand the scope and sensitivity of our tests of GR and our probes of the nature of BHs when analyzing data from O3b and future observing runs.

\acknowledgments
Analyses in this paper made use of
\textsc{NumPy} \cite{Numpy2020},
\textsc{SciPy} \cite{2020SciPy-NMeth},
\textsc{Astropy} \cite{astropy:2013,astropy:2018},
\textsc{IPython} \cite{ipython},
\textsc{qnm} \cite{Stein:2019mop},
\textsc{PESummary} \cite{Hoy:2020vys},
and \textsc{GWpy} \cite{gwpy};
plots were produced with
\textsc{Matplotlib} \cite{matplotlib}, and
\textsc{Seaborn} \cite{seaborn}.
Posteriors were sampled with 
\textsc{Stan} \cite{JSSv076i01},
\textsc{CPNest} \cite{CPNest},
\textsc{PyMultinest} \cite{Feroz:2009,Feroz:2013},
and \linf{} \cite{Veitch:2014wba}.
The authors gratefully acknowledge the support of the United States
National Science Foundation (NSF) for the construction and operation of the
LIGO Laboratory and Advanced LIGO as well as the Science and Technology Facilities Council (STFC) of the
United Kingdom, the Max-Planck-Society (MPS), and the State of
Niedersachsen/Germany for support of the construction of Advanced LIGO 
and construction and operation of the GEO600 detector. 
Additional support for Advanced LIGO was provided by the Australian Research Council.
The authors gratefully acknowledge the Italian Istituto Nazionale di Fisica Nucleare (INFN),  
the French Centre National de la Recherche Scientifique (CNRS) and
the Netherlands Organization for Scientific Research, 
for the construction and operation of the Virgo detector
and the creation and support  of the EGO consortium. 
The authors also gratefully acknowledge research support from these agencies as well as by 
the Council of Scientific and Industrial Research of India, 
the Department of Science and Technology, India,
the Science \& Engineering Research Board (SERB), India,
the Ministry of Human Resource Development, India,
the Spanish Agencia Estatal de Investigaci\'on,
the Vicepresid\`encia i Conselleria d'Innovaci\'o, Recerca i Turisme and the Conselleria d'Educaci\'o i Universitat del Govern de les Illes Balears,
the Conselleria d'Innovaci\'o, Universitats, Ci\`encia i Societat Digital de la Generalitat Valenciana and
the CERCA Programme Generalitat de Catalunya, Spain,
the National Science Centre of Poland and the Foundation for Polish Science (FNP),
the Swiss National Science Foundation (SNSF),
the Russian Foundation for Basic Research, 
the Russian Science Foundation,
the European Commission,
the European Regional Development Funds (ERDF),
the Royal Society, 
the Scottish Funding Council, 
the Scottish Universities Physics Alliance, 
the Hungarian Scientific Research Fund (OTKA),
the French Lyon Institute of Origins (LIO),
the Belgian Fonds de la Recherche Scientifique (FRS-FNRS), 
Actions de Recherche Concertées (ARC) and
Fonds Wetenschappelijk Onderzoek – Vlaanderen (FWO), Belgium,
the Paris \^{I}le-de-France Region, 
the National Research, Development and Innovation Office Hungary (NKFIH), 
the National Research Foundation of Korea,
the Natural Science and Engineering Research Council Canada,
Canadian Foundation for Innovation (CFI),
the Brazilian Ministry of Science, Technology, Innovations, and Communications,
the International Center for Theoretical Physics South American Institute for Fundamental Research (ICTP-SAIFR), 
the Research Grants Council of Hong Kong,
the National Natural Science Foundation of China (NSFC),
the Leverhulme Trust, 
the Research Corporation, 
the Ministry of Science and Technology (MOST), Taiwan
and
the Kavli Foundation.
The authors gratefully acknowledge the support of the NSF, STFC, INFN and CNRS for provision of computational resources.

{\it We would like to thank all of the essential workers who put their health at risk during the COVID-19 pandemic, without whom we would not have been able to complete this work.}

\vspace{5mm}

\appendix
\section{Residuals $p$-value uncertainty}
\label{app:res}
The light-blue band in Fig.~\ref{fig:res:pp} represents the 90\%-credible band on the cumulative distribution of $p$-values from the residuals analysis (Sec.~\ref{sec:res}).
This incorporates two types of uncertainty~\cite{2018arXiv180406788T}:
\begin{enumerate}
\item uncertainty in the true $p$-value for any specific event, due to the finite number of noise instantiations used to compute the background $\rhores$;
\item uncertainty in the fraction of events yielding a $p$-value below any given benchmark, due to the finite number of events observed.
\end{enumerate}
These two types of ignorance translate into uncertainty in the abscissa and ordinate values in Fig.~\ref{fig:res:pp}, respectively.
We compute the corresponding credible band as explained below.

The true (unknown) $p$-value for a given event is estimated by counting the number of noise instances $n$ that yield an $\rhores$ greater than or equal to the on-source threshold $\rhores^{\rm thr}$, out of a total $N=\resN{}$ trials.
We denote the true $p$-value by $p = P(\rhores^{\rm thr} \leq \rhores)$, and the estimate from finite noise instances as $\hat{p} = n/N$. 
For a given true value of $p$, the expected likelihood of observed $\hat{p}$ will be given by the binomial distribution,
\begin{equation} \label{eq:res:binomial}
P(n,\, N \mid p) = \binom{N}{n}\, p^n \left(1-p\right)^{N-n}\, ,
\end{equation}
by definition of the $p$-value.
Under the null hypothesis, we expect $p$ to be uniformly distributed, so we may set this as our prior and obtain a posterior distribution on $p$ functionally identical to the likelihood.
With $p$ as the variable, this is a beta distribution, 
\begin{equation} \label{eq:res:beta}
p \sim \mathrm{Beta}(n+1, N-n+1)\, ,
\end{equation}
which has mean $\left\langle p \right\rangle = (n+1)/(N+2) \approx \hat{p}$.
The central blue line in Fig.~\ref{fig:res:pp} corresponds to $\hat{p}$, rather than $\left\langle p \right\rangle$, but the two are effectively equivalent.

To produce the credible band in Fig.~\ref{fig:res:pp}, we further need to understand the expected distribution of $\hat{p}$'s for a set of $N_e = \ResNevents{}$ events.
To do this, we produce a large number of synthetic $p$-value sets by drawing each of the $N_e$ elements from Eq.~\eqref{eq:res:beta}, with $n$ and $N$ corresponding to the measured values for each event.
Each individual simulation produces a PP curve akin to the central line in Fig.~\ref{fig:res:pp}.
These curves are contained within the light blue band 90\% of the time.

\section{Inspiral-merger-ringdown consistency test systematics}
\label{app:imr}
\begin{figure}
	\begin{center}
	\includegraphics[width=3.5in]{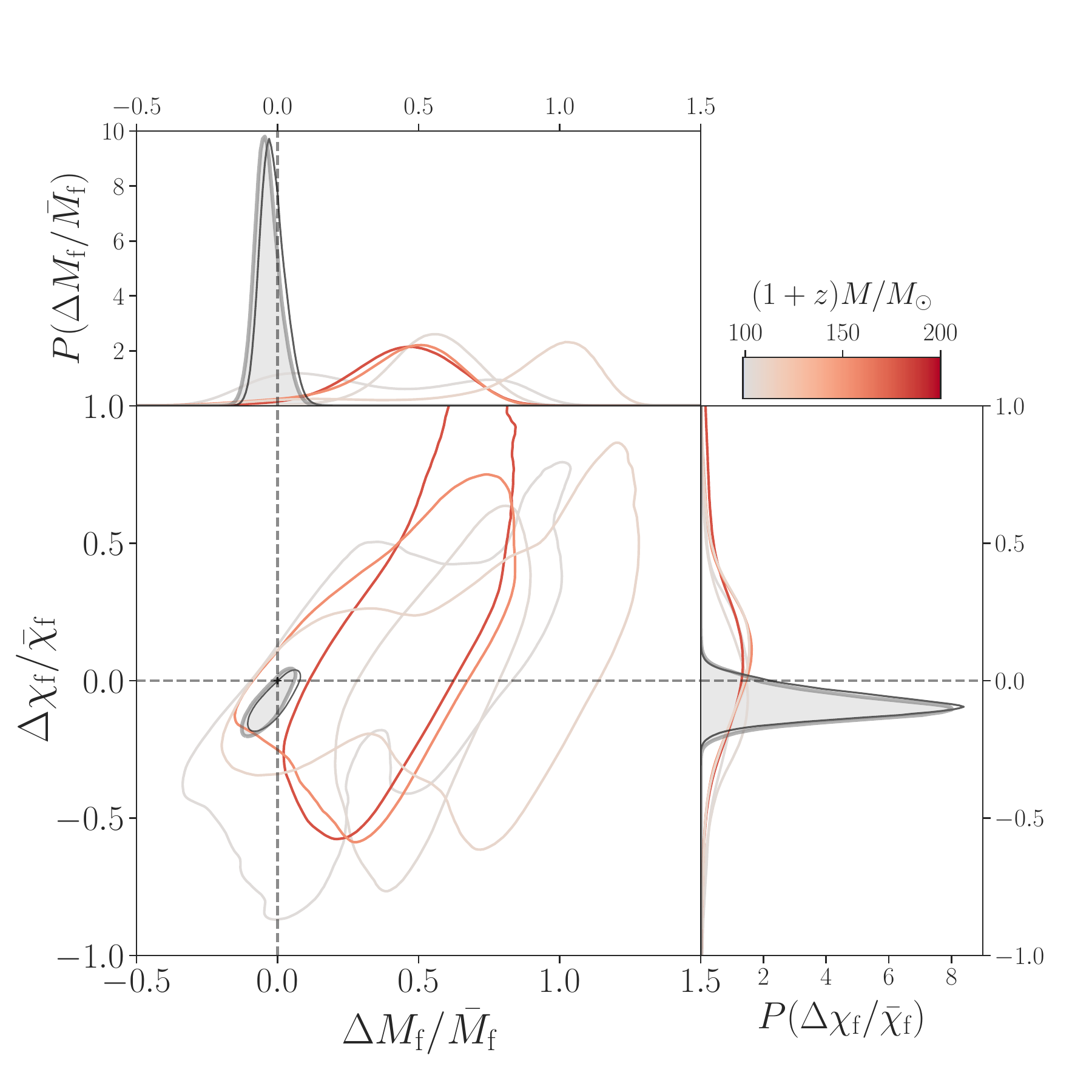}
	\end{center}
	\caption{As in Fig.~\ref{fig:imr_test_posteriors} of the main text, but for the events excluded for having median $(1+z) M  > 100 M_{\odot}$ (Table~\ref{tab:imr_test_params}). These events present a systematic bias in $\dMf$. The gray distribution corresponds to the same joint posterior as in Fig.~\ref{fig:imr_test_posteriors}, while the thin black one is obtained if the heavy events are also included.
  }
	\label{fig:imr_test_posteriors_heavy}
\end{figure}

\subsection{Redshifted total mass}

From the study of simulated signals, it is known that the IMR consistency test of Sec.~\ref{sec:imr} may be strongly biased for heavy BBHs.
This is because sources with high redshifted mass lead to short signals in the detectors and do not contain sufficient information about the inspiral regime.
For this reason, most of the results discussed in the main text (namely, Figs.~\ref{fig:imr_test_posteriors} and \ref{fig:imr_hier}) imposed a criterion on the median redshifted total mass so that $(1+z)M < 100\, M_{\odot}$.
Here we discuss the results for the events that did not make that cut.

Excessively high redshifted masses can lead to strong systematic biases in $\dMf$.
This is evident in Fig.~\ref{fig:imr_test_posteriors_heavy}, which is the equivalent of Fig.~\ref{fig:imr_test_posteriors} for the heavy events with median $(1+z)M > 100\, M_{\odot}$ that we excluded in the main text.
In spite of this, the joint posterior obtained by multiplying the individual results is hardly affected by the inclusion of the biased events (cf.\ gray and black distributions in Fig.~\ref{fig:imr_test_posteriors_heavy}).
This is because the joint posterior is driven by the individual events whose distributions have the narrowest support: the deviations towards high \dMf{} get washed out, and the combined result thus fails to identify that a significant fraction of the signals do not conform to the null hypothesis.

\begin{figure}
	\begin{center}
	\includegraphics[width=\columnwidth]{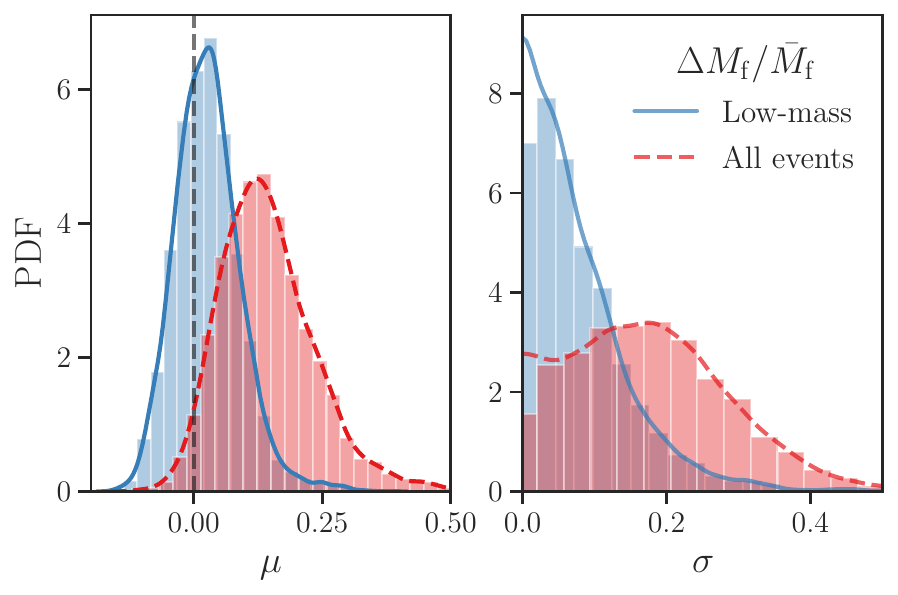}
	\end{center}
	\caption{Marginal posteriors for the hyperdistribution mean $\mu$ and standard deviation $\sigma$ for the \dMf{} measurements in GWTC-2. If the biased events with median $(1+z)M > 100\, M_{\odot}$ are included (red) the analysis mildly suggests a deviation from the null-hypothesis ($\mu=\sigma=0$); as expected, this goes away if the heavy events are excluded (blue).
  The nominal blue posteriors correspond to the population distribution presented in Fig.~\ref{fig:imr_hier}.
  }
	\label{fig:imr_hier_hyper}
\end{figure}

The hierarchical results are, on the other hand, sensitive to this sort of effect.
This can be seen most clearly in the posterior for the \dMf{} hyperdistribution mean $\mu$ and standard deviation $\sigma$, as defined in Sec.~\ref{sec:inference:populations}.
Figure~\ref{fig:imr_hier_hyper} shows the marginal distributions for these parameters as obtained when including (excluding) the events with $(1+z)M > 100\, M_{\odot}$ in red (blue).
The subpopulation of biased events manifests itself in anomalous distributions for the hyperparameters that disfavor $\mu=\sigma=0$.
Removing the heavy events, which are known to be biased, restores support for $\mu=\sigma=0$, and yields the %
nominal observed distribution shown in Fig.~\ref{fig:imr_hier}.

\subsection{Waveform modeling}

\begin{figure}[tb] 
	\begin{center}
	\includegraphics[width=3.5in]{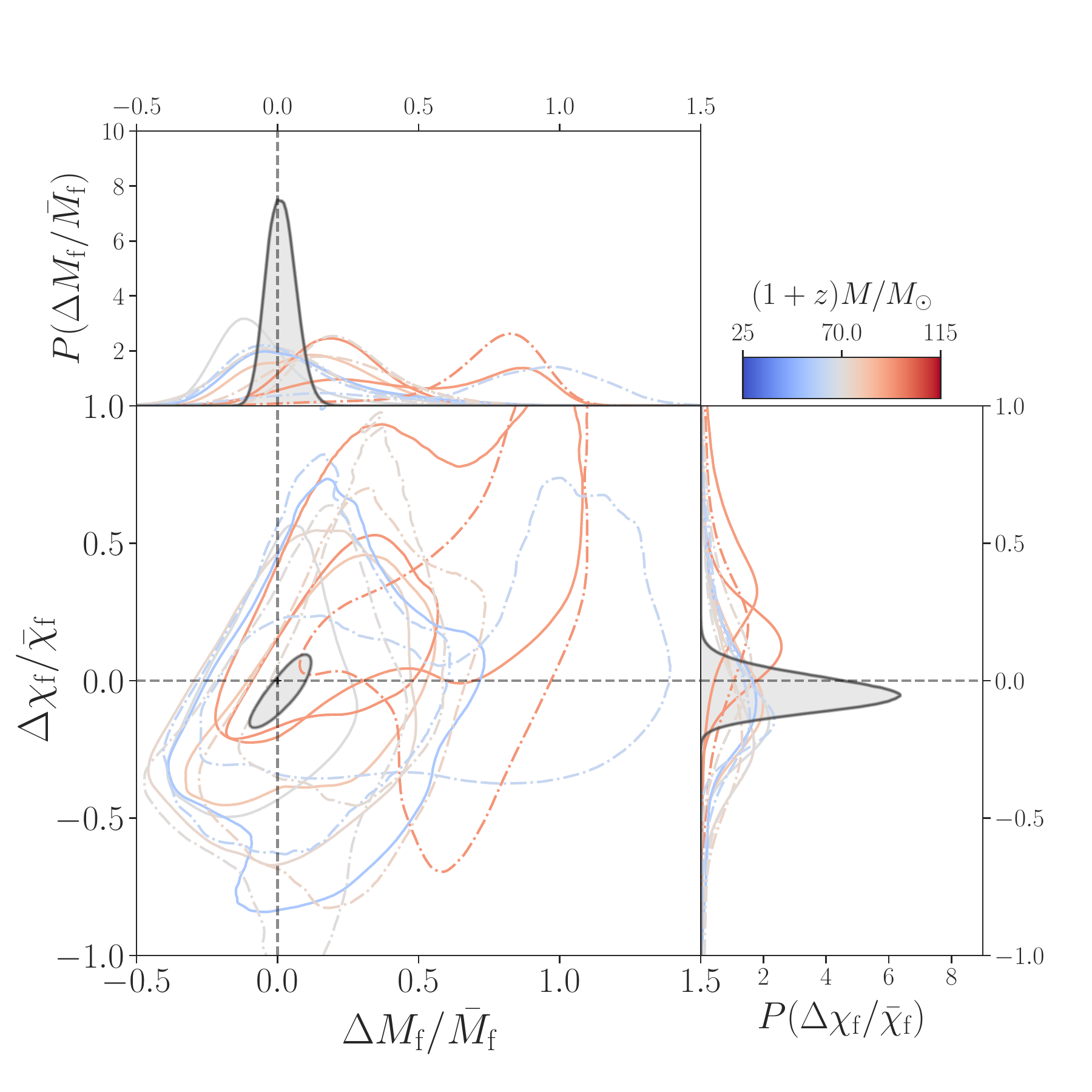}
	\end{center}
	\caption{As per Fig.~\ref{fig:imr_test_posteriors} but using the non-precessing \SEOBROM{} waveform model.
  Posteriors for the heavier events in Fig.~\ref{fig:imr_hier_hyper} are not shown here, but are included in the data release for this paper \cite{GWTC2:TGR:release}. Results for \protect\NAME{GW190412A}{} and \protect\NAME{GW190814A}{} are not included due to the relative importance of HMs, as discussed in Sec.~\ref{sec:imr}. 
  }
	\label{fig:imr_test_posteriors_seob}
\end{figure}

In order to gauge systematic errors arising from imperfect waveform modeling, we perform the IMR consistency test using both \IMRP{} and \SEOBROM{}. Although \SEOBROM{} is a non-precessing waveform approximant, we find that the posteriors are in broad agreement with no qualitative differences between the results (Fig.~\ref{fig:imr_test_posteriors_seob}). Assuming that the fractional deviations take the same value for all events, at $90\%$ credibility we find $\dMf = \ImrGWTCTWO{DMFGWTC2SEOB}$ and $\dchif = \ImrGWTCTWO{DCHIFGWTC2SEOB}$ when using \SEOBROM{}, consistent with the GR values. 

The differences in individual posteriors are expected due to differing physics and modeling of the final state between the approximants. For the two events in the IMR test where HMs are important, \protect\NAME{GW190412A}{}~\cite{GW190412} and \protect\NAME{GW190814A}{}~\cite{GW190814}, we use \IMRPHM{} as the preferred waveform approximant. As systematic errors are demonstrably larger when neglecting HMs for these two events, they are excluded when constructing the joint posteriors for the \SEOBROM{} analysis. 

\section{Impact of higher moments on parametrized tests}
\label{app:par}
\begin{figure*}[ht!]
\centering
\includegraphics[width=0.95\textwidth]{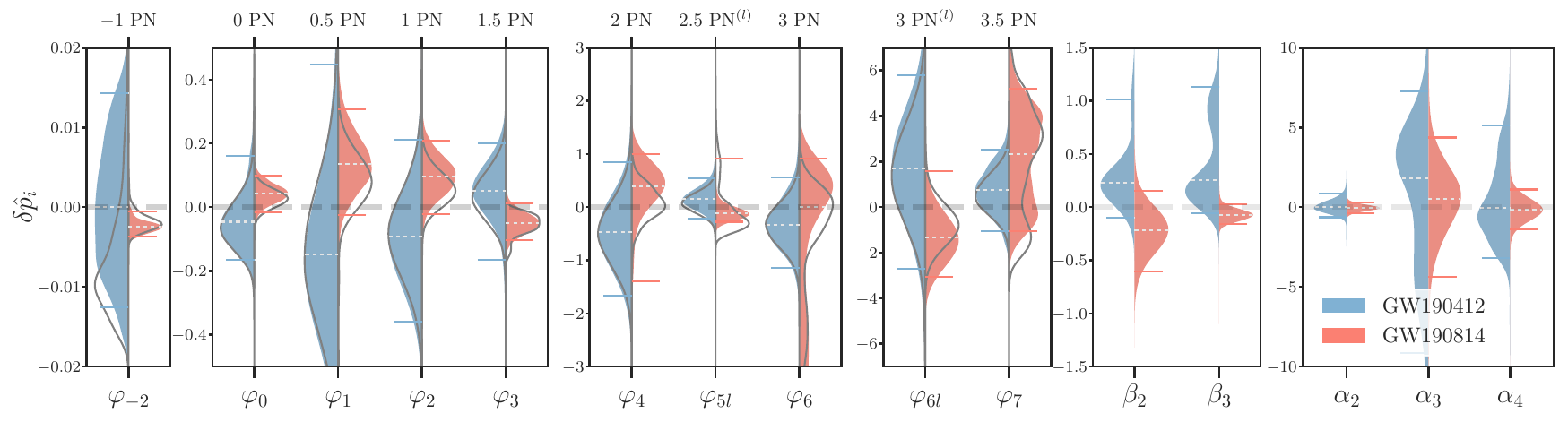}
\caption{Posteriors for parametrized violations of GR inferred using \IMRPHM{} and \SEOBHMROM{} (black solid lines). The horizontal solid lines indicate the $90\%$ credible intervals and the white dashed line marks the median. The horizontal dashed line at $\delta\hat p_i = 0$ denote the GR values. Posteriors for \protect\NAME{GW190412A}{} are shown in blue and for \protect\NAME{GW190814A}{} in red.}
\label{fig:tig:phenompv3hm_posteriors}
\end{figure*}

For the tests detailed in Sec.~\ref{sec:par}, the majority of events were analyzed using \IMRP{} and \SEOBROM{}, which only model the dominant $\ell = 2$ moments of the radiation and neglect subdominant spherical harmonic multipoles.  However, two of the BBHs considered in our analysis, \protect\NAME{GW190412A}{} \cite{GW190412} and \protect\NAME{GW190814A}{} \cite{GW190814}, have asymmetric component masses and detailed investigations show strong evidence for the presence of HMs. Using approximants that only capture the dominant $\ell = 2$ multipole moments could therefore lead to systematic errors and biases that present as false deviations of GR. In order to mitigate such systematics, we analyzed both these events using \IMRPHM{}, a precessing waveform approximant incorporating HMs, and \SEOBHMROM{}, an aligned-spin approximant with HMs, as described in Sec.~\ref{sec:inference}.

In Fig.~\ref{fig:tig:phenompv3hm_posteriors} we show the marginalized 1D posteriors for the parametrized violations of GR using \IMRPHM{} and \SEOBHMROM{}. As this is the first time that constraints are obtained using approximants with HMs, we explicitly show the marginalized 1D posteriors for the deviation coefficients. As mentioned in the main text, it is not necessarily surprising that we find some events for which the GR values fall in the tail of the posterior, as is the case for \protect\NAME{GW190814A}{}. The fact that this takes place for several \protect\NAME{GW190814A}{} coefficients is also not necessarily abnormal, since these are not statistically independent measurements. In addition, due to the way in which the parametrized tests are implemented, certain regions of the parameter space can lead to unphysical and pathological features in the waveform, potentially leading to multimodal posteriors and poor convergence of the posterior samples. Such features are observed in the $\delta\varphi_6$ and $\delta\varphi_7$ posteriors for the \IMRPHM{} analysis of \protect\NAME{GW190814A}, as in Fig.~\ref{fig:tig:phenompv3hm_posteriors}, and pathologies were found to occur when $\delta\varphi_6$ ($\delta\varphi_7$) becomes too negative (positive). We urge caution about the use and interpretation of these two coefficients in further studies, but find that these \protect\NAME{GW190814A}{} results do not have any notable impact on the combined posteriors and the resulting hierarchical analysis. \protect\NAME{GW190814A}{} is highly asymmetric and occurs in a region of the parameter space in which parametrized tests have not been systematically studied. For future analyses, detailed studies across the parameter space will be important in characterizing systematic errors, biases, and waveform pathologies as well as their impact on parameter estimation.

\bibliography{cbc-group,software}

\clearpage

\iftoggle{endauthorlist}{
 \let\author\myauthor
 \let\affiliation\myaffiliation
 \let\maketitle\mymaketitle
 \title{Authors}
 \pacs{}

 \newpage
 \maketitle
}

\end{document}